\documentclass[fleqn]{mnras}
\usepackage{epsfig}
\usepackage{threeparttable}
\usepackage{multicol}
\usepackage{multirow}
\usepackage{aas_macros,graphicx,times,multirow,amsmath,amssymb,color,longtable}
\usepackage[T1]{fontenc}
\usepackage{pdflscape}

\title[Systematic study of magnetar outbursts]{Systematic study of magnetar outbursts}
\author[F. Coti Zelati et al.]
{Francesco Coti Zelati,$^{1,2,3,4}$\thanks{E-mail: cotizelati@ice.csic.es} Nanda Rea,$^{1,2}$ Jos\'e~A. Pons,$^{5}$ Sergio Campana$^{3}$ \newauthor and Paolo Esposito$^{2}$\\
$^{1}$ Institute of Space Sciences (IEEC--CSIC), Campus UAB, Carrer de Can Magrans, E-08193, Barcelona, Spain\\
$^{2}$ Anton Pannekoek Institute for Astronomy, University of Amsterdam, Postbus 94249, NL-1090-GE Amsterdam, The Netherlands\\
$^{3}$ INAF -- Osservatorio Astronomico di Brera, via Bianchi 46, I-23807 Merate (LC), Italy\\
$^{4}$ Dipartimento di Scienza e Alta Tecnologia, Universit\`a dell'Insubria, via Valleggio 11, I-22100 Como, Italy\\
$^{5}$ Departament de F\'isica Aplicada, Universitat d'Alacant, Ap. Correus 99, E-03080 Alacant, Spain\\
}

\date{Accepted 2017 October 11. Received 2017 October 11; in original form 2017 August 1}
\pubyear{2018}

\def\ltsima{$\; \buildrel < \over \sim \;$}
\def\lsim{\lower.5ex\hbox{\ltsima}}
\def\gtsima{$\; \buildrel $\geq$ \over \sim \;$}
\def\gsim{\lower.5ex\hbox{\gtsima}}

\def\nh{\hbox{$N_{\rm H}$}}

\def\lum {\mbox{erg\,s$^{-1}$}}

\def\aa {1E\,1547$-$5408}
\def\xte{XTE\,J1810$-$197}
\def\wes{CXOU\,J164710.2$-$455216}
\def\sgrd{SGR\,1627$-$41}
\def\sgre{SGR\,0501$+$4516}
\def\lowba{SGR\,0418$+$5729}
\def\3xmm{3XMM\,J1852$+$0033}
\def\lowbb{Swift\,J1822.3$-$1606}
\def\galcen{SGR\,1745$-$2900}
\def\sgrm{SGR\,1935$+$2154}
\def\antgli{1E\,2259$+$586}
\def\psrr{PSR\,J1846$-$0258}
\def\cco{1E\,161348$-$5055}
\def\psr{PSR\,J1119$-$6127}

\newcommand{\cxo}{{\em Chandra}}
\newcommand{\ros}{{\em ROSAT}}
\newcommand{\asca}{{\em ASCA}}

\newcommand{\ein}{{\em Einstein}}
\newcommand{\xmm}{{\em XMM--Newton}}
\newcommand{\nustar}{{\em NuSTAR}}
\newcommand{\bsax}{{\em BeppoSAX}}
\newcommand{\suzaku}{{\em Suzaku}}
\newcommand{\rxte}{{\em RXTE}}
\newcommand{\swift}{{\em Swift}}
\newcommand{\fermi}{{\em Fermi}}
\newcommand{\integral}{{\em INTEGRAL}}

\begin{document}
\label{firstpage}
\pagerange{\pageref{firstpage}--\pageref{lastpage}}
\maketitle

\begin{abstract}
We present the results of the systematic study of all magnetar outbursts observed to date, through a reanalysis 
of data acquired in about 1100 X-ray observations. We track the temporal evolution of the outbursts soft X-ray spectral properties 
and the luminosities of the single spectral components as well as of the total emission. We model empirically all outburst light 
curves, and estimate the characteristic decay time-scales as well as the energetics involved. We investigate the link between 
different parameters (e.g., the luminosity at the peak of the outburst and in quiescence, the maximum luminosity increase, the 
decay time-scale and energy of the outburst, the neutron star surface dipolar magnetic field and characteristic age, etc.), and 
unveil several correlations among these quantities. We discuss our results in the context of the internal crustal heating and 
twisted bundle models for magnetar outbursts. This study is complemented by the Magnetar Outburst Online Catalogue 
(\url{http://magnetars.ice.csic.es}), an interactive data base where the user can plot any combination of the parameters derived 
in this work, and download all data.
\end{abstract}


\section{Introduction}

Magnetars are strongly magnetized (up to $B \sim 10^{14}-10^{15}$~G) isolated X-ray pulsars  
with luminosities $L_X\sim10^{31}-10^{36}$ \lum. They rotate at comparatively long periods 
($P\sim0.3-12$\,s) with respect to the general pulsar population, and are typically characterized 
by large secular spin-down rates ($\dot{P}\sim10^{-15}$ to $10^{-10}$~s~s$^{-1}$). According to 
the magnetar scenario, their emission is ultimately powered by the decay and the instability of 
their ultra-strong magnetic field (e.g. Duncan \& Thompson 1992; Paczy\'nski 1992; Thompson 
\& Duncan 1993, 1995, 1996, 2001; see Turolla, Zane \& Watts 2015 and Kaspi \& Beloborodov 
2017 for recent reviews). The hallmark of magnetars is the unpredictable and highly variable 
bursting/flaring activity in the X-/gamma-ray energy range, which encompasses a wide interval 
of time-scales (from a few milliseconds up to tens of seconds) and luminosities 
($10^{39}-10^{47}$~\lum\ at the peak; Turolla et al. 2015). The bursting episodes are often 
accompanied by large and rapid enhancements of the persistent X-ray emission (typically by a 
factor of $\sim10-1000$), which then decline and attain the quiescent level on a time-scale ranging 
from a few weeks up to several years. We will refer to these phases as outbursts, to distinguish 
from the bursting/flaring activity (see Rea \& Esposito 2011, for an observational review).


At the moment of writing (2017 July), 26 isolated X-ray pulsars have unambiguously shown 
magnetar-like activity, including the rotation-powered pulsars \psrr\ and \psr\ (Gavriil et al. 2008; 
Kumar \& Safi-Harb 2008; Kuiper \& Hermsen 2009; Archibald et al. 2016a; G{\"o}{\u g}{\"u}{\c s} 
et al. 2016; Archibald et al. 2017a), the low-field magnetars \lowba\ and \lowbb\ (e.g., 
Rea et al. 2010, 2012a), and the central compact object \cco\ (D'A\`i et al. 2016; Rea et al. 2016). 
These discoveries demonstrate how magnetar activity might have a larger spread within the neutron 
star population. 

The soft X-ray ($\lesssim 10$~keV) emission of magnetars is typically well described by a combination 
of a thermal component (a blackbody with temperature $kT\sim 0.3-0.9$~keV) plus a power law with 
photon index $\Gamma \sim 2-4$, commonly interpreted in terms of repeated resonant cyclotron up-scattering 
of thermal photons from the star surface on to charged particles flowing in a twisted magnetosphere (e.g., 
Thompson, Lyutikov \& Kulkarni 2002; Nobili, Turolla \& Zane 2008a,b). In some cases, a multiple-blackbody 
model provides an adequate description as well, and it is usually ascribed to thermal emission from regions 
of different temperature and size on the star surface (e.g., Tiengo, Esposito \& Mereghetti 2008; Alford \& 
Halpern 2016). 

In the last decades and especially following the advent of the new generation of imaging 
instruments on board \swift, \cxo\ and \xmm, several magnetar outbursts were monitored 
in the X-rays, leading to a number of unexpected breakthroughs which have changed our 
understanding of these strongly magnetized neutron stars (Turolla et al. 2015; Kaspi \& 
Beloborodov 2017). The large field of view (FoV) and the fast response of the \swift\ satellite proved 
(and still prove) to be key ingredients to spot the bursting/flaring activity of magnetars and precisely 
track spectral variations since the very first active phases and on time-scales ranging from days 
to months. \cxo\ and \xmm\ have revealed to be of paramount importance to characterize 
adequately the X-ray emission of faint outbursts particularly at later stages, thanks to dedicated 
follow-up observational programmes and the large collecting area of their instruments. In some 
cases, the monitoring campaigns covered the whole outburst evolution, and disclosed the source 
quiescent level. Although the cooling pattern varies significantly from outburst to outburst, the 
spectral softening throughout the decay seems an ubiquitous characteristic for these events (Rea 
\& Esposito 2011).

\subsection{Magnetar outbursts: mechanisms}
 
Although it is widely accepted that magnetar outbursts are attributable to some form 
of heat deposition in a restricted region of the star surface which then cools, the 
mechanism responsible for their activation, as well as the energy supply responsible 
for sustaining their long-term emission, still remain somewhat elusive. They are 
probably triggered by local internal magnetic stresses strong enough to deform irreversibly
part of the stellar crust, possibly in the form of a prolonged avalanche of plastic 
failures (Li, Levin \& Beloborodov 2016). An additional contribution may be provided 
by magnetospheric Alfv\'en waves created during flaring activity (Parfrey, Beloborodov \& 
Hui 2013). According to Li \& Beloborodov (2015), these waves are impulsively transmitted 
inside the star, and induce a strong oscillating plastic flow in the crust that subsists for 
a few ms, after which the waves are damped.

Regardless of the triggering mechanism, the plastic flows induced in the crust lead 
to transient thermoplastic waves that move the crust, convert mechanically its magnetic 
energy into heat and relieve the stresses (Beloborodov \& Levin 2014). A fraction of the 
deposited heat is then conducted up to the surface and radiated, producing a delayed 
thermal afterglow emission that can be sustained up to a few years, also depending on 
the flare rate (see also Beloborodov \& Li 2016). The crustal cooling time-scale chiefly 
depends on the thermal properties of the outer crust, the depth at which the energy is 
released and the neutrino emission processes operating in the crust (Pons \& Rea 
2012; Li et al. 2016). Moreover, the crustal displacements implant a strong external 
magnetic twist, presumably confined to a bundle of current-carrying closed field lines 
anchored in the crust. Additional heating of the surface layers is then produced as the 
currents flowing along the field lines of the twisted bundle impact upon the star (e.g. 
Thompson et al. 2002; Beloborodov \& Thompson 2007; Beloborodov 2009). As the 
energy reservoir stored in the star interior is progressively depleted, the twist must 
decay to support its currents. Consequently, the spatial extent of the bundle gets 
gradually more and more limited, the area on the star surface hit by the charges shrinks 
and the luminosity decreases. The time-scale of the resistive untwisting can be of the 
order of a few years if the crustal motions take place at high latitudes and the footpoints 
of the bundle are positioned close to the magnetic poles (Beloborodov 2009). 

Both heating mechanisms -- internal and external -- are likely at work during outbursts.

\subsection{Motivation of the study and plan of the paper}

Although several detailed studies were conducted for each of these events, an overall 
systematic and homogeneous analysis of the spectral properties of these stars, from the very 
first active phases of their outbursts throughout their decays, is still missing. A systematic 
reanalysis of all data sets is required to compare properly these properties, model accurately 
the outbursts cooling curves in a consistent way and unveil possible correlations among 
different parameters such as maximum luminosity, quiescent luminosity, luminosity increase 
during the outburst, energetics, decay time-scale, magnetic field, rotational energy loss rate and age. 

This paper presents the results of the X-ray spectral modelling for 23 magnetar outbursts from 17 
different sources using all the available data acquired by the \swift, \cxo\ and \xmm\ X-ray observatories, 
as well as data collected in a handful of observations by the instruments aboard \bsax, {\em Roentgen 
Satellite} (\ros) and \rxte. This sums up to about 1100 observations, for a total dead-time corrected on 
source exposure time of more than 12~Ms.
The paper is structured as follows: in Section~\ref{transients} we introduce the sample of 
magnetars considered in this study, and the monitoring campaigns that were activated 
following the detection of their outbursts. In Section~\ref{reduction} we describe the data 
reduction and extraction procedures. In Section~\ref{analysis} we report details on the 
spectral analysis. In Section~\ref{curves} we exploit the results of our analysis to extract 
the light curves for each outburst and estimate the outburst energetics and decay time-scale. 
In Section~\ref{maxvsquiescent} we report on accurate estimates of peak and quiescent 
luminosities of magnetars, including those showing only subtle variability on top of their 
persistent emission. In Section~\ref{correlations} we present the results of a search for 
possible (anti)correlations between several different parameters. In Section~\ref{discussion} 
we discuss the results of our study. A brief description of the Magnetar Outburst Online Catalogue 
(MOOC) follows in Section~\ref{mooc}. The results of the detailed modelling of the outbursts 
evolution with physically motivated models will be presented in a forthcoming work.

\section{The sample}
\label{transients}

This section summarizes the properties of the 17 magnetars that so far have undergone 
at least one outburst. The sources are listed according to the chronological order of their (first) 
outburst activation, except for the three sources \psr, \psrr\ and \cco, which are described at the 
end of the section. Details about the prompt and follow-up X-ray observations used in this work 
are reported in a series of tables in Appendix~\ref{journal}. In the following, all the values reported 
for the magnetic field are computed using the spin-down formula for force-free magnetospheres by 
Spitkovsky (2006), and assuming an aligned rotator. They refer to the dipolar component of the 
magnetic field at the polar caps (this is a factor of $\sim2$ larger than the value computed at the equator).

\subsection{\sgrd}

\sgrd\ was discovered on 1998 June 15 (Kouveliotou et al. 1998), when three 
consecutive bursts were detected by the Burst and Transient Source Experiment 
(BATSE) aboard the {\em Compton Gamma Ray Observatory}. More than 100 
bursts were recorded from the same location within the subsequent 6 weeks, and 
the X-ray counterpart was identified 2 months later by the narrow field instruments
on board \bsax\ (Woods et al. 1999). The burst detections marked the onset of an 
outburst, which gradually recovered the quiescent level over the course of the ensuing 
decade (see Table~\ref{tab:sgr1627_1998}).

On 2008 May 28 the Burst Alert Telescope (BAT; Barthelmy et al. 2005) aboard \swift\ 
triggered on dozens of bursts from \sgrd\ (Palmer et al. 2008). A conspicuous enhancement 
of the persistent X-ray flux was measured (a factor of about 100 larger with respect to 
3 months and a half before), and the magnetar nature of the source was incontrovertibly 
settled with the detection of 2.59-s X-ray pulsations in \xmm\ data sets (with $\dot{P} 
\sim 1.9 \times 10^{-11}$~s~s$^{-1}$; Esposito et al. 2009b). Table~\ref{tab:sgr1627_2008} 
reports the log of the X-ray observations carried out after the second outburst.
We assume a distance of 11~kpc throughout the paper.

\subsection{\antgli}

After more than two decades of rather persistent X-ray emission since its discovery at the 
centre of the supernova remnant (SNR) G109.1-1.0 (CTB~109) in 1979 December (Fahlman 
\& Gregory 1981), the 6.98-s X-ray pulsar \antgli\ attracted attention on 2002 June 18, when 
more than 80 bursts were detected within 3~h of observing time by the {\em Rossi X-ray 
Timing Explorer} (\rxte), and the persistent flux rose by a factor of $\sim10$ compared to 
the quiescent level (Kaspi et al. 2003). Eight \xmm\ observations were carried out to study 
the subsequent evolution of the outburst (see Table~\ref{tab:1e2259_2002}).

Nearly 10 yr later, on 2012 April 21, the Gamma-ray Burst Monitor (GBM) on board 
\fermi\ triggered on a single 40-ms long event (Foley et al. 2012), which was accompanied 
by an increase in the soft X-ray flux (as observed about a week later by the X-ray Telescope 
(XRT) on board \swift; see Table~\ref{tab:1e2259_2012} for a journal comprising this and all 
the follow-up observations of the first $\sim 1400$~d since the outburst onset). We assume 
a distance of 3.2~kpc throughout the paper.

\subsection{\xte}
\label{xte}

Originally a soft and faint X-ray source serendipitously recorded by the \ros\ during four observations 
between 1991 and 1993, the transient nature of \xte\ was disclosed in 2003, when the \rxte\ detected 
it at an X-ray flux a factor about 100 larger with respect to the pre-outburst level. X-ray pulsations 
were measured at a period of 5.54~s (Ibrahim et al. 2004). Radio pulsations at the spin period were 
detected in 2006 (about 3 yr later), a property never observed before in any other magnetar, 
which definitely proved that pulsed radio emission could be produced even in sources with 
magnetar-strength fields (Camilo et al. 2006). Although the initial phases of the outburst 
were missed, \xte\ has been studied in great detail over the last 12 yr, especially with the 
\xmm\ observatory and up to the return to quiescence (see Table~\ref{tab:xte})\footnote{The 
source was observed also with \cxo\ for 12 times and with \swift\ for 5 times. We focus 
here on the \xmm\ pointings alone, because they provide a good coverage of the whole outburst 
evolution down to the quiescent level, as well as the spectra with the largest counting statistics. 
Note that a recent \swift\ XRT observation performed in 2017 February caught the source again 
at the historical quiescent flux.}.
We assume a distance of 3.5~kpc throughout the paper.

\subsection{SGR\,1806$-$20}
\label{sgr1806}

Initially catalogued as a classical $\gamma$-ray burst (GRB\,790107) based on observations 
by the \emph{Konus} experiment (Mazets et al. 1981) and other all-sky monitors of the 
interplanetary network (Laros et al. 1986), SGR\,1806$-$20 was recognized to be a member 
of a distinct class of astrophysical transients after the detection of more than 100 bursts of soft 
$\gamma$-rays between 1979 and 1986 (Laros et al. 1987). Two observations were carried out 
by the \emph{Advanced Satellite for Cosmology and Astrophysics} (\asca) soon after an intense 
bursting activity in 1993 October (as unveiled by BATSE), leading to the identification of a previously 
uncatalogued, persistent, point-like X-ray counterpart (Murakami et al. 1994; Sonobe et al. 1994). 
The spin period, $\sim 7.5$~s, was measured in 1996 November by means of five \rxte\ observations 
that were performed following another reactivation of the source (Kouveliotou et al. 1998).

SGR\,1806$-$20 experienced an exceptionally intense flare on 2004 December 27 with a peak 
luminosity of a few 10$^{47}$~\lum\ (for a distance of 8.7~kpc and under the assumption of 
isotropic emission; Hurley et al. 2005; Palmer et al. 2005), which then decayed by a factor of 
$\sim50$ per cent, and stabilized at an approximately steady level over the subsequent 7 
yr (Younes, Kouveliotou \& Kaspi 2015). Table~\ref{tab:sgr1806} reports a log of all 10 \xmm\ 
observations tracking the post-flare evolution (no \swift\ observations were performed during the 
first 2 months of the outburst). We assume a distance of 8.7~kpc throughout the paper.

\subsection{\wes}

\wes\ was discovered in 2005 during an X-ray survey of the young cluster of massive stars 
Westerlund~1, and tentatively identified as a magnetar candidate based on the value of its 
spin period, 10.61~s, and the X-ray spectral properties (Muno et al. 2006). The case was 
clinched the following year, when a rather intense burst lasting about 20~ms was fortuitously 
detected by the \swift\ BAT from the direction of the source, on 2006 September 21 (Krimm et 
al. 2006). This episode was indeed associated with an abrupt enhancement of the X-ray flux, 
which marked the onset of a magnetar-like outburst. Table~\ref{tab:cxou1647} reports a summary 
of all follow-up X-ray observations. 

The source underwent another weaker outburst on 2011 September 19, when four more sporadic 
bursts were detected from the source position (Baumgartner et al. 2011; Rodr\'iguez Castillo et 
al. 2014). Table~\ref{tab:cxou1647_2011} lists the few X-ray observations of this outburst.
We assume a distance of 4~kpc throughout the paper.

\subsection{\sgre}

\sgre\ joined the magnetar family on 2008 August 22, after the \swift\ BAT detection of a 
series of short bursts of soft $\gamma$-rays ($<100$~keV; Barthelmy et al. 2008) and the 
discovery of pulsations at a period of 5.76~s from the X-ray counterpart (G{\"o}{\u g}{\"u}{\c s} et 
al. 2008). The source continued to be active over the following 36 h, showing a total of 
about 30 bursts. It was soon recognized that the bursting activity was related to the onset of an 
outburst, and several X-ray observations were promptly undertaken (see Table~\ref{tab:sgr0501}).
We assume a distance of 1.5~kpc throughout the paper. 

\subsection{\aa}

Discovered by the \ein\ satellite on 1980 March 2 during a search for X-ray counterparts 
of unidentified $\gamma$-ray sources (Lamb \& Markert 1981), \aa\ (aka SGR\,1550$-$5418) 
was later suspected to be a magnetar candidate based on its X-ray spectral properties, 
the observed long-term X-ray variability between 1980 and 2006, and its putative association 
with the SNR G327.24$-$0.13 (Gelfand \& Gaensler 2007). The `smoking gun' in favour of this 
classification came with the measurement of 2.07-s pulsations from the radio counterpart 
(Camilo et al. 2007), later confirmed also in the X-rays (Halpern et al. 2008).

On 2008 October 3, the \swift\ BAT triggered on and localized a short burst from a position 
consistent with that of \aa\ (Krimm et al. 2008). \swift\ executed a prompt slew, and the XRT 
started observing the field only 99 s after the BAT trigger, catching the source at a flux a factor 
about 20 above that in quiescence (see Table~\ref{tab:1e1547_08} for the log of all the 
follow-up X-ray observations).

No further bursts were reported until 2009 January 22, when the source resumed a new state of 
extreme bursting activity (Connaughton \& Briggs 2009; Gronwall et al. 2009), culminating in a 
storm of more than 200 soft $\gamma$-ray bursts recorded by the \emph{International Gamma-Ray 
Astrophysics Laboratory} (\integral) in a few hours (Mereghetti et al. 2009), and characterized by 
a considerable increase in the persistent X-ray flux. The source was repeatedly observed in the 
X-rays after the burst trigger (especially with \swift), leading to one of the most intensive samplings 
of a magnetar outburst ever performed (see Table~\ref{tab:1e1547_09})\footnote{\swift\ and \xmm\ 
observations performed from 2007 June to October caught the magnetar while recovering from 
another outburst likely occurred prior to 2007 June (Halpern et al. 2008). We do not include the 
analysis of this outburst in this study owing to the unknown epoch of the episode onset and the 
sparse X-ray coverage. Our analysis of the 2009 event is limited to the first 1000~d of the outburst, 
but the source is currently being observed by \swift. However, a preliminary extraction of the long-term 
light curve with the \swift\ online tool (see below), reveals an extremely slow decay which is consistent 
with the extrapolation of our long-term light curve, giving no significant differences in the estimate of 
the total energetics and decay time-scale.}. 
We assume a distance of 4.5~kpc throughout the paper.

\subsection{\lowba}

\lowba\ was discovered after the detection of a couple of short hard X-ray bursts on 2009 June 5 with 
\fermi\ GBM and other instruments sensitive to the hard X-ray range (van der Horst et al. 2010). 
Coherent X-ray pulsations were observed at a period of 9.1~s 5 d later during an \rxte\ pointing 
(G{\"o}{\u g}{\"u}{\c s} et al. 2009). Since then, \swift, \cxo\ and \xmm\ observed the field of the new 
source for a total of 39 pointings (see Table~\ref{tab:sgr0418}). It took more than 3 yr of 
continuous monitoring to establish unambiguously the first derivative of the spin period, making this 
source the magnetar with the lowest inferred surface dipolar magnetic field known to date, $\sim 1.2 
\times 10^{13}$~G (Rea et al. 2013a). We assume a distance of 2~kpc throughout the paper.

\subsection{SGR\,1833$-$0832}

SGR\,1833$-$0832 was discovered on 2010 March 19, when the \swift\ BAT triggered on and localized 
a short ($<$1~s) hard X-ray burst in a region close to the Galactic plane (Gelbord et al. 2010; 
G{\"o}{\u g}{\"u}{\c s} et al. 2010a) and the fast slew of the XRT promptly detected a previously unnoticed 
7.57-s X-ray pulsator (G{\"o}{\u g}{\"u}{\c s} et al. 2010a; Esposito et al. 2011). Starting right after its 
discovery, \swift\ and \xmm\ pointed their instruments towards the source multiple times for the first 
$\sim$160~d of the outburst decay (see Table~\ref{tab:sgr1833}).
We assume an arbitrary distance of 10~kpc throughout the paper.

\subsection{\lowbb}

On 2011 July 14, the detection of a magnetar-like burst by the \swift\ BAT and of an 
associated bright and persistent XRT counterpart heralded the existence of a new 
magnetar, \lowbb\ (Cummings et al. 2011), with a spin period of 8.43~s 
(G{\"o}{\u g}{\"u}{\c s} et al. 2011a). \lowbb\ was densely monitored in the X-rays 
until 2012 November 17, covering a time span of $\sim$1.3 yr (see Table~\ref{tab:swift1822}).
With an estimated surface dipolar magnetic field of $\sim 6.8 \times 10^{13}$~G 
(Rodr\'iguez Castillo et al. 2016, and references therein), it also belongs to the 
sub-class of the so called `low-$\dot{P}$ magnetars'.
We assume a distance of 1.6~kpc throughout the paper.

\subsection{Swift\,J1834.9$-$0846}
\label{swift1834}

The BAT aboard \swift\ was triggered by a short SGR-like burst on 2011 August 7 
(D'Elia et al. 2011). This episode was not isolated: a second burst from the same 
direction on the sky was recorded by the \fermi\ GBM approximately 3.3~h later 
(Guiriec et al. 2011), and another similar event triggered the BAT again on 
August 29 (Hoversten et al. 2011). The magnetar nature of this newly discovered 
source was nailed down with the discovery of pulsations at 2.48~s from the X-ray 
counterpart (G{\"o}{\u g}{\"u}{\c s} \&  Kouveliotou 2011). \swift, \cxo\ and \xmm\ observed 
this new SGR for a total of 25 times since the first burst detection (see Table~\ref{tab:swift1834}).

Swift\,J1834.9$-$0846 represents a unique case among magnetars. It is indeed 
embedded in a patch of diffuse X-ray emission with a complex spatial structure 
consisting of a symmetric component within $\sim 50$ arcsec around the magnetar, 
and an asymmetric component strechted towards the south--west of the point source 
and extending up to $\sim 150$ arcsec. The former was interpreted as a halo 
created by the scattering of X-rays by intervening dust (dust-scattering halo; Kargaltsev 
et al. 2012; Esposito et al. 2013). The latter was attributed to a magnetar-powered 
wind nebula based on its highly absorbed power law-like X-ray spectrum, the flux 
constancy and the absence of statistically significant variations in the spectral shape 
over a time span of 9 yr, between 2005 and 2014 (Younes et al. 2016). 
Swift\,J1834.9$-$0846 would then provide the first observational evidence for the 
existence of wind nebulae around magnetically powered pulsars (see also Granot et 
al. 2017; Torres 2017). 
We assume a distance of 4.2~kpc throughout the paper.

\subsection{1E\,1048.1$-$5937}
\label{sgrn}	

The discovery of 1E\,1048.1$-$5937 dates back to 1979 July 13, when \ein\ detected 
6.44-s pulsed X-ray emission from a point-like source in the Carina Nebula (Seward, 
Charles \& Smale 1986). With five long-term outbursts shown to date, this source holds 
the record as the most prolific outbursting magnetar hitherto known. The first three flux 
enhancements were observed in 2001, 2002 and 2007 by \rxte, which monitored this source 
about twice per month from 1999 February to 2011 December (see Dib \& Kaspi 2014 and 
references therein). An additional flux increase was observed in 2011, and the subsequent 
evolution was the object of a prolonged monitoring campaign with \swift, to which two \cxo\ 
and one \xmm\ observations have to be added (see Table~\ref{tab:1e1048_2011} for the 
observations of the first $\sim 1000$ d of the outburst decay). The last outburst from this 
source dates back to 2016 July 23 (Archibald et al. 2016b), and its evolution was again densely 
monitored thanks to the ongoing \swift\ campaign (see Table~\ref{tab:1e1048_2016}). The 
outbursts are remarkably periodic, with a recurrence time of about 1800~d (Archibald et al. 
2015). In this study we focus on the last two outbursts.
We assume a distance of 9~kpc throughout the paper.

\subsection{\galcen}

At a projected separation of $\sim0.1$~pc from the supermassive black hole at the Centre of 
the Milky Way, Sagittarius~A$^*$ (hereafter Sgr~A$^*$), the magnetar \galcen\ is the 
closest neutron star to a black hole ever observed, and it spins at a period of about 
3.76~s (e.g., Coti Zelati et al. 2015a, 2017). According to numerical simulations and to 
the recently detected proper motion, it is likely in a bound orbit around Sgr~A$^*$ (Rea 
et al. 2013b; Bower et al. 2015).

\galcen\ is the object of an ongoing intensive monitoring campaign by \cxo\ (see 
Table~\ref{tab:sgr1745}), still more than 3 yr after the detection of the first 
$\sim 30$~ms long soft gamma-ray burst from the source on 2013 April 25 (Kennea 
et al. 2013a). 
We assume a distance of 8.3~kpc throughout the paper.

\subsection{\sgrm}

The most recent addition to the magnetar class is represented by \sgrm, whose existence 
was announced on 2014 July 5 once more through the detection of low-Galactic latitude short 
bursts by \swift\ BAT (Stamatikos et al. 2014). A deep follow-up observation carried out by 
\cxo\ enabled to determine its spin period (3.24~s; Israel et al. 2014), and the post-outburst 
behaviour was then observed with \swift, \cxo\ and \xmm. On 2015 February 22 the BAT 
triggered on another burst from the source (D'Avanzo et al. 2015), which led to further monitoring 
through 14 observations with \swift\ and two with \xmm. Another 50-ms long burst was detected 
in 2015 December by \integral\ in the soft gamma rays (Mereghetti et al. 2015), albeit no 
concurrent increase in the X-ray emission over the long-term behaviour was observed (Coti Zelati 
et al. 2015b). The source reactivated once more on 2016 May 16 (Barthelmy et al. 2016), and 
bursting activity was observed over the following $\sim 5$ d. Some of these flux enhancements 
were recently studied in detail by Younes et al. (2017). See Table~\ref{tab:sgr1935} for the log of 
the observations. We assume a distance of 9~kpc throughout the paper.

\subsection{\psr}
\label{psr}

The 0.4-s radio pulsar \psr\ was discovered in the \emph{Parkes} multibeam 1.4-GHz survey (Camilo 
et al. 2000), and it is likely associated with the SNR G292.2$-$0.5 (Crawford et al. 2001). 
The dipolar surface magnetic field implied by the timing parameters is about $8.2 \times10^{13}$~G, 
among the highest known among radio pulsars. On 2016 July 27 and 28 two magnetar-like bursts 
signalled the onset of an outburst from this source (Archibald et al. 2016a; Kennea et al. 2016; Younes, 
Kouveliotou \& Roberts 2016). Table~\ref{tab:psr1119} lists the follow-up \swift\ observations analysed in 
this work. Interestingly, simultaneous radio and X-ray observations about 1 month after the outburst 
onset revealed a significant anticorrelation between the emission in the two bands: the rotation-powered 
radio emission switched off during periods of multiple magnetar-like X-ray bursts (Archibald et al. 2017a).  
We assume a distance of 8.4~kpc throughout the paper.

\subsection{\psrr}
\label{psr1846}

\psrr\ is a young (<1~kyr) rotation-powered pulsar located at the centre of the SNR Kesteven 75 (Gotthelf 
et al. 2000). It rotates at a period of $\sim326$~ms (Livingstone et al. 2011a) and is endowed with a surface 
dipolar magnetic field of $\sim1\times10^{14}$~G, which is higher than the vast majority of rotation-powered 
pulsars. On 2006 June 8 several magnetar-like X-ray bursts were detected in the time series of the \rxte\ data 
sets, and a sudden X-ray outburst took place (Gavriil et al. 2008; see also Kumar \& Safi-Harb 2008; Kuiper 
\& Hermsen 2009). The source then returned to the quiescent state in about 6 weeks. In this study we will 
adopt the values estimated by Gavriil et al. (2008) for the total energy released during the outburst, as well as 
the time-scale of the decay (see Table~\ref{tab:decays}).
We assume a distance of 6~kpc throughout the paper.

\subsection{\cco}
\label{cco}

The source \cco\ near the geometrical centre of the SNR RCW\,103 defied any interpretation for more 
than two decades because of its puzzling phenomenology (in particular, a periodicity at 6.67~h and the 
lack of an optical/infrared counterpart; De Luca et al. 2006, 2008). On 2016 June 22, the \swift\ BAT 
detected a magnetar-like burst from \cco, also coincident with a large long-term X-ray outburst (D'A\`i 
et al. 2016). The long-term light curve of the source from 1999 to 2016 July was already extracted by 
Rea et al. (2016; see in particular their fig.~2) in a way completely consistent with the procedure reported 
in this work for the other magnetar outbursts, and shows that the source experienced another major outburst 
in 2000 February. In the following, we will thus refer to that publication when quoting our estimates for the 
energetics and decay time-scale for the first outburst. On the other hand, the \swift\ XRT monitoring campaign 
of this object is ongoing on a monthly cadence and we are currently tracking the decay of the second outburst 
to refine the time-scale and energetics of this episode. The outburst is showing a slower evolution with respect 
to that we predicted in Rea et al. (2016), and in the following we will consider our updated values for the 
energetics and time-scales (up to mid-July 2017; see Table~\ref{tab:decays}). We assume a distance of 3.3~kpc 
throughout the paper.

\section{Data reduction and extraction}
\label{reduction}

This section describes the standard procedures employed to extract the scientific products 
(source and background spectra) and create or assign the response and auxiliary files starting 
from the raw \swift, \xmm\ and \cxo\ data files publicly available. In addition to these data sets, 
we also looked at other few observations carried out with the Medium-Energy Concentrator 
Spectrometer (MECS; Boella et al. 1997) on board \bsax, the \ros\ Position Sensitive Proportional 
Counter (PSPC; Pfeffermann et al. 1987), and the Proportional Counter Array (PCA; Jahoda 
et al. 2006) instrument of the \rxte. In particular, we focused on the data concerning the quiescent 
stages (pre-outburst observations), or the very early phases, of the outbursts. These data sets 
revealed to be crucial to estimate fluxes and luminosities for the magnetar \xte\ during quiescence 
or for other magnetars  (i.e., \sgrd\ during its 1998 event and \lowba) at the very early stages of the 
outburst decay, and were reduced and analysed as described by Esposito et al. (2008, 2010a) and 
Rea et al. (2009, 2012a).

\subsection{\swift\ data}
\label{swift}

XRT (Burrows et al. 2005) on board the \swift\, 
satellite uses a front-illuminated charge-coupled device (CCD) detector 
sensitive to photons with energies between 0.2 and 10~keV, with an 
effective area of about 110~cm$^2$ at 1.5~keV. Two readout modes are 
now available: photon counting (PC) and windowed timing (WT). In the former, 
the entire CCD is read every $\sim2.5$~s, whereas in the latter 10 rows are 
compressed in one, and only the central 200 (out of 600) columns are read out. 
One-dimensional imaging is preserved, achieving a time resolution of 
$\sim1.7$~ms and thus providing a larger dynamic range of measurable source 
intensities (see Hill et al. 2004 for a detailed description of the XRT readout modes).

We processed the data with standard screening criteria (see Capalbi et al. 2005) 
and generated exposure maps with the task \textsc{xrtpipeline} (version 0.13.3) 
from the \textsc{ftools} package (Blackburn 1995), using the spacecraft attitude file. 
We selected events with grades 0--12 and 0 for the PC and WT data\footnote{Because 
of issues in the modelling of the response matrix files, spectra of heavily absorbed 
sources ($N_{\rm H} \gtrsim 10^{22}$~cm$^{-2}$) occasionally are known to exhibit a 
bump and/or turn-up at low energy (typically below 1~keV) in WT mode for events with 
grades $\ge1$. See \url{http://www.swift.ac.uk/analysis/xrt/digest_cal.php}.}, respectively, 
and extracted the source and background spectra using \textsc{xselect} (v. 2.4). We 
accumulated the source counts from a circular region centred on the source position and 
with a radius of 20 pixels (one XRT pixel corresponds to about 2.36 arcsec). Noteworthy 
exceptions are represented by the magnetar Swift\,J1834.9$-$0846 and the source \cco, 
for which we opted for a circle of radius 6 and 10 pixels, respectively, to minimize the 
contribution from the surrounding diffuse emission (see Section~\ref{swift1834}). To estimate 
the background in the PC-mode data, we extracted the events within an annulus centred 
on the source position with inner and outer radius of 40 and 80 pixels, respectively (12 and 
19 pixels for Swift\,J1834.9$-$0846, 10 and 20 pixels for \cco). For the observations targeting 
the 2009 outburst of \aa\ we considered instead a circle as far as possible from the source, 
to reduce the contamination by the three expanding dust scattering X-ray rings (see Tiengo 
et al. 2010). For the WT-mode data we adopted a region far from the target and of the same 
size as that used for the source.

For all the observations we built exposure-corrected and background-subtracted 
light curves using \textsc{xrtlccorr} and \textsc{lcmath} (the latter accounting also for 
different areas of the source and background extraction regions). We binned them with 
different time resolutions, and removed possible bursts/flares episodes by applying 
intensity filters to the event lists. This procedure aims at minimizing flux overestimates, 
and avoiding possible spectral distorsions induced by the bursting emission, which is 
typically harder than that of the underlying continuum. 

In case an observation in PC mode suffered from photon pile-up (typically this occurs 
when the source net count rate exceeds $\sim 0.6$ counts~s$^{-1}$), we determined 
the extent of the piled-up region as follows. First, we modelled the wings of the radial 
profile of the source point-spread function (at a distance $>15$ arcsec from the centre) 
with a King function reproducing the PSF of the XRT (Moretti et al. 2005). We then 
extrapolated the model back to the core of the PSF, and compared it to the data points. 
The region where the observed PSF lies underneath the extrapolation of the King function 
was then excluded from our analysis\footnote{See \url{http://www.swift.ac.uk/analysis/xrt/xrtpileup.php}.}. 

We created the observation-specific ancillary response files with \textsc{xrtmkarf} 
(v. 0.6.3), thereby correcting for the loss of counts due to hot columns and bad pixels, 
and accounting for different extraction regions, telescope vignetting and PSF corrections. 
We then assigned the appropriate redistribution matrix available in the \textsc{heasarc} 
calibration data base, and excluded bad spectral channels (at energy $< 0.3$~keV). We 
co-added individual spectra and responses for contiguous observations with very few counts 
and that were carried out with the same observing mode, to improve the statistics quality 
and increase the signal-to-noise ratio\footnote{Ancillary response files were weighted by the 
net number of counts of the source in each observation.}. For extensively monitored outbursts 
we also constructed the long-term 0.3--10~keV count rate light curves (using the online \swift\ 
XRT data products generator; see Evans et al. 2009 for details), to gauge the decay time-scales 
(see Appendix~\ref{xrtcurves}).

\subsection{\xmm\ data}

The \xmm\ satellite carries three co-aligned X-ray telescopes, each with an European Photon 
Imaging Camera (EPIC) imaging spectrometer at the focus. Two of the EPIC spectrometers 
use Metal Oxide Semiconductor CCD arrays (MOS cameras; Turner et al. 2001) and one uses 
pn CCDs (pn camera; Str\"uder  et al. 2001). They all cover the 0.1--15~keV energy range with 
an effective area of about 500 cm$^2$ for each MOS and 1400 cm$^2$ for the pn at 1.5~keV. 
In this work we shall consider only data acquired with the pn camera, which provides the 
spectra with the highest counting statistics owing to its larger effective area.

The pn camera can operate in different modes. In full frame mode (FF; 73.4-ms time resolution), 
all pixels of the 12 CCDs are read out simultaneously and the full FoV is covered. 
In large window mode (LW; 47.7-ms time resolution), only half of the area in all CCDs is read out 
and in small window mode (SW; 5.7-ms time resolution) just part of one single CCD is used to collect 
data. The pn can also operate in timing mode, where data from a predefined area on one CCD chip 
are collapsed into a one-dimensional row to be read every 30~$\mu$s.

We retrieved the raw observation data files from the \xmm\ Science Archive, and processed them to 
produce calibrated, concatenated photon event lists using the \textsc{epproc} tool of the \xmm\ Science 
Analysis System (\textsc{sas} v. 15.0; Gabriel et al. 2004) and the most up to date calibration files available 
(XMM-CCF-REL-332). For each observation we built a light curve of single pixel events (\textsc{pattern} = 0) 
for the entire FoV, and discarded episodes (if any) of strong soft-proton flares of solar origin using intensity filters.
We then estimated the amount of residual contamination in each event file by comparing 
the area-corrected count rates in the in- and out-of-FoV regions of the detector\footnote{We 
used the script provided by the \xmm\ EPIC Background working group available at 
\url{http://www.cosmos.esa.int/web/xmm-newton/epic-scripts\#flare}.}, and verified that it 
was negligible or low in all cases (here `negligible' and `low' are defined following De Luca 
\& Molendi 2004). We extracted the source photons from a circular region centred on the 
source position and with a typical radius of 20--30 arcsec, depending on the source brightness, 
the presence of closeby sources and the distance from the edge of the CCD. The background 
was extracted from a circle located on the same CCD, and the position and size of the region 
were determined so as to guarantee similar low-energy noise subtraction and avoid detector 
areas possibly contaminated by out-of-time events from the source or too near to the CCD 
edges (we used the \textsc{ebkgreg} tool, which typically yielded larger radii for the cases 
where the source was particularly faint, e.g. \sgrd\ or \lowba\ close to the quiescent level). 
The case of Swift\,J1834.9$-$0846 stands apart owing to the surrounding extended emission 
(see Section~\ref{swift1834}), and the photon counts were collected within similar regions as 
those adopted by Younes et al. (2016).

We built background-subtracted and exposure-corrected light curves with different time 
binnings using the \textsc{epiclccorr} task, which also corrects the time series for any 
relevant instrumental effect such as bad pixels, chip gaps, PSF variation, vignetting, quantum 
efficiency and dead time, and accounts for the different sizes of the source and background 
extraction regions. We then removed possible source flaring episodes by applying ad hoc 
intensity filters on the light curves. 

We estimated the potential impact of pile-up by comparing the observed event pattern 
distribution as a function of energy with the theoretical prediction in the 0.3--10~keV 
energy interval, by means of the \textsc{epatplot} task. For piled-up sources, we selected 
the most suitable annular extraction region for the source counts via an iterative procedure, 
by excising larger and larger portions of the inner core of the source PSF until a match 
was achieved between the observed and expected distributions at the 1$\sigma$ 
confidence level (c.l.) for both single and double pixel events.

We employed the standard filtering procedure in the extraction of the scientific products, 
retaining only single and double pixel events optimally calibrated for spectral analysis 
(\textsc{pattern} $\leq4$), and excluding border pixels and columns with higher offset for 
which the pattern type and the total energy are known with significantly lower precision 
(\textsc{flag} = 0). We calculated the area of source and background regions using the 
\textsc{backscale} tool, and generated the redistribution matrices and effective area files 
with \textsc{rmfgen} and \textsc{arfgen}, respectively. We used the \textsc{epispeccombine} 
task to co-add the spectra and average the response files of closeby observations carried 
out with the same instrumental setup (i.e. same observing mode and optical blocking filter 
in front of the pn CCD) and with a scarce number of counts, to obtain a reasonable 
number of spectral bins for a meaningful spectral analysis.

\subsection{\cxo\ data}
\label{sec:cxo}

The {\em Chandra X-Ray Observatory} includes two focal plane 
instruments: the Advanced CCD Imaging Spectrometer (ACIS; 
Garmire et al. 2003) and the High Resolution Camera (HRC; 
Zombeck et al. 1995). The ACIS operates in the 0.2--10~keV energy 
range with an effective area of about 340~cm$^2$ at 1~keV. It consists 
of an imaging (ACIS-I) and a spectroscopic (ACIS-S) CCD arrays. 
The HRC covers the 0.1--10~keV interval with an effective area of 
about 225 cm$^2$ at 1~keV and comprises the HRC-I and the HRC-S
detectors. The former optimized for wide-field imaging, and the latter 
designed for spectroscopy\footnote{Observations performed with the 
HRC-I were not analysed because this camera provides only a limited 
energy resolution on the detected photons.}. 

The ACIS detectors enable two modes of data acquisition: the timed 
exposure (TE) mode, and the continuous clocking (CC) mode. In the 
former, each chip is exposed for a nominal time of 3.241~s (or a 
sub-multiple, if only a sub-array of a chip is being read-out). In the latter, 
data are transferred from the imaging array to the frame store array every 
2.85~ms, at the expense of one dimension of spatial information.
 
We analysed the data following the standard analysis threads for a point-like source 
with the \cxo\ Interactive Analysis of Observations software (\textsc{ciao}, v. 4.8; Fruscione 
et al. 2006) and the calibration files stored in the \cxo\ \textsc{caldb} (v. 4.7.1). Only 
non-dispersed (zeroth-order) spectra were extracted for observations where 
a grating array was used. We used the \textsc{chandra$_{-}$repro} script 
to reprocess the data and generate new `level 2' events files with the latest 
time-dependent gain, charge transfer inefficiency correction, and sub-pixel 
adjustments. For TE-mode data and on-axis targets, we collected the source 
photons from a circular region around the source position with a radius of 2 
arcsec. An important outlier is \galcen\ amid the Galactic Centre, for which 
the counts were accumulated within a 1.5-arcsec radius circular region. A 
larger radius would have included too many counts from Sgr~A$^*$ 
(see Coti Zelati et al. 2015a, 2017 for details). The \cxo\ PSF exhibits 
significant variations in size and shape across the focal plane. Therefore, for 
the few cases where the target of interest was located far from the position of 
the aim point, we proceeded as follows. First, we accurately measured the 
coordinates of the source centroid by applying the \textsc{ciao} source detection 
algorithm \textsc{wavdetect} (Freeman et al. 2002) to the exposure-corrected image. 
We adopted the default `Ricker' wavelet (`Mexican Hat' wavelet) functions with 
scales ranging from 1 to 16 pixels with a $\sqrt{2}$ step size and the default value 
for the source pixel threshold ($\textsc{sigthresh}=10^{-6}$). We then calculated 
the off-axis angle from the pointing direction, and used the \textsc{ciao} tool 
\textsc{psfsize$_-$srcs} to estimate the radius of the 90 per cent encircled counts 
fraction at 3~keV. In all cases the background was extracted from an annulus centred 
on the source location. For observations with the ACIS set in CC mode, source events 
were instead collected through a rectangular region of dimension 4 arcsec along 
the readout direction of the CCD. Background events were extracted within two similar 
boxes oriented along the image strip, symmetrically placed with respect to the target 
and sufficiently far from the position of the source, to minimize the contribution from the 
PSF wings. 
 
We filtered the data for flares from particle-induced background (e.g., Markevitch 
et al. 2003) by running the \textsc{deflare} routine on the lightcurves, and estimated 
the impact of photon pile-up in the TE-mode observations using the \textsc{pileup$_{-}$map} 
tool. On the other hand, the fast readout of the ACIS in the CC-mode ensured in all 
cases that the corresponding spectra were not affected by pile-up. Because of the 
sharp \cxo\ PSF, discarding photons in the core of the PSF to correct for pile-up effects 
results in a significant loss of counts. Spectral distortions were then mitigated directly 
in the spectral modelling, as described in Section~\ref{analysis}. 

We created the source and background spectra, the associated redistribution 
matrices and ancillary response files using the \textsc{specextract} 
script\footnote{Ancillary response files are automatically corrected to account 
for continuous degradation in the ACIS CCD quantum efficiency.}. Spectra and 
auxiliary and response files for contiguous observations with low counting statistics 
were combined using the \textsc{combine$_{-}$spectra} script.

\section{Spectral analysis}
\label{analysis}

We generally grouped the background-subtracted spectra to have at least 20 counts in 
each spectral bin using \textsc{grppha}, to allow for fitting using the $\chi^2$ statistics. 
For the spectra with the largest number of counts (typically those extracted from \xmm\ 
and \cxo\ observations, but in some cases also from \swift\ pointings at the earliest stages 
of the most powerful outbursts), we adopted a higher grouping minimum and the optimal 
binning prescription of Kaastra \& Bleeker (2016)\footnote{See \url{http://cms.unige.ch/isdc/ferrigno/developed-code}.}. 
For the spectra with too few counts for the $\chi^2$-fitting, we opted to group the data to 
a lower degree (or even not to group them in the case of the \swift\ XRT spectra of \sgrd\ 
and Swift\,J1834.9$-$0846), and use the Cash statistics ($C$-statistics; Cash 1979).

We performed the spectral analysis separately for the \swift, \cxo\ and \xmm\ data, owing 
to known cross-calibration uncertainties (e.g. Tsujimoto et al. 2011) and their remarkably 
different effective areas and energy dependence, which translate into different counting 
statistics and therefore best-fitting models in most cases (the larger the statistics available, 
the larger the number of spectral components required to properly fit the data). 

For the spectral modelling we employed the \textsc{xspec} spectral fitting package (v. 12.9.1; 
Arnaud 1996), and applied the Levenberg--Marquardt minimization algorithm (Press et al. 1992). 
We restricted our analysis to the energy interval whereby the calibration of the spectral responses 
is best known, i.e. 0.3--10~keV for \swift\ XRT and \xmm\ EPIC (with some exceptions for the XRT 
WT-mode data; see below), 0.3--8~keV for \cxo, 1.8--10~keV for \bsax\ MECS, 0.1--2.4~keV for 
\ros\ PSPC and 3--10~keV for \rxte\ PCA. For the faintest outbursts (e.g. those of \wes\ and the 
2008 event from \aa) and heavily absorbed sources (e.g. SGR\,1833$-$0832\ and Swift\,J1834.9$-$0846), 
we further limited our study to photons with energy above $1 - 2$~keV, owing to the few available 
counts at lower energy. On the other hand, the spectra of \lowba\ softened significantly as the source 
approached the quiescent phase. The few photons at energy $\gtrsim 3$~keV were overwhelmed 
by the background and hence discarded. In some cases, spectra acquired by \swift\ and with 
the XRT configured in WT mode exhibited some residual bumps due to calibration uncertainties 
below $\sim 1$~keV. Because these features would yield a misleading (systematically 
underestimated) value for the absorption column density, we decided to filter out the spectral 
channels at low energy ($<0.8$~keV).
 
\subsection{Spectral models}

For the continuum emission we tested a set of different single and double-component empirical models: 
a blackbody (\textsc{bbodyrad}; BB), a power law (PL), a blackbody plus a power law (BB+PL), the 
superposition of two blackbodies (2BB) and resonant cyclotron scattering models. In particular, we 
applied the \textsc{ntz} model developed by Nobili et al. (2008a,b), which is based on three-dimensional 
Monte Carlo simulations. The topology of the magnetic field is assumed to be a globally twisted, force-free 
dipole in the model, and its parameters are the surface temperature (assumed to be the same over the 
whole surface), the bulk motion velocity of the charged particles in the magnetosphere (assumed constant
through the magnetosphere), the twist angle and a normalization constant. This model has the same 
number of free parameters as the empirical two-component models mentioned above (2BB and BB+PL). 
In the cases of \xte\ and the 2002 outburst of \antgli, the higher statistics quality available from \xmm\ 
observations allowed us to probe more complicated models, such as the sum of three thermal components 
(3BB). Because the internal calibration accuracy of the pn CCD for on-axis sources is estimated to be 
better than 2 per cent at the 1$\sigma$ c.l. (Smith 2016\footnote{See 
\url{http://xmm2.esac.esa.int/docs/documents/CAL-TN-0018.pdf}.}), we added an extra 2 per cent systematic 
error term to each spectral channel in these cases, as also recommended by the online threads. We then 
assessed the number of required spectral components by means of the Fisher test (e.g. Bevington 1969), 
setting a minimum threshold of 3$\sigma$ (99.7 per cent) for the statistical significance of the improvement 
in the fit.

If pile-up was detected in a \cxo\ observation (typically at the early stages of the outburst), the multiplicative 
pile-up model of Davis (2001) was included, as implemented in \textsc{xspec}. Following the prescriptions 
reported in `{\em The Chandra ABC Guide to Pile-up}'\footnote{See \url{http://cxc.harvard.edu/ciao/download/doc/pileup_abc.pdf}.}, 
the only parameters allowed to vary were the grade-migration parameter and the fraction of events within the 
central, piled up, portion of the source PSF.

The photoelectric absorption by the interstellar medium along the line of sight was described through the 
Tuebingen--Boulder model (\textsc{TBabs} in \textsc{xspec}), and we adopted the photoionization cross-sections 
from Verner et al. (1996) and the chemical abundances from Wilms, Allen \& McCray (2000). The choice of these 
abundances typically translates into values for the column density about 30 per cent larger than those estimated 
assuming the solar abundance tables from Anders \& Grevesse (1989). 

For \galcen\ the \textsc{fgcdust} model was also included to correct for the effects of scattering of X-ray 
photons on interstellar dust grains located along the line of sight towards the source (likely in the Galactic 
disc and a few kpc away from the Galactic Centre according to Jin et al. 2017; see also Coti Zelati et al. 2017). 
For SGR\,1833$-$0832 and Swift\,J1834.9$-$0846, i.e. the most absorbed sources of our sample besides
\galcen\ (see Table~\ref{tab:results}), we tested the inclusion of the \textsc{xscat} model (Smith, Valencic \& 
Corrales 2016) to account for the effect of dust scattering of spreading the photons along the line of sight 
around the source, an effect that is more relevant for the most heavily absorbed objects\footnote{We assumed 
different models for the dust composition and grain size distribution (see Mathis, Rumpl \& Nordsieck 
1977; Weingartner \& Draine 2001; Zubko et al. 2004).}.

Although both the adoption of different chemical abundances and the correction for dust scattering opacity yield 
some differences in the values for the hydrogen column density and hence the unabsorbed fluxes, they provide 
only a secondary source of systematic error on the estimate of the luminosities compared to the uncertainties on 
the sources distances (see Section~\ref{curves}). Furthermore, we checked that they did not translate into 
significantly different decay patterns and estimates for the total energy released during the outburst.

\subsection{Spectral fits}

For each outburst, we started by fitting together the absorbed BB+PL and 2BB models to the spectra acquired 
by \swift\ XRT\footnote{For \galcen, we considered instead \cxo\ data alone, because only the exquisitely sharp 
PSF of the ACIS instrument enables to single out the magnetar counterpart in the crowded region of the Galactic 
Centre. See Coti Zelati et al. (2015a, 2017) for details.}  (with the exception of \xte, the 2008 outburst of \sgrd, \lowba, 
SGR\,1833$-$0832, Swift\,J1834.9$-$0846, the 2011 outburst of \wes, \sgrm\ and \psr, for which a single absorbed 
blackbody model provided an acceptable fit across the entire data set). All parameters of the BB and PL components 
were left free to vary from observation to observation. The absorption column density was left free to vary as well, 
but with the request to be the same at all stages of the outburst evolution. For extensively monitored sources (i.e., 
\sgre, \aa\ during the 2009 outburst, \lowbb, 1E\,1048.1$-$5937, \antgli\ during the 2012 outburst, \sgrm\ and \psr), 
the joint modelling was performed on groups of 20 spectra, to reduce the time-scale of the convergence of the fit 
and of the computation of parameter uncertainties. 

In most cases, spectra of observations carried out at late stages of the outburst were described adequately 
by an absorbed blackbody alone and the addition of a second component was not statistically required. 
However, we decided to retain the second component in the spectral fits, and freeze its pivotal parameter (the 
power law photon index in the BB+PL model or the temperature of the second, hotter, blackbody in the 2BB 
model) to the value inferred for the spectrum of the last pointing where the second component is significantly 
detected. Alternatively, this parameter was tied up between all these data sets. For both alternative strategies, 
the normalizations of the spectral components were left free to vary. We then derived stringent upper limits on 
the contribution of the additional spectral component, and verified that the fits to the single spectra yielded values 
for the parameters consistent with those inferred from the joint modelling. 

The above-mentioned fitting procedure was subsequently repeated for the \xmm\ and \cxo\ data sets. 
Table~\ref{tab:results} reports the best-fitting models. Appendix~\ref{goodspectra} reports a series of 
figures ( Fig.~\ref{fig:spectra_outbursts}) showing a set of high-quality X-ray spectra and the best-fitting 
empirical models for several outbursts that were repeatedly monitored by the \xmm\ or \cxo\ observatories. 
 
For the cases where the $C$-statistics was employed, we evaluated the quality of the fit by Monte Carlo simulations. 
We used the \textsc{goodness} command within \textsc{xspec} to simulate a total of 1000 spectra (based on a 
Gaussian distribution of parameters centred on the best-fitting model parameters and with Gaussian width set by 
the 1$\sigma$ uncertainties on the parameters), and determined the percentage of simulations having a $C$-statistics 
value much lower or higher than that obtained from the best fit of the data.

\section{Light curves}
\label{curves}

For each fitted spectrum, we calculated the absorbed flux for the total source emission, as well as the unabsorbed 
flux and the luminosity for the single spectral components and for the total source emission (all in the 0.3--10~keV 
energy range). Unabsorbed fluxes were calculated using the convolution model \textsc{cflux}, and converted to 
luminosities (as measured by an observer at infinity) assuming isotropic emission and the most reliable value for 
the distance of the source (see Section~\ref{transients} and Table\,\ref{tab:timing}). All the uncertainties are quoted 
at the 1$\sigma$ c.l. for a single parameter of interest ($\Delta \chi^2 = 1$; Lampton, Margon \& Bowyer 1976) 
throughout this work, whereas upper limits are reported at the 3$\sigma$ c.l. For each outburst, we checked that the 
unabsorbed fluxes inferred for observations carried out with different instruments approximately at the same epoch 
were consistent with each other within the cross-correlation uncertainties.
 
The determination of unabsorbed fluxes from models comprising absorbed power law components is known to 
overestimate the source flux by a large factor, owing to the divergence of the power law component at low energy. 
We hence considered the results obtained from the \textsc{ntz} model (Nobili et al. 2008a,b) to estimate bolometric 
(0.01--100~keV) luminosities for the cases where a power law spectral component was required in the spectral fits. 
In all cases the bolometric fluxes were determined after having defined dummy response matrices with the 
\textsc{dummyrsp} command in \textsc{xspec}. An uncertainty of 15 per cent was assigned to each flux. We note that, 
in some cases, an additional spectral component was observed in the hard X-rays at the outburst peak, which then 
became undetectable within the following few weeks. However, the paucity of hard X-ray monitorings of magnetar 
outbursts prevents a proper study of the appearance, disappearance, and total energetics of this component over the 
whole class. If we consider all the hard X-ray observations of magnetar outbursts performed so far by \integral, \suzaku\ 
and \nustar\ (for \sgre, SGR\,1806$-$20, \aa, \galcen, \sgrm, \psr, \cco; see Esposito et al. 2007; Rea et al. 2009; Enoto 
et al. 2010a,b; Kuiper et al. 2012; Kaspi al. 2014; Archibald et al. 2016a; Rea et al. 2016; Younes et al. 2017; see Enoto 
et al. 2017 for a review of all these cases), our values for the bolometric fluxes neglecting this component are 
underestimated (only close to the outburst peak though) by a factor $\lesssim20$ per cent in all cases but SGR\,1806$-$20 
and \aa. For these two magnetars we might be underestimating our bolometric flux at the outburst peak by a factor of $\sim2-3$, 
however the lack of a proper monitoring of the hard X-ray component precludes an accurate modelling of the time evolution 
of the hard X-ray emission.

All cooling curves are shown in a series of figures in Appendix~\ref{out_lcurves} (see Figs~\ref{fig:sgr1627_1998_parameters_outbursts}--\ref{fig:psr1119_parameters_outbursts}), 
and include the evolution of the absorbed fluxes, of the 0.3--10~keV luminosities (for the single spectral components 
and for the total emission) and of the bolometric luminosities. 

\begin{figure*}
\begin{center}
\includegraphics[width=1.69\columnwidth]{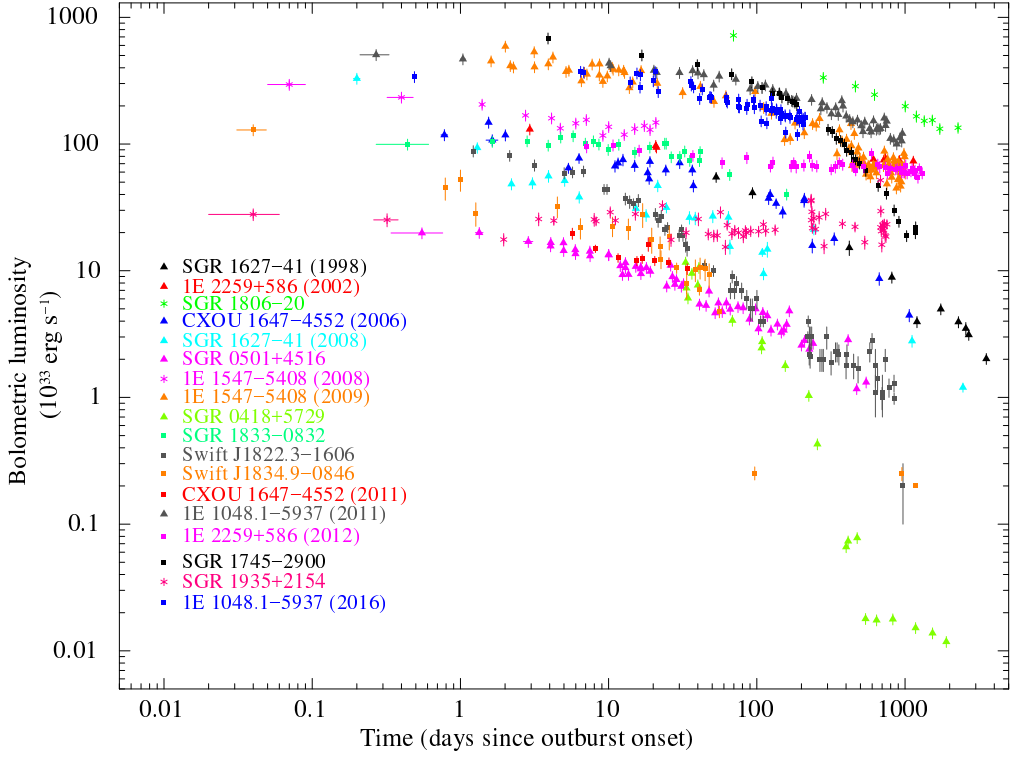}
\vspace{-10pt}
\caption{Temporal evolution of the bolometric (0.01--100~keV) luminosities for all outbursts re-analysed in this work. 
The distances assumed are those quoted in Table~\ref{tab:timing}.}
\label{fig:lum_plot}
\end{center}
\end{figure*}

Fig.~\ref{fig:lum_plot} shows the temporal decays of the  bolometric luminosities for all outbursts. We refer 
each curve to the epoch of the outburst onset, defined as the time of the first burst detection from the source (mostly 
with \swift\ BAT or \fermi\ GBM), or of the giant flare in the case of SGR\,1806$-$20 (note however that the source flux 
already doubled during the first half of 2004 with respect to the quiescent level; Mereghetti et al. 2005). For \xte\ and 
1E\,1048.1$-$5937, for which no bursts were detected prior to their outbursts (see Sections~\ref{xte} and ~\ref{sgrn}, 
respectively), we adopted the epoch of the observation where an increase in the X-ray flux was first measured as the 
reference epoch. 

\subsection{Phenomenological modelling}

We modelled the decays of the X-ray luminosities of the single spectral components and of the total  bolometric 
luminosities using a constant (dubbed $L_q$ and representing the quiescent level; see Table\,\ref{tab:quiescence}) plus 
one or more exponential functions (dubbed \textsc{exp}, \textsc{2exp}, and \textsc{3exp} in the following) of the form 
\begin{equation}
L(t)=L_q + \sum_{i=1}^j A_i\times \rm{exp}(-$$t$$/\tau_i)~,
\end{equation} 
with $j\leq3$ ($\tau_i$ denotes the $e$-folding time). The number of required exponential functions was evaluated by 
means of the $F$-test, i.e., an additional exponential was included only if it yielded an improvement in the fit of at least 
$3\sigma$. A superexponential function of the form 
\begin{equation}
L(t)=L_q + B\times \rm{exp}[-($$t$$/\tau)^\alpha]
\end{equation}
was also tested as an alternative to the double-exponential model in several cases, leading however to an extrapolated 
luminosity at the very early stages of the outburst systematically overestimated compared to that obtained with the 
double-exponential function. We assume conservatively that the peak luminosity attains a value not so different from that 
measured in the earliest observation available, and favour the double-exponential model in the following.

$L_q$ was fixed at the quiescent value or, in cases of non-detections, constrained to be lower than the upper limit 
(see Table~\ref{tab:quiescence} for our estimates of the quiescent bolometric luminosities). There are however two
exceptions for the modelling of the  bolometric light curves: in the case of the 2008 outburst of \aa\ and of the 1998 
outburst of \sgrd, the sources did not reach the historical quiescent level while recovering from the outburst. In these 
cases the constant term was held fixed at the quiescent value reached after that particular outburst (which is larger 
than the historical minimum reported in Table~\ref{tab:quiescence}). 

Although other alternative phenomenological models such as broken power laws or those consisting in the combination 
of one or multiple linear and power law terms could satisfactorily reproduce the decays of several outbursts, we opted to
fit exponential functions to all light curves, to allow a direct comparison of the decay time-scales (i.e., the $\tau$ parameter) 
among different outbursts. On the other hand, the estimate of the outburst energetics is not sensitive to 
the model used to fit the luminosity decay.

\subsection{Outburst energetics}

\begin{figure*}
\begin{center}
\includegraphics[width=1.69\columnwidth]{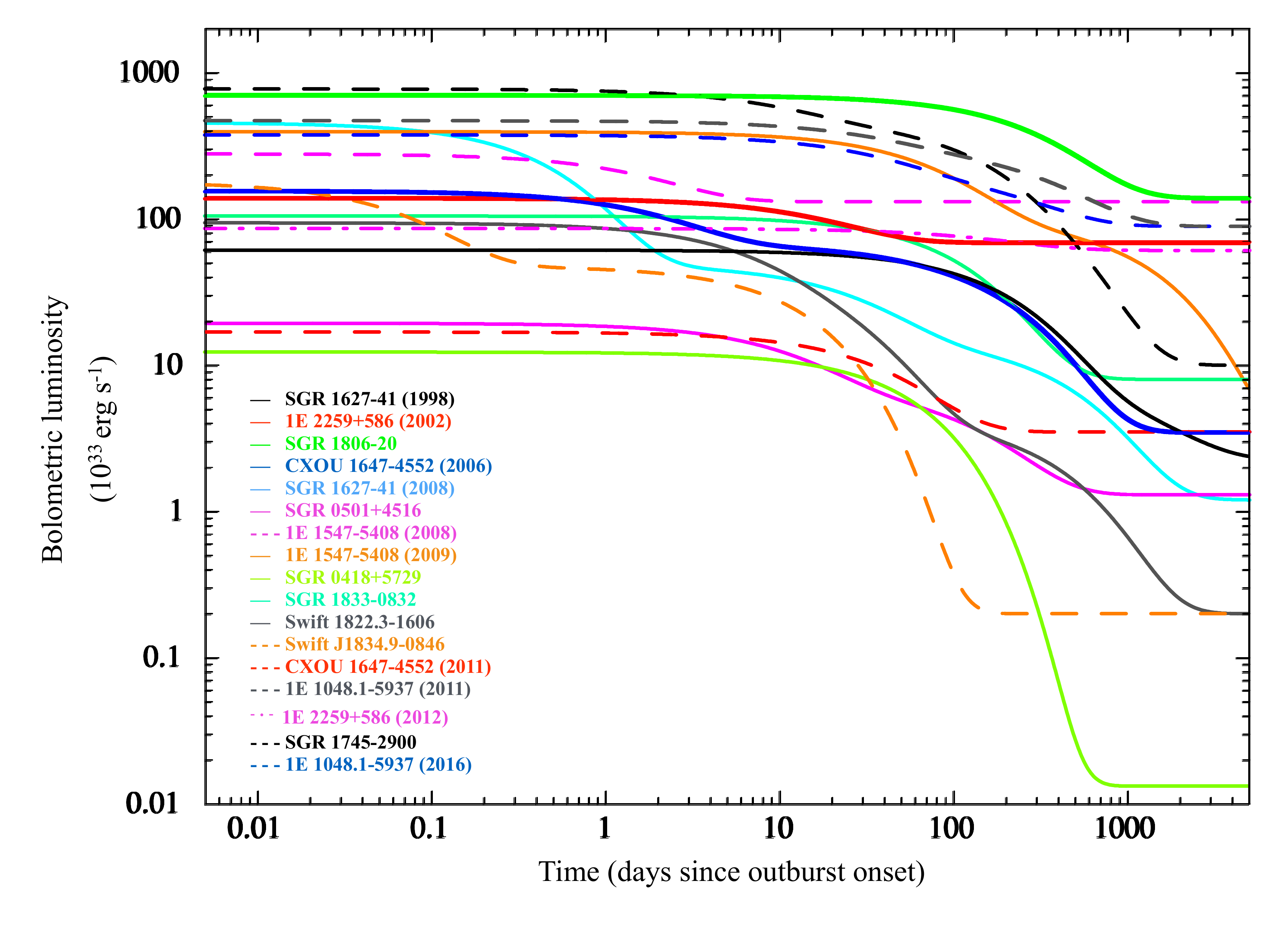}
\vspace{-10pt}
\caption{Models describing the temporal evolution of the bolometric (0.01--100~keV) luminosities for all outbursts re-analysed in this work.}
\label{fig:lum_plot_models}
\end{center}
\end{figure*}

We estimated the total outburst impulsive energetics by integrating the best-fitting model for the bolometric light curves 
over the whole duration of the event, and extrapolating it to the quiescent value for the cases where the observational 
campaign was not extended enough to follow completely the return to the pre-outburst state: 
\begin{equation}
E=\int_{0}^{t_{q}} L_{{\rm bol}}(t) dt~,
\end{equation} 
where $t_{q}$ is the epoch of the recovery of the quiescent state expressed as time since the outburst onset. 
For the sources that are still recovering from their outbursts (i.e., \aa, \galcen, 1E\,1048.1$-$5937, \psr\ and \cco), 
we assumed no changes in the decay pattern down to quiescence for the estimate of the time-scales and energetics 
(in these cases the derived decay time-scale and the total outburst energetics should be considered as upper limits). 
Fig.~\ref{fig:lum_plot_models} shows the best-fitting models (see also the right-hand panels of 
Figs~\ref{fig:sgr1627_1998_parameters_outbursts}--\ref{fig:psr1119_parameters_outbursts} for the individual cases), 
and Table~\ref{tab:decays} reports the corresponding parameters. The assumed 15 per cent error on each bolometric 
value is likely an underestimate (the largest uncertainty arising from the poorly constrained distance of the source in 
almost all cases). For some extensively monitored outbursts we verified that the choice of larger uncertainties on these 
values yielded no significant alteration of the decay pattern and of our estimates for the characteristic time-scales and 
the amount of energy released during the event.

\section{Peak and quiescent luminosities}
\label{maxvsquiescent}

Fig.~\ref{fig:flux_deltaf} shows the maximum luminosity increase as a function of the quiescent (steady) 
X-ray luminosity for all magnetars that so far have displayed substantial enhancements and/or variability in 
their X-ray emission. To have a more complete sample, we have also included SGR\,1900$+$14, 4U\,0142$+$61 
and 1E\,1841$-$045. In fact, although extensive X-ray observations in the \swift, \xmm\ and \cxo\ era did not detect 
major X-ray outbursts from these targets, re-brightenings or subtle variations around their persistent activity have 
been nevertheless reported throughout the last 15 yr. SGR\,1900$+$14 exhibited a giant flare in 1998 (Hurley 
et al. 1999), and re-brightened in the X-rays on two occasions, 2001 April and 2006 March (G{\"o}{\u g}{\"u}{\c s} et 
al. 2011b and references therein). Its flux decline was monitored by \cxo\ and \xmm\ until 2008, and after both 
episodes the source reached the same minimum flux level, which we identify as the bona fide quiescent one (see 
Fig.~\ref{fig:flux_outbursts} for the \swift\ XRT light curve). 4U\,0142$+$61 showed repeated low-level variability 
on top of its persistent emission on at least two occasions, in 2011 July and 2015 February (Archibald et al. 2017b)\footnote{An 
additional magnetar-like burst in 2017 July was recently reported by Hamburg (2017).}. 1E\,1841$-$045 also 
showed sporadic bursting/flaring activity between 2010 May and 2011 July (a total of nine bursts were indeed recorded 
by \swift\ BAT and \fermi\ GBM), and some deviation from the source historical persistent X-ray flux has been noticed 
between 2008 and 2011 (Lin et al. 2011). The source has been subsequently monitored by \swift\ XRT about 145 times 
until the end of 2017 April. Finally, a recent analysis by Scholz et al. (2014) showed that the flux of magnetar 
1RXS\,J170849.0$-$4009 remained constant within uncertainties between 2003 and 2013, in contradiction with what 
reported by G\"{o}tz et al. (2007). In particular, the maximum variability for the X-ray flux is constrained to be lower than 
10 per cent over this decade. We also verified that the source flux remained approximately steady between 2013 April 
and 2017 May by visually inspecting the long-term X-ray light curve generated using all the \swift\ XRT observations 
carried out during this period (which also covers the epoch of the detection of a magnetar-like burst, on 2017 February 
17; see Archibald, Scholz \& Kaspi 2017). In light of these characteristics, we decided not to include this source in our sample.

For each magnetar, spectra relative to the first observation following the outburst onset were used to measure accurately 
fluxes and luminosities at the very early phases of the outburst. For the sources showing low-level variability (see above), 
we extracted and fitted the spectrum relative to the observation where the source is found at the highest flux ever. 
Table~\ref{tab:outburst} lists the inferred values. The quiescent 0.3--10~keV fluxes and luminosities for all magnetars 
monitored so far are reported in Table~\ref{tab:quiescence}. Bolometric luminosities are also quoted. We also calculated the 
flux during pre-outburst observations (if available), and considered the lowest value historically to estimate the quiescent level. 
For the sources where only low-level variability has been reported, we focused on the observations with high counting statistics 
to derive the persistent flux.

In all cases, the spectra were fitted using thermal models (i.e., one or multiple blackbody components) or the \textsc{ntz} 
model, to avoid possible overestimates in the values for the fluxes introduced when fitting a power law model to the data. 
For \ros\ data, we extrapolated the 0.3--2.4~keV fluxes using the \textsc{dummyrsp} tool. For the cases where the source 
is not detected, we applied the \textsc{srcflux} task of \textsc{ciao} (for the \cxo\ observations of Swift\,J1834.9$-$0846 and 
\galcen) and the \textsc{eupper} tool of \textsc{sas} (for the \xmm\ observation of SGR\,1833$-$0832) to derive 3$\sigma$ 
upper limits on the net count rates at the source position (the background was estimated locally). We found values of 
$2 \times 10^{-4}$, $1.1\times 10^{-3}$ and $7\times 10^{-4}$ counts s$^{-1}$ for Swift\,J1834.9$-$0846, \galcen\ and 
SGR\,1833$-$0832, respectively. We then assumed a blackbody spectral model with $kT = 0.3$~keV (similarly to what 
observed in other quiescent magnetars; see e.g. table~3 by Olausen \& Kaspi 2014), and the same column density derived 
from the joint spectral fits of the outburst decay, to infer upper limits on the fluxes with the Portable, Interactive Multi-Mission 
Simulator (\textsc{pimms}, v. 4.8; Mukai 1993).

\section{Search for correlations}
\label{correlations}

Our systematic analysis allows us to search for correlations between different parameters for all sources of our 
sample and their outbursts, in particular between parameters measured in this work (e.g., quiescent luminosity, 
maximum luminosity, maximum luminosity increase, outburst energetics and time-scale) and the timing-inferred 
parameters (e.g., surface dipolar magnetic field, rotational energy loss rate, characteristic age). 

Table~\ref{tab:timing} reports the most up-to-date values for the spin period and the spin-down rate for our sample of 
magnetars, and for the other sources we included in our correlation study (see below). We list the strength of the 
surface dipolar component of the magnetic field at the pole, the spin-down luminosity and characteristic age (all 
estimated assuming simple magnetic dipole braking in vacuo, the initial spin period to be much smaller than the 
current value, and no variation of the magnetic field in time). Several magnetars displayed a high level of timing 
noise in their rotational evolution across the outburst decay, and significant deviations from simple spin-down were 
often detected. For these cases we assumed a long-term average value for the spin-down rate to infer the 
characteristic parameters, following Olausen \& Kaspi (2014). 

Table~\ref{tab:correlations} reports the significance for the (anti)correlations among several different combinations of parameters.
The significance was evaluated from the two-sided null-hypothesis probability ($p$-value) obtained from the Spearman and Kendall 
$\tau$ rank correlation tests. We did not include upper limit measurements in our computations, but verified that the reported upper 
limits were consistent with the observed trend for all cases where a significant ($>2\sigma$) correlation or anticorrelation was observed. 
The table also reports on the shape of the (anti)correlation. The power law index was estimated for each case via a power law regression 
test based on the least squares fitting method (only for the cases where the significance for the correlation was above $2\sigma$). 
The table also indicates whether the observed/unobserved correlation/anticorrelation fits either the internal crustal cooling model 
(Perna \& Pons 2011; Pons \& Perna 2011; Pons \& Rea 2012) or the untwisting bundle model (Beloborodov 2009) proposed to 
account for the evolution of magnetar outbursts (see also Section\,\ref{discussion}).

All plots are shown in Figs~\ref{fig:flux_deltaf}--\ref{fig:energy_deltaflux}. In all figures, the black triangles denote all magnetars 
of our sample; black squares represent the high-field rotation-powered pulsars that underwent an outburst (i.e., \psr\ and \psrr) and 
the grey cross denotes \cco. The year of outburst onset is indicated in parentheses for sources that underwent more than one 
luminosity enhancement. To have a more complete sample, we included also the other few magnetars (black stars), the central 
compact objects (grey crosses), the rotation-powered pulsars clearly showing a thermal component in their spectra (red diamonds) 
and the X-ray dim isolated neutron stars (orange crosses) already reported by Vigan\`o et al. (2013; see in particular their table~1 for 
a list and their section~2 for the criteria adopted to select the sample). \psr\ and the magnetars \xte, \aa\ and \galcen, for which radio 
pulsed emission was detected (Camilo et al. 2006, 2007; Eatough et al. 2013), are marked by black circles. Upper and lower 
limits are indicated by black arrowheads.

\section{Discussion}
\label{discussion}

We carried out the first systematic study of all sources experiencing magnetar-like outbursts up to the end of 2016, and for which extensive 
X-ray monitoring campaigns of their outbursts are available. We re-analysed in a coherent way about 1100 X-ray observations, adopting the 
same assumptions and spectral models throughout the whole sample. This work allows us to study possible correlations and anticorrelations 
between several different combinations of parameters, and put the results in the context of the models proposed to explain the triggering 
mechanism and evolution of magnetar outbursts.

\subsection{On the relation between the outburst luminosity increase and the quiescent luminosity}

A few years ago, Pons \& Rea (2012) showed how magnetars with low quiescent luminosities ($L_q\sim10^{31}-10^{33}$ \lum) experience large 
luminosity increases during an outburst, whereas the brightest sources in quiescence ($L_q\sim10^{34}-10^{35}$ \lum) undergo only subtle 
enhancements in luminosity. This discovery clarified that the distinction between `transient' and `persistent' sources within the magnetar 
population is deceptive, and only dependent on the initial quiescent luminosity of each source.

\begin{figure*}
\begin{center}
\includegraphics[width=2.1\columnwidth]{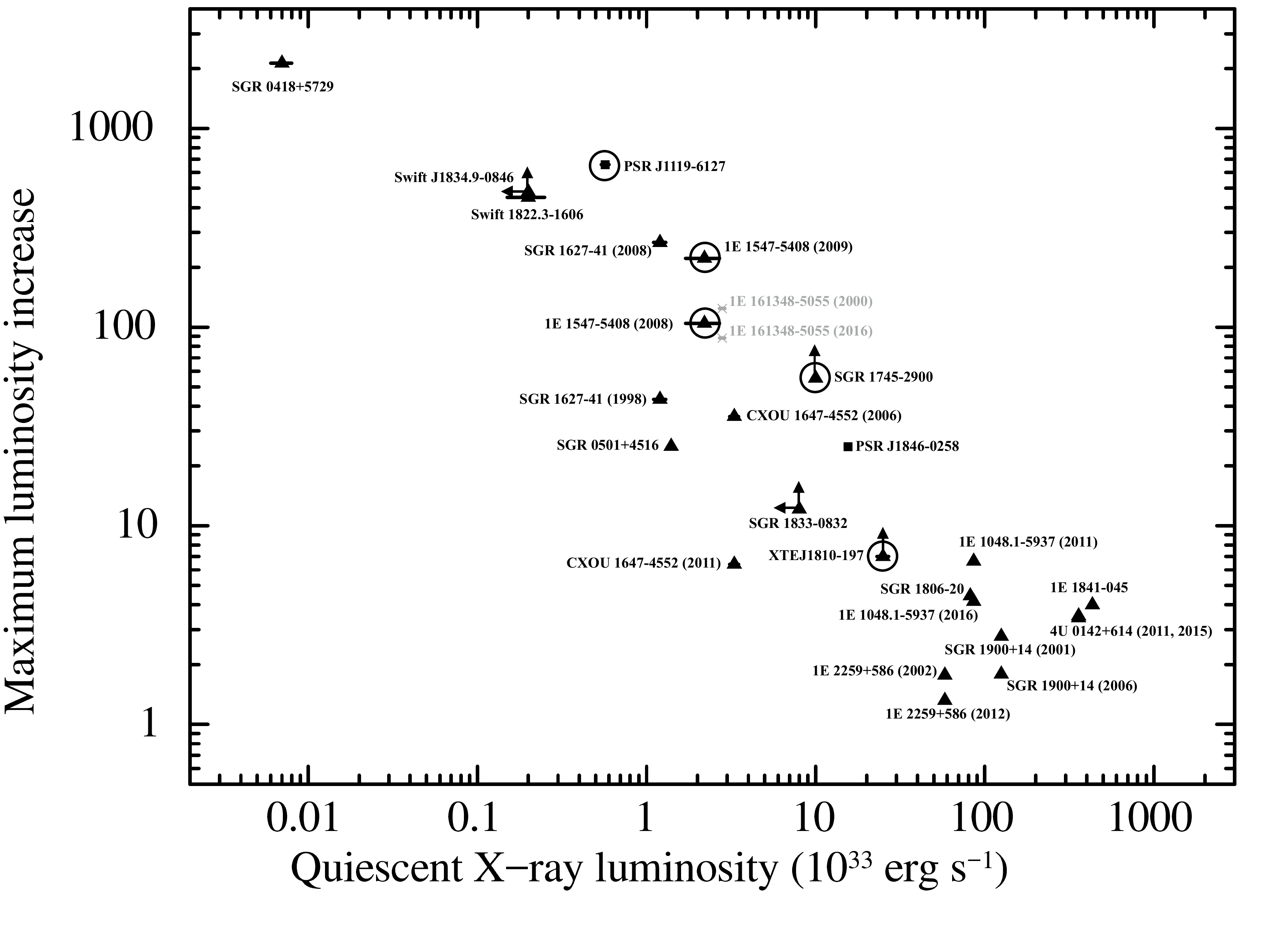}
\vspace{-1cm}
\caption{Maximum X-ray luminosity increase versus quiescent X-ray luminosity.}
\label{fig:flux_deltaf}
\end{center}
\end{figure*}

The anticorrelation between magnetars quiescent luminosities and their luminosity increases is observed at a significance 
of 5.7$\sigma$ (according to the Spearman test; see Table\,\ref{fig:flux_deltaf} and Fig.~\ref{fig:flux_deltaf}), and suggests the existence of a limiting luminosity of 
$\sim10^{36}$ \lum\ for magnetar outbursts (regardless of the quiescent level of the source). This result was interpreted in the framework of the 
internal crustal heating model as the observational manifestation of the self-regulating effect resulting from the strong temperature-dependence 
of the neutrino emissivity (Pons \& Rea 2012): the surface photon luminosity for injected energies larger than $\sim10^{43}$\,erg reaches a limiting 
value of $\sim10^{36}$ \lum , because the crust is so hot that most of the energy is released in the form of neutrinos before reaching the star surface. 
The observed anticorrelation is expected also in the untwisting magnetospheric bundle model, where the maximum theoretically 
predicted luminosity could be somewhat higher, a few 10$^{36}$ \lum\, for the generous case of a twist with $\psi \sim 1$~rad extended to a large 
part of the magnetospheric volume. The generally lower values observed for the peak luminosity are interpreted, in this model, as a consequence 
of the limited size of the current bundle and the twist (Beloborodov 2009).

We used our updated sample (see Fig.~\ref{fig:flux_deltaf}) to gauge the general trend of this anticorrelation via a power law regression test:
\begin{equation}
\Delta L_X \equiv \frac{L_{X,peak}}{L_{X,q}}  \propto L_{X,q} ^{-0.7}~. 
\end{equation}

We observed a similar trend when considering fluxes, suggesting a weak dependence on the sources distance.

We note that, although there is no observational bias in detecting large luminosity increases in sources with a high quiescent luminosity (see the empty 
regions on the top-right corners of Fig.~\ref{fig:flux_deltaf}), the lack of detections of weak outbursts in magnetars with a low quiescent level (see the 
empty regions on the bottom-left corners of Fig.~\ref{fig:flux_deltaf}) might follow from the lack of sufficient sensibility of the current all sky X-ray monitors 
in detecting relatively subtle outbursts in low-luminosity sources.

We also point out that, throughout this study, the epoch of the outburst onset was defined as the time of the first burst detection from the source (mostly 
with \swift\ BAT or \fermi\ GBM), or of the giant flare in the case of SGR\,1806$-$20. This is a somewhat arbitrary choice, because the increase of the 
persistent flux during the time interval preceding the detection of magnetars bursting/flaring activity is usually missed by X-ray instruments. In some cases (e.g., 
\wes\ and \galcen; see Muno et al. 2007 and Kennea et al. 2013b, respectively), the time-scale for the flux rise was constrained to be shorter than a couple 
of days, but this might not be necessarily the case for all magnetars. However, given the large sample, and the clear trend observed over several orders 
of magnitude, we do not expect to measure significantly different values for the outburst peak luminosity.

Different estimates on the time-scale of the luminosity increase were proposed in the past years. The internal crustal cooling models by Pons \& Rea 
(2012) show that the internal heat wave takes some time to propagate from the location in the crust where the energy is injected up to the surface 
layers. Therefore, the luminosity increase is not instantaneous but relatively fast, and might range from a few hours up to a few days depending on the 
depth of the region where heat is released. On the other hand, simplified one-dimensional models show that the time-scale of magnetospheric twisting 
by a large thermoplastic wave (corresponding to the rise time of the outburst) can span from days to weeks (Li et al. 2016). Within the large uncertainties, 
both models are compatible with a typical rise time of a few days.

\subsection{On the quiescent luminosity versus the spin-down luminosity and the dipolar magnetic field}

\begin{figure*}
\begin{center}
\includegraphics[width=1.75\columnwidth]{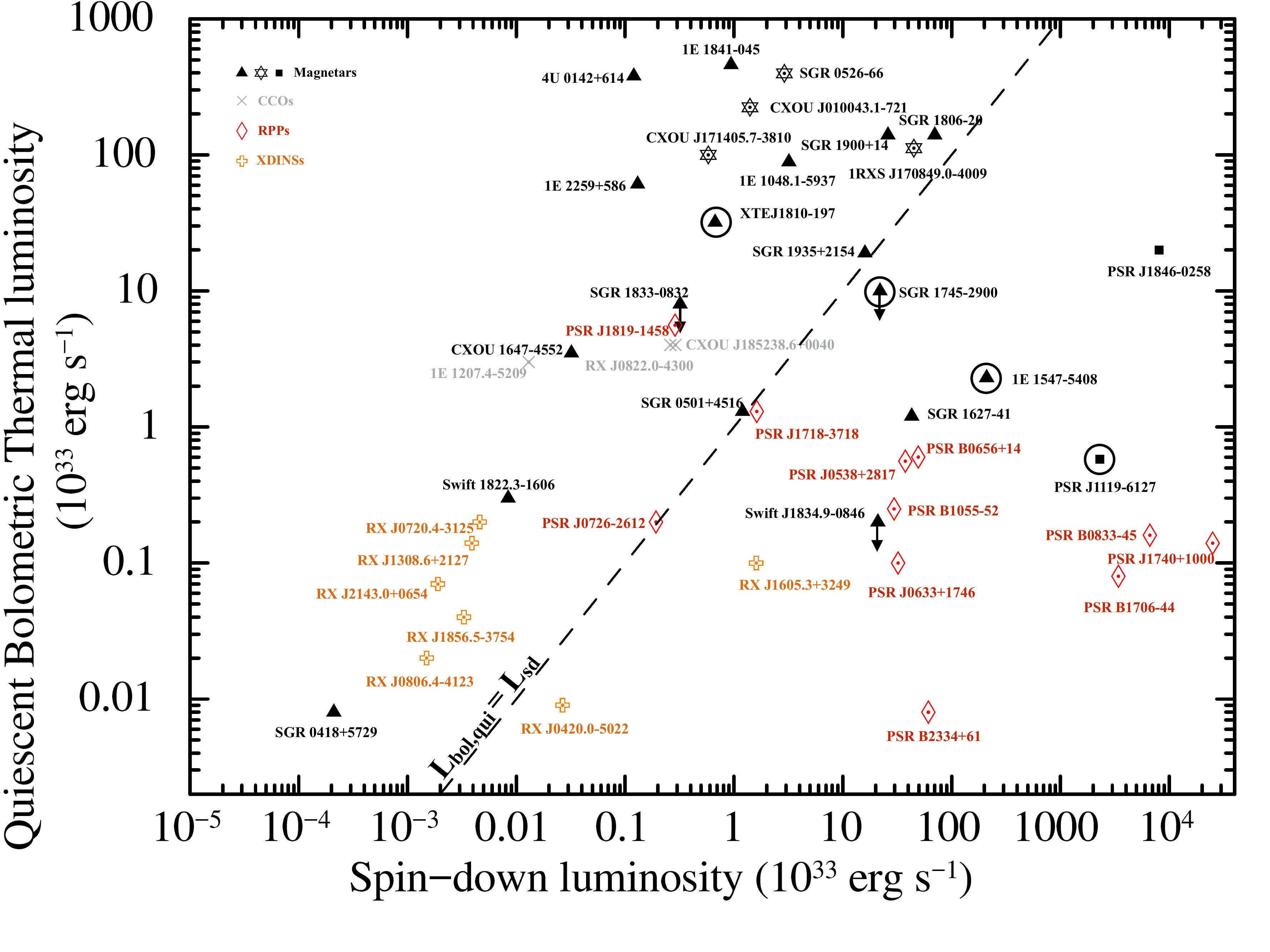}
\includegraphics[width=1.75\columnwidth]{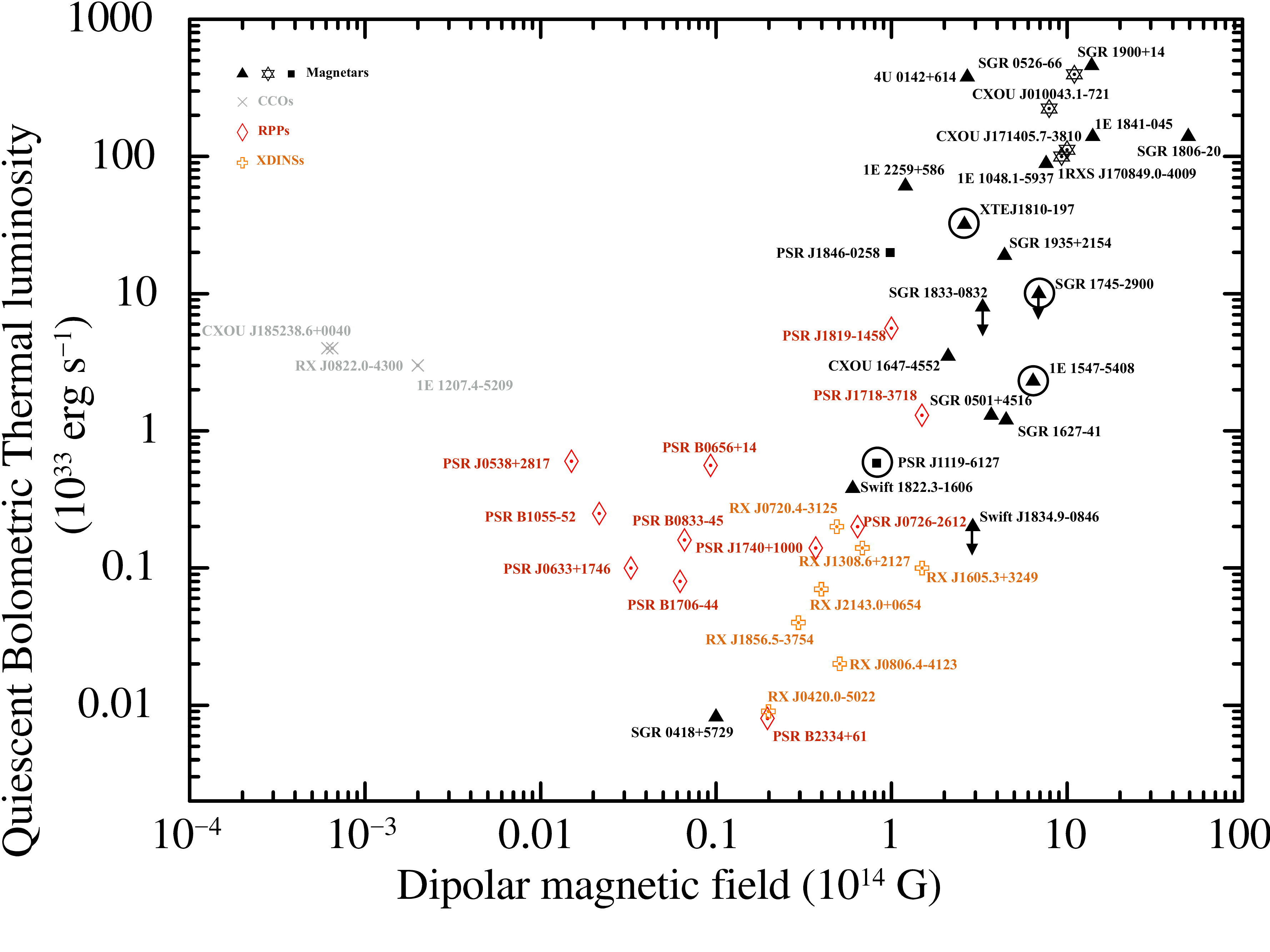}
\vspace{-0.5cm}
\caption{\emph{Top panel}: quiescent bolometric luminosity relative to the thermal component versus the rotational energy loss rate for all isolated X-ray pulsars with clear 
thermal emission. The black dashed line marks the region on the diagram where the  bolometric luminosity equals the spin-down luminosity. \emph{Bottom panel}: quiescent 
bolometric luminosity relative to the thermal component versus the dipolar component of the magnetic field. In both figures, black triangles refer to the `canonical' magnetars 
of our sample, black stars indicate magnetars that did not experience outburst activity, black squares denote the rotation-powered pulsars with high magnetic field that showed 
magnetar-like activity, light grey crosses are the central compact objects, red diamond the rotation-powered pulsars selected by Vigan\`o et al. (2013) and orange crosses refer 
to the X-ray dim isolated neutron stars.}
\label{fig:b_lqui}
\end{center}
\end{figure*}

\begin{figure*}
\begin{center}
\includegraphics[width=1.04\columnwidth]{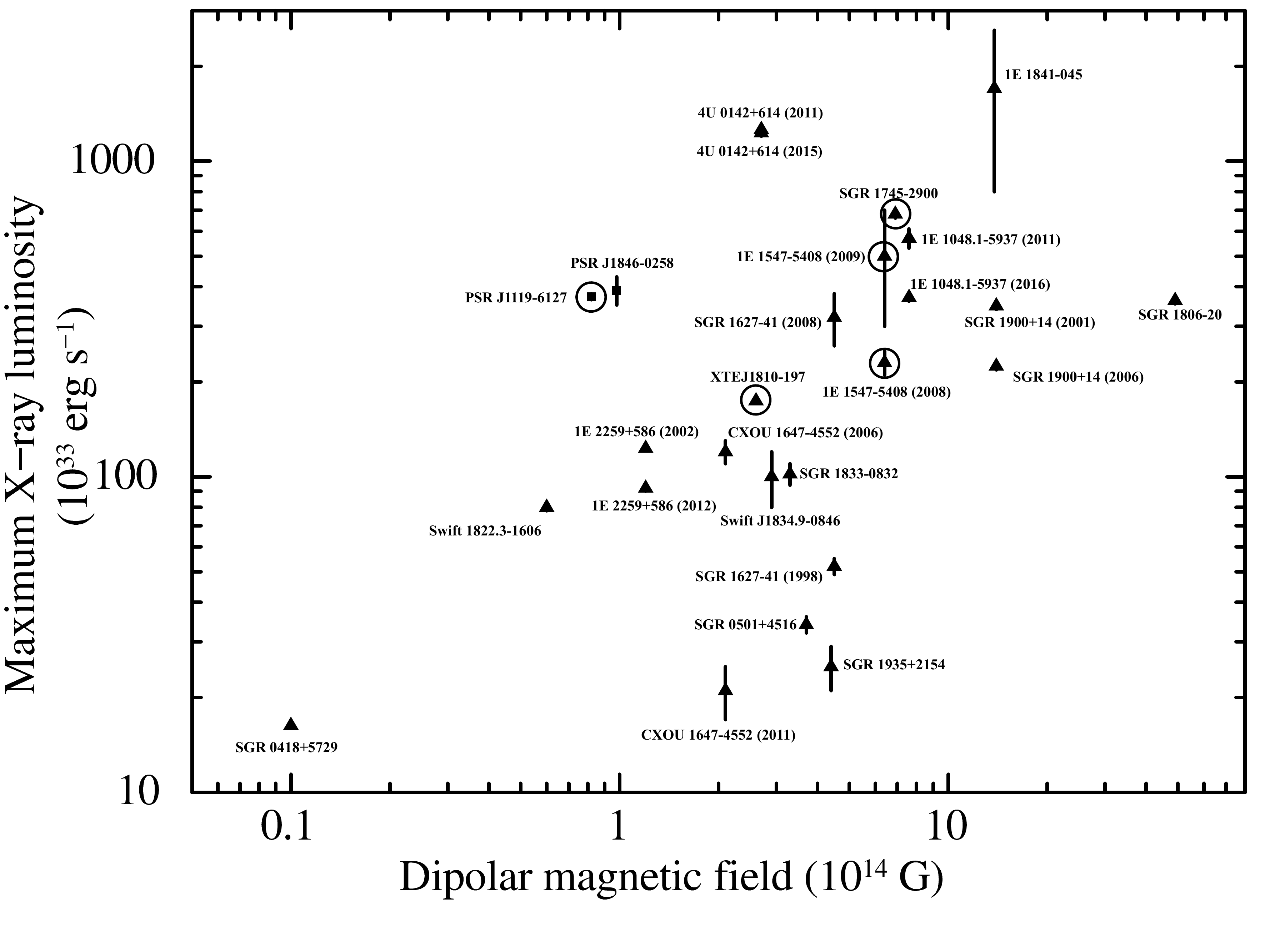}
\includegraphics[width=1.04\columnwidth]{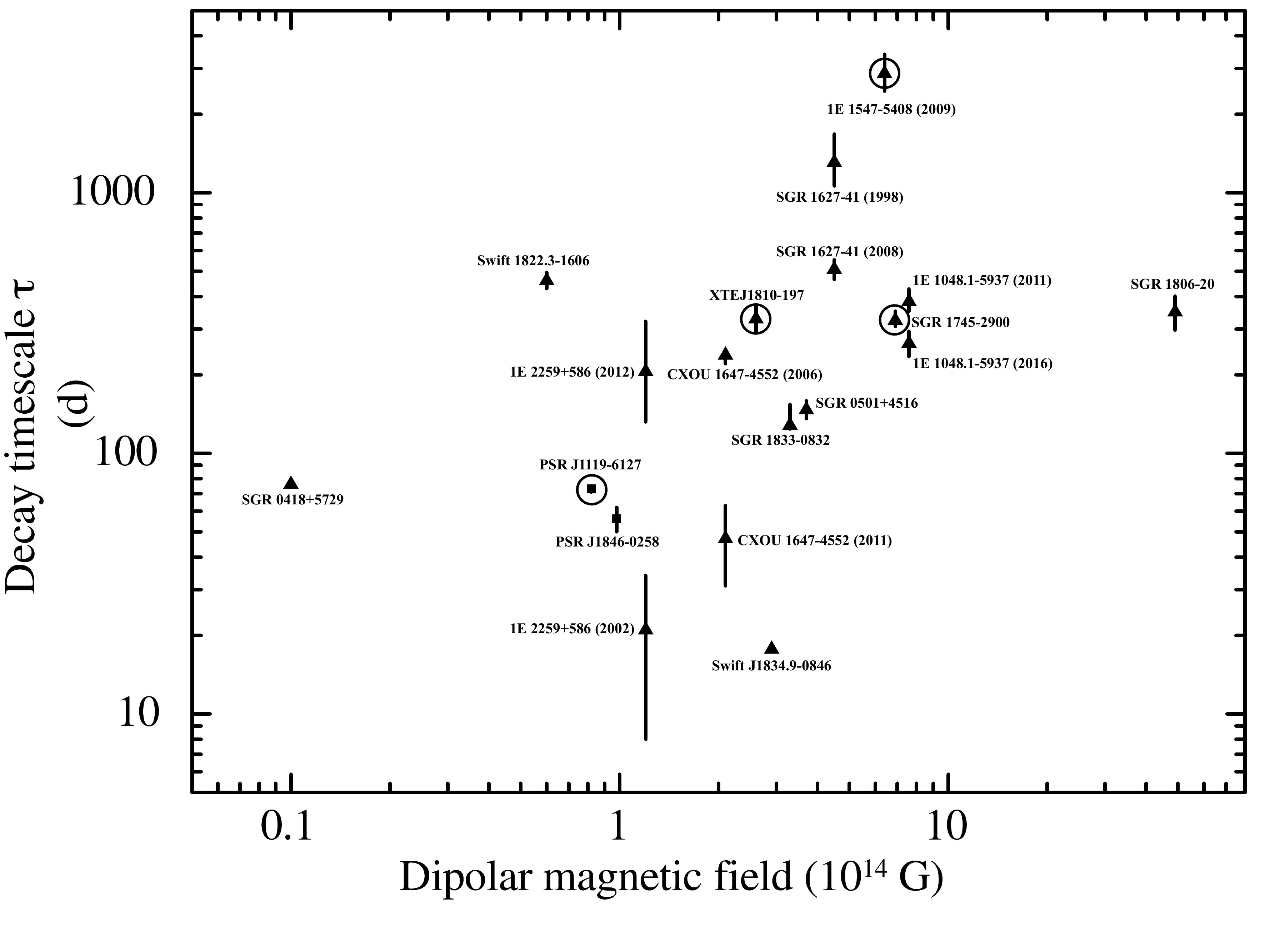}
\vspace{-0.2cm}
\caption{\emph{Left-hand panel}: maximum X-ray luminosity as a function of the dipolar component of the magnetic field. \emph{Right-hand panel}: decay time-scale as a function
of the dipolar component of the magnetic field.}
\label{fig:b_lpeak}
\end{center}
\end{figure*}

\begin{figure*}
\begin{center}
\hbox{
\includegraphics[width=1.04\columnwidth]{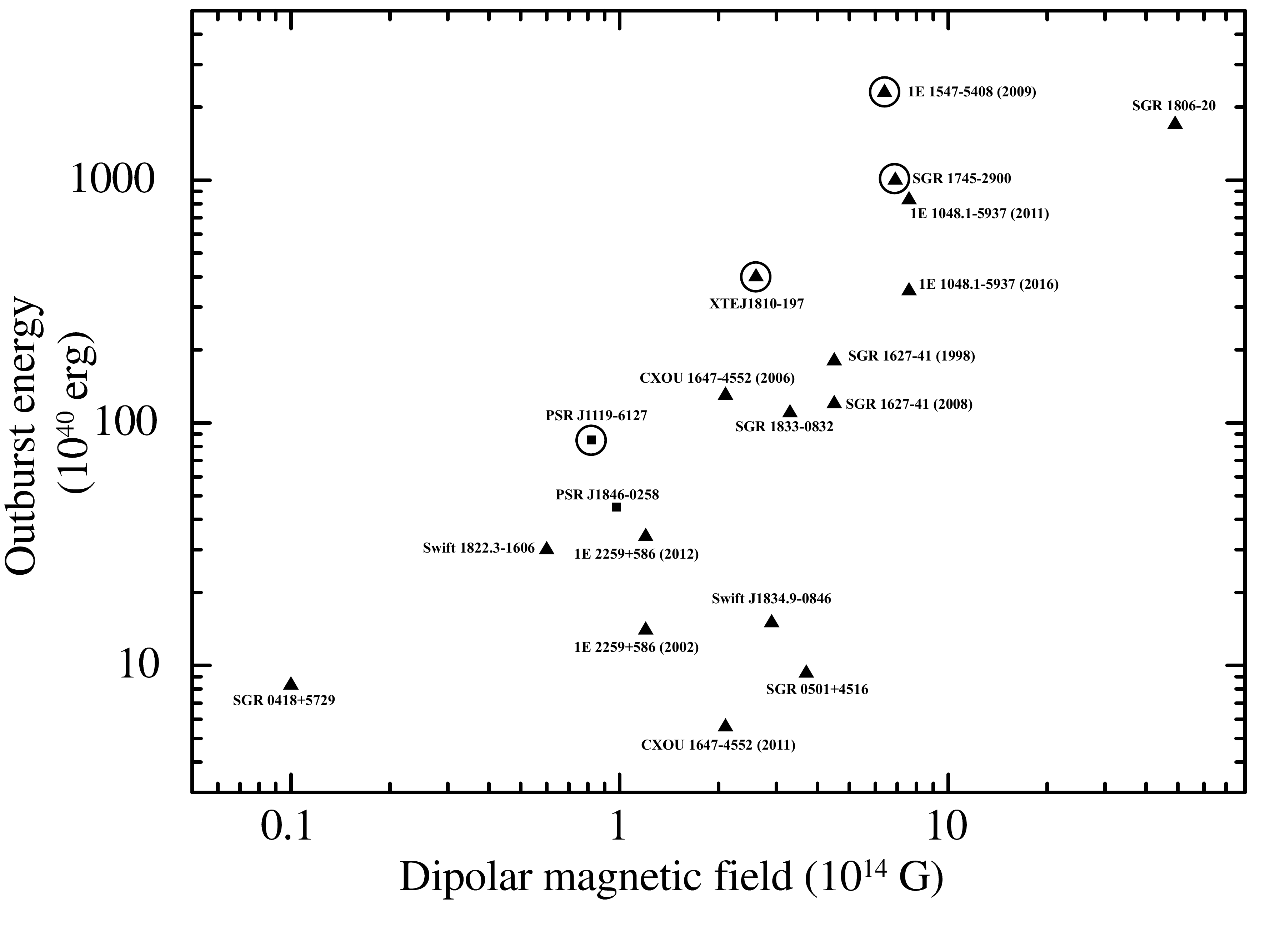}
\includegraphics[width=1.04\columnwidth]{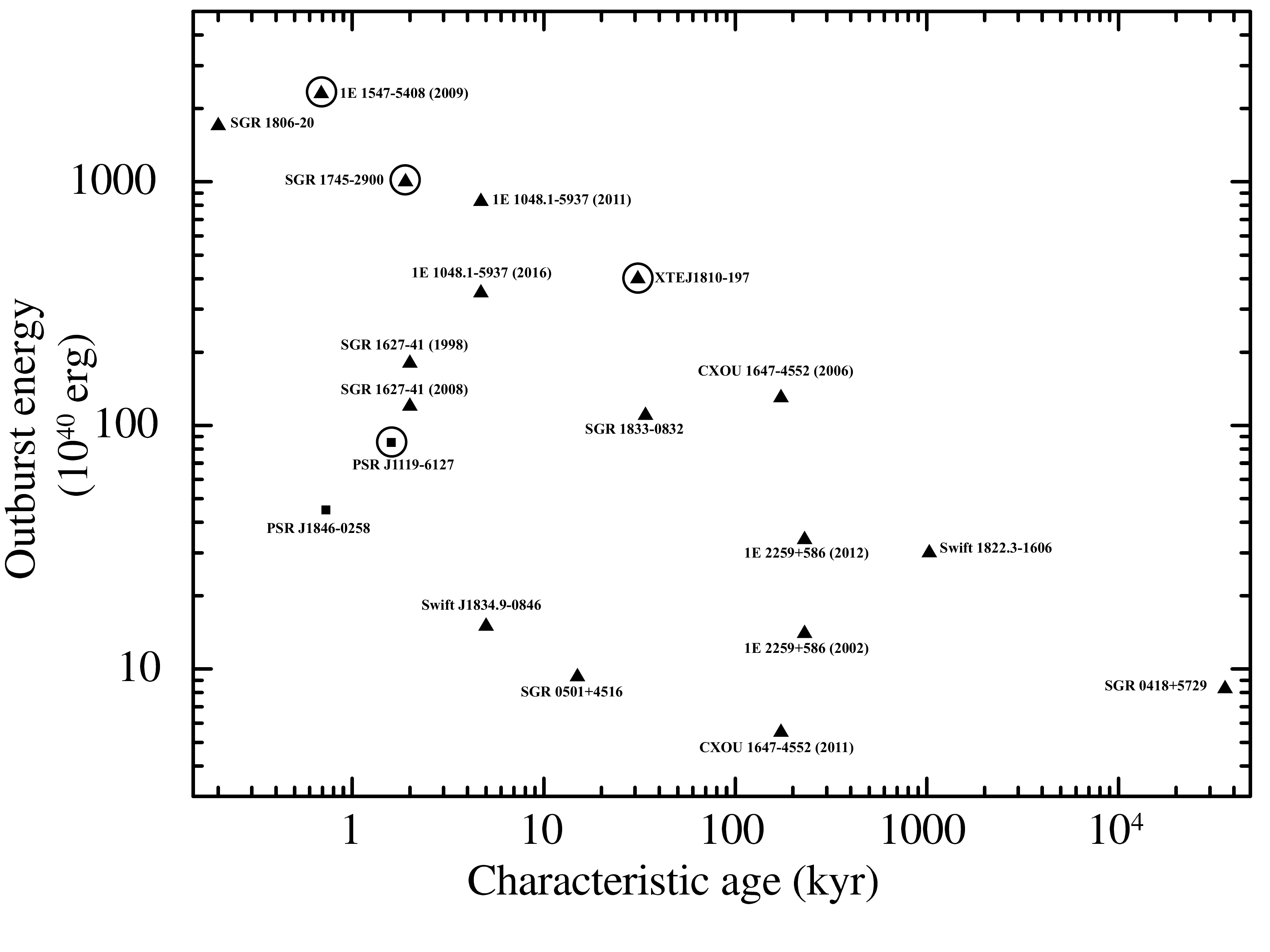}}
\hbox{
\includegraphics[width=1.04\columnwidth]{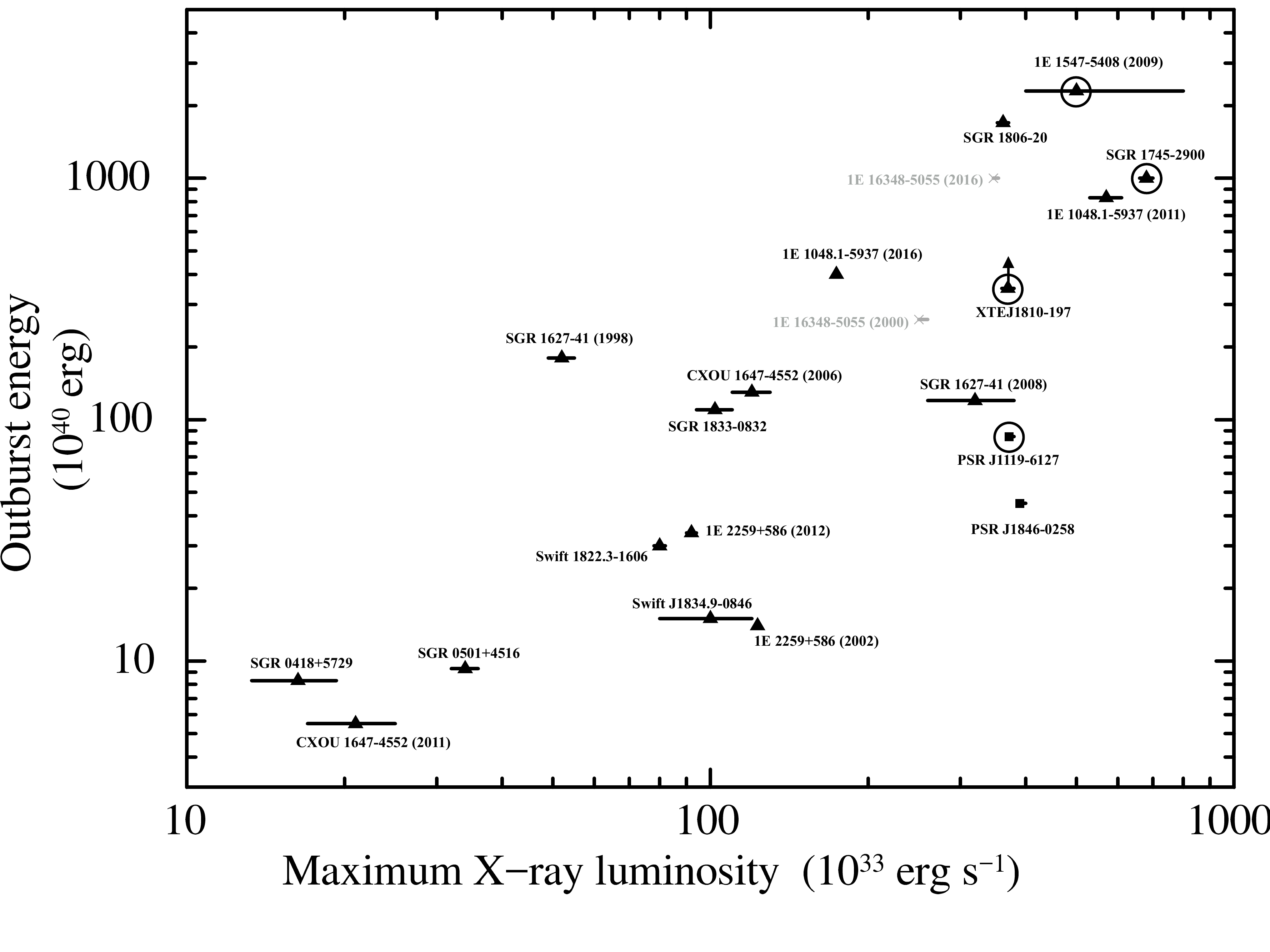}
\includegraphics[width=1.04\columnwidth]{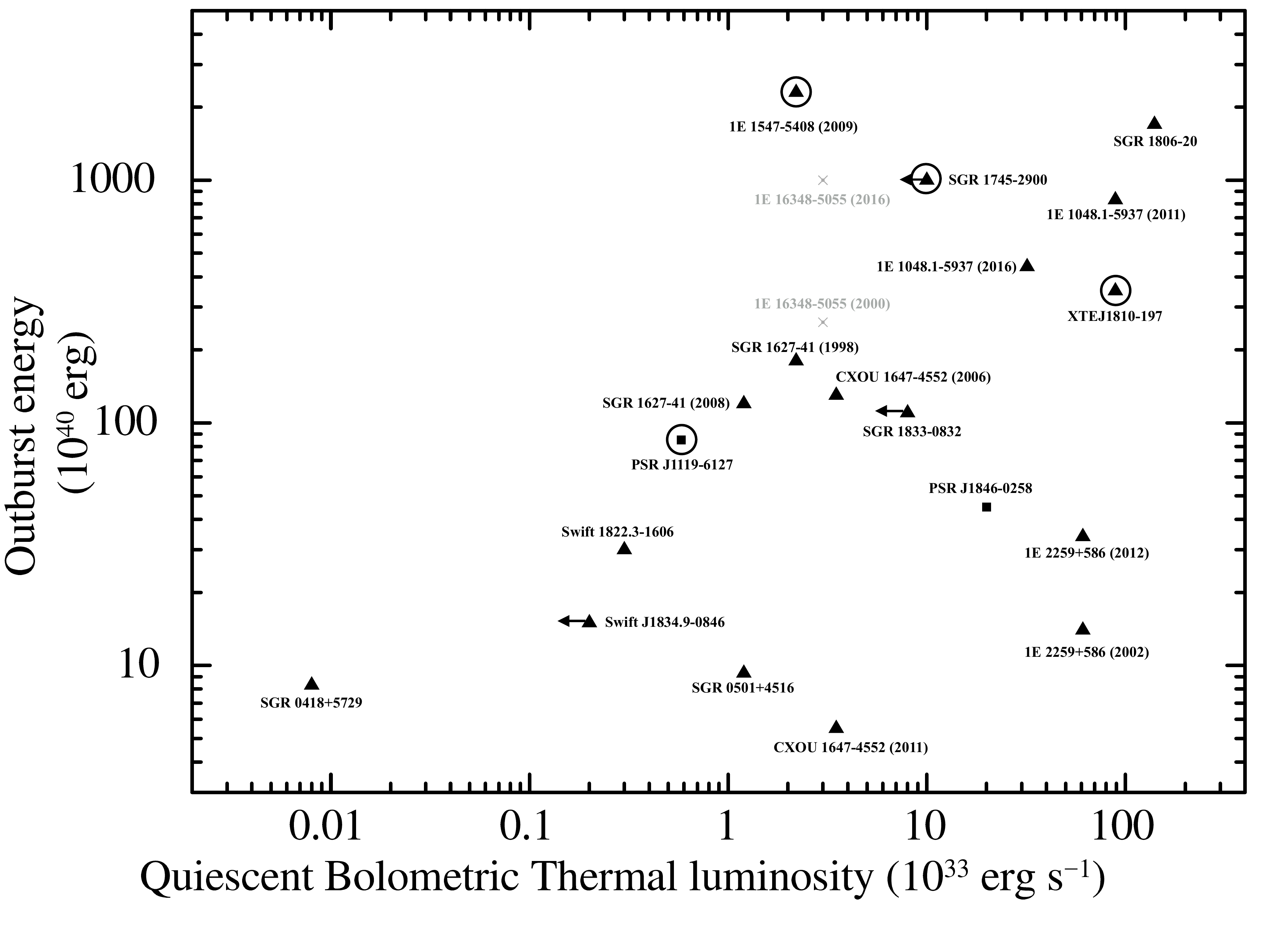}}
\caption{\emph{Top panels}: total energy released during the outburst as a function of the dipolar component of the magnetic field (left), and total energy released during the 
outburst as a function of the characteristic age (right). \emph{Bottom panels}: total energy released during the outburst versus maximum X-ray luminosity at the peak of the 
outburst (left), and total energy released during the outburst as a function of the quiescent  bolometric luminosity relative to the thermal component (right).}
\label{fig:b_energy}
\label{fig:lpeak_energy}
\end{center}
\end{figure*}

\begin{figure*}
\begin{center}
\includegraphics[width=1.04\columnwidth]{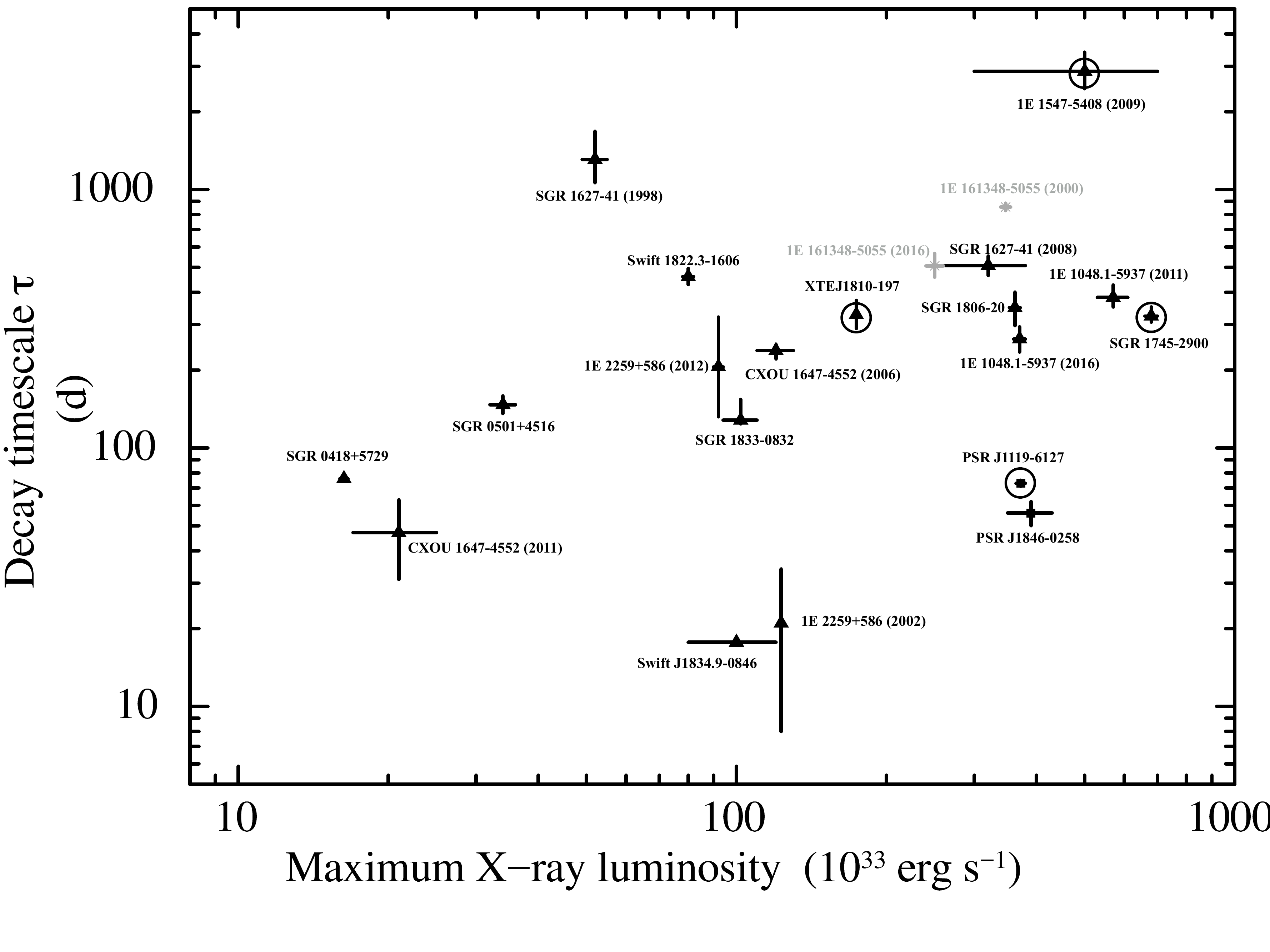}
\includegraphics[width=1.04\columnwidth]{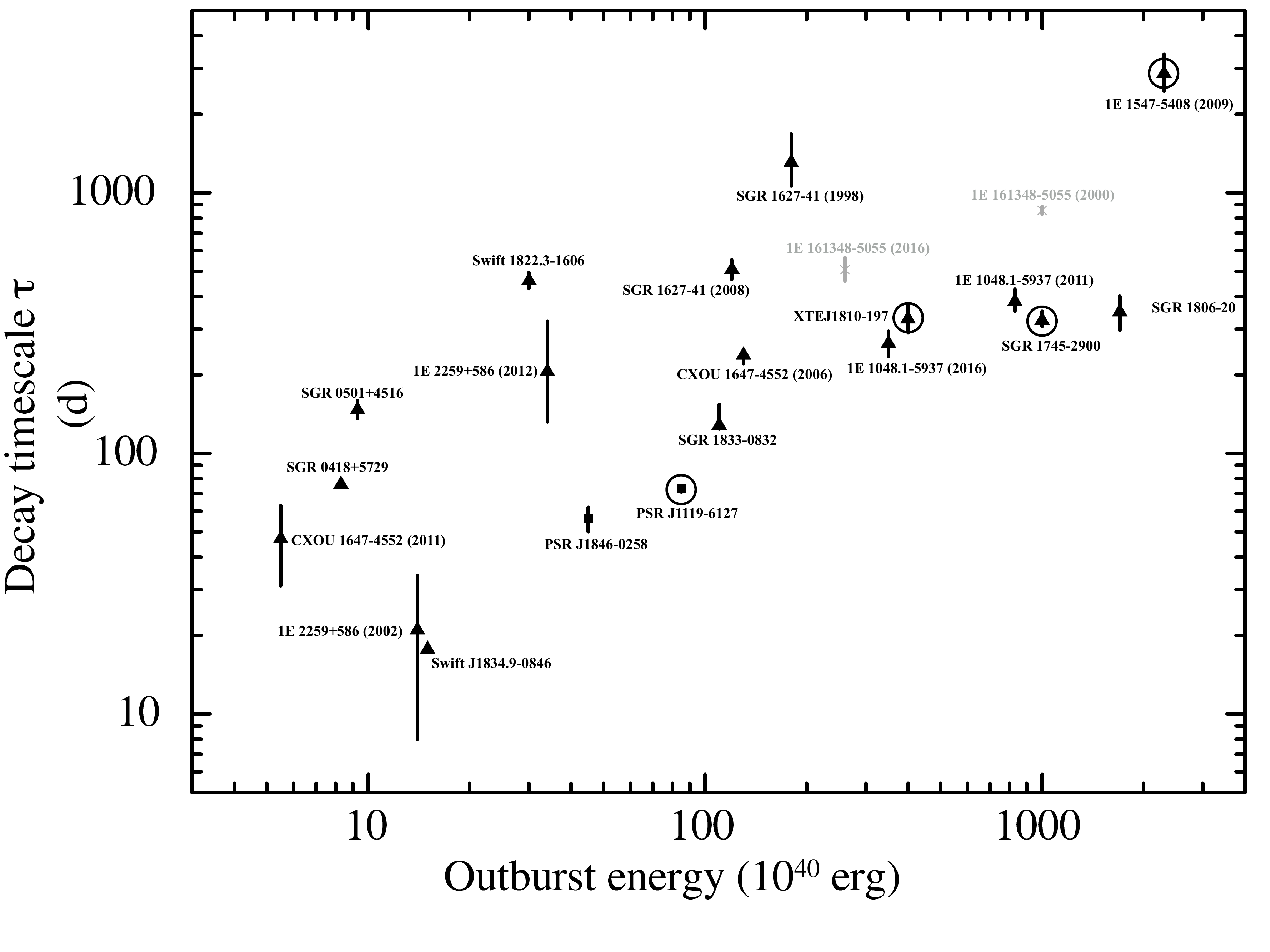}
\caption{\emph{Left-hand panel}: decay time-scale as a function of the maximum X-ray luminosity at the peak of the outburst. \emph{Right-hand panel}: decay time-scale as a 
function of the total energy released during the outburst.}
\label{fig:lpeak_decay}
\end{center}
\end{figure*}

\begin{figure*}
\begin{center}
\includegraphics[width=1.04\columnwidth]{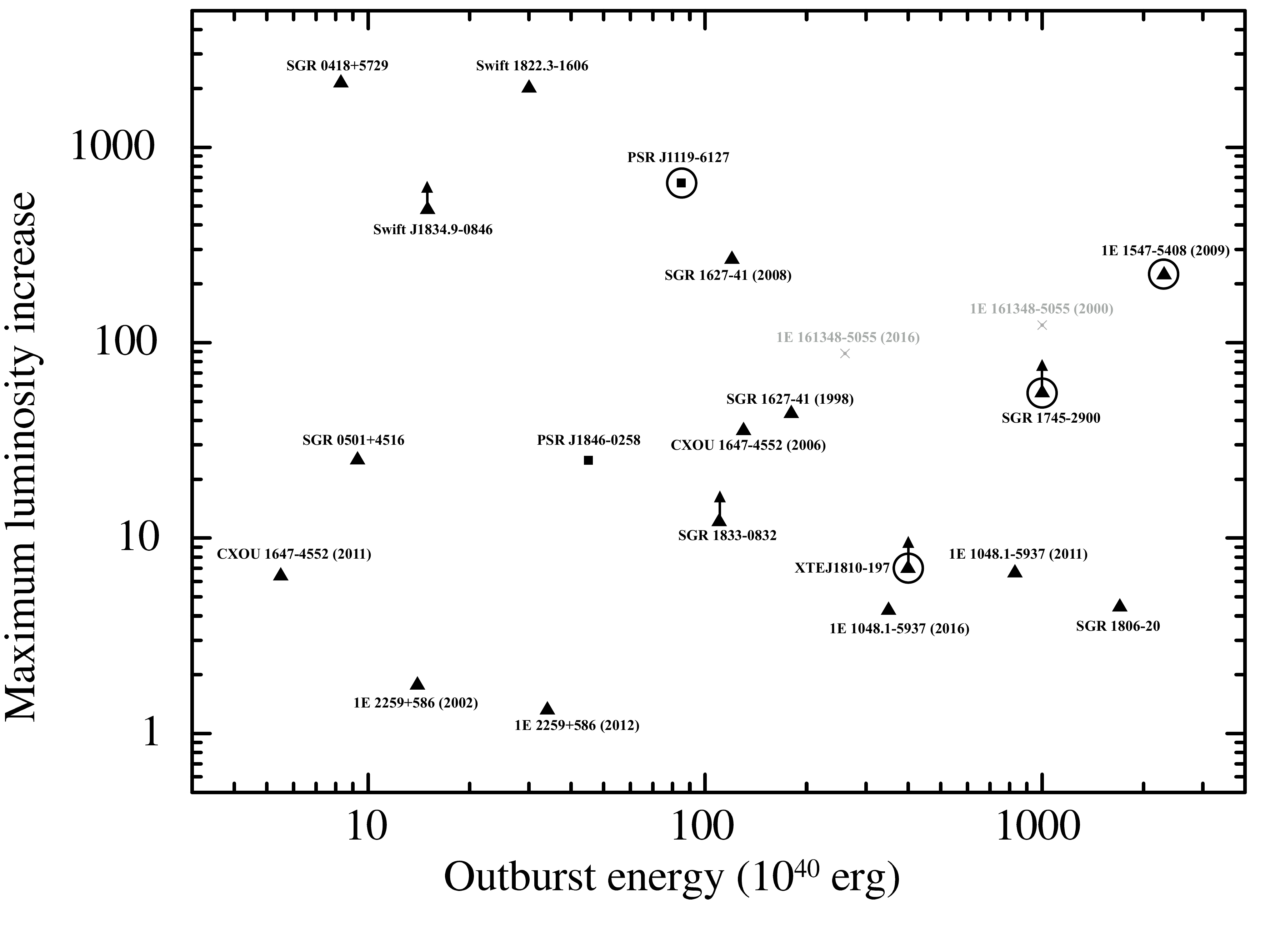}
\includegraphics[width=1.04\columnwidth]{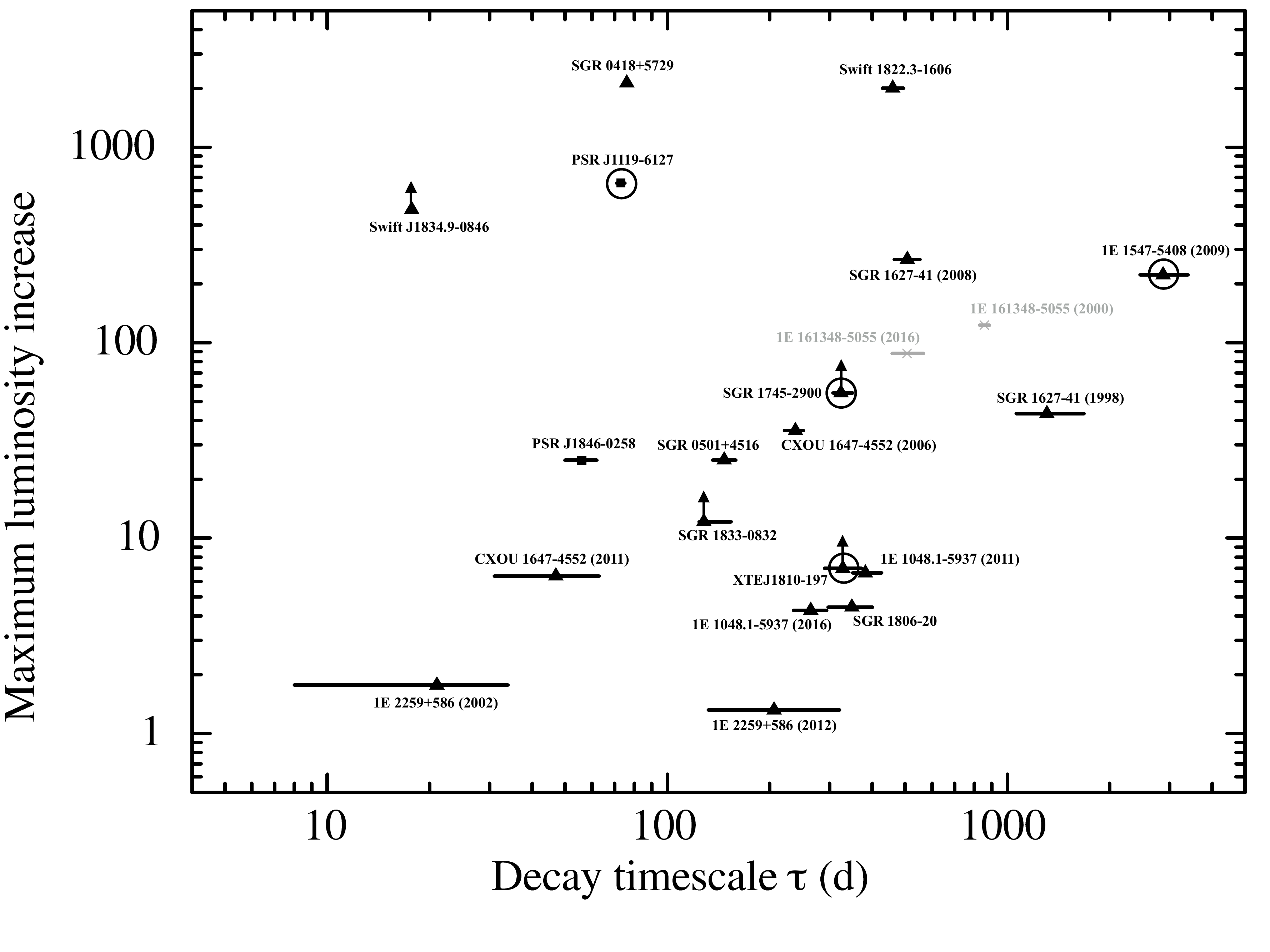}
\caption{\emph{Left-hand panel}: maximum luminosity increase as a function of the total energy released during the outburst. \emph{Right-hand panel}: maximum luminosity increase as a function 
of the decay time-scale.}
\label{fig:energy_deltaflux}
\end{center}
\end{figure*}

The top panel of Fig.~\ref{fig:b_lqui} reports the quiescent thermal bolometric luminosity ($L_{bol,q}$) of magnetars and of the other classes of 
isolated X-ray pulsars as a function of their spin-down luminosity ($\dot{E}_{\rm {rot}}$). The dashed line represents  $L_{bol,q}=\dot{E}_{\rm {rot}}$. 
The emission of all sources lying above the dashed line must be ultimately powered by magnetic energy. On the other hand, the emission of all sources 
located below the dashed line might be entirely rotation-powered, or switch between magnetar-like and rotation-powered emission. An interesting case is 
represented by the magnetar \xte, whose steady, quiescent, luminosity, $L_{bol,q} \sim 4 \times 10^{34}$ \lum\ attained in the past $\sim5$ yr (see 
Fig.~\ref{fig:xte1810_parameters_outbursts}) is a factor of $\sim60$ larger than its spin-down luminosity $\dot{E}_{\rm {rot}} \sim 6.7 \times 10^{32}$ \lum, 
accurately estimated from timing analysis of X-ray data taken over the last 12 yr (see Table~\ref{tab:timing}). This result contradicts the prediction put 
forward by Rea et al. (2012b), according to which a magnetar with $L_{X,q} \gtrsim \dot{E}_{\rm {rot}}$ is expected to be radio quiet, regardless of its 
possible X-ray outburst activity. 

The bottom panel of Fig.~\ref{fig:b_lqui} shows the quiescent thermal bolometric luminosity as a function of the surface dipolar magnetic field. 
We observe a significant correlation (3.2$\sigma$ according to the Spearman test) when including all sources belonging to the different classes 
considered in this study (the correlation is 3.9$\sigma$ after excluding the central compact objects). This correlation is naturally explained in terms 
of magnetic field decay and Joule heating  (Pons, Miralles \& Geppert 2009; Vigan\`o et al. 2013). The central compact objects clearly depart from 
the general trend. The peculiar behaviour of these objects might be explained in the framework of the `hidden magnetic field' scenario: hypercritical 
accretion on to the neutron star surface during the initial stages of the star life can bury a magnetic field of a few 10$^{13}$~G into the inner crust, 
yielding a strength for the external magnetic field that is significantly lower than the internal `hidden' magnetic field. The large luminosity observed 
for these objects is most probably due to the toroidal and higher order mulipolar components of the magnetic field trapped inside the crust (Geppert, 
Page \& Zannias 1999; Ho 2011; Shabaltas \& Lai 2012; Vigan\`o \& Pons 2012; Torres-Forn\'e et al. 2016). The magnetic field will eventually re-emerge, after 
a few thousands of years, settling on a value comparable to that at birth. If this picture is correct, we would expect a `shift' of the central compact objects 
towards the right in the quiescent luminosity versus dipolar magnetic field diagram, as the CCOs get older. Some of the rotation-powered pulsars also 
depart slightly from the observed trend (e.g., PSR\,J0538$+$2817, PSR\,B1055$-$52 and PSR\,J0633$+$1746). This might be possibly due to an 
additional contribution to the surface heating from slamming particles on to the stellar surface, as typically observed for pulsars with a high rotational 
energy loss rate. 

We investigated the shape of the correlation via a power law regression test, and found
\begin{equation}
L_{bol,q} \propto B_{p,dip}^{2}
\end{equation} 
(see Table~\ref{tab:correlations}). This is in agreement with the dependence reported by Pons et al. (2007) using a reduced sample of sources.

\subsection{On the dipolar magnetic field versus the outburst properties}

We also investigated possible correlations between the strength of the surface dipolar magnetic field and all the outburst parameters derived in this 
work. There is no significant correlation between the magnetic field and either the maximum luminosity or the decay time-scale. Furthermore, in a few 
cases the same source was observed to undergo two different outbursts with distinct properties (see Fig.~\ref{fig:b_lpeak}).

The correlation between the magnetic field and the outburst energetics is more evident (3.4$\sigma$ according to the Spearman test; see 
Fig.~\ref{fig:b_energy}), and supports the idea that the energy reservoir of the outbursts is mainly provided by the dissipation of the magnetic field. 
The two variables are linearly related (i.e., $E \propto B_{p,dip}$). We observe a sort of limiting energy as a function of age. Young magnetars tend 
to experience more energetic outbursts than older magnetars, a characteristic that can be explained simply in terms of field decay. The expected 
energetics distribution was estimated by Perna \& Pons (2011), who did not find a significant dependence of the energy of the events with age, but 
the fact that magnetic field decay limits the energy budget available for old magnetars, compared to young sources. They also estimated the 
recurrence times between consecutive outbursts, and found as a general trend that the older the object, the longer the average recurrence time.

\subsection{On the outburst energy versus other properties}

The outburst energy correlates with the peak luminosity reached during the outburst (at a significance of 4.0$\sigma$ according to the Spearman test), 
but not with the quiescent X-ray luminosity (<2$\sigma$; see Fig.~\ref{fig:lpeak_energy}). These results suggest that a larger luminosity at the peak 
of the outburst results in a larger energy released during the entire outburst event, regardless of the quiescent level of the source, and reflect similar 
decay patterns for magnetar outbursts. This is  expected in both internal crustal cooling and untwisting bundle scenarios, since it only reflects the 
normalization of the decay curve.

The energetics correlates significantly with the decay time-scale (at a significance of 3.9$\sigma$ according to the Spearman test): the longer the 
outburst, the more energetic (Fig.~\ref{fig:lpeak_decay}). This suggests again that the decay pattern is similar from outburst to outburst. For 
example, we never observe a magnetar undergoing a rather weak outburst and then returning to quiescence over an extremely long time interval, 
or a magnetar showing an extremely powerful outburst and then rapidly decaying back to quiescence.

\section{The Magnetar Outburst Online Catalogue}
\label{mooc}

All the key parameters derived for the magnetar outbursts presented in this study, as well as the reduced spectral files, are available at the MOOC 
(\url{http://magnetars.ice.csic.es}). We have also included all important parameters for the other thermally emitting isolated X-ray pulsars (see 
Table\,\ref{tab:timing}; see also Vigan\`o et al. 2013), to allow a direct comparison between the different classes of isolated neutron stars.

The webpage consists of three distinct sections: {\tt Sources}, {\tt Analysis} and {\tt Download}. In the {\tt Sources} section, the user can plot any 
combination of the parameters for all thermally emitting isolated X-ray pulsars. In the {\tt Analysis} section, the user can plot the light curves for all magnetar 
outbursts, as well as any combination of the parameters characterizing these events.

In both the {\tt Sources} and {\tt Analysis} sections, a detailed description of all parameters is provided, and the user can download all values of the
plotted parameters in the form of a \emph{csv} table. Furthermore, restricted ranges of values can be selected and plotted using the `Filter' task. 
The user can also create mathematical functions linking different parameters via the `Create Function Field' tool, and download the resulting plot as 
an image or an ascii file.

Finally, in the {\tt Download} section, the user can download the fits files relative to all the observations of magnetar outbursts analysed in this study, 
i.e. the source and background average spectra, the redistribution matrix files and the auxiliary response files. Each file is named according to the 
following general scheme: `source name$_-$name of the satellite$_-$type of file$_-$obsID.fits', where `type of file' is either src$_-$spectrum, 
bg$_-$spectrum, rmf or arf. The user can perform a spectral analysis of the data by uploading these files in the \textsc{xspec} spectral fitting package.

The webpage will be updated periodically and expanded as new outbursts are observed.

\section*{Acknowledgements}
We are indebted to Martin Folger and Santiago Serrano Elorduy from the Institute of Space Sciences 
(IEEC--CSIC) for designing the Magnetar Outburst Online Catalogue and implementing several plotting 
and analysis online tools. FCZ acknowledges Alice Borghese, Niccol\`{o} Bucciantini and Jason Hessels 
for useful suggestions that helped improving a preliminary version of the manuscript, Alexander Kaminker 
and Stefano Carignano for useful discussions, and Chichuan Jin for providing the \cxo\ ACIS-S version of 
the dust scattering model for Galactic Centre X-ray sources. We thank the referee for comments.
The scientific results reported in this study are based on observations obtained with the 
{\em Chandra X-ray Observatory}, \xmm, \swift\ and \rxte. \xmm\ is an ESA science mission 
with instruments and contributions directly funded by ESA Member States and the National 
Aeronautics and Space Administration (NASA). \swift\ is a NASA/UK/ASI mission. \rxte\ is 
a NASA mission. This research has made extensive use of software provided by the 
{\em Chandra X-ray Center} [operated for and on behalf of NASA by the Smithsonian Astrophysical 
Observatory (SAO) under contract NAS8--03060] in the application package \textsc{ciao}. The 
\xmm\ \textsc{sas} is developed and maintained by the Science Operations Centre at the European 
Space Astronomy Centre. We made use of data supplied by the UK \swift\ Science Data Centre 
at the University of Leicester and of the XRT Data Analysis Software (XRTDAS) developed under 
the responsibility of the ASI Science Data Center (ASDC), Italy. We also used softwares and tools 
provided by the High Energy Astrophysics Science Archive Research Center (HEASARC) Online 
Service, which is a service of the Astrophysics Science Division at NASA/GSFC and the High 
Energy Astrophysics Division of SAO. We made use of the McGill Online Magnetar Catalog 
(www.physics.mcgill.ca/$\sim$pulsar/magnetar/main.html) and NASA's Astrophysics Data System Bibliographic 
Services. FCZ, NR and PE acknowledge funding in the framework of the Netherlands Organization for 
Scientific Research (NWO) Vidi award number 639.042.321 (PI: N.~Rea) and the European COST Action 
MP1304 (NewCOMPSTAR). FCZ and NR are also supported by grants AYA2015-71042-P and SGR2014-1073. 
JAP acknowledges support by grants AYA2015-66899-C2-2-P and PROMETEOII-2014-069.

\begin{table*}
\caption{Spectral fitting results of magnetar outbursts. The year of the outburst onset is indicated in parentheses. 
\textsc{ntz} denotes the resonant cyclotron scattering code by Nobili et al. (2008a,b), and was applied only in the cases 
where a power law component was needed when fitting `empirical' models to the data. To account for interstellar 
absorption, we adopted the \textsc{tbabs} model, cross-sections from Verner et al. (1996) and abundances from 
Wilms et al. (2000). The hydrogen column density was tied up among all the observations targeting a specific 
source and the associated uncertainty is quoted at the 1$\sigma$ c.l.}
\resizebox{1.72\columnwidth}{!}{
\begin{tabular}{cccccc}
\hline
Source    					& Observatory (\# obs)		& Best-fitting model		& \nh	\ (emp)					& \nh	\ (\textsc{ntz})				& Reference\\
						& 						&					&\multicolumn{2}{c}{($10^{22}$ cm$^{-2}$)}  					& \\
\hline
\multirow{3}{*}{\sgrd\ (1998)}	& \bsax\ (4)				& BB					& $6\pm1$			 		& --						& \multirow{3}{*}{Table~\ref{tab:sgr1627_1998}} 	\\
						& \xmm\ (2) 				& BB					& 6$_{-1}^{+2}$			 		& --						& 	\\
						& \cxo\ (4)					& BB					& 10$_{-1}^{+2}$				& --						&	\\
\hline
\antgli\ (2002)				& \xmm\ (8) 				& 3BB				& $0.816 \pm 0.007$  			& --						& Table~\ref{tab:1e2259_2002}	\\
\hline
\multirow{2}{*}{\xte\ (2003)}	& \multirow{2}{*}{\xmm\ (22)}	& 3BB (1--11)  			 & \multirow{2}{*}{$1.22\pm0.02$}	 & --						&  \multirow{2}{*}{Table~\ref{tab:xte}} 	\\
						& 						& 2BB (11--22) 			 &  		 					& --						&  	\\
\hline
SGR\,1806$-$20 (2004)		& \xmm\ (10) 				& 2BB				 & $8.5\pm0.1$   				& --						& Table~\ref{tab:sgr1806}	\\
\hline
\multirow{3}{*}{\wes\ (2006)}	& \swift\ (18)				& BB+PL				& $3.06\pm0.08$ 				 & 2.43$_{-0.03}^{+0.04}$		& \multirow{3}{*}{Table~\ref{tab:cxou1647}} 	 \\
						& \xmm\ (5) $^{\rm a}$ 		& BB+PL				& 3.01 (fixed) 					 & 2.39 (fixed)				&	\\
						& \cxo\ (5)					& BB+PL				& $3.01\pm0.04$ 				 & 2.39$_{-0.01}^{+0.02}$		&	\\
\hline
\multirow{3}{*}{\sgrd\ (2008)}	& \swift\ (21)				& BB					& $9\pm2$ 					& --						& \multirow{3}{*}{Table~\ref{tab:sgr1627_2008}} 		\\
						& \xmm\ (2)				& 2BB				& 10$_{-2}^{+3}$				& --						&	\\
						& \cxo\ (4)					& BB					& $10\pm2$ 					& --						&	\\
\hline
\multirow{5}{*}{\sgre\ (2008)}	& \multirow{3}{*}{\swift\ (62) $^{\rm b}$}	& BB+PL (1--20)  & 1.319 (fixed) 				& 0.708$_{-0.006}^{+0.007}$	& \multirow{5}{*}{Table~\ref{tab:sgr0501}}	\\ 
		 				& 						& BB+PL (21--40)		& 1.319 (fixed)					& $0.71\pm0.03$ 			& \\ 
	 	      				&						& BB+PL (41--62)		& 1.319 (fixed)					& 0.708 (fixed)				& \\	
		     				& \xmm\ (6)				& BB+PL				& $1.319\pm0.009$ 				& $0.705\pm0.004$ 			& \\
						& \cxo\ (1)					& BB+PL				& $1.33\pm0.03$				& $0.85\pm0.01$			& \\
\hline
\multirow{2}{*}{\aa\ (2008)}	& \swift\ (15)				& BB+PL				& $4.9\pm0.1$					& 4.65$_{-0.07}^{+0.05}$		& \multirow{2}{*}{Table~\ref{tab:1e1547_08}} \\ 
						& \cxo\ (5)					& BB+PL				& $5.1\pm0.2$					& $4.83\pm0.06$			&	\\
\hline
\multirow{7}{*}{\aa\ (2009)} 	& \multirow{5}{*}{\swift\ (97)}	& BB+PL (1--20)		& 4.91$_{-0.13}^{+0.03}$  		& $4.59\pm0.09$			& \multirow{7}{*}{Table~\ref{tab:1e1547_09}}	\\	
		 				&						& BB+PL (21--40)		& 4.91 (fixed)					& 4.59 (fixed)				&	\\	
						& 						& BB+PL (41--60) 		& 4.91 (fixed)					& 4.59 (fixed)				& \\ 
						&						& BB+PL (61--80)		& 4.91 (fixed)					& 4.59 (fixed)				&	\\	
		 				&						& BB+PL (81--97)		& 4.91 (fixed)					& 4.59 (fixed)				&	\\	
		  				& \xmm\ (2) $^{\rm a}$		& BB+PL				& 4.9 (fixed)					& 4.65 (fixed) 				&	\\
						& \cxo\ (3)					& BB+PL				& $5.0\pm0.1$ 					& 4.71$_{-0.07}^{+0.10}$ 		&	\\
\hline
\multirow{4}{*}{\lowba\ (2009)}	& \swift\ (24)				& BB					& 0.57 (fixed)			 		& --						& \multirow{4}{*}{Table~\ref{tab:sgr0418}}	\\
						& \multirow{2}{*}{\xmm\ (11)}	& 2BB (1)				& 0.57$_{-0.03}^{+0.04}$	 		& --						&	\\
						& 						& BB (2--11)			& 0.57 (fixed)					& --						&	\\
						& \cxo\ (4)					& BB					& 0.57 (fixed) 	 				& --						&	\\
\hline
\multirow{3}{*}{SGR\,1833$-$0832 (2010)}	& \swift\ (27)	& BB					& $13.1\pm0.9$ 		 		& --						& \multirow{3}{*}{Table~\ref{tab:sgr1833}}	\\	
						& \xmm\ (3)				& BB					& $15.5\pm0.4$		 		& --						&	\\
						& \cxo\ (1)					& BB					& $13.7\pm0.9$		 		& --						&	\\
\hline
\multirow{5}{*}{\lowbb\ (2011)}	& \multirow{3}{*}{\swift\ (60)}	& BB+PL (1--20)		& 0.68 (fixed)	 				& 0.289 (fixed)				& \multirow{5}{*}{Table~\ref{tab:swift1822}}			\\	
 				 		&						& BB+PL (21--40)		& 0.68 (fixed)			 		& 0.289 (fixed)				& 	\\
						&						& BB+PL (40--60)		& 0.68 (fixed)		 			& 0.289 (fixed)				&	\\	
						& \xmm\ (5)				& BB+PL				& $0.68\pm0.01$		 		& $0.289\pm0.004$			&	\\
						& \cxo\ (5)					& BB+PL				& $0.62\pm0.02$		 		& $0.283\pm0.005$			&	\\
\hline	
\multirow{3}{*}{Swift\,J1834.9$-$0846 (2011)}	& \swift\ (19)			& BB					& 19 (fixed) 		 	 & --						& \multirow{3}{*}{Table~\ref{tab:swift1834}}	\\
						& \xmm\ (3)						& BB					& $15\pm1$			 & --						&	\\	
						& \cxo\ (4)							& BB					& $19\pm1$			 & --						&	\\
\hline						
\multirow{3}{*}{\wes\ (2011)}	& \swift\ (7)						& BB					& $2.6\pm0.3$ 			 & --						& \multirow{3}{*}{Table~\ref{tab:cxou1647_2011}}\\
						& \xmm\ (1)						& 2BB				& $2.5\pm0.1$			 & --						&	\\	
						& \cxo\ (1)							& BB					& $2.8\pm0.1$ 			 & --						&	\\
\hline
\multirow{3}{*}{1E\,1048.1$-$5937 (2011)}  & \multirow{3}{*}{\swift\ (55)}	& BB (1--20)			& $0.61 \pm 0.02$ 		 & --						& \multirow{3}{*}{Table~\ref{tab:1e1048_2011}}	\\
								  & 						& BB (21--40) 			& 0.61 (fixed)			 & --						& 	\\
								  & 						& BB (40--55) 			& 0.56 (fixed)			 & --						& 	\\ 	     
\hline
\multirow{2}{*}{\antgli\ (2012)}	& \multirow{2}{*}{\swift\ (44)}			& 2BB (1--20)			&$0.38\pm0.01$ 	 	 & --						& \multirow{2}{*}{Table~\ref{tab:1e2259_2012}}	\\
						& 								& 2BB (21--44) 			& 0.38 (fixed)			 & --						& 	\\ 
\hline
\galcen\ (2013)				& \cxo\ (35)						& BB					& $18.7\pm0.1$		 & --						& Table~\ref{tab:sgr1745}		\\	
\hline
\multirow{4}{*}{\sgrm\ (2014)}	& \multirow{2}{*}{\swift\ (45)}			& BB	 (1--20)			& $2.3\pm0.2$ 			 & --						& \multirow{4}{*}{Table~\ref{tab:sgr1935}}	\\
		 				&								& BB (21--45)			& 2.3 (fixed)			 & --						&	\\
						& \xmm\ (9)						& 2BB				& $2.37 \pm 0.07$		 & --						&	\\
						& \cxo\ (3)							& 2BB				& $2.8 \pm 0.1$ 		 &						&	\\
\hline	
\multirow{3}{*}{1E\,1048.1$-$5937 (2016)}  & \multirow{3}{*}{\swift\ (60)}	& BB (1--20)			& $0.56 \pm 0.04$ 		 & --						& \multirow{3}{*}{Table~\ref{tab:1e1048_2016}}	\\
								  & 						& BB (21--40) 			& $0.59 \pm 0.04$		 & --						& 	\\
								  & 						& BB (40--60) 			& 0.56 (fixed)			 & --						& 	\\ 
\hline
\hline
\multirow{2}{*}{\psr\ (2016)}	& \multirow{2}{*}{\swift\ (35)}			& BB (1--20)			& $0.69\pm0.05$		 & --						& \multirow{2}{*}{Table~\ref{tab:psr1119}}		\\	
					        & 								& BB (21--36) 			& 0.69 (fixed)			 & --						& 	\\					
\hline								  
\cco\ (1999, 2016)			& \cxo\ (25), \xmm\ (2), \swift\ (129)		& 2BB				& $2.05\pm0.05$	 	 & --						& Rea et al. (2016)		\\	 	 
\hline
\end{tabular}
}
 \begin{tablenotes}
\item \emph{Notes}. $^a$The absorption column density was fixed to a value compatible with that inferred from the fits of the data sets from the other X-ray instruments, because a significant excess in the fit 
residuals was detected below about 1~keV independently on the choice of the background region and of the adopted spectral model (see e.g. Bernardini et al. 2009 for this issue). We obtained acceptable 
fits in all cases.
\item $^b$The absorption column density was fixed to the value obtained from the fit to the \xmm\ spectra, because the XRT was operating in WT in all cases and bumps of instrumental origin were 
present at $\sim 0.8-1$~keV (see the text). We obtained acceptable fits in all cases.
 \end{tablenotes}
\label{tab:results}
\end{table*}

\clearpage

\begin{table*}
\caption{Results of the empirical modelling of the outburst decays for the 0.3--10~keV luminosities of the single spectral components (BB, PL, BB1, BB2, BB3) and for the total bolometric luminosities. 
The cooling curves were fitted with one or multiple exponential functions plus a constant (see the text for details). Uncertainties on the best-fitting parameters are quoted at the 1$\sigma$ c.l. for a single 
parameter of interest. The total outburst energy is also reported. The values for \psrr\ and the first outburst of \cco\ are taken from Gavriil et al. (2008), Rea et al. (2016), respectively. The values for the 
second outburst of \cco\ are taken from the ongoing \swift\ XRT monitoring campaign.}
\begin{tabular}{lccccc}
\hline
Source 			  				& Component 			& Best-fitting decay model	& $\tau$					& $\tau_1$ / $\tau_2$ / $\tau_3$  	& $E$	\\
								& 					& 					& (d)						& (d)		  					& (erg)				 \\
\hline
\sgrd\ (1998)						& BB/bol				& \textsc{2exp}			& --						& 234$_{-38}^{+37}$ / 1307$_{-245}^{+373}$	& $2 \times 10^{42}$		 \\ 
\hline			
\multirow{3}{*}{\antgli\ (2002)}			& BB1				& \textsc{exp}			& $1.41 \pm 0.05$			& --							& --		 \\ 	
								& BB2				& \textsc{exp}			& 47$_{-16}^{+40}$			& --							& --		 \\ 		
								& bol					& \textsc{exp}			& $21 \pm 13$				& --							& $10^{41}$	 \\ 
\hline
\multirow{3}{*}{\xte}					& BB1				& \textsc{exp}			& 376$_{-58}^{+72}$			& --							& --		 \\ 	
								& BB2				& \textsc{exp}			& 372$_{-29}^{+33}$			& --							& --		 \\ 
								& bol					& \textsc{exp}			& 328$_{-38}^{+44}$			& --							& $4 \times 10^{42}$	 \\ 
\hline								
SGR\,1806$-$20					& bol					& \textsc{exp}			& $349 \pm 52$			& --							& $2 \times 10^{43}$	 \\ 				
\hline
\multirow{3}{*}{CXOU\,J1647$-$4552 (2006)}  & BB				& \textsc{3exp}			& --						& $2.9 \pm 0.7$ / 91$_{-27}^{+54}$ / 225$_{-57}^{+32}$							& --		 \\ 
								& PL					& \textsc{2exp}			& --						& $3 \pm 1$ / 458$_{-60}^{+64}$							& --		 \\
								& bol					& \textsc{3exp}			& --						& 2.4$_{-0.6}^{+0.8}$ / $53 \pm 3$ / 238$_{-17}^{+13}$  	& $10^{42}$		 \\
\hline
\sgrd\ (2008)						& BB/bol				& \textsc{3exp}			& --						& 0.56$_{-0.06}^{+0.07}$ / 31$_{-4}^{+5}$  / 508$_{-43}^{+45}$  	& $10^{42}$		 \\
\hline
\multirow{3}{*}{\sgre}					& BB					& \textsc{exp}			& $33 \pm 2$				& --							& --		 \\ 
								& PL					& \textsc{2exp}			& --						& 9$_{-2}^{+3}$ / 345$_{-51}^{+68}$	& --		 \\ 
								& bol					& \textsc{2exp}			& --						& $13 \pm 2$ / 147$_{-11}^{+12}$	 & $9 \times 10^{40}$		 \\ 
\hline
\multirow{3}{*}{\aa\ (2009)}			& BB					& \textsc{2exp}			& --						& $4.8^{+0.7}_{-0.6}$ / $1131^{+156}_{-120}$							& --		 \\  
								& PL					& \textsc{exp}			& $364 \pm 15$			& --							& --		 \\ 
								& bol					& \textsc{3exp}			& --						& $3\pm1$ / $109\pm8$  / $2870^{+528}_{-416}$	& $2.4 \times 10^{43}$		 \\ 

\hline
\lowba							& BB/bol				& \textsc{exp}			& $76 \pm 1$				& --							& $8 \times 10^{40}$		 \\ 		
\hline
SGR\,1833$-$0832					& BB/bol				& \textsc{exp}			& 128$_{-4}^{+26}$			& --							& $10^{42}$		 \\ 	
\hline
\multirow{3}{*}{\lowbb}				& BB					& \textsc{3exp}			& --						& 0.78$_{-0.3}^{+0.4}$ / 16.7$_{-0.9}^{+1.0}$ / 207$_{-11}^{+12}$			& --		 \\  
								& PL					& \textsc{2exp}			& --						& $14.6 \pm 0.8$ / 817$_{-47}^{+54}$ 	& --		 \\ 
								& bol					& \textsc{3exp}			& --						& $7 \pm 2$ / 28$_{-3}^{+4}$ / 460$_{-31}^{+35}$ 	 & $3 \times 10^{41}$  \\ 
\hline
Swift\,J1834.9$-$0846			 	& BB/bol				& \textsc{2exp}			& --						& $0.08 \pm 0.01$  / $17.7 \pm 0.4$   & $2 \times 10^{41}$		 \\ 	
\hline
CXOU\,J1647$-$4552 (2011)		 	& BB/bol				& \textsc{exp}			& $47 \pm 16$				& --							& $6 \times 10^{40}$		 \\ 	
\hline
1E\,1048.1$-$5937 (2011)				& BB/bol				& \textsc{2exp}			& --						& 39$_{-16}^{+26}$  / 382$_{-31}^{+45}$ 		& $8 \times 10^{42}$		 \\ 	
\hline
\multirow{3}{*}{\antgli\ (2012)}			& BB1				& \textsc{exp}			& 79$_{-35}^{+59}$			& --							& --		 \\ 
								& BB2				& \textsc{exp}			& 33.7$_{-8}^{+9}$			& --							& --		 \\ 
								& bol					& \textsc{exp}			& 206$_{-74}^{+115}$		& --							& $3 \times 10^{41}$	 \\ 	
\hline
\galcen							& BB/bol				& \textsc{2exp}			& --						& 81$_{-20}^{+6}$ / 324$_{-17}^{+27}$ 		& $10^{43}$		 \\ 	
\hline			
1E\,1048.1$-$5937 (2016)				 & BB/bol				& \textsc{2exp}			& --						& 42$_{-6}^{+8}$  / 264$_{-29}^{+30}$			& $4 \times 10^{42}$		 \\ 
\hline
\hline
\psr								 & bol				& \textsc{3exp}			& -- 						& $0.25 \pm 0.06$ / $18 \pm 2$ / $73 \pm 2$	  & $8.5\times10^{41}$ \\						
\hline
\psrr								 & bol				& \textsc{exp}			& $56 \pm 6$ 				& --			    				& $4.5\times10^{41}$ \\						
\hline
\cco\ (2000)						 & bol				& \textsc{2exp}			&  --						& 110$_{-15}^{+13}$  / 856$_{-27}^{+29}$	  	& $10^{43}$ \\						
\hline	
\cco\ (2016)						 & bol				& \textsc{2exp}			& -- 						& 0.5$_{-0.1}^{+0.2}$	/ 507$_{-49}^{+59}$	    & $2.6\times10^{42}$ \\					
\hline					
\end{tabular}
\label{tab:decays}
\end{table*}

\clearpage

\begin{table*}
\caption{Maximum fluxes and luminosities (0.3--10~keV) for magnetars showing major outbursts or variations in their persistent emission. 
The table is ordered according to the chronological order of the outburst episodes, and the cases of the peculiar high $B$-field pulsars 
and the CCO \cco\ are reported below the double horizontal line. Uncertainties are reported at the 1$\sigma$ c.l.}
\resizebox{1.9\columnwidth}{!}{
\begin{tabular}{ccccccc}
\hline
Source    			& Date 				& Observatory				& Obs ID			& Exposure		& Abs/Unabs flux					& $L_{X,p}$  			\\
				& 					&						&				& (ks)			& (erg~cm$^{-2}$~s$^{-1}$)			& (erg~s$^{-1}$)	 \\
\hline
\sgrd\			& 1998 Aug 07 			& \bsax\ 					& 70566001		& 44.9 			& $(2.4\pm0.1) \times 10^{-12}$ 		& $(5.2\pm0.3) \times 10^{34}$ \\ 
				&					&						&				&				& $(3.6\pm0.2) \times 10^{-12}$		& \\
SGR\,1900$+$14	& 2001 Apr 22			& \cxo\					& 2458			& 20.1			& $(1.02\pm0.02) \times 10^{-11}$ 		& $(3.48\pm0.09) \times 10^{35}$ \\
				&					&						&				&				& $(1.86\pm0.05) \times 10^{-11}$ 		& \\						
\antgli\			& 2002 Jun 21   		& \xmm\ 					& 0155350301 		& 18.4			& $(5.87\pm0.01) \times 10^{-11}$ 		& $(1.23\pm0.01) \times 10^{35}$		\\ 
				&					&						&				&				& $(1.006\pm0.009) \times 10^{-10}$		& \\	
\xte				& 2003 Sep 08  		& \xmm\ 					& 0161360301 		& 6.6 			& $(3.84\pm0.02) \times 10^{-11}$ 		& $(1.74\pm0.01) \times 10^{35}$		\\ 
				&					&						&				&				& $(1.19\pm0.01) \times 10^{-10}$		& \\						 
SGR\,1806$-$20	& 2004 Oct 06			& \xmm\					& 0164561101		& 12.9 			& $(2.68\pm0.02) \times 10^{-11}$		& $(3.62\pm0.09) \times 10^{35}$ \\	
				&					&						&				&				& $(4.0\pm0.1) \times 10^{-11}$			& \\	
SGR\,1900$+$14	& 2006 Mar 29			& \cxo\					& 6709			& 40.0			& $(6.10\pm0.09) \times 10^{-12}$ 		& $(2.24\pm0.06) \times 10^{35}$ \\
				&					&						&				&				& $(1.20\pm0.03) \times 10^{-11}$ 		& \\		
CXOU\,J1647$-$4552  & 2006 Sep 21 		& \swift\ 					& 00030806001 	& 7.7  			& $(3.36\pm0.08) \times 10^{-11}$  		& $(1.2\pm0.1) \times 10^{35}$		\\ 
				&					&						&				&				& $(6.1\pm0.5) \times 10^{-11}$			& \\	
\sgrd\			& 2008 May 28 			& \swift\					& 00312579001	& 2.0 			& $(1.2\pm0.2) \times 10^{-11}$			& $(3.2\pm0.6) \times 10^{35}$ \\
				&					& 						& 				& 				& $(2.2\pm0.4) \times 10^{-11}$			& \\	
\sgre\			& 2008 Aug 23 			& \xmm\ 					& 0560191501		& 33.8 			& $(4.03\pm0.01) \times 10^{-11}$  		& $(3.4\pm0.2) \times 10^{34}$		\\ 
				&					&						&				&				& $(1.28\pm0.08) \times 10^{-10}$		& \\  
\aa\				& 2008 Oct 03  			& \swift\  					& 00330353000 	& 4.1  			& $(6.2\pm0.2) \times 10^{-11}$ 		& $(2.3\pm0.2) \times 10^{35}$		\\ 
				&					&						&				&				& $(9.4\pm0.7) \times 10^{-11}$			& \\
\aa\				& 2009 Jan 23			& \swift\  					& 00340923000 	& 1.7 			& $(8.2\pm0.5) \times 10^{-11}$ 		& $5_{-1}^{+3}\times 10^{35}$		\\ 
				&					&						&				&				& $2.0_{-0.4}^{+1.4} \times 10^{-10}$	& \\
\lowba\			& 2009 Jun 11  			& \rxte\ 					& 94048--03--01--00 	& 5.2				& $(3.31\pm0.06) \times 10^{-11}$ 		& $(1.63\pm0.03) \times 10^{34}$		\\ 
				&					&						&				&				& $(3.41\pm0.06) \times 10^{-11}$		& \\			
SGR\,1833$-$0832$^{\rm a}$  & 2010 Mar 20  	& \swift\ 					& 00416485000	& 29.0  			& $(4.0\pm0.2) \times 10^{-12}$  		& $(1.02\pm0.08) \times 10^{35}$		\\ 
				&					&						&				&				& $(8.5\pm0.7) \times 10^{-12}$		& \\	
1E\,1841$-$045	& 2011 Jul 02			& \swift\					& 00456505000	& 1.4 			& $(2.0\pm0.2) \times 10^{-11}$			& $(1.7\pm0.9) \times 10^{36}$	\\ 
				&					&						&				&				& $(2\pm1) \times 10^{-10}$			& 					 \\	
\lowbb\			& 2011 Jul 16 			& \swift\  					& 00032033001	& 1.6				& $(2.35\pm0.04) \times 10^{-10}$ 		& $(8.0\pm0.2) \times 10^{34}$		\\ 
				&					&						&				&				& $(2.61\pm0.05) \times 10^{-10}$		& \\
4U\,0142$+$61		& 2011 Jul 29			& \swift\					& 00458345000 	& 3.9 			& $(6.7\pm0.3) \times 10^{-10}$ 		& $(1.23 \pm 0.03) \times 10^{36}$ 	\\
				&					&						&				&				& $(7.9\pm0.2) \times 10^{-10}$ 		&	\\
Swift\,J1834.9$-$0846 & 2011 Aug 07		& \swift\ 					& 00458907000 	& 1.5 			& $(3.2\pm0.6) \times 10^{-11}$  		& $(1.0\pm0.2) \times 10^{35}$		\\ 
				&					&						&				&				& $(4.8\pm0.8) \times 10^{-11}$			& \\		
CXOU\,J1647$-$4552  & 2011 Sep 25		& \swift\ 					& 00030806020	& 3.1 			& $(6.5\pm0.5) \times 10^{-12}$  		& $(2.1\pm0.4) \times 10^{34}$		\\ 
				&					&						&				&				& $(1.1\pm0.2) \times 10^{-11}$			& \\	
1E\,1048.1$-$5937	& 2011 Dec 31 			& \swift\ 					& 00031220066 	& 2.0				& $(4.6\pm0.3) \times 10^{-11}$  		& $(5.7\pm0.4) \times 10^{35}$		\\ 
				&					&						&				&				& $(5.9\pm0.4) \times 10^{-11}$			& \\
\antgli\			& 2012 Apr 28   		& \swift\ 					& 00032035021 	& 3.9				& $(5.7\pm0.1) \times 10^{-11}$  		& $(9.2\pm0.2) \times 10^{34}$		\\ 
				&					&						&				&				& $(7.5\pm0.2) \times 10^{-11}$			& \\
\galcen\			& 2013 Apr 29 			& \cxo$^{\rm b}$ 			& 14701 			& 9.7  			& $\sim 1.8 \times 10^{-11}$	 		& $\sim 6.8 \times 10^{35}$		\\ 
				&					&						&				&				& $\sim 8.3 \times 10^{-11}$			& \\			
\sgrm$^{\rm c}$		& 2014 Jul 05 			& \swift\ 					& 00603488000 	& 3.4 			& $(1.7\pm0.2) \times 10^{-12}$  		& $(2.5\pm0.4) \times 10^{34}$	\\ 
				&					&						&				&				& $(2.6\pm0.4) \times 10^{-12}$		& \\
4U\,0142$+$61		& 2015 Feb 28			& \swift\					& 00632888000 	& 0.5 			& $(6.5\pm0.3) \times 10^{-10}$	 	& $(1.26 \pm 0.02) \times 10^{36}$ 	\\
				&					&						&				&				& $(8.1\pm0.1) \times 10^{-10}$ 		&	\\
1E\,1048.1$-$5937	& 2016 Jul 29  			& \swift\ 					& 00032923249  	& 1.4 			& $(3.16\pm0.2) \times 10^{-11}$  		& $(3.7\pm0.1) \times 10^{35}$		\\ 
				&					&						&				&				& $(3.8\pm0.1) \times 10^{-11}$			& \\
\hline
\hline
\psr\				& 2016 Jul 28			& \swift\					& 00706396000	& 2.2				& $(4.1\pm0.2)\times10^{-11}$			& $(3.72\pm0.08)\times10^{35}$ \\	
				&					&						&				&				& $(4.4\pm0.1)\times10^{-11}$			& \\	
\hline
\psrr$^{\rm d}$		& 2006 Jun 08			& \rxte\					& 92012--01--14-00	& 20.1			& $\sim1.2\times10^{-11}$				& $\sim3.9\times10^{35}$ \\	
				&					&						&				&				& $\sim9\times10^{-11}$				& \\						
\hline
\cco\				& 2000 Feb 08			& \cxo\					& 970			& 18.9			& $(6.5\pm1.6)\times10^{-11}$			& $(3.47\pm0.08)\times10^{35}$ \\	
				&					&						&				&				& $(2.66\pm0.06)\times10^{-10}$							& \\	
\hline
\cco\				& 2016 Jun 22			& \swift\					& 00030389032	& 0.6				& $(1.4\pm0.1)\times10^{-10}$			& $(2.5\pm0.1)\times10^{35}$ \\	
				&					& 						&				&				& $(1.9\pm0.1)\times10^{-10}$			& \\	
\hline				
\end{tabular}
}
\begin{tablenotes}
\item \emph{Notes}. $^a$A distance of 10~kpc was assumed.
\item $^b$The field around the source has been previously observed by \swift. In order to avoid contamination by nearby active X-ray sources, we consider here the first \cxo\ observation, which was carried out with the HRC. The flux was then estimated by assuming a blackbody model at 0.9~keV and the column density inferred from the joint fits of all \cxo\ data sets, i.e. $\nh = 1.87 \times 10^{23}$~cm$^{-2}$.
\item $^c$A distance of 9~kpc was assumed.
\item $^d$We used \textsc{pimms} to convert the absorbed 2--10~keV pulsed flux reported by Kuiper \& Hermsen (2009) into unabsorbed 0.3--10~keV fluxes and luminosities. 
\end{tablenotes}
\label{tab:outburst}
\end{table*}

\begin{table*}
\caption{Quiescent fluxes and luminosities (0.3--10~keV) for magnetars showing major outbursts or variations in their persistent emission. Bolometric luminosities are also listed. 
Magnetars are ordered as in Table~\ref{tab:outburst}. Uncertainties are reported at the 1$\sigma$ c.l., upper limits at the 3$\sigma$ c.l.}
\centering
\resizebox{2.1\columnwidth}{!}{
\begin{tabular}{cccccccc}
\hline
Source    			& Date 				& Observatory				& Obs ID			& Exposure		& Abs/Unabs flux					& $L_{X,q}$  					& $L_{bol,q}$		\\
				& 					&						&				& (ks)			& (erg~cm$^{-2}$~s$^{-1}$)			& (erg~s$^{-1}$)				& (erg~s$^{-1}$) \\
\hline
\sgrd\			& 2015 Feb 18			& \xmm\ 					& 0742650101		& 19.1			& $4.2^{+0.2}_{-2.1} \times 10^{-14}$ 	& $(1.2 \pm 0.1) \times 10^{33}$	& $\sim 1.2  \times 10^{33}$	\\ 
				&					&						&				&				& $(8\pm1) \times 10^{-14}$			& 							& 	\\
SGR\,1900$+$14	& 2005 Sep 22			& \xmm\					& 0305580201		& 19.1			& $(3.92\pm0.05) \times 10^{-12}$ 		& $(1.25\pm0.04) \times 10^{35}$	& $\sim 1.4 \times 10^{35}$ 	\\
				&					&						&				&				& $(6.7\pm0.2) \times 10^{-12}$ 		&							&	 \\	
\antgli$^{\rm a}$	& 2014 Nov 04 -- 2015 Nov 17 	& \swift\  				& 00032035087--114  & 40.1			& $(3.55\pm0.02) \times 10^{-11}$  		& $(5.8 \pm 0.3) \times 10^{34}$	& $\sim 6.1 \times 10^{34}$	\\
				&					&						&				&				& $(4.74\pm0.03) \times 10^{-11}$		&							&  	\\
\xte\				& 1993 Apr 03			& \ros/PSPC  				& RP900399N00	& 5.3				& $\sim 5.3\times 10^{-13}$			& $\sim 2.5 \times 10^{34}$		& $\sim 3.2 \times 10^{34}$\\
				&					&						&				&				& $\sim 1.7\times 10^{-11}$			& 							& 	\\	
SGR\,1806$-$20	& 2011 Mar 23			& \xmm\					& 0654230401		& 22.4			& $(5.49\pm0.07) \times 10^{-12}$		& $(8.2 \pm 0.3) \times 10^{34}$ 	& $\sim 1.4 \times 10^{35}$ \\	
				&					&						&				&				& $(9.0\pm0.3) \times 10^{-12}$		& 							& \\	
CXOU\,J1647$-$4552 & 2009 Aug 24		& \xmm\					& 0604380101		& 38.2			& $(8.0\pm0.2) \times 10^{-13}$  		& $(3.3 \pm 0.2) \times 10^{33}$ 	& $\sim 3.5 \times 10^{33}$ 	\\	
				&					&						&				&				& $(1.72\pm0.08) \times 10^{-12}$		& 							& 	\\			
4U\,0142$+$61		& 2004 Jul 24			& \xmm\					& 0206670201		& 21.9			& $(1.215\pm0.002) \times 10^{-10}$ 	& $(3.58\pm0.05) \times 10^{35}$	& $\sim 3.8 \times 10^{35}$   \\
				&					&						&				&				& $(2.309\pm0.003) \times 10^{-10}$ 	&							&   \\
\sgre\			& 2009 Dec 07 --2010 Feb 21	& \swift\ 				& 0032117465--68	& 25.1			& $(2.5\pm0.1) \times 10^{-12}$ 		& $(1.2\pm0.8) \times 10^{33}$		& $\sim 1.3  \times 10^{33}$	\\	
				&					&						&				&				& $(4.4\pm0.3) \times 10^{-12}$		& 							&		\\	  
\aa\ 				& 2006 Jul 01  			& \cxo\ 					& 7287			& 9.5				& $(3.2\pm0.3) \times 10^{-13}$  		& $(2.2 \pm 0.5) \times 10^{33}$	& $\sim 2.3 \times 10^{33}$	\\
				&					&						&				&				& $(9\pm2) \times 10^{-13}$			&							&	 \\	
\lowba\			& 2014 Aug 13--18		& \xmm\					& 0741970201--401	& 108.1  			& $(1.01\pm0.06) \times 10^{-14}$  		& $(7 \pm 1) \times 10^{30}$		& $\sim 8  \times 10^{30}$	\\
				&					&						&				&				& $(1.6\pm0.2) \times 10^{-14}$		& 							&		\\				
SGR\,1833$-$0832$^{\rm b}$	& 2006 Sep 16	& \xmm\ 					& 0400910101		& 8.3				& $<6 \times 10^{-14}$  				& $<8 \times 10^{33}$  			& $< 8 \times 10^{33}$	\\
				&					&						&				&				& $<7 \times 10^{-13}$				&							&	 \\			
1E\,1841$-$045	& 2000 Jul 29 			& \cxo$^{\rm c}$			& 730			& 10.5			& $(2.33\pm0.03) \times 10^{-11}$		& $(4.32\pm0.03) \times 10^{35}$	& $\sim 4.6 \times 10^{35}$\\ 
				&					&						&				&				& $(5.00\pm0.04) \times 10^{-11}$		& 							& \\
\lowbb\			& 2014 Mar 08			& \xmm\					& 0722520101		& 40.3			& $(2.3\pm0.8)\times 10^{-13}$			& $(2.0\pm0.5)\times 10^{32}$		& $\sim 2.3 \times 10^{32}$ \\
				&					&						&				&				& $(6.5\pm1.0)\times 10^{-13}$			& 							&	\\
Swift\,J1834.9$-$0846 & 2009 Jun 06 		& \cxo\					& 10126			& 46.6 			& $<1\times 10^{-14}$  				& $<2 \times 10^{32}$ 			& $< 2 \times 10^{32}$ \\
				&					&						&				&				& $<1\times 10^{-13}$				& 							&	\\			
1E\,1048.1$-$5937	& 2011 Aug 06			& \xmm\	 			 	& 0654870101		& 21.9			& $(5.56\pm0.04) \times 10^{-12}$ 		& $(8.6 \pm 0.2) \times 10^{34}$	& $\sim 8.9 \times 10^{34}$ \\
				&					&						&				&				& $(8.9\pm0.2) \times 10^{-12}$		& 							&	\\
\galcen\			& 1999 Sep 21 -- 2012 Oct 29 & \cxo\				& 129 obs $^{\rm d}$	& 4808.6			& $<2\times 10^{-14}$ 				& $<1 \times 10^{34}$			& $< 1 \times 10^{34}$ \\ \vspace{0.15cm}
				& 					&						&				&				& $<1.5\times 10^{-12}$				&							& \\				
\sgrm			$^{\rm e}$	& 2014 Oct 04 	& \xmm\ 					& 0722412701		& 16.1			& $(8.6\pm0.2) \times 10^{-13}$  		& $(1.6\pm0.1) \times 10^{34}$ 	& $\sim 1.9 \times 10^{34}$\\ 
				&					&						&				&				& $(1.7\pm0.2) \times 10^{-12}$		& 							&	\\
\hline
\hline
\psr\				& 2004 Oct 31			& \cxo\					& 4676			& 60.5			& $(4.8\pm0.6) \times 10^{-14}$		& $(5.7\pm0.3) \times 10^{32}$ 	& $\sim5.8\times10^{32}$ \\	
				&					&						&				&				& $(6.7\pm0.4) \times 10^{-14}$		& \\	
\hline
\psrr				& 2000 Oct 15			& \cxo\					& 748			& 37.3			& $(3.2\pm0.2)\times 10^{-12}$			& $(1.55\pm0.04)\times10^{34}$ 	& $\sim2\times10^{34}$ \\	
				&					&						&				&				& $(3.6\pm0.1)\times 10^{-12}$			& \\						
\hline
\cco\				& 1999 Sep 26			& \cxo\					& 0123			& 13.4			& $(9.8\pm0.6) \times 10^{-13}$		& $(2.8\pm0.1)\times10^{33}$           & $\sim3\times10^{33}$ \\	
				&					&						&				&				& $(2.15\pm0.09) \times 10^{-12}$		& \\	
\hline							
\end{tabular}
}
\begin{tablenotes}
\item \emph{Notes}. $^a$The steady level of the source is slightly lower after the 2012 outburst (as measured with \swift) compared to that after the 2002 outburst (as measured with \xmm), but they are however consistent 
with each other within the uncertainties. We then consider the more precise value derived from the \xmm\ data sets.
\item $^b$A distance of 10~kpc is assumed.
\item $^c$The field around the source has been observed three times by \cxo\ (two with the ACIS set in TE mode and one in CC mode). We consider here the CC-mode observation to minimize pile-up issues.
\item $^d$See \url{http://www.sgra-star.com} for the 2012 \cxo\ X-ray Visionary Project for HETGS Observations of Sgr~A* (see e.g. table~1 by Neilsen et al. 2013 for the log of the observations).
\item $^e$A distance of 9~kpc was assumed.
\end{tablenotes}
\label{tab:quiescence}
\end{table*}

\begin{table*}
\caption{Distances, timing properties and timing-inferred parameters for all magnetars, high magnetic field pulsars, central compact objects, rotation-powered pulsars and 
X-ray dim isolated neutron stars included in our correlation study (see \url{http://magnetars.ice.csic.es}). Sources that underwent major and extensively monitored magnetar-like 
outbursts are marked in bold.}
\begin{threeparttable}
\resizebox{2.1\columnwidth}{!}{
\begin{tabular}{rcccccccc}
\hline
Source    				& Class					& $D$	& $P$	& $\dot{P}$				& $B_{\rm{p, dip}}$$^{\rm a}$	&$\dot{E}_{\rm {rot}}$$^{\rm b}$ 	& $\tau$$^{\rm c}_c$		& Reference		 \\
					&						& (kpc)	& (s)		& (10$^{-11}$~s~s$^{-1}$)		& (10$^{14}$~G)			& (erg~s$^{-1}$)				& (kyr)				&				\\
\hline
{\bf \sgrd$^{\rm d}$}		& \multirow{21}{*}{Magnetars}	& 11 		& 2.59	& 1.9						& 4.5						& $4.3 \times 10^{34}$			& 2 					& Esposito et al. (2009a)		     \\		
{\bf \antgli} 			& 						& 3.2 	& 6.98 	& 0.048					& 1.2 					& $1.3 \times 10^{32}$			& 230				& Dib \& Kaspi (2014)		      \\
{\bf \xte}				& 						& 3.5 	& 5.54 	& 0.283					& 2.6						& $6.7 \times 10^{32}$			& 31					& Camilo et al. (2016)		       \\
{\bf SGR\,1806$-$20}       	& 						& 8.7     	& 7.55      & 76.95                                      & 49                                           & $7.0 \times 10^{34}$                       & 0.2                			& Younes et al. (2015)                       \\
{\bf \wes}	 			& 						& 4	 	& 10.61	& 0.097					& 2.1			 			& $3.2 \times 10^{31}$			& 173				& Rodr\'iguez Castillo et al. (2014)     \\
{\bf \sgre} 	 			& 						& 1.5 	& 5.76	& 0.594					& 3.7						& $1.2 \times 10^{33}$			& 15					& Camero et al. (2014)	       \\ 
{\bf \aa} 				& 						& 4.5 	& 2.07	& 4.77					& 6.4						& $2.1 \times 10^{35}$			& 0.7 				& Dib et al. (2012)		        \\	
{\bf \lowba} 			& 						& 2		& 9.08	& 0.0004					& 0.1					& $2.1 \times 10^{29}$			& $\sim 36000$			& Rea et al. (2013a)				\\	
{\bf SGR\,1833$-$0832$^{\rm e}$} 	& 				& 10		& 7.57	& 0.35					& 3.3						& $3.2 \times 10^{32}$			& 34					& Esposito et al. (2011)		\\	
{\bf \lowbb} 	    		& 						& 1.6 	& 8.44	& 0.013					& 0.7					& $8.4 \times 10^{30}$			& 1030				& Rodr\'iguez Castillo et al. (2016)	\\	
{\bf Swift\,J1834.9$-$0846} & 						& 4.2  	& 2.48	& 0.806					& 2.9						& $2.1 \times 10^{34}$			& 5					& Esposito et al. (2013)		\\			
{\bf 1E\,1048.1$-$5937}	& 						& 9 		& 6.46	& 2.18					& 7.6						& $3.2 \times 10^{33}$			& 4.7					& Dib \& Kaspi (2014)		\\			     
{\bf \galcen} 			& 						& 8.3 	& 3.76	& 3.06					& 6.9						& $2.2 \times 10^{34}$			& 1.9 				& Coti Zelati et al. (2017)		\\	
{\bf \sgrm} 			& 						& 9 		& 3.24	& 1.43					& 4.4						& $1.6 \times 10^{34}$			& 3.6					& Israel et al. (2016)			\\
SGR\,1900$+$14		& 						& 12.5	& 5.20	& 9.2						& 14.0					& $2.6 \times 10^{34}$			& 0.9					& Olausen \& Kaspi (2014)	\\	
4U\,0142$+$614		& 						& 3.6	 	& 8.69	& 0.20					& 2.7						& $1.3 \times 10^{32}$			& 69					& Olausen \& Kaspi (2014)	\\
1E\,1841$-$045		& 						& 8.5	  	& 11.79	& 4.09					& 13.8					& $9.9 \times 10^{33}$			& 4.6					& Olausen \& Kaspi (2014)	\\
1RXS\,J170849.0$-$4009	& 						& 3.8 	& 11.01	& 1.95					& 9.3						& $5.8 \times 10^{32}$			& 9.1					& Olausen \& Kaspi (2014)	\\
CXOU\,J010043.1$-$721	& 						& 62.4	& 8.02	& 1.88					& 7.9						& $1.4 \times 10^{33}$			& 6.8					& Olausen \& Kaspi (2014)	\\
CXOU\,J171405.7$-$3810 & 						& 13.2	& 3.83	& 6.40					& 10.0 					& $4.5 \times 10^{34}$			& 0.95				& Olausen \& Kaspi (2014)	\\
SGR\,0526$-$66		& 						& 49.7	& 8.05	& 3.8						& 11.0					& $2.9 \times 10^{33}$			& 3.4					& Olausen \& Kaspi (2014)	\\	 
\hline
{\bf PSR\,J1119$-$6127}	& \multirow{5}{*}{High-$B$ pulsars} & 8.4 	& 0.41	& 0.4						& 0.82					& $2.5 \times 10^{36}$			& 1.6					& Vigan\`o et al. (2013) \\
{\bf PSR\,J1846$-$0258}	&						& 6.0		& 0.33	& 0.71					& 0.98					& $8.1 \times 10^{36}$			& 0.7					& Vigan\`o et al. (2013) \\
PSR\,J0726$-$2612		&  						& 1.0 	& 3.44	& 0.03					& 0.64					& $2.5 \times 10^{32}$ 			& 190				& Vigan\`o et al. (2013) \\
PSR\,J1819$-$1458		&						& 3.6 	& 4.26	& 0.057					& 1.0						& $3.2 \times 10^{32}$			& 120				& Vigan\`o et al. (2013) \\
PSR\,J1718$-$3718		&						& 4.5 	& 3.38	& 0.16					& 1.5						& $1.6 \times 10^{33}$			& 33					& Vigan\`o et al. (2013) \\
\hline
{\bf 1E\,161348$-$5055}	& \multirow{4}{*}{CCOs}		& 3.3		& 24030	& < 70					& < 2600					& $<2 \times 10^{24}$			& > 540				& Rea et al. (2016) \\
CXOU\,J185238.6$+$0040 & 						& 7.1 	& 0.105	& 0.00000087				& 0.00061					& $3.2 \times 10^{32}$   			& $\sim 190000$		& Vigan\`o et al. (2013) \\						
1E\,1207.4$-$5209		&						& 2.1 	& 0.424	& 0.00000220				& 0.00196					& $1.2 \times 10^{31}$   			& $\sim 302000	$		& Vigan\`o et al. (2013) \\
RX\,J0822$-$4300		&						& 2.2 	& 0.112	& 0.00000093				& 0.00058					& $1.9 \times 10^{32}$  			& $\sim 254000$		& Vigan\`o et al. (2013) \\
\hline
RX\,J0420.0$-$502		& \multirow{7}{*}{XDINSs}		& 0.34	& 3.45	& 0.004					& 0.2						& $2.5 \times 10^{31}$			& $\sim 2000$			& Vigan\`o et al. (2013) \\
RX\,J1856.5$-$375		&						& 0.12	& 7.06	& 0.003					& 0.3						& $3.2 \times 10^{30}$			& $\sim 3800$			& Vigan\`o et al. (2013) \\
RX\,J2143.0$+$065		&						& 0.43	& 9.43	& 0.004					& 0.4						& $2.0 \times 10^{30}$			& $\sim 3600$			& Vigan\`o et al. (2013) \\
RX\,J0720.4$-$312		&						& 0.29	& 8.39	& 0.007					& 0.5						& $4.7 \times 10^{30}$			& $\sim 1900$			& Vigan\`o et al. (2013) \\
RX\,J0806.4$-$412		&						& 0.25	& 11.37	& 0.0055					& 0.5						& $1.6 \times 10^{30}$			& $\sim 3300$			& Vigan\`o et al. (2013) \\
RX\,J1308.6$+$212		&						& 0.50	& 10.31	& 0.01					& 0.7						& $4.0 \times 10^{30}$			& $\sim 1500$			& Vigan\`o et al. (2013) \\
RX\,J1605.3$+$324		&						& 0.35	& 3.39	& 0.16					& 1.5						& $1.6 \times 10^{33}$			& 34					& Vigan\`o et al. (2013) \\
\hline
PSR\,J0538$+$281		& \multirow{8}{*}{RPPs}		& 1.3 	& 0.143	& 0.0005					& 0.015					& $5.0 \times 10^{34}$			& 620				& Vigan\`o et al. (2013) \\
PSR\,B1055$-$52		&						& 0.73	& 0.197	& 0.0006					& 0.02					& $3.2 \times 10^{34}$			& 540				& Vigan\`o et al. (2013) \\
PSR\,J0633$+$174		&						& 0.25	& 0.237 	& 0.001					& 0.03					& $3.2 \times 10^{34}$			& 340				& Vigan\`o et al. (2013) \\
PSR\,B1706$-$44		&						& 2.6 	& 0.102	& 0.009					& 0.06					& $3.2 \times 10^{36}$			& 17 					& Vigan\`o et al. (2013) \\
PSR\,B0833$-$45		&						& 0.28	& 0.089	& 0.01					& 0.07					& $6.3 \times 10^{36}$			& 11					& Vigan\`o et al. (2013) \\
PSR\,B0656$+$14		&						& 0.28	& 0.385	& 0.0055					& 0.09					& $4.0 \times 10^{34}$ 			& 110				& Vigan\`o et al. (2013) \\
PSR\,B2334$+$61		&						& 3.1 	& 0.495	& 0.02					& 0.2						& $6.3 \times 10^{34}$			& 41					& Vigan\`o et al. (2013) \\
PSR\,J1740$+$100		&						& 1.4 	& 0.154	& 0.2						& 0.4						& $2.5 \times 10^{37}$			& 1.2					& Vigan\`o et al. (2013) \\
\hline	
\end{tabular}
}
\end{threeparttable}
\begin{tablenotes}
\item \emph{Notes}. $^a$Assuming a force-free magnetosphere and an aligned rotator, a star radius $R = 10$~km and moment of inertia $I = 10^{45}$~g~cm$^2$, the dipolar component of 
the surface magnetic field at the polar caps is given by $B_{\rm{p, dip}} \sim 2 \cdot (3c^3 IP\dot{P}/8\pi^2 R^6)^{1/2} \sim 6.4 \times 10^{19} (P\dot{P})^{1/2}$~G. Relativistic 
magnetohydrodynamic simulations of pulsar magnetospheres have shown that the estimate offered by this formula is correct within a factor of $\sim2-3$ (Spitkovsky 2006).
\item $^b$With the same assumptions, the rotational energy loss is given by $\dot{E}_{\rm {rot}} = 4\pi^2 I\dot{P} P^{-3} \sim 3.9 \times 10^{46} \dot{P}P^{-3}$~erg~s$^{-1}$.
\item $^c$With the same assumptions and assuming that the spin period at birth was much smaller than the current value, the characteristic age is given by $\tau_c=P/2\dot{P}$.
\item $^d$The spin period and its derivative were detected only following the 2008 re-activation of the source. We assume the same spin period derivative also for the 1998 
outburst, and consider the same values for $B_{\rm{p, dip}}$, $\dot{E}_{\rm {rot}}$ and $\tau_c$ in our searches for correlations.
\item $^e$The value for the distance is assumed.
\end{tablenotes}
\label{tab:timing}
\end{table*}

\begin{table*}
\caption{Results of the search for (anti)correlations between different parameters. Letters in parentheses indicate the case of a correlation ($c$) or an anticorrelation ($a$). The 
decay-time-scale is defined as the $e$-folding parameter ($\tau$) and it refers to the larger value (the parameter $\tau_2$ in Table~\ref{tab:decays}) for the cases where the outburst 
decay curve was modelled by more than one exponential function. Values for the significance are not reported if below 2$\sigma$. The yes / no flag in the last column indicates 
if a correlation or anticorrelation is predicted by either the internal crustal cooling or the untwisting bundle models for the evolution of magnetar outbursts (see the text for more details).} 
\resizebox{2.1\columnwidth}{!}{
\begin{tabular}{lccccc}
\hline
First parameter    					& Second parameter						& Corr (c) or anticorr (a),  		& PL index	& Reference figure 				& Correlation expected?		\\
								& 									& Significance ($\sigma$) for		&			& 							& Internal cooling / 	\\
								&									& Spearman / Kendall $\tau$ tests 	&			&							& untwisting bundle	\\
\hline	
Quiescent X-ray luminosity			& Maximum luminosity increase			& (a)	, 5.7 / 4.9				&  -0.7				& Fig.~\ref{fig:flux_deltaf}			& Yes / yes	\\
Spin-down luminosity				& Quiescent bolometric luminosity			& --	  					& --					& Fig.~\ref{fig:b_lqui}				& No/no 	\\
Dipolar magnetic field				& Quiescent bolometric luminosity			& (c)	,  3.2 / 2.9 			&  2.0				& Fig.~\ref{fig:b_lqui}				& Does not apply 	\\
Dipolar magnetic field				& Maximum luminosity					& (c)	,  2.5 / 2.4				&  0.5				& Fig.~\ref{fig:b_lpeak}			& Yes/yes	\\
Dipolar magnetic field				& Decay time-scale						& --	   					& --					& Fig.~\ref{fig:b_lpeak}			& Yes/yes	\\
Dipolar magnetic field				& Outburst energy						& (c)	,  3.7 / 3.3 			&  1.0				& Fig.~\ref{fig:b_energy}			& Yes / yes	\\
Characteristic age					& Outburst energy						& (a)	,  3.3	/ 3.0				&  -0.4				& Fig.~\ref{fig:b_energy}			& Yes / ?	\\
Maximum luminosity					& Outburst energy						& (c) ,  4.0 / 3.7				&  1.4				& Fig.~\ref{fig:lpeak_energy}		& Yes / yes	\\
Quiescent bolometric luminosity		& Outburst energy						& --	   					& -- 					& Fig.~\ref{fig:lpeak_energy}		& No / no	\\
Maximum luminosity					& Decay time-scale						& --	   					& --					& Fig.~\ref{fig:lpeak_decay}		& No / no	\\
Outburst energy					& Decay time-scale						& (c)	,   3.9 / 3.6	 		&  0.5				& Fig.~\ref{fig:lpeak_decay}		& Yes / yes	\\
Outburst energy					& Maximum luminosity increase			& --	    					& --					& Fig.~\ref{fig:energy_deltaflux}		& No/no 	\\	
Decay time-scale					& Maximum luminosity increase			& --	    					& --					& Fig.~\ref{fig:energy_deltaflux}		& No/no 	\\	
\hline
\end{tabular}
}
\label{tab:correlations}
\end{table*}

\clearpage

\clearpage

\appendix

\begin{table*}
\section{Journal of observations}
\label{journal}

This Section provides a log of all observations carried out by the X-ray instruments on board the 
\swift, \xmm, \cxo\ and \bsax\ satellites, relative to the magnetar outbursts that were analysed in our 
study. The tables are reported following the chronological order of the outburst onsets (spanning a time 
interval of $\sim 18$ yr, from 1998 to 2016). We give references to previous papers where part of the 
listed observations have been already analysed. For every single observation, each table lists:

\vspace{0.5cm}

\begin{enumerate}

\item the X-ray instrument (legend: XRT = X-ray Telescope on board \swift; EPN = pn CCD 
of the EPIC camera on board \xmm; ACIS-S = Advanced CCD Imaging Spectrometer spectroscopic 
array on board \cxo; ACIS-I = Advanced CCD Imaging Spectrometer imaging array on board \cxo; 
MECS = Medium-Energy Concentrator Spectrometer on board \bsax); 

\item the operating mode of the X-ray instrument (legend: PC = photon counting; WT = windowed 
timing; FF = full frame; LW = large window: SW = small window; TE = timed exposure; CC = continuous 
clocking);

\item the mid point of the observation expressed in modified Julian date (MJD); 

\item the time (in units of days) elapsed from the outburst onset, which is defined as the epoch when 
the first burst was detected in the hard X-/soft $\gamma$-rays from the target of interest as reported 
by the `Gamma-ray Burst Coordinates Network' (\url{http://gcn.gsfc.nasa.gov/gcn3_archive.html}) and/or 
`The Astronomer's Telegram' website (\url{http://www.astronomerstelegram.org/}). Two exceptions to 
this definition are represented by the magnetars \xte\ and 1E\,1048.1$-$5937, as discussed in 
Sections~\ref{xte} and ~\ref{sgrn}, respectively;

\item the exposure time after filtering for intrinsic source flares and bursts and, 
in the case of the \xmm\ and \cxo\ data sets, also for particle background flaring;

\item the background-subtracted count rate of the source (in units of counts s$^{-1}$) in the 
0.3--10~keV energy band. Count rates are corrected for point spread function and vignetting 
effects, but not for pile-up.

\end{enumerate}
\end{table*}

\begin{table*}
\centering
\caption{Log of all X-ray observations of \sgrd\ following the 1998 June outburst.
The outburst onset occurred on MJD 50\,979.109 (Kouveliotou 1998). Two additional 
\xmm\ observations (obs ID: 0204500201, 0204500301, pn in FF mode) were not included 
because the source was detected at an off-axis angle of about 9.6 arcmin in these cases, 
leading to a too poor statistics for a meaningful spectral analysis. Part of these observations 
were already analysed by Woods et al. (1999), Kouveliotou et al. (2003), Mereghetti 
et al. (2006) and Esposito et al. (2008, 2009a, 2009b).} 
\label{tab:sgr1627_1998}
\vspace{10pt}
\begin{tabular}{ccccccc} \hline
Instrument     	& Mode	& Obs. ID 			&  Mid point of observation	& Time since outburst onset 	& Exposure  		& Source net count rate \\
         		&          	&       			&  (MJD)					& (d)     			     	& (ks)   			& (counts s$^{-1}$)\\ 
\hline 
\bsax/MECS$^{a}$	& -- 	& 70566001		& 51\,032.25	  			& 53.14	 				& 44.9 	 		& $0.0200\pm0.0008$\\
\bsax/MECS$^{a}$	& -- 	& 70566002		& 51\,072.51		   		& 93.41	 				& 30.4	 		& $0.0153\pm0.0008$\\
\bsax/MECS$^{a}$	& --	& 70821005		& 51\,399.77		 		& 420.66	 				& 80.4	 		& $0.0060\pm0.0005$\\
\bsax/MECS$^{a}$	& -- 	& 70821001		& 51\,793.58				& 814.47	 	 			& 61.3 	 		& $0.0034\pm0.0005$\\
\cxo\ ACIS-S	& TE  	& 1981			& 52\,182.50   				& 1203.39	 				& 48.9 			& $0.0049\pm0.0003$\\
\cxo\ ACIS-S	& TE  	& 3877			& 52\,722.33	  			& 1743.22					& 25.7 			& $0.0059\pm0.0005$\\
\xmm\ EPN	& SW 	& 0202560101		& 53\,270.98 				& 2291.87	 				& 35.9	  		& $0.023\pm0.002$\\
\cxo\ ACIS-S	& TE  	& 5573			& 53\,549.28	  			& 2570.17					& 9.8	   			& $0.0042\pm0.0007$\\
\cxo\ ACIS-S	& TE  	& 5574			& 53\,668.54				& 2689.43					& 10.0			& $0.0038\pm0.0006$\\
\xmm\ EPN	& FF 	& 0502140101		& 54\,509.28  				& 3530.17	 				& 47.5	 		& $0.0083\pm0.0008$\\
\hline
\end{tabular}
\begin{tablenotes}
\item \emph{Note}. $^a$Data acquired by the LECS were not considered, owing to the large value for the column density towards the source direction (the LECS is best calibrated 
for spectral analysis in the 0.1--4~keV range). 
\end{tablenotes}
\end{table*}

\begin{table*}
\centering
\caption{Log of all X-ray observations of \antgli\ following the 2002 June outburst.
The outburst onset occurred on MJD 52\,443.66 (Kaspi et al. 2003). An additional 
\cxo\ observation (obs ID: 6730, ACIS-S in TE mode) was not included owing to the 
combination of severe pile-up and extended emission (due to both the SNR surrounding 
the source and a halo from dust scattering) beyond a radial distance of about 4 arcsec. 
Part of these observations were already analysed by Woods et al. (2004) and Zhu et al. (2008).} 
\label{tab:1e2259_2002}
\vspace{10pt}
\begin{tabular}{ccccccc} \hline
Instrument     	& Mode	& Obs. ID 			&  Mid point of observation	& Time since outburst onset 	& Exposure  		& Source net count rate \\
         		&          	&       			&  (MJD)					& (d)     			     	& (ks)   			& (counts s$^{-1}$)\\ 
\hline 
\xmm\ EPN 	& SW 	& 0155350301		& 52\,446.60   				& 2.94  					& 18.4   			& $17.92\pm0.03$\\	
\xmm\ EPN 	& FF 	& 0057540201 		& 52\,464.45 				& 20.79 					& 5.7   			& $4.52\pm0.03$\\	
\xmm\ EPN 	& FF 	& 0057540301 		& 52\,464.68 				&  21.02					& 10.1   			& $4.89\pm0.02$\\	
\xmm\ EPN 	& SW	& 0203550301		& 53\,055.63				& 611.97  					& 3.8   			& $10.47\pm0.05$\\	
\xmm\ EPN 	& SW	& 0203550601		& 53\,162.70				& 719.04 					& 4.9   			& $10.33\pm0.05$\\	
\xmm\ EPN 	& SW	& 0203550401		& 53\,178.66				& 735.00 					& 3.6  			& $10.41\pm0.05$\\	
\xmm\ EPN 	& SW 	& 0203550501		& 53\,358.04				& 914.38 					& 3.6  			& $10.22\pm0.05$\\	
\xmm\ EPN 	& SW 	& 0203550701		& 53\,580.00				& 1136.34 				& 3.5  			& $9.98\pm0.05$\\	
\hline
\end{tabular}
\end{table*}

\begin{table*}
\centering
\caption{Log of all \xmm\ observations of \xte\ following the 2003 outburst.
The outburst onset was missed and is constrained to be in the range MJD 
52\,595--52\,662 (Ibrahim et al. 2004). Part of these observations were already 
analysed by Rea et al. (2004), Gotthelf et al. (2004), Halpern \& Gotthelf (2005), 
Gotthelf \& Halpern (2005), Bernardini et al. (2009, 2011a), Alford \& Halpern (2016), 
Camilo et al. (2016) and Pintore et al. (2016).} 
\label{tab:xte}
\vspace{10pt}
\begin{tabular}{ccccccc} 
\hline
Instrument     	& Mode	& Obs. ID 			&  Mid point of observation	& Time since outburst onset 	& Exposure  	& Source net count rate \\
         		&          	&       			&  (MJD)					& (d)     			     	& (ks)   		& (counts s$^{-1}$)\\ 
\hline 
\xmm\ EPN 	& SW	& 0161360301		& 52\,890.61				& 229--296				& 6.6  		& $10.04\pm0.04$\\	
\xmm\ EPN 	& LW 	& 0161360501		& 53\,075.59				& 414--481 				& 2.5			& $5.50\pm0.05$\\
\xmm\ EPN 	& LW 	& 0164560601		& 53\,266.66				& 605--672 				& 21.5 		& $3.57\pm0.01$\\
\xmm\ EPN 	& LW 	& 0301270501		& 53\,448.23				& 786--853 				& 32.3 		& $1.832\pm0.008$\\
\xmm\ EPN 	& LW 	& 0301270401		& 53\,633.68				& 972--1039 				& 28.3 		& $0.919\pm0.006$\\
\xmm\ EPN 	& LW 	& 0301270301		& 53\,807.03 				& 1145--1212 				& 21.3		& $0.641\pm0.006$\\
\xmm\ EPN 	& LW 	& 0406800601		& 54\,002.34				& 1340--1407 				& 39.8		& $0.493\pm0.004$\\
\xmm\ EPN 	& LW 	& 0406800701		& 54\,166.12				& 1504--1571 				& 36.7		& $0.459\pm0.004$\\
\xmm\ EPN 	& LW 	& 0504650201		& 54\,359.48				& 1698--1765 				& 67.0		& $0.455\pm0.003$\\
\xmm\ EPN 	& LW	& 0552800201		& 54\,896.02				& 2234--2301  				& 30.3  		& $0.418\pm0.004$\\
\xmm\ EPN 	& LW	 & 0605990201		& 55\,079.73 				& 2418--2485 				& 17.9  		& $0.423\pm0.005$\\
\xmm\ EPN 	& LW	 & 0605990301		& 55\,081.66				& 2420--2487 				& 16.3  		& $0.418\pm0.005$\\
\xmm\ EPN 	& LW	 & 0605990401		& 55\,097.77				& 2436--2503 				& 11.1 		& $0.419\pm0.006$\\
\xmm\ EPN 	& LW	 & 0605990501		& 55\,295.23 				& 2633--2700 				& 3.5  		& $0.43\pm0.01$\\
\xmm\ EPN 	& LW 	 & 0605990601		& 55\,444.73 				& 2783--2850				& 8.4  		& $0.428\pm0.007$\\
\xmm\ EPN 	& LW 	& 0671060101		& 55\,654.19				& 2992--3059 				& 16.0  		& $0.430\pm0.005$\\
\xmm\ EPN 	& LW	& 0671060201		& 55\,813.46				& 3152--3219 				& 12.0  		& $0.418\pm0.006$\\
\xmm\ EPN 	& LW 	& 0691070301		& 56\,177.07 				& 3515--3582 				& 14.6 		& $0.423\pm0.005$\\
\xmm\ EPN 	& LW 	 & 0691070401		& 56\,354.29 				& 3692--3759 				& 7.4  		& $0.421\pm0.008$\\
\xmm\ EPN 	& LW 	 & 0720780201		& 56\,540.98				& 3879--3946 				& 18.0  		& $0.429\pm0.005$\\
\xmm\ EPN 	& LW 	& 0720780301		& 56\,721.10  				& 4059--4126 				& 19.3  		& $0.429\pm0.005$\\
\hline
\end{tabular}
\end{table*}

\begin{table*}
\centering
\caption{Log of all \xmm\ observations of SGR\,1806$-$20 following the 2004 December giant 
flare. We assume that the outburst onset occurred in concomitance with the pinnacle of the 
giant flare, i.e., on MJD 53\,366.89613426 (e.g., Palmer et al. 2004). Part of these observations 
were already analysed by Tiengo et al. (2005), Esposito et al. (2007), Mereghetti, Esposito \& 
Tiengo (2007) and Younes et al. (2015).} 
\label{tab:sgr1806}
\vspace{10pt}
\begin{tabular}{ccccccc} \hline
Instrument     	& Mode	& Obs. ID 			&  Mid point of observation		& Time since outburst onset 	& Exposure  	& Source net count rate \\
         		&          	&       			&  (MJD)						& (d)     			     	& (ks)   		& (counts s$^{-1}$)\\ 
\hline 
\xmm\ EPN	& SW	& 0164561301		& 53\,436.49					& 69.59					& 10.5		& $1.56 \pm 0.01$	\\
\xmm\ EPN	& SW 	& 0164561401		& 53\,647.62 					& 280.72					& 22.8 		& $1.101 \pm 0.007$	\\
\xmm\ EPN	& SW 	& 0406600301		& 53\,829.45					& 462.55					& 20.5 		& $0.876 \pm 0.007$	\\
\xmm\ EPN	& SW 	& 0406600401		& 53\,988.61 					& 621.71					& 22.3 		& $0.911 \pm 0.007$	\\
\xmm\ EPN	& SW 	& 0502170301		& 54\,369.82					& 1002.92					& 21.3		& $0.736 \pm 0.006$	\\
\xmm\ EPN	& SW	& 0502170401		& 54\,558.74 					& 1191.84					& 22.7		& $0.600 \pm 0.005$	\\
\xmm\ EPN	& FF		& 0554600301		& 54\,714.36 					& 1347.46					& 25.9 		& $0.560 \pm 0.005$	\\
\xmm\ EPN	& FF  	& 0554600401 		& 54\,893.89					& 1526.99					& 22.6		& $0.516 \pm 0.005$	\\
\xmm\ EPN	& FF		& 0604090201		& 55\,081.97					& 1715.07					& 23.2 		& $0.474 \pm 0.005$	\\
\xmm\ EPN	& FF		& 0654230401		& 55\,643.69					& 2276.79					& 22.4		& $0.423 \pm 0.004$	\\
\hline
\end{tabular}
\end{table*}

\begin{table*}
\centering
\caption{Log of all X-ray observations of \wes\ following the 2006 September outburst.
The outburst onset  occurred on MJD 53\,999.06587963 (Krimm et al. 2006). Two \swift\ 
observations (obs. ID 00030806003 and 00030806004) were carried with the XRT in both 
the PC and WT modes (indeed, count rates above 1 counts s$^{-1}$ cause an automated 
shift of the PC to the WT mode, to prevent heavy pile-up). We then considered only the data 
from the mode which resulted in the largest counting statistics. Part of these observations 
were already analysed by Israel et al. (2007), Woods et al. (2011), An et al. (2013) and 
Rodr\'iguez Castillo et al. (2014).} 
\label{tab:cxou1647}
\vspace{10pt}
\begin{tabular}{ccccccc} \hline
Instrument     	& Mode	& Obs. ID 			&  Mid point of observation		& Time since outburst onset 	& Exposure  		& Source net count rate \\
         		&          	&       			&  (MJD)						& (d)     			     	& (ks)   			& (counts s$^{-1}$)\\ 
\hline 
\swift\ XRT 	& PC		& 00030806001 	& 53\,999.85					& 0.78 					& 7.7 			& $0.523\pm0.008$\\ 
\swift\ XRT 	& WT	& 00030806002 	& 54\,000.62					& 1.55 					& 0.7 			& $0.60\pm0.03$\\ 
\xmm\ EPN 	& FF	 	& 0311792001 		& 54\,000.70					& 1.64 					& 26.2 			& $3.80 \pm 0.01$\\ 
\swift\ XRT 	& WT	& 00030806003 	& 54\,001.07					& 2.01 					& 4.9 			& $0.49\pm0.01$\\ 
\swift\ XRT 	& PC		& 00030806004 	& 54\,004.41					& 5.35	 				& 2.5				& $0.30\pm0.01$\\
\cxo\ ACIS-S	& CC 	& 6724 			& 54\,005.38					& 6.31 					& 15.1 			& $1.59 \pm 0.01$\\ 
\cxo\ ACIS-S	& CC 	& 6725 			& 54\,010.11					& 11.04 					& 20.1 			& $1.301\pm0.009$\\ 
\swift\ XRT 	& WT	& 00030806006 	& 54\,010.64					& 11.57 					& 2.0 			& $0.37\pm0.01$\\ 
\swift\ XRT 	& WT	& 00030806007 	& 54\,011.56					& 12.49 					& 2.0 			& $0.36\pm0.01$\\ 
\swift\ XRT 	& WT	& 00030806008 	& 54\,014.05					& 14.99 					& 2.1 			& $0.36\pm0.01$\\ 
\cxo\ ACIS-S	& CC 	& 6726 			& 54\,017.42					& 18.35 					& 25.1 			& $ 1.221 \pm 0.007$\\ 
\swift\ XRT 	& WT	& 00030806009 	& 54\,017.85					& 18.78 					& 3.5 			& $0.286\pm0.009$\\ 
\swift\ XRT 	& WT	& 00030806010 	& 54\,018.05					& 18.99 					& 2.8 			& $0.33\pm0.01$\\ 
\swift\ XRT 	& WT	& 00030806011 	& 54\,023.38					& 24.32 					& 5.6 			& $0.303\pm0.008$\\ 
\swift\ XRT 	& WT	& 00030806012 	& 54\,029.24					& 30.17 					& 5.5 			& $0.276\pm0.007$\\ 
\swift\ XRT 	& WT	& 00030806013 	& 54\,035.76					& 36.69 					& 2.8 			& $0.29\pm0.01$\\ 
\cxo\ ACIS-S	& CC 	& 8455 			& 54\,036.39					& 37.32 					& 15.1 			& $0.981 \pm 0.008$ \\ 
\swift\ XRT 	& WT	& 00030806014 	& 54\,119.21					& 120.15 					& 2.0 			& $0.21\pm0.01$\\ 
\swift\ XRT 	& WT	& 00030806015 	& 54\,122.13					& 123.06 					& 3.8 			& $0.219\pm0.008$\\ 
\cxo\ ACIS-S	& CC 	& 8506 			& 54\,133.92					& 134.86 					& 20.1 			& $0.683 \pm 0.006$\\ 
\xmm\ EPN	& LW 	& 0410580601 		& 54\,148.47					& 149.41 					& 17.3 			& $1.224 \pm 0.009$\\ 
\swift\ XRT 	& WT	& 00030806016 	& 54\,207.38					& 208.32 					& 4.3 			& $0.175\pm0.007$\\ 
\swift\ XRT 	& WT	& 00030806017 	& 54\,208.23					& 209.16 					& 2.2 			& $0.183\pm0.009$\\ 
\swift\ XRT 	& PC		& 00030806018 	& 54\,235.53					& 236.46 					& 2.6				& $0.079\pm0.006$\\ 
\swift\ XRT 	& PC		& 00030806019 	& 54\,237.53					& 238.47 					& 1.1 			& $0.13\pm0.01$\\ 
\xmm\ EPN	& LW  	& 0505290201		& 54\,331.59					& 332.52 					& 28.4			& $0.714 \pm 0.005$\\
\xmm\ EPN	& LW  	& 0555350101		& 54\,698.68 					& 699.61 					& 28.4			& $0.292 \pm 0.003$\\
\xmm\ EPN	& LW  	& 0604380101		& 55\,067.57					& 1068.50	 				& 38.2			& $0.163 \pm 0.002$\\  
\hline
\end{tabular}
\end{table*}

\begin{table*}
\centering
\caption{Log of all X-ray observations of \sgrd\ following the 2008 May outburst. The 
outburst onset  occurred on MJD 54\,614.34841435 (Palmer et al. 2008). Part of these 
observations were already analysed by Esposito et al. (2008, 2009a, 2009b) and An 
et al. (2012).}
\label{tab:sgr1627_2008}
\vspace{10pt}
\begin{tabular}{ccccccc} \hline
Instrument     	& Mode	& Obs. ID 			&  Mid point of observation	& Time since outburst onset 	& Exposure  		& Source net count rate \\
         		&          	&       			&  (MJD)					& (d)     			     	& (ks)   			& (counts s$^{-1}$)\\ 
\hline			
\swift\ XRT	& PC		& 00312579001 	& 54\,614.55 				& 0.20					& 2.0 			& $0.058 \pm 0.006$\\
\swift\ XRT	& PC		& 00312579002 	& 54\,615.65 				& 1.30					& 2.0 			& $0.017 \pm 0.003$\\
\swift\ XRT	& PC		& 00312579003 	& 54\,616.55 				& 2.21					& 1.9 			& $0.007 \pm 0.002$\\
\swift\ XRT	& PC		& 00312579004 	& 54\,617.41  				& 3.06					& 1.8 			& $0.009 \pm 0.003$\\
\swift\ XRT	& PC		& 00312579005 	& 54\,618.27 				& 3.93					& 2.0 			& $0.010 \pm 0.002$\\
\swift\ XRT	& PC		& 00312579006 	& 54\,619.40 				& 5.05					& 2.1 			& $0.009 \pm 0.002$\\
\cxo\ ACIS-S	& CC	& 9126 			& 54\,620.68 				& 6.33 					& 40.0			& $0.029 \pm 0.002$\\
\swift\ XRT	& PC		& 00312579007 	& 54\,623.48 				& 9.13					& 0.6 			& $<$0.01$^{\rm a}$\\
\swift\ XRT	& PC		& 00312579008 	& 54\,626.65 				& 12.30					& 0.3 			& $<$0.02$^{\rm a}$\\
\swift\ XRT	& PC		& 00312579009 	& 54\,629.62  				& 15.27					& 1.9 			& $0.006 \pm 0.002$\\
\swift\ XRT	& PC		& 00312579010 	& 54\,632.26 				& 17.91					& 3.8 			& $0.005 \pm 0.001$\\
\swift\ XRT	& PC		& 00312579011 	& 54\,635.78 				& 21.43					& 2.3 			& $0.010 \pm 0.002$\\
\swift\ XRT	& PC		& 00312579012 	& 54\,638.88 				& 24.54					& 5.2 			& $0.007 \pm 0.001$\\
\swift\ XRT	& PC		& 00312579013 	& 54\,649.47 				& 35.12					& 1.5 			& $0.004 \pm 0.002$\\
\swift\ XRT	& PC		& 00312579014 	& 54\,652.78 				& 38.43					& 5.6 			& $0.005 \pm 0.001$\\
\swift\ XRT	& PC		& 00312579015 	& 54\,664.76 				& 50.41  					& 7.0				& $0.0047 \pm 0.0009$\\		
\swift\ XRT	& PC		& 00312579016 	& 54\,678.64 				& 64.29  					& 5.2			 	& $0.006 \pm 0.001$\\		
\swift\ XRT	& PC		& 00312579017 	& 54\,680.14 				& 65.80  					& 1.7				& $0.002 \pm 0.001$\\			
\swift\ XRT	& PC		& 00312579018 	& 54\,723.19 			 	& 108.84					& 3.2				& $0.003 \pm 0.001$\\				
\swift\ XRT	& PC		& 00312579019 	& 54\,724.26 				& 109.91  					& 0.6  			& $<$0.007$^{\rm a}$\\				
\swift\ XRT	& PC		& 00312579020 	& 54\,725.19 			 	& 110.84  					& 3.5				& $ 0.002 \pm 0.001$\\				
\swift\ XRT	& PC 	& 00312579022 	& 54\,732.09 				& 117.75  					& 3.2  			& $ 0.005 \pm 0.001$\\				
\xmm\ EPN	& FF		& 0560180401 		& 54\,734.76				& 120.41 					& 94.7			& $0.0226 \pm 0.0005$\\
\cxo\ ACIS-S	& TE		& 10519 			& 54\,856.86				& 242.51					& 6.6 			& $0.018 \pm 0.002$	\\
\cxo\ ACIS-I	& TE		& 12528$^{b}$ 		& 55\,728.17				& 1113.83					& 19.0			& $0.0026 \pm 0.0004$	\\
\cxo\ ACIS-I	& TE		& 12529$^{b}$ 		& 55\,728.41				& 1114.06					& 19.0			& $0.0027 \pm 0.0004$ 	\\
\xmm\ EPN	& FF		& 0742650101 		& 57\,071.56				& 2457.21					& 19.1			& $0.0046 \pm 0.0007$\\
\hline
\end{tabular}
\begin{tablenotes}
\item \emph{Notes}. $^a$The upper limit is quoted at the 3$\sigma$ c.l., and is derived by applying the prescription for low number statistics given by Gehrels (1986). 
The corresponding upper limits on the fluxes and luminosities were estimated by assuming an absorbed blackbody spectral model with the same parameters as those of 
the spectra of the closeby observations.
\item $^b$The spectral files and responses of these observations were combined to improve the fit statistics. 
\end{tablenotes}
\end{table*}

\begin{table*}
\centering
\caption{Log of all X-ray observations of \sgre\ following the 2008 August outburst. 
The outburst onset  occurred on MJD 54\,700.52915509 (Barthelmy et al. 2008). All 
reported count rates are not corrected for pile-up. One \cxo\ observation was performed 
with the HRC-I (obs. ID 9131) and was not included in our analysis. Part of these 
observations were already analysed by Rea et al. (2009), G{\"o}{\u g}{\"u}{\c s} et al. 
(2010b) and Camero et al. (2014).}
\label{tab:sgr0501}
\vspace{10pt}
\begin{tabular}{ccccccc} \hline
Instrument     	& Mode	& Obs. ID 			&  Mid point of observation	& Time since outburst onset 	& Exposure  		& Source net count rate \\
         		&          	&       			&  (MJD)					& (d)     			     	& (ks)   			& (counts s$^{-1}$)\\ 
\hline	
\swift\ XRT	& PC 	& 00321174000		& 54\,701.08				& 0.55					& 37.1			& $0.738 \pm 0.004$		\\		
\xmm\ EPN	& SW  	& 0560191501		& 54\,701.33				& 0.80	 				& 33.8			& $ 8.48\pm0.02$		\\
\swift\ XRT	& PC 	& 00321174001		& 54\,701.87				& 1.34					& 14.6			& $0.651 \pm 0.007$		\\	
\cxo\ ACIS-S	& CC	& 10164			& 54\,703.40				& 2.87					& 36.5			& $4.51 \pm 0.01$		\\	
\swift\ XRT	& WT 	& 00321174003		& 54\,704.59				& 4.06					& 9.0				& $ 0.80 \pm 0.01$		\\		
\swift\ XRT	& WT 	& 00321174004		& 54\,705.60				& 4.07					& 16.3			& $ 0.615 \pm 0.006$	\\		
\swift\ XRT	& WT 	& 00321174005		& 54\,705.48				& 4.96					& 10.6			& $ 0.780 \pm 0.009$	\\		
\swift\ XRT	& WT 	& 00321174006		& 54\,705.50				& 4.97					& 14.5			& $ 0.762 \pm 0.006$	\\		
\swift\ XRT	& WT 	& 00321174007		& 54\,706.49				& 5.96					& 25.0			& $ 0.756 \pm 0.006 $	\\		
\swift\ XRT	& WT 	& 00321174008		& 54\,706.51				& 5.98					& 7.9				& $ 0.697 \pm 0.009$	\\		
\xmm\ EPN	& SW  	& 0552971101		& 54\,707.44				& 6.91	 				& 17.1			& $7.12\pm0.02$ 		\\ 
\swift\ XRT	& WT 	& 00321174009		& 54\,708.00				& 7.47					& 43.5			& $ 0.766 \pm 0.004$	\\		
\swift\ XRT	& WT 	& 00321174010		& 54\,708.02				& 7.49					& 20.7			& $ 0.669 \pm 0.006$	\\		
\xmm\ EPN	& SW  	& 0552971201		& 54\,709.57				& 9.04	  				& 7.2				& $6.68\pm0.03$		\\ 
\swift\ XRT	& WT 	& 00321174011		& 54\,710.55				& 10.02					& 67.6			& $ 0.728 \pm 0.003$	\\		
\swift\ XRT	& WT 	& 00321174012		& 54\,710.52				& 10.00					& 32.2			& $ 0.556 \pm 0.004$	\\		
\xmm\ EPN	& SW  	& 0552971301		& 54\,711.54				& 11.01	  				& 14.3			& $6.05\pm0.02$		\\
\swift\ XRT	& WT 	& 00321174013		& 54\,712.52				& 11.99					& 6.1				& $ 0.71 \pm 0.01$		\\		
\swift\ XRT	& WT 	& 00321174014		& 54\,712.93				& 12.40					& 2.0				& $ 0.58 \pm 0.02$		\\		
\swift\ XRT	& WT 	& 00321174017		& 54\,713.51				& 12.98					& 2.8				& $ 0.70 \pm 0.02$		\\		
\swift\ XRT	& WT 	& 00321174018		& 54\,713.52				& 12.99					& 16.7			& $ 0.695 \pm 0.006$	\\		
\swift\ XRT	& WT 	& 00321174019		& 54\,714.13				& 13.61					& 2.1				& $ 0.65 \pm 0.02 $		\\		
\swift\ XRT	& WT 	& 00321174020		& 54\,715.78				& 15.25					& 1.3				& $ 0.36 \pm 0.02$		\\		
\swift\ XRT	& WT 	& 00321174021		& 54\,717.02				& 16.50					& 4.6				& $ 0.62 \pm 0.01$		\\		
\swift\ XRT	& WT 	& 00321174022		& 54\,717.00				& 16.47					& 44.0			& $ 0.640 \pm 0.004$	\\		
\swift\ XRT	& WT 	& 00321174023		& 54\,717.01				& 16.48					& 14.7			& $ 0.642 \pm 0.007$	\\		
\swift\ XRT	& WT 	& 00321174024		& 54\,718.50				& 17.97					& 1.8				& $ 0.53 \pm 0.02$		\\		
\swift\ XRT	& WT 	& 00321174025		& 54\,719.66				& 19.13					& 1.1				& $ 0.58 \pm 0.02$		\\		
\swift\ XRT	& WT 	& 00321174026		& 54\,725.19				& 24.66					& 1.0				& $ 0.39 \pm 0.02$		\\		
\swift\ XRT	& WT 	& 00321174027		& 54\,726.75				& 26.22					& 1.0				& $ 0.53 \pm 0.02 $		\\		
\swift\ XRT	& WT 	& 00321174028		& 54\,727.47				& 26.94					& 1.7				& $ 0.47\pm 0.02$		\\		
\swift\ XRT	& WT 	& 00321174029		& 54\,728.52				& 27.99					& 1.6				& $ 0.43 \pm 0.02 $		\\		
\swift\ XRT	& WT 	& 00321174030		& 54\,729.19				& 28.66					& 1.3				& $ 0.45 \pm 0.02 $		\\	
\swift\ XRT	& WT 	& 00321174032		& 54\,731.49				& 30.96					& 1.7				& $ 0.50 \pm 0.02$		\\		
\swift\ XRT	& WT 	& 00321174033		& 54\,732.56				& 32.04					& 1.1				&  $ 0.46 \pm 0.02$		\\
\xmm\ EPN	& LW  	& 0552971401		& 54\,739.29				& 38.76	  				& 28.1			& $3.22\pm0.01$		\\ 	
\swift\ XRT	& WT 	& 00321174036		& 54\,741.57				& 41.04					& 0.9				& $ 0.29 \pm 0.02$		\\		
\swift\ XRT	& WT 	& 00321174037		& 54\,745.96				& 45.43					& 1.5				& $ 0.36 \pm 0.02$ 		\\		
\swift\ XRT	& WT 	& 00321174038		& 54\,748.09				& 47.57					& 1.0				& $ 0.26 \pm 0.02$		\\		
\swift\ XRT	& WT 	& 00321174039		& 54\,752.83				& 52.30					& 2.5				& $ 0.31 \pm 0.01$		\\		
\swift\ XRT	& WT 	& 00321174040		& 54\,755.54				& 55.01					& 3.8				& $ 0.36 \pm 0.01$		\\		
\swift\ XRT	& WT 	& 00321174041		& 54\,758.95				& 58.42					& 4.1				& $ 0.326 \pm 0.009$	\\		
\swift\ XRT	& WT 	& 00321174042		& 54\,762.86				& 62.34					& 3.7				& $ 0.35 \pm 0.01$		\\		
\swift\ XRT	& WT 	& 00321174043		& 54\,766.61				& 66.09					& 4.2				& $ 0.284 \pm 0.008$	\\		
\swift\ XRT	& WT 	& 00321174044		& 54\,775.16				& 74.63					& 2.9				& $ 0.31 \pm 0.01$		\\		
\swift\ XRT	& WT 	& 00321174045		& 54\,781.36				& 80.84					& 3.5				& $ 0.292 \pm 0.009$	\\		
\swift\ XRT	& WT 	& 00321174046		& 54\,789.69				& 89.16					& 3.3				& $ 0.254 \pm 0.009$	\\		
\swift\ XRT	& WT 	& 00321174047		& 54\,798.63				& 98.11					& 3.2				& $ 0.29 \pm 0.01$		\\	
\swift\ XRT	& WT 	& 00321174048		& 54\,803.13				& 102.60					& 2.4				& $ 0.25 \pm 0.01$		\\		
\swift\ XRT	& WT 	& 00321174049		& 54\,810.22				& 109.70					& 3.5				& $ 0.264 \pm 0.009$	\\		
\swift\ XRT	& WT 	& 00321174050		& 54\,818.20				& 117.68					& 3.6				& $ 0.242 \pm 0.008$	\\
\swift\ XRT	& WT 	& 00321174051		& 54\,825.07				& 124.55					& 4.5				& $ 0.227 \pm 0.007$	\\		
\swift\ XRT	& WT 	& 00321174053		& 54\,838.18				& 137.65					& 4.4				& $ 0.237 \pm 0.007$	\\		
\swift\ XRT	& WT 	& 00321174054		& 54\,846.86				& 146.33					& 2.8				& $ 0.224 \pm 0.009$	\\		
\swift\ XRT	& WT 	& 00321174055		& 54\,852.51				& 151.98					& 4.5				& $ 0.251 \pm 0.008$	\\		
\swift\ XRT	& WT 	& 00321174056		& 54\,859.51				& 158.98					& 3.3				& $ 0.215 \pm 0.008$	\\				
\hline
\end{tabular}
\end{table*}

\setcounter{table}{6}
\begin{table*}
  \caption{ -- \emph{continued}}
\label{tab:sgr0501}
\vspace{10pt}
\begin{tabular}{ccccccc} \hline
Instrument     	& Mode	& Obs. ID 			&  Mid point of observation	& Time since outburst onset 	& Exposure  		& Source net count rate \\
         		&          	&       			&  (MJD)					& (d)     			     	& (ks)   			& (counts s$^{-1}$)\\ 
\hline
\swift\ XRT	& WT 	& 00321174057		& 54\,866.12				& 165.60					& 3.3				& $ 0.253 \pm 0.009$		\\		
\swift\ XRT	& WT 	& 00321174058		& 54\,900.33				& 199.80					& 4.6				& $ 0.183 \pm 0.006$		\\		
\swift\ XRT	& WT 	& 00321174059		& 54\,910.38				& 209.85					& 6.0				& $ 0.177 \pm 0.005$		\\		
\swift\ XRT	& WT 	& 00321174060		& 54\,926.39				& 225.87					& 3.9				& $ 0.177 \pm 0.007$		\\		
\swift\ XRT	& WT 	& 00321174061		& 54\,940.56				& 240.04					& 5.2				& $ 0.126 \pm 0.005$		\\	
\xmm\ EPN	& SW 	& 0604220101		& 55\,073.93				& 373.40					& 24.6			& $0.224\pm0.003$\\
\swift\ XRT	& PC 	& 00321174062		& 55\,111.81				& 411.29					& 3.3				& $ 0.074 \pm 0.005$		\\		
\swift\ XRT	& PC 	& 00321174063		& 55\,114.81				& 414.29					& 3.9				& $ 0.077 \pm 0.004$		\\		
\swift\ XRT	& PC 	& 00321174064		& 55\,115.29				& 414.76					& 6.0				& $ 0.100 \pm 0.004$		\\		
\swift\ XRT	& PC 	& 00321174065		& 55\,172.27				& 471.75					& 13.4			& $ 0.072 \pm 0.002$		\\		
\swift\ XRT	& PC 	& 00321174066		& 55\,245.45				& 544.92					& 2.3				& $ 0.073 \pm 0.006$		\\		
\swift\ XRT	& PC 	& 00321174067		& 55\,246.89				& 546.36					& 4.4				& $ 0.069 \pm 0.004$		\\		
\swift\ XRT	& PC 	& 00321174068		& 55\,248.62				& 548.10					& 5.0				& $ 0.065 \pm 0.004$		\\			
\hline
\end{tabular}
\end{table*}

\begin{table*}
\centering
\caption{Log of all X-ray observations of \aa\ following the 2008 October outburst.
The outburst onset  occurred on MJD 54\,742.39453704 (Krimm et al. 2008).
All reported count rates are not corrected for pile-up. Part of these observations 
were already analysed by Israel et al. (2010), Ng et al. (2011), Dib et al. (2012) 
and Kuiper et al. (2012).}
\label{tab:1e1547_08}
\vspace{10pt}
\begin{tabular}{ccccccc} \hline
Instrument     	& Mode	& Obs. ID 			&  Mid point of observation	& Time since outburst onset 	& Exposure  		& Source net count rate \\
         		&          	&       			&  (MJD)					& (d)     			     	& (ks)   			& (counts s$^{-1}$)\\ 
\hline			
\swift\ XRT	& PC 	& 00330353000 	& 54\,742.46  				& 0.07  					& 4.1 			& $0.60 \pm 0.01$	\\
\swift\ XRT	& WT	& 00330353001 	& 54\,742.80				& 0.40					& 14.2 			& $0.539 \pm 0.006$ 	\\
\swift\ XRT	& WT	& 00330353002  	& 54\,743.79 				& 1.40					& 4.8 			& $0.45 \pm 0.01$  	\\
\swift\ XRT 	& WT 	& 00330353004  	& 54\,745.14 				& 2.75					& 10.5 			& $0.389 \pm 0.006$  	\\
\swift\ XRT	& WT 	& 00330353005  	& 54\,746.51 				& 4.12					& 7.7 			& $0.407 \pm 0.007$	\\
\cxo\ ACIS-S	& CC 	& 8811			& 54\,746.60 		 		& 4.21 					& 12.1 			& $1.39 \pm 0.01$ \\
\swift\ XRT	& WT 	& 00330353006  	& 54\,747.11 				& 4.72					& 4.5 			& $0.380 \pm 0.009$	\\
\swift\ XRT	& WT 	& 00330353007  	& 54\,748.39 				& 6.00					& 3.7  			& $0.314 \pm 0.009$	\\
\swift\ XRT	& WT 	& 00330353008 	& 54\,749.46 				& 7.07					& 3.9 			& $0.328 \pm 0.009$	\\
\cxo\ ACIS-S	& CC 	& 8812			& 54\,749.50 				& 7.11					& 15.1 			& $1.209 \pm 0.009$ \\
\swift\ XRT	& WT 	& 00330353010  	& 54\,751.53				& 9.14 					& 3.7				& $0.309 \pm 0.009$	\\
\swift\ XRT	& WT 	& 00330353011  	& 54\,752.40				& 10.01					& 3.4 			& $0.31 \pm 0.01$	\\
\swift\ XRT	& WT 	& 00330353012  	& 54\,755.14				& 12.75					& 4.0 			& $0.303 \pm 0.009$	\\
\swift\ XRT	& WT 	& 00330353013  	& 54\,757.38				& 14.99					& 5.0  			& $0.352 \pm 0.008$	\\
\cxo\ ACIS-S	& CC 	& 8813			& 54\,757.60 				& 15.21 					& 10.1 			& $1.20 \pm 0.01$ \\
\swift\ XRT	& WT 	& 00330353014  	& 54\,759.80				& 17.41					& 3.9 			& $0.322 \pm 0.009$	\\
\cxo\ ACIS-S	& CC 	& 10792			& 54\,760.80				& 18.41 					& 10.1 			& $1.11 \pm 0.01$ \\
\swift\ XRT	& WT 	& 00330353015 	& 54\,761.43				& 19.04					& 3.9  			& $0.35 \pm 0.01$	\\
\swift\ XRT	& WT 	& 00330353016  	& 54\,763.21				& 20.82					& 3.6 			& $0.34 \pm 0.01$	\\
\cxo\ ACIS-S	& CC 	& 8814			& 54\,765.10 				& 22.71					& 23.1 			& $1.032 \pm 0.007$\\
\hline	
\end{tabular}
\end{table*}

\begin{table*}
\centering
\caption{Log of all X-ray observations of \aa\ following the 2009 January outburst. 
The outburst onset  occurred on MJD 54\,853.03740937 (Connaughton \& Briggs 
2009). All reported count rates are not corrected for pile-up. We did not include the 
following \swift\ observations: obs. ID 0003095041, because the source PSF falls on a 
column of bad pixels in this case; obs. ID 00030956050, 00090404025 and 00091032011, 
owing to the low number of net source counts (about 80 counts in the former two cases 
and 40 counts in the latter case). Part of these observations were already analysed by 
Bernardini et al. (2011b), Ng et al. (2011), Scholz \& Kaspi (2011), Dib et al. (2012) and 
Kuiper et al. (2012).}
\label{tab:1e1547_09}
\vspace{10pt}
\begin{tabular}{ccccccc} \hline
Instrument     	& Mode	& Obs. ID 			&  Mid point of observation	& Time since outburst onset 	& Exposure  		& Source net count rate \\
         		&          	&       			&  (MJD)					& (d)     			     	& (ks)   			& (counts s$^{-1}$)\\ 
\hline	
\swift\ XRT	 & PC 	& 00340923000 	& 54\,854.65  				& 1.61					& 1.7				& $ 0.99 \pm 0.02$\\
\cxo/HETG	 & CC 	& 10185			& 54\,855.04				& 2.00 					& 10.1 			& $ 1.04 \pm 0.01$\\
\swift\ XRT	 & WT	& 00090007026  	& 54\,855.05  				& 2.01 					& 8.2				& $ 1.18 \pm 0.01$\\
\swift\ XRT	 & PC 	& 00340986000  	& 54\,855.21 				& 2.17					& 2.9 			& $ 0.77 \pm 0.02$	\\
\swift\ XRT	 & PC	& 00030956031  	& 54\,855.32 				& 2.28					& 2.5 			& $ 0.69 \pm 0.02$\\
\swift\ XRT	 & WT	& 00090007027 	& 54\,856.19 				& 3.15					& 3.3 			& $ 1.10 \pm 0.02$\\
\swift\ XRT	 & PC  	& 00341055000  	& 54\,856.20 				& 3.16					& 4.0 			& $ 0.68 \pm 0.01$\\
\cxo\ ACIS-S	 & CC 	 & 10186 			& 54\,856.73				& 3.70 					& 12.1 			& $ 3.33 \pm 0.02$\\
\swift\ XRT	 & PC 	& 00341114000  	& 54\,856.96 				& 3.92					& 4.6 			& $ 0.64 \pm 0.01$\\
\swift\ XRT	 & WT 	& 00090007028  	& 54\,857.25 				& 4.21					& 3.5 			& $ 0.99 \pm 0.02$\\
\swift\ XRT	 & PC 	& 00030956032  	& 54\,858.20 				& 5.16					& 6.2 			& $ 0.67 \pm 0.01$\\
\swift\ XRT	 & WT 	& 00090007029  	& 54\,858.43 				& 5.39					& 1.8 			& $ 0.86 \pm 0.02$\\
\swift\ XRT	 & PC 	& 00030956033  	& 54\,859.59 				& 6.55					& 5.0 			& $ 0.57 \pm 0.01$\\
\swift\ XRT	 & WT 	& 00090007030  	& 54\,859.91				& 6.87					& 1.9 			& $ 0.55 \pm 0.02$\\
\swift\ XRT	 & PC 	& 00030956034  	& 54\,860.18				& 7.14					& 5.9 			& $ 0.63 \pm 0.01$\\
\swift\ XRT	 & WT 	& 00090007031  	& 54\,860.78 				& 7.74					& 2.1 		 	& $ 0.83 \pm 0.02$\\
\cxo\ ACIS-S	 & CC 	 & 10187			& 54\,860.84 				& 7.80 					& 13.1 			& $2.69 \pm 0.01$\\
\swift\ XRT	 & WT 	& 00090007032  	& 54\,861.69 				& 8.65					& 2.9 			& $ 0.74 \pm 0.02$\\
\swift\ XRT	 & WT 	& 00030956035  	& 54\,861.72 				& 8.68					& 3.0 			& $ 0.80 \pm 0.02$\\
\swift\ XRT	 & WT	& 00030956036  	& 54\,862.18 				& 9.14					& 3.0 			& $ 0.72 \pm 0.02$\\
\swift\ XRT	 & WT 	& 00090007033  	& 54\,862.83 				& 9.79					& 2.5 			& $ 0.79 \pm 0.02$\\
\swift\ XRT	 & WT 	& 00030956037  	& 54\,863.61 				& 10.57					& 2.0 			& $ 0.68 \pm 0.02$\\
\swift\ XRT	 & WT 	& 00090007034  	& 54\,863.76 				& 10.72					& 2.0 			& $ 0.76 \pm 0.02$\\
\swift\ XRT	 & PC 	& 00030956038  	& 54\,865.66 				& 12.62					& 5.9 			& $ 0.474 \pm 0.009$\\
\swift\ XRT	 & PC 	& 00341965000	& 54\,865.84				& 12.80					& 0.9				& $ 0.68 \pm 0.03$\\
\xmm\ EPN	 & FF	& 0560181101		& 54\,866.09				& 13.06					& 48.9			& $ 4.99 \pm 0.01$\\
\swift\ XRT	 & PC 	& 00030956039  	& 54\,866.84 				& 13.80					& 6.1  			& $ 0.56 \pm 0.01$\\
\swift\ XRT	 & WT 	& 00030956040  	& 54\,867.57 				& 14.53					& 6.1  			& $ 0.60 \pm 0.01$\\
\cxo\ ACIS-S	 & CC 	& 10188			& 54\,868.68  				& 15.64 					& 14.3 			& $2.35 \pm 0.01$\\
\swift\ XRT	 & WT 	& 00030956042  	& 54\,869.59 				& 16.55					& 1.6 			& $ 0.68 \pm 0.02$\\
\swift\ XRT	 & WT 	& 00090007036  	& 54\,874.32 				& 21.28					& 4.6 			& $ 0.47 \pm 0.01$\\
\swift\ XRT	 & WT 	& 00090007037  	& 54\,884.63 				& 31.59					& 4.6 			& $ 0.55 \pm 0.01$\\
\swift\ XRT	 & WT 	& 00090007038  	& 54\,894.61 				& 41.57					& 3.9 			& $ 0.45 \pm 0.01$\\
\swift\ XRT	 & WT 	& 00090007039  	& 54\,904.45 				& 51.41					& 4.0 			& $ 0.46 \pm 0.01$\\
\swift\ XRT	 & WT 	& 00090007040  	& 54\,914.90 				& 61.86					& 4.2 		 	& $ 0.288 \pm 0.008$\\
\swift\ XRT	 & WT  	& 00030956043  	& 54\,950.47 				& 97.43					& 1.7 			& $ 0.44 \pm 0.02$  \\
\swift\ XRT	 & WT  	& 00030956044  	& 54\,964.32		 		& 111.28					& 1.8	 			& $ 0.43 \pm 0.02$  \\
\swift\ XRT	 & WT  	& 00030956045  	& 54\,978.52 				& 125.48					& 2.2 			& $ 0.36 \pm 0.01$  \\
\swift\ XRT	 & PC  	& 00030956047  	& 55\,006.88		 		&153.84					& 1.8 	 		& $ 0.29 \pm 0.01$  \\
\swift\ XRT	 & PC  	& 00030956048  	& 55\,020.80		 		&167.76					& 2.5 	 		& $ 0.29 \pm 0.01$  \\
\swift\ XRT	 & PC  	& 00030956049  	& 55\,034.12		 		&181.08					& 1.7	 			& $ 0.33 \pm 0.01$  \\
\swift\ XRT	 & PC  	& 00030956051  	& 55\,062.11		 		& 209.07					& 2.4	 			& $ 0.30 \pm 0.01$  \\
\swift\ XRT	 & WT  	& 00030956053  	& 55\,090.52 				& 237.48					& 1.5 			& $ 0.32 \pm 0.01$  \\
\swift\ XRT	 & WT 	& 00030956054  	& 55\,104.63 				& 251.59					& 3.3  			& $ 0.35 \pm 0.01$	\\
\swift\ XRT	 & WT 	& 00030956055   	& 55\,118.09 				& 265.05					& 2.0 			& $ 0.25 \pm 0.01$  \\
\swift\ XRT	 & WT  	& 00030956056  	& 55\,200.35	 			& 347.31					& 1.9	 			& $ 0.28 \pm 0.01$  \\
\swift\ XRT	 & WT  	& 00030956057  	& 55\,214.10	 			& 361.06					& 2.0	 			& $ 0.25 \pm 0.01 $  \\
\swift\ XRT	 & WT  	& 00030956058  	& 55\,228.87		 		& 375.83					& 2.0	 			& $0.26 \pm 0.01$  \\
\xmm\ EPN	 & LW 	& 0604880101 		& 55\,237.44				& 384.41					& 39.4 			& $1.819 \pm 0.007$ \\	
\swift\ XRT	 & WT  	& 00030956059  	& 55\,256.06	 			& 403.02					& 2.0 	 		& $0.25 \pm 0.01$  \\
\swift\ XRT	 & WT  	& 00030956060  	& 55\,270.86	 			& 417.82					& 1.9 	 		& $0.23 \pm 0.01$  \\
\swift\ XRT	 & WT  	& 00030956061  	& 55\,284.96	 			& 431.92					& 2.0	 			& $0.28 \pm 0.01$  \\
\hline
\end{tabular}
\end{table*}


\setcounter{table}{8}
\begin{table*}
  \caption{ -- \emph{continued}}
\label{tab:1e1547_09}
\vspace{10pt}
\begin{tabular}{ccccccc} \hline
Instrument     	& Mode	& Obs. ID 			&  Mid point of observation	& Time since outburst onset 	& Exposure  		& Source net count rate \\
         		&          	&       			&  (MJD)					& (d)     			     	& (ks)   			& (counts s$^{-1}$)\\ 
\hline
\swift\ XRT	 & PC  	& 00090404001  	& 55\,287.82	 			& 434.78					& 0.9  	 		& $0.14 \pm 0.01$  \\
\swift\ XRT	 & PC  	& 00090404002  	& 55\,291.37	 			& 438.33					& 3.6  			& $0.180 \pm 0.007$  \\
\swift\ XRT	 & PC  	& 00090404003  	& 55\,298.72	 			& 445.68					& 5.7 	 		& $0.156 \pm 0.005$  \\
\swift\ XRT	 & PC  	& 00090404004  	& 55\,307.90	 			& 454.86					& 3.5 	 		& $0.195 \pm 0.008$  \\
\swift\ XRT	 & PC  	& 00090404005  	& 55\,317.22	 			& 464.18					& 2.3	 			& $0.166 \pm 0.009$  \\
\swift\ XRT	 & PC  	& 00090404006  	& 55\,327.63	 			& 474.59					& 3.0	 			& $0.171 \pm 0.008$  \\
\swift\ XRT	 & PC  	& 00090404007  	& 55\,337.91	 			& 484.87					& 3.0	 			& $0.188 \pm 0.008$  \\
\swift\ XRT	 & PC  	& 00090404008  	& 55\,347.50	 			& 494.46					& 2.9		 		& $0.187 \pm 0.008$  \\
\swift\ XRT	 & WT  	& 00030956062  	& 55\,357.31	 			& 504.27					& 2.8	 			& $0.25 \pm 0.01$  \\
\swift\ XRT	 & PC  	& 00090404010  	& 55\,377.39	 			& 524.35					& 2.8	 			& $0.181 \pm 0.008$  \\
\swift\ XRT	 & PC  	& 00090404011  	& 55\,387.59	 			& 534.55					& 3.3 	 		& $0.148 \pm 0.007$  \\
\swift\ XRT	 & PC  	& 00090404012  	& 55\,398.77	 			& 545.73					& 3.2	 			& $0.167 \pm 0.007$  \\
\swift\ XRT	 & PC  	& 00090404013  	& 55\,407.22		 		& 554.18					& 3.6	 			& $0.162 \pm 0.007$  \\
\swift\ XRT	 & PC  	& 00090404014  	& 55\,417.41		 		& 564.37					& 3.0	 			& $0.193 \pm 0.008$  \\
\swift\ XRT	 & PC  	& 00090404015  	& 55\,427.38	 			& 574.34					& 2.9 	 		& $0.189 \pm 0.008$  \\
\swift\ XRT	 & PC  	& 00090404016  	& 55\,436.25	 			& 583.21					& 3.6 	 		& $0.171 \pm 0.007$  \\
\swift\ XRT	 & PC  	& 00090404017  	& 55\,447.08		 		& 594.04					& 3.0	 			& $0.144 \pm 0.007$  \\
\swift\ XRT	 & PC  	& 00090404018  	& 55\,457.53	 			& 604.49					& 3.1	 			& $0.180 \pm 0.008$  \\
\swift\ XRT	 & PC  	& 00090404019  	& 55\,467.09	 			& 614.05					& 3.5 	 		& $0.191 \pm 0.007$  \\
\swift\ XRT	 & PC  	& 00090404020  	& 55\,477.26		 		& 624.22					& 3.5	 			& $0.159 \pm 0.007$  \\
\swift\ XRT	 & PC  	& 00090404021  	& 55\,487.14	 			& 634.10					& 3.1	 			& $0.172 \pm 0.007$  \\
\swift\ XRT	 & PC  	& 00090404022  	& 55\,493.86		 		& 640.82					& 2.8 	 		& $0.186 \pm 0.008$  \\
\swift\ XRT	 & PC  	& 00090404023  	& 55\,567.69				& 714.65					& 3.2				& $ 0.154 \pm 0.007$ \\
\swift\ XRT	 & PC  	& 00090404024  	& 55\,578.14				& 725.10					& 3.0				& $ 0.179 \pm 0.008$	\\
\swift\ XRT	 & PC  	& 00090404026  	& 55\,607.45				& 754.41					& 2.8				& $ 0.107 \pm 0.006$	\\
\swift\ XRT	 & PC  	& 00090404027  	& 55\,617.27				& 764.23					& 3.2				& $ 0.134 \pm 0.007$	\\
\swift\ XRT	 & PC  	& 00090404028  	& 55\,627.91				& 774.88					& 3.3				& $ 0.175 \pm 0.007$	\\
\swift\ XRT	 & PC  	& 00090404029  	& 55\,637.86				& 784.82					& 2.3				& $ 0.148 \pm 0.008$	\\
\swift\ XRT	 & PC  	& 00090404030  	& 55\,647.43				& 794.39					& 2.9				& $ 0.153 \pm 0.007$	\\
\swift\ XRT	 & PC  	& 00091032001 	& 55\,656.33				& 803.29					& 2.5 			& $ 0.128 \pm 0.007$	\\
\swift\ XRT	 & PC  	& 00091032002 	& 55\,666.55				& 813.51					& 3.1 			& $ 0.135 \pm 0.007$	\\
\swift\ XRT	 & PC  	& 00091032003 	& 55\,676.45				& 823.41					& 3.3				& $ 0.141 \pm 0.007$	\\
\swift\ XRT	 & PC  	& 00091032004 	& 55\,687.09				& 834.05					& 2.7				& $ 0.132 \pm 0.007$	\\
\swift\ XRT	 & PC  	& 00091032005 	& 55\,696.83				& 843.79					& 3.2 			& $ 0.148 \pm 0.007$	\\
\swift\ XRT	 & PC  	& 00091032006 	& 55\,706.57				& 853.54					& 2.7 			& $ 0.151 \pm 0.008$	\\
\swift\ XRT	 & PC  	& 00091032007 	& 55\,716.46				& 863.42					& 2.8 			& $ 0.138 \pm 0.007$	\\
\swift\ XRT	 & PC  	& 00091032008 	& 55\,726.62				& 873.59					& 3.0 			& $ 0.093 \pm 0.006$	\\
\swift\ XRT	 & PC  	& 00091032009 	& 55\,736.73				& 883.69					& 2.3				& $ 0.156 \pm 0.008$	\\
\swift\ XRT	 & PC  	& 00091032011  	& 55\,749.45				& 896.41					& 1.4				& $ 0.16 \pm 0.01$		\\
\swift\ XRT	 & PC  	& 00091032012 	& 55\,756.27				& 903.23					& 3.0				& $ 0.154 \pm 0.007$	\\
\swift\ XRT	 & PC  	& 00091032013 	& 55\,766.60				& 913.56					& 2.9 			& $ 0.161 \pm 0.008$	\\
\swift\ XRT	 & PC  	& 00091032015 	& 55\,780.15				& 927.11					& 1.5				& $ 0.14 \pm 0.01$		\\
\swift\ XRT	 & PC  	& 00091032016 	& 55\,786.43				& 933.39					& 1.7				& $ 0.146 \pm 0.009$	\\
\swift\ XRT	 & PC  	& 00091032017 	& 55\,790.71				& 937.68					& 3.3				& $ 0.164 \pm 0.007$	\\
\swift\ XRT	 & PC  	& 00091032018 	& 55\,796.84				& 943.80					& 2.6 			& $ 0.157 \pm 0.008$	\\
\swift\ XRT	 & PC  	& 00091032019 	& 55\,806.41				& 953.37					& 2.9 			& $ 0.150 \pm 0.007$	\\
\swift\ XRT	 & PC  	& 00091032020 	& 55\,816.45				& 963.41					& 3.2				& $ 0.153 \pm 0.007$	\\
\swift\ XRT	 & WT  	& 00091032021 	& 55\,826.23				& 973.20					& 2.6 			& $ 0.25 \pm 0.01$		\\
\swift\ XRT	 & WT  	& 00091032022 	& 55\,836.27				& 983.23					& 3.1 			& $ 0.214 \pm 0.008$	\\
\swift\ XRT	 & PC  	& 00091032023 	& 55\,846.88				& 993.84					& 3.0				& $ 0.146 \pm 0.007$	\\
\swift\ XRT	 & PC  	& 00091032024 	& 55\,856.40				& 1003.36					& 2.7 			& $ 0.129 \pm 0.007$	\\
\hline
\end{tabular}
\end{table*}


\begin{table*}
\centering
\caption{Log of all X-ray observations of \lowba\ following the 2009 June outburst. 
The outburst onset  occurred on MJD 54\,987.86167685 (van der Horst et al. 2009). 
Part of these observations were already analysed by Esposito et al. (2010a) and Rea 
et al. (2010, 2013a).}
\label{tab:sgr0418}
\vspace{10pt}
\begin{tabular}{ccccccc} \hline
Instrument     	& Mode	& Obs. ID 				&  Mid point of observation	& Time since outburst onset 	& Exposure  		& Source net count rate \\
         		&          	&       				&  (MJD)					& (d)     			     	& (ks)   			& (counts s$^{-1}$)\\ 
\hline
\swift\ XRT  	& PC		& 00031422001 		& 55\,020.91 				& 33.05					& 2.9 			& $0.241 \pm 0.009$  \\
\swift\ XRT 	& PC 	& 00031422002 		& 55\,021.41				& 33.54					& 10.6 			& $0.249 \pm 0.005$ \\
\swift\ XRT 	& PC 	& 00031422003 		& 55\,022.15    				& 34.29					& 5.6 			& $0.188 \pm 0.006$  \\
\swift\ XRT 	& WT 	& 00031422004  		& 55\,024.32   				& 36.46					& 7.1 			& $0.286 \pm 0.006$ \\
\swift\ XRT	& WT  	& 00031422006   		& 55\,027.81				& 39.95					& 7.7 			& $0.321 \pm 0.007$ \\
\swift\ XRT 	& WT  	& 00031422007     		& 55\,028.51				& 40.65					& 16.4 			& $0.276 \pm 0.004$ \\
\xmm\ EPN  	& SW	& 0610000601   		& 55\,056.26				& 68.40					& 45.0 			& $1.480 \pm 0.006$ \\
\swift\ XRT  	& PC  	& 00031422008 		& 55\,095.42   				& 107.56					& 9.4 			& $0.067 \pm 0.003$ \\
\swift\ XRT  	& PC  	& 00031422009  		& 55\,096.50  				& 108.64					& 7.6 			& $0.077 \pm 0.003$ \\
\swift\ XRT   	& PC 	& 00031422010 		& 55\,143.49 				& 155.63					& 15.1			& $0.046 \pm 0.002$ \\
\swift\ XRT  	& PC 	& 00031422011$^{a1}$   	& 55\,210.51  				& 222.64					& 3.6 			& $0.020 \pm 0.002$ \\
\swift\ XRT  	& PC 	& 00031422012$^{a1}$   	& 55\,211.66  				& 223.79					& 3.6 			& $0.015 \pm 0.002$ \\
\swift\ XRT  	& PC 	& 00031422013$^{a1}$   	& 55\,212.45  				& 224.59					&  4.0 			& $0.021 \pm 0.002$ \\
\swift\ XRT  	& PC 	& 00031422014$^{a1}$   	& 55\,213.36 				& 225.50					&  3.7 			& $0.024 \pm 0.003$ \\
\swift\ XRT  	& PC 	& 00031422015$^{a2}$   	& 55\,241.84  				& 253.97					&  4.5 			& $0.020 \pm 0.002$ \\
\swift\ XRT  	& PC 	& 00031422016$^{a2}$   	& 55\,242.84  				& 254.97					&  4.5 			& $0.017 \pm 0.002$ \\
\swift\ XRT  	& PC 	& 00031422017$^{a2}$   	& 55\,243.30 				& 255.44					&  4.5 			& $0.017 \pm 0.002$ \\
\swift\ XRT  	& PC 	& 00031422018$^{a2}$   	& 55\,244.68  				& 256.82					&  4.6 			& $0.016 \pm 0.002$ \\
\swift\ XRT  	& PC 	& 00031422019$^{a2}$   	& 55\,245.81 				& 257.95					&  3.4 			& $0.018 \pm 0.002$ \\
\swift\ XRT  	& PC 	& 00031422020$^{a2}$   	& 55\,246.25  				& 258.39					& 3.2 			& $0.017 \pm 0.002$ \\
\swift\ XRT  	& PC 	& 00031422021$^{a3}$   	& 55\,386.63  				& 398.77					&  3.6 			& $0.005 \pm 0.001$ \\
\swift\ XRT  	& PC 	& 00031422022$^{a3}$   	& 55\,387.87  				& 400.00					&  5.1 			& $0.0023 \pm 0.0007$ \\
\swift\ XRT  	& PC 	& 00031422023$^{a3}$   	& 55\,388.30   				& 400.44					& 5.0 			& $0.0026 \pm 0.0008$ \\
\swift\ XRT  	& PC 	& 00031422024$^{a3}$   	& 55\,389.07   				& 401.21					& 5.4 			& $0.0032 \pm 0.0008$ \\
\swift\ XRT  	& PC 	& 00031422025$^{a3}$   	& 55\,390.14    				& 402.28					& 4.8 			& $0.0014 \pm 0.0007$ \\
\cxo\ ACIS-S 	& TE		& 12312 				& 55\,400.81				& 412.95					& 27.2 			& $0.0170 \pm 0.0008$ \\
\xmm\ EPN 	& FF		& 0605852201 			& 55\,463.31				& 475.45					& 8.6 			& $0.040 \pm 0.002$ \\
\cxo\ ACIS-S 	& TE		& 13148 				& 55\,529.43				& 541.57					& 27.2 			& $0.0045 \pm 0.0004$ \\
\xmm\ EPN 	& LW	& 0672670201			& 55\,630.34				& 642.48					& 11.6 			& $0.007 \pm 0.001$ \\
\cxo\ ACIS-S 	& TE 	& 13235 				& 55\,762.56				& 774.70					& 69.8 			& $0.0034 \pm 0.0002$ \\
\xmm\ EPN 	& LW	& 0672670401$^{a4}$ 	& 55\,813.84				& 825.98				 	& 25.9 			& $0.0062 \pm 0.0006$ \\
\xmm\ EPN	& LW	& 0672670501$^{a4}$	& 55\,816.19				& 828.33					& 28.7 			& $0.0070 \pm 0.0006$ \\
\cxo\ ACIS-S 	& TE		& 13236 				& 55\,891.94				& 904.08					& 68.0			& $0.0029 \pm 0.0002$ \\
\xmm\ EPN  	& LW	& 0693100101 			& 56\,165.04				& 1177.18					& 54.3 			& $0.0060 \pm 0.0004$ \\
\xmm\ EPN  	& LW	& 0723810101$^{a5}$	& 56\,520.00				& 1532.14					& 32.7			& $0.0045 \pm 0.0005$\\
\xmm\ EPN  	& LW	& 0723810201$^{a5}$	& 56\,522.19				& 1534.33					& 35.3			& $0.0053 \pm 0.0005$\\
\xmm\ EPN  	& LW	& 0741970201$^{a6}$	& 56\,883.20				& 1895.34					& 36.0			& $0.0046 \pm 0.0004$\\
\xmm\ EPN  	& LW	& 0741970301$^{a6}$	& 56\,885.20				& 1897.33					& 41.2			& $0.0044 \pm 0.0004$\\
\xmm\ EPN  	& LW	& 0741970401$^{a6}$	& 56\,887.20				& 1899.34					& 30.9			& $0.0038 \pm 0.0004$\\
\hline
\end{tabular}
\begin{tablenotes}
\item \emph{Note}. $^a$The spectral files and responses of these observations were combined to improve the fit statistics. 
\end{tablenotes}
\end{table*}

\begin{table*}
\centering
\caption{Log of all X-ray observations of SGR\,1833$-$0832\ following the 2010 March 
outburst. The outburst onset  occurred on MJD 55\,274.77418981 (Gelbord et al. 2010). 
Part of these observations were already analysed by G{\"o}{\u g}{\"u}{\c s} et al. (2010a) 
and Esposito et al. (2011).}
\label{tab:sgr1833}
\vspace{10pt}
\begin{tabular}{ccccccc} \hline
Instrument     	& Mode	& Obs. ID 				&  Mid point of observation	& Time since outburst onset 	& Exposure  		& Source net count rate \\
         		&          	&       				&  (MJD)					& (d)     			     	& (ks)   			& (counts s$^{-1}$)\\ 
\hline
\swift\ XRT 	& PC		& 00416485000 		& 55\,275.21 				& 0.44					& 29.0			& $0.031 \pm 0.001$	\\
\swift\ XRT 	& PC		& 00416485001 		& 55\,276.41 				& 1.64					& 10.7			& $0.036 \pm 0.002$	\\
\swift\ XRT 	& WT	& 00416485002		& 55\,276.90 				& 2.13					& 9.9				& $0.121 \pm 0.004$	\\
\swift\ XRT 	& PC		& 00416485003 		& 55\,277.60 				& 2.82					& 13.3			& $0.038 \pm 0.002$	\\
\cxo\ ACIS-I	& TE		& 11114				& 55\,278.27 				& 3.49					& 33.1			& $0.100 \pm 0.002$	\\
\xmm\ EPN   	& FF		& 0605851901 			& 55\,278.67 				& 3.90					& 18.5			& $0.320 \pm 0.004$ \\
\swift\ XRT 	& PC		& 00416485004 		& 55\,278.68 				& 3.91					& 12.8			& $0.032 \pm 0.002$	\\
\swift\ XRT 	& PC		& 00416485005 		& 55\,279.52 				& 4.75					& 10.3			& $0.034 \pm 0.002$	\\
\swift\ XRT 	& PC		& 00416485006 		& 55\,280.53 				& 5.76					& 9.9				& $0.036 \pm 0.002$	\\
\swift\ XRT 	& PC		& 00416485007 		& 55\,281.73				& 6.96					& 10.0			& $0.032 \pm 0.002$	\\
\swift\ XRT 	& PC		& 00416485008 		& 55\,282.66 				& 7.89					& 9.8				& $0.036 \pm 0.002$	\\
\swift\ XRT 	& PC		& 00416485009 		& 55\,283.38 				& 8.60					& 10.9			& $0.034 \pm 0.002$	\\
\swift\ XRT 	& PC		& 00416485010 		& 55\,284.71 				& 9.93					& 9.5				& $0.032 \pm 0.002$	\\
\swift\ XRT 	& PC		& 00416485011 		& 55\,286.49 				& 11.72					& 7.9				& $0.030 \pm 0.002$	\\
\xmm\ EPN  	& FF		& 0605852001 			& 55\,288.63 				& 13.86					& 18.2			& $0.317 \pm 0.005$	\\
\swift\ XRT 	& PC		& 00416485012 		& 55\,289.80 				& 15.03					& 10.0			& $0.026 \pm 0.002$	\\
\swift\ XRT 	& PC		& 00416485013 		& 55\,293.75 				& 18.98					& 10.1			& $0.031 \pm 0.002$	\\
\swift\ XRT 	& PC		& 00416485014 		& 55\,298.67 				& 23.90					& 5.1				& $0.026 \pm 0.002$	\\
\swift\ XRT 	& PC		& 50041648015 		& 55\,299.24 				& 24.46					& 4.0				& $0.029 \pm 0.003$	\\
\xmm\ EPN   	& FF		& 0605852101 			& 55\,299.30 				& 24.53					& 14.6			& $0.266 \pm 0.004$	\\
\swift\ XRT 	& PC		& 00416485016 		& 55\,301.24 				& 26.46					& 9.4				& $0.026 \pm 0.002$	\\
\swift\ XRT 	& PC		& 00416485017 		& 55\,304.96 				& 30.19					& 8.8				& $0.025 \pm 0.002$	\\
\swift\ XRT 	& PC		& 00416485018 		& 55\,307.36 				& 32.59					& 10.3			& $0.025 \pm 0.002$	\\
\swift\ XRT 	& PC		& 00416485019 		& 55\,310.05 				& 35.28					& 7.6				& $0.027 \pm 0.002$	\\
\swift\ XRT 	& PC		& 00416485020 		& 55\,315.82 				& 41.05					& 5.5				& $0.026 \pm 0.002$	\\
\swift\ XRT 	& PC		& 00416485021 		& 55\,316.16 				& 41.39					& 4.4				& $0.030 \pm 0.003$	\\
\swift\ XRT 	& PC		& 00416485022 		& 55\,340.39 				& 65.62					& 18.0			& $0.019 \pm 0.001$	\\
\swift\ XRT 	& PC		& 00416485023$^{a}$	& 55\,432.30 				& 157.52					& 5.3				& $0.012 \pm 0.002$	\\
\swift\ XRT 	& PC		& 00416485024$^{a}$ 	& 55\,433.67 				& 158.90					& 2.2				& $0.009 \pm 0.002$	\\
\swift\ XRT 	& PC		& 00416485025$^{a}$ 	& 55\,434.51 				& 159.73					& 9.9				& $0.010 \pm 0.001$	\\
\swift\ XRT 	& PC		& 00416485026$^{a}$ 	& 55\,435.28 				& 160.51					& 2.5				& $0.011 \pm 0.002$	\\			
\hline
\end{tabular}
\begin{tablenotes}
\item \emph{Note}. $^a$The spectral files and responses of these observations were combined to improve the fit statistics. 
\end{tablenotes}
\end{table*}


\begin{table*}
\centering
\caption{Log of all X-ray observations of \lowbb\ following the 2011 July outburst. The 
outburst onset occurred on MJD 55\,756.53318403 (Cummings et al. 2011). We did 
not include the following \swift\ observations: obs. ID 00032033038, because the source 
was located at the edge of the detector and only a few counts were collected from 
the source; obs. ID 00032033041, because the source PSF falls on a column of bad pixels. 
Moreover, one \cxo\ observation was performed with the HRC-I (obs. ID 13511), with 
no sufficient spectral information, and thus was not included in our analysis as well. Part 
of these observations were already analysed by Livingstone et al. (2011b), Rea et al. (2012a), 
Scholz et al. (2012), Scholz, Kaspi \& Cumming (2014) and Rodr\'iguez Castillo et al. (2016).}
\label{tab:swift1822}
\vspace{10pt}
\begin{tabular}{ccccccc} \hline
Instrument     	& Mode	& Obs. ID 			&  Mid point of observation	& Time since outburst onset 	& Exposure  		& Source net count rate \\
         		&          	&       			&  (MJD)					& (d)     			     	& (ks)   			& (counts s$^{-1}$)\\ 
\hline	
\swift\ XRT 	& PC		& 00032033001 	& 55\,757.75 				& 1.2	2					& 1.6				& $ 2.18 \pm 0.04$ \\				
\swift\ XRT 	& WT	& 00032033002 	& 55\,758.68 				& 2.1	5					& 2.0				& $ 5.16 \pm 0.05$	 \\
\swift\ XRT 	& WT	& 00032033003 	& 55\,759.69 				& 3.1	6					& 2.0				& $ 4.29 \pm 0.05$	\\
\swift\ XRT 	& WT	& 00032033005 	& 55\,761.54 				& 5.01					& 0.5				& $ 3.98 \pm $	0.09 \\
\swift\ XRT 	& WT	& 00032033006 	& 55\,762.24 				& 5.71					& 1.8				& $ 3.78 \pm 0.05$	 \\
\swift\ XRT 	& WT	& 00032033007 	& 55\,763.30 				& 6.77					& 1.6				& $ 3.46 \pm 0.05$	\\
\swift\ XRT 	& WT	& 00032033008 	& 55\,765.85 				& 9.32					& 2.2				& $ 2.10 \pm 0.03$	 \\
\swift\ XRT 	& WT	& 00032033009  	& 55\,766.28 				& 9.75					& 1.7				& $ 2.98 \pm 0.04$	 \\
\cxo\ ACIS-S 	& CC	& 12612 			& 55\,769.28 				& 12.75					& 15.0 			& $11.64 \pm 0.03$	\\
\swift\ XRT 	& WT	& 00032033010 	& 55\,769.50 				& 12.97					& 2.1				& $ 2.54 \pm 0.03$	 \\
\swift\ XRT 	& WT	& 00032033011  	& 55\,770.40 				& 13.87					& 2.1 			& $ 2.44 \pm 0.03$	\\
\swift\ XRT 	& WT	& 00032033012 	& 55\,771.23 				& 14.70					& 2.1 			& $ 2.38 \pm 0.03$	\\		
\swift\ XRT 	& WT	& 00032033013  	& 55\,772.40 				& 15.87					& 2.1				& $ 2.13 \pm 0.03$	\\
\cxo\ ACIS-S	& CC	& 12613  			& 55\,777.22 				& 20.68					& 13.6			& $ 7.45 \pm 0.02$	 \\
\swift\ XRT 	& WT	& 00032051001  	& 55\,778.11 				& 21.58					& 1.7				& $ 1.74 \pm 0.03$	 \\
\swift\ XRT 	& WT	& 00032051002  	& 55\,779.19 				& 22.66					& 1.7 			& $ 1.66 \pm 0.03$	\\
\swift\ XRT 	& WT	& 00032051003 	& 55\,780.50 				& 23.97					& 2.3				& $ 1.59 \pm 0.03$	 \\
\swift\ XRT 	& WT	& 00032051004  	& 55\,781.50 				& 24.97					& 2.3				& $ 1.57 \pm 0.03$	 \\
\swift\ XRT 	& WT	& 00032051005  	& 55\,786.42 				& 29.89					& 2.2 			& $ 1.28 \pm 0.02$	\\
\swift\ XRT 	& WT	& 00032051006  	& 55\,787.59  				& 31.06					& 2.2 			& $ 1.29 \pm 0.02$	\\
\swift\ XRT 	& WT	& 00032051007  	& 55\,788.26  				& 31.73					& 2.3				& $ 1.25 \pm 0.02$	 \\
\swift\ XRT 	& WT	& 00032051008  	& 55\,789.66  				& 33.13					& 2.2				& $ 1.17 \pm 0.02$	 \\	
\swift\ XRT 	& WT	& 00032051009  	& 55\,790.36  				& 33.83					& 2.2				& $ 1.07 \pm 0.02$	 \\
\swift\ XRT 	& WT	& 00032033015  	& 55\,800.86  				& 44.33 					& 2.9				& $ 0.85 \pm 0.02$	\\
\swift\ XRT 	& WT	& 00032033016  	& 55\,807.49  				& 50.96					& 2.4				& $ 0.77 \pm 0.02$	\\
\cxo\ ACIS-S	& CC	& 12614 			& 55\,822.80 				& 66.26					& 10.0			& $ 2.68 \pm 0.02$	 \\
\swift\ XRT 	& PC		& 00032033017	& 55\,822.83  				& 66.30					& 4.9				& $ 0.45 \pm 0.01$	\\
\swift\ XRT 	& WT	& 00032033018  	& 55\,824.71  				& 68.18					& 1.5				& $ 0.60 \pm 0.02$	\\
\xmm\ EPN 	& LW	& 0672281801 		& 55\,827.25 				& 70.72					& 9.9				& $5.03 \pm 0.02$	 \\
\swift\ XRT 	& WT	& 00032033019  	& 55\,829.45  				& 72.92					& 2.3				& $ 0.61 \pm 0.02$	\\
\swift\ XRT 	& WT	& 00032033020  	& 55\,835.54  				& 79.01					& 2.6				& $ 0.53 \pm 0.01$	\\
\swift\ XRT 	& WT	& 00032033021  	& 55\,842.06  				& 85.53					& 4.2				& $ 0.44 \pm 0.01$	\\
\xmm\ EPN 	& LW	& 0672282701 		& 55\,847.02 				& 90.49 					& 24.0			& $ 3.71  \pm 0.01$	 \\
\swift\ XRT 	& WT	& 00032033022  	& 55\,849.62  				& 93.09 					& 3.4				& $ 0.40 \pm 0.01$	\\
\swift\ XRT	& WT	& 00032033023	& 55\,856.58				& 100.05					& 2.2				& $ 0.37 \pm 0.01$	\\
\swift\ XRT 	& PC		& 00032033024  	& 55\,862.59  				& 106.06					& 10.2			& $ 0.263 \pm 0.005$	\\
\cxo\ ACIS-S	& CC	& 12615  			& 55\,867.18				& 110.65					& 16.2			& $ 1.47 \pm 0.01$	 \\
\swift\ XRT 	& PC		& 00032033025  	& 55\,977.17				& 220.64					& 6.2 			& $ 0.152 \pm 0.005$	\\
\swift\ XRT 	& WT	& 00032033026  	& 55\,978.53 				& 222.00					& 10.2			& $ 0.198 \pm 0.005$	 \\
\swift\ XRT 	& PC	 	& 00032033027  	& 55\,981.99  				& 225.46					& 11.0 			& $ 0.137 \pm 0.004$	\\
\swift\ XRT 	& WT	& 00032033028  	& 55\,982.96   				& 226.43					& 6.6				& $ 0.194 \pm 0.006$	\\
\swift\ XRT 	& WT	& 00032033029  	& 55\,985.18 				& 228.65					& 7.0				& $ 0.201 \pm 0.006$	 \\
\swift\ XRT 	& WT	& 00032033030  	& 55\,985.55  				& 229.02					& 7.0				& $ 0.195 \pm 0.006$	 \\
\swift\ XRT 	& WT	& 00032033031  	& 55\,991.09 				& 234.56					& 6.8				& $ 0.193 \pm 0.006$	 \\
\xmm\ EPN 	& LW	& 0672282901		& 56\,023.12 				& 266.59					& 23.0			& $ 1.421 \pm 0.008$	\\
\swift\ XRT 	& WT	& 00032033032  	& 56\,031.14 				& 274.61					& 4.2 			& $ 0.236 \pm 0.008$	\\
\cxo\ ACIS-S	& CC	& 14330 		  	& 56\,037.09				& 280.56 					& 20.0 			& $ 0.663 \pm 0.006$	\\
\swift\ XRT  	& WT	& 00032033033 	& 56\,052.66  				& 296.13 					& 5.1				& $ 0.242 \pm 0.007$	\\
\swift\ XRT  	& WT	& 00032033034 	& 56\,073.25				& 316.72 					& 4.9				& $ 0.200 \pm 0.007$	\\
\swift\ XRT  	& WT	& 00032033035 	& 56\,095.59				& 339.67					& 5.6				& $ 0.179 \pm 0.006$	\\
\swift\ XRT  	& WT	& 00032033036 	& 56\,104.55				& 348.02 					& 6.2				& $ 0.167 \pm 0.005$	\\
\swift\ XRT  	& WT	& 00032033037 	& 56\,114.30				& 357.77					& 6.8				& $ 0.146 \pm 0.005$	 \\
\swift\ XRT  	& WT	& 00032033039 	& 56\,156.20 				& 399.67 					& 4.9				& $ 0.205 \pm 0.007$	\\
\swift\ XRT  	& WT	& 00032033040 	& 56\,161.70				& 405.17 					& 5.0				& $ 0.214 \pm 0.007$	 \\
\hline
\end{tabular}
\end{table*}

\setcounter{table}{11}
\begin{table*}
  \caption{ -- \emph{continued}}
\label{tab:swift1822}
\vspace{10pt}
\begin{tabular}{ccccccc} \hline
Instrument     	& Mode	& Obs. ID 			&  Mid point of observation	& Time since outburst onset 	& Exposure  	& Source net count rate \\
         		&          	&       			&  (MJD)					& (d)     			     	& (ks)   		& (counts s$^{-1}$)\\ 
\hline
\xmm\ EPN	& LW	& 0672283001		& 56\,178.85				& 422.32					& 20.2		& $ 0.950 \pm 0.007$	\\
\swift\ XRT  	& WT	& 00032033042 	& 56\,206.01				& 449.48					& 5.0			& $ 0.147 \pm 0.006$	 \\
\swift\ XRT  	& WT	& 00032033043 	& 56\,238.71				& 482.18					& 4.9 		& $ 0.117 \pm 0.005$	\\
\swift\ XRT	& WT	& 00032051010 	& 56\,334.76				& 578.23					& 9.5 		& $ 0.114 \pm 0.004$	\\
\swift\ XRT	& WT	& 00032051011 	& 56\,355.40				& 598.87					& 8.2			& $ 0.142 \pm 0.004$	\\
\swift\ XRT	& WT	& 00032051012 	& 56\,386.91				& 630.38					& 2.7			& $ 0.157 \pm 0.008$	\\
\swift\ XRT	& WT	& 00032051013 	& 56\,387.28				& 630.75					& 7.7 		& $ 0.180 \pm 0.005$	\\
\swift\ XRT	& WT	& 00032051014 	& 56\,409.09				& 652.55					& 8.4			& $ 0.105 \pm 0.004$	\\
\swift\ XRT	& WT	& 00032051016 	& 56\,456.35				& 699.82					& 8.8			& $ 0.141 \pm 0.004$	\\
\swift\ XRT	& WT	& 00032051017 	& 56\,459.66				& 703.13					& 7.1 		& $ 0.104 \pm 0.004$	\\
\swift\ XRT	& WT	& 00032051018 	& 56\,490.67				& 734.14					& 4.0 		& $ 0.142 \pm 0.006$	\\
\swift\ XRT	& WT	& 00032051019 	& 56\,491.52				& 734.98					& 18.0 		& $ 0.165 \pm 0.003$	\\
\swift\ XRT	& WT	& 00032051020 	& 56\,536.11				& 779.58					& 13.8 		& $ 0.132 \pm 0.003$	\\
\swift\ XRT	& WT	& 00032051021 	& 56\,598.52 				& 841.99					& 6.3 		& $ 0.119 \pm 0.005$	\\
\swift\ XRT	& WT	& 00032051022 	& 56\,599.99				& 843.46					& 9.1 		& $ 0.128 \pm 0.004$	\\
\xmm\ EPN 	& FF		& 0722520101 		& 56\,724.58				& 968.05					& 40.3		& $ 0.139 \pm 0.002$	\\
\hline
\end{tabular}
\end{table*}

\begin{table*}
\centering
\caption{Log of all X-ray observations of Swift\,J1834.9$-$0846\ following the 2011 
August outburst. The outburst onset  occurred on MJD 55\,780.83178241 (D'Elia et 
al. 2011). We did not include two \swift XRT WT-mode observations carried out within 
6~h since the outburst onset (obs ID 00458907001, 00458907002) because they lasted 
only 91 and 141~s, respectively, and provided a low number of counts for a 
meaningful spectral analysis. Part of these observations were already analysed by 
Kargaltsev et al. (2012) and Esposito et al. (2013).}
\label{tab:swift1834}
\vspace{10pt}
\begin{tabular}{ccccccc} \hline
Instrument     	& Mode	& Obs. ID 			&  Mid point of observation	& Time since outburst onset 	& Exposure  		& Source net count rate \\
         		&          	&       			&  (MJD)					& (d)     			     	& (ks)   			& (counts s$^{-1}$)\\ 
\hline
\swift\ XRT 	& PC 	& 00458907000 	& 55\,780.87 				& 0.04					& 1.5				& $0.14 \pm 0.01$ \\
\swift\ XRT 	& WT 	& 00458907003 	& 55\,781.62 				& 0.79					& 1.7				& $0.05 \pm 0.006$ \\
\swift\ XRT 	& WT 	& 00458907004 	& 55\,781.84 				& 1.00					& 1.0				& $0.051 \pm 0.008$ \\
\swift\ XRT 	& WT 	& 00458907006 	& 55\,782.10 				& 1.27					& 2.7				& $0.053 \pm 0.005$ \\
\swift\ XRT	& WT 	& 00458907007 	& 55\,785.34 				& 4.51					& 5.7				& $0.046 \pm 0.003$ \\
\swift\ XRT 	& WT 	& 00458907008 	& 55\,787.28 				& 6.45					& 5.4				& $0.033 \pm 0.003$ \\
\swift\ XRT 	& WT 	& 00458907009 	& 55\,791.46  				& 10.63					& 8.0				& $0.041 \pm 0.002$ \\
\swift\ XRT 	& WT 	& 00458907010 	& 55\,794.43  				& 13.60					& 2.5				& $0.038 \pm 0.004$ \\
\cxo\ ACIS-S 	& TE		& 14329 			& 55\,795.74  				& 14.91					& 13.0			& $0.071 \pm 0.002$ \\
\swift\ XRT 	& WT 	& 00458907011 	& 55\,797.81  				& 16.98					& 0.9				& $0.033 \pm 0.006$ \\
\swift\ XRT 	& WT 	& 00458907012 	& 55\,800.35 				& 19.52					& 1.9				& $0.029 \pm 0.004$ \\
\swift\ XRT 	& PC 	& 00501752000 	& 55\,803.06  				& 22.23					& 2.6				& $0.010 \pm 0.002$ \\
\swift\ XRT 	& WT 	& 00458907013 	& 55\,803.38  				& 22.54				         & 2.2			& $0.029 \pm 0.004$ \\
\swift\ XRT 	& PC 	& 00458907014 	& 55\,806.47  				& 25.64					& 2.1				& $0.016 \pm 0.003$ \\
\cxo\ ACIS-S 	& TE		& 14055 			& 55\,809.59  				& 28.76					& 16.3			& $0.056 \pm 0.002$\\
\swift\ XRT 	& PC 	& 00458907016 	& 55\,814.45  				& 33.61					& 2.0				& $0.007 \pm 0.002$ \\
\swift\ XRT 	& WT 	& 00032097001 	& 55\,819.28  				& 38.45					& 9.1				& $0.025 \pm 0.002$\\
\xmm\ EPN 	& FF		& 0679380201 		& 55\,821.80  				& 40.96					& 23.7			& $0.116 \pm 0.002$ \\
\swift\ XRT 	& WT 	& 00032097002 	& 55\,822.27  				& 41.43					& 10.4			& $0.035 \pm 0.002$\\
\swift\ XRT 	& WT 	& 00032097003 	& 55\,825.60  				& 44.77					& 7.7				& $0.033 \pm 0.002$\\
\swift\ XRT 	& WT 	& 00032097004 	& 55\,828.52  				& 47.69					& 8.1				& $0.028 \pm 0.002$\\
\cxo\ ACIS-S 	& TE		& 14056 			& 55\,836.71  				& 55.88					& 24.5			& $0.023 \pm 0.001$\\
\cxo\ ACIS-S 	& TE		& 14057 			& 55\,877.60  				& 96.77					& 37.6			& $0.0014 \pm 0.0002$\\
\xmm\ EPN 	& FF		& 0723270101		& 56\,733.38   				& 952.54					& 58.0 			& $0.0052 \pm 0.0005$ \\
\xmm\ EPN 	& FF		& 0743020201		& 56\,946.78   				& 1165.95					& 50.3 			& $<$0.003$^{\rm a}$ \\
\hline
\end{tabular}
\begin{tablenotes}
\item \emph{Note}. $^a$The upper limit is quoted at the 3$\sigma$ c.l., and is derived by applying the \textsc{eupper} task of \textsc{sas}. The corresponding 
upper limits on the fluxes and luminosities were estimated by assuming an absorbed blackbody spectral model with the same parameters as those of the spectrum of 
the penultimate \xmm\ observation (obs ID: 0723270101).
\end{tablenotes}{}
\end{table*}

\begin{table*}
\centering
\caption{Log of all X-ray observations of \wes\ following the 2011 September outburst.
The outburst onset  occurred on MJD 55\,823.88623843 (Baumgartner et al. 2011).
Part of these observations were already analysed by An et al. (2013) and Rodr\'iguez 
Castillo et al. (2014).} 
\label{tab:cxou1647_2011}
\vspace{10pt}
\begin{tabular}{ccccccc} \hline
Instrument     	& Mode	& Obs. ID 			&  Mid point of observation		& Time since outburst onset 	& Exposure  		& Source net count rate \\
         		     	&          	&       			&  (MJD)						& (d)     			     	& (ks)   			& (counts s$^{-1}$)\\ 
\hline 
\swift\ XRT 	& PC		&  00030806020 	& 55\,829.58					& 5.69 					& 3.1				& $0.103\pm0.006$\\
\xmm\ EPN	& LW  	& 0679380501		& 55\,832.03					& 8.14  					& 15.5			& $0.770 \pm 0.007$\\
\swift\ XRT 	& PC		& 00030806022 	& 55\,835.46					& 11.57 					& 4.3				& $0.063\pm0.004$\\ 
\swift\ XRT 	& PC		& 00030806023 	& 55\,839.41					& 15.52 					& 3.6				& $0.058\pm0.004$\\ 
\swift\ XRT 	& PC		& 00030806024 	& 55\,840.72					& 16.83					& 3.7				& $0.065\pm0.004$\\ 
\swift\ XRT 	& PC		& 00030806025 	& 55\,842.40					& 18.51 					& 3.9				& $0.063\pm0.004$\\ 
\swift\ XRT 	& PC		& 00030806026 	& 55\,844.37					& 20.48 					& 4.0 			& $0.064\pm0.004$\\ 
\swift\ XRT 	& PC		& 00030806027 	& 55\,849.42 					& 25.53 					& 8.8 			& $0.064\pm0.003$\\ 
\cxo\ ACIS-S	& TE   	& 14360			& 55\,857.78					& 33.90  					& 19.1			& $0.244 \pm 0.004$\\
\hline
\end{tabular}
\end{table*}

\begin{table*}
\centering
\caption{Log of the X-ray observations of 1E\,1048.1$-$5937\ following the 2011 December 
outburst. No burst signalling the outburst onset was detected in this case. The outburst onset 
is thus considered to be occurred on MJD 55\,926, when an increase in the X-ray flux was 
measured (the previous \swift\ observation was carried out on MJD 55\,877.20). Part of these 
observations were already analysed by Archibald et al. (2015). 
We focus here on the observations covering the first $\sim 1000$~d of the outburst. We did 
not include the \swift\ observation 00031220126, because the source PSF falls on a column of bad 
pixels in this case.} 
\label{tab:1e1048_2011}
\vspace{10pt}
\begin{tabular}{ccccccc} \hline
Instrument     	& Mode	& Obs. ID 			&  Mid point of observation		& Time since outburst onset 	& Exposure  	& Source net count rate \\
         		&          	&       			&  (MJD)						& (d)     			     	& (ks)   		& (counts s$^{-1}$)\\ 
\hline 
\swift\ XRT	& WT	& 00031220066	& 55\,926.27					& 0.27					& 2.0			& $0.96  \pm 0.02$	\\
\swift\ XRT	& WT	& 00031220067	& 55\,927.04					& 1.04					& 2.0			& $0.91  \pm 0.02$	\\
\swift\ XRT	& WT	& 00031220068	& 55\,936.07					& 10.07					& 2.2			& $0.88  \pm 0.02$	\\
\swift\ XRT	& WT	& 00031220069	& 55\,936.32					& 10.32					& 1.1			& $0.85  \pm 0.03$	\\
\swift\ XRT	& WT	& 00031220070	& 55\,937.04					& 11.04					& 1.3			& $0.66  \pm 0.02$	\\
\swift\ XRT	& WT	& 00031220071	& 55\,946.19					& 20.19					& 3.2			& $0.74  \pm 0.02$	\\
\swift\ XRT	& WT	& 00031220073	& 55\,947.12					& 21.12					& 2.2			& $0.64  \pm 0.02$	\\
\swift\ XRT	& WT	& 00031220074	& 55\,956.05					& 30.05					& 2.0			& $0.72  \pm 0.02$	\\
\swift\ XRT	& WT	& 00031220077	& 55\,962.84					& 36.84					& 6.3			& $0.72  \pm 0.01$	\\
\swift\ XRT	& WT	& 00031220078	& 55\,966.09					& 40.09					& 2.3			& $0.74  \pm 0.02$	\\
\swift\ XRT	& WT	& 00031220081	& 55\,969.86					& 43.86					& 7.4			& $0.73  \pm 0.01$	\\
\swift\ XRT	& WT	& 00031220085	& 55\,977.14					& 51.14					& 1.9			& $0.46  \pm 0.02$	\\
\cxo\ ACIS-S 	& CC	& 14139			& 55\,980.95					& 54.95					& 6.1			& $3.09  \pm 0.02$	\\
\swift\ XRT	& WT	& 00031220086	& 55\,981.75					& 55.75					& 5.0 		& $0.73  \pm 0.01$	\\
\swift\ XRT	& WT	& 00031220090	& 55\,998.15					& 72.15					& 1.8			& $0.58  \pm 0.02$	\\
\swift\ XRT	& WT	& 00031220095	& 56\,016.12					& 90.12					& 1.9			& $0.43  \pm 0.02$	\\
\swift\ XRT	& WT	& 00031220098	& 56\,027.44					& 101.44					& 2.0			& $0.61  \pm 0.02$	\\
\cxo\ ACIS-S 	& CC	& 14140			& 56\,027.56					& 101.56					& 12.0		& $2.68  \pm 0.02$	\\
\swift\ XRT	& WT	& 00031220099	& 56\,040.21					& 114.21					& 1.3			& $0.60  \pm 0.02$	\\
\swift\ XRT	& WT	& 00031220102	& 56\,054.10					& 128.10					& 2.2			& $0.59  \pm 0.02$	\\
\swift\ XRT	& WT	& 00031220105	& 56\,068.25					& 142.25 					& 0.6			& $0.60  \pm $	0.03\\
\swift\ XRT	& WT	& 00031220110		& 56\,083.25					& 157.25					& 1.9			& $0.59  \pm 0.02$	\\
\swift\ XRT	& WT	& 00031220113		& 56\,097.24					& 171.24					& 1.6			& $0.55  \pm 0.02$	\\
\swift\ XRT	& WT	& 00031220116		& 56\,111.24					& 185.24					& 2.1			& $0.51  \pm 0.02$	\\
\swift\ XRT	& WT	& 00031220133	& 56\,160.05					& 234.05					& 1.6			& $0.53  \pm 0.02$	\\
\swift\ XRT	& WT	& 00031220147	& 56\,196.04					& 270.04					& 1.6			& $0.39  \pm 0.02$	\\
\swift\ XRT	& WT	& 00031220148	& 56\,202.05					& 276.05					& 1.5			& $0.41  \pm 0.02$	\\
\swift\ XRT	& WT	& 00031220158	& 56\,223.27					& 297.27					& 1.5			& $0.44  \pm 0.02$	\\
\swift\ XRT	& WT	& 00031220167	& 56\,244.28					& 318.28					& 1.8			& $0.47  \pm 0.02$	\\
\swift\ XRT	& WT	& 00031220177	& 56\,266.34					& 340.34					& 1.5			& $0.39  \pm 0.02$	\\
\swift\ XRT	& WT	& 00031220181	& 56\,280.31					& 354.31					& 1.5			& $0.39  \pm 0.02$	\\
\swift\ XRT	& WT	& 00031220189	& 56\,300.28					& 374.28					& 1.3			& $0.37  \pm 0.02$	\\
\swift\ XRT	& WT	& 00031220192	& 56\,307.12					& 381.12					& 1.2			& $0.37  \pm 0.02$	\\
\swift\ XRT	& WT	& 00031220201	& 56\,327.30					& 401.30					& 1.2			& $0.46  \pm 0.02$	\\
\swift\ XRT	& WT	& 00031220211		& 56\,349.39					& 423.39					& 1.4			& $0.42  \pm 0.02$	\\
\swift\ XRT	& WT	& 00031220220	& 56\,370.20					& 444.20					& 1.6			& $0.40  \pm 0.02$	\\
\swift\ XRT	& WT	& 00031220224	& 56\,403.25					& 477.25					& 1.0			& $0.35  \pm 0.02$	\\
\swift\ XRT	& WT	& 00031220231	& 56\,433.07					& 507.07 					& 1.2			& $0.39  \pm 0.02$	\\
\swift\ XRT	& WT	& 00031220234	& 56\,446.39					& 520.39					& 1.1			& $0.37  \pm 0.02$	\\
\swift\ XRT	& WT	& 00031220238	& 56\,460.46					& 534.46					& 0.8			& $0.38  \pm 0.02$	\\
\xmm\ EPN 	& FF		& 0723330101		& 56\,496.13					& 570.13					& 48.3		& $2.395  \pm 0.007$ \\
\swift\ XRT	& WT	& 00031220246	& 56\,502.12					& 576.12					& 1.5			& $0.36  \pm 0.02$	\\
\swift\ XRT	& WT	& 00031220249	& 56\,516.07					& 590.07 					& 1.4			& $0.32  \pm 0.02$	\\
\swift\ XRT	& WT	& 00032923002	& 56\,538.55					& 612.55					& 1.4			& $0.37  \pm 0.02$	\\
\swift\ XRT	& WT	& 00032923008	& 56\,566.38					& 640.38					& 1.3			& $0.33  \pm 0.02$	\\
\swift\ XRT	& WT	& 00032923012	& 56\,581.48					& 655.48					& 1.5			& $0.35  \pm 0.02$	\\
\swift\ XRT	& WT	& 00032923014	& 56\,594.45					& 668.45					& 1.6			& $0.37  \pm 0.02$	\\
\swift\ XRT	& WT	& 00032923016	& 56\,609.42					& 683.42 					& 1.4			& $0.36  \pm 0.02$	\\
\swift\ XRT	& WT	& 00032923023	& 56\,650.12					& 724.12 					& 1.4			& $0.28  \pm 0.01$	\\
\swift\ XRT	& WT	& 00032923026	& 56\,664.07					& 738.07					& 1.4			& $0.30  \pm 0.01$	\\
\swift\ XRT	& WT	& 00032923029	& 56\,678.27					& 752.27 					& 0.8			& $0.20  \pm 0.02$	\\
\swift\ XRT	& WT	& 00032923039	& 56\,721.45					& 795.45					& 1.7			& $0.28  \pm 0.01$	\\
\swift\ XRT	& WT	& 00032923048	& 56\,763.87					& 837.87					& 1.4			& $0.31  \pm 0.02$	\\
\swift\ XRT	& WT	& 00032923058	& 56\,805.72					& 879.72					& 1.3			& $0.32  \pm 0.02$	\\
\swift\ XRT	& WT	& 00032923069	& 56\,848.78					& 922.78					& 1.8			& $0.31  \pm 0.01$	\\
\swift\ XRT	& WT	& 00032923078	& 56\,875.20					& 949.20					& 1.5			& $0.29  \pm 0.01$	\\
\swift\ XRT	& WT	& 00032923082	& 56\,885.53					& 959.53					& 1.2			& $0.27  \pm 0.02$	\\
\hline
\end{tabular}
\end{table*}

\clearpage

\begin{table*}
\centering
\caption{Log of X-ray observations of \antgli\ following the 2012 April outburst and up to the 
return to quiescence. The outburst onset occurred on MJD 56\,038.34564479 (Foley et al. 
2012). One \cxo\ observation was performed with the HRC-I (obs. ID 15265), and was not 
included in our analysis. Part of these observations were already analysed by Archibald et al. 
(2013).} 
\label{tab:1e2259_2012}
\vspace{10pt}
\begin{tabular}{ccccccc} \hline
Instrument     	& Mode	& Obs. ID 			&  Mid point of observation		& Time since outburst onset 	& Exposure  		& Source net count rate \\
         		&          	&       			&  (MJD)						& (d)     			     	& (ks)   			& (counts s$^{-1}$)\\ 
\hline 
\swift\ XRT	& WT	& 00032035021 	& 56\,045.45					& 7.10  					& 3.9  			& $1.74\pm0.02$\\
\swift\ XRT	& WT	& 00032035022 	& 56\,049.06					& 10.71 					& 6.6   			& $1.68\pm0.02$\\
\swift\ XRT	& WT	& 00032035023 	& 56\,054.41 					& 16.07 					& 3.9   			& $1.65\pm0.02$\\
\swift\ XRT	& WT	& 00032035026 	& 56\,075.37					& 37.03 					& 3.9 			& $1.15\pm0.02$\\
\swift\ XRT	& WT	& 00032035029 	& 56\,096.55					& 58.20 					& 1.1  			& $1.47\pm0.04$\\
\swift\ XRT	& WT	& 00032035033 	& 56\,124.02					& 85.68 					& 0.9   			& $0.99\pm0.03$\\
\swift\ XRT	& WT	& 00032035036 	& 56\,161.80					& 123.46 					& 1.9  			& $1.44\pm0.03$\\
\swift\ XRT	& WT	& 00032035037 	& 56\,166.77					& 128.43 					& 1.0   			& $1.38\pm0.04$\\
\swift\ XRT	& WT	& 00032035040 	& 56\,208.62					& 170.28 					& 2.9  			& $0.86\pm0.02$\\
\swift\ XRT	& WT	& 00032035041 	& 56\,215.64					& 177.30 					& 2.9  			& $1.31\pm0.02$\\
\swift\ XRT	& WT	& 00032035042 	& 56\,222.54					& 184.20 					& 2.7  			& $1.21\pm0.02$\\
\swift\ XRT	& WT	& 00032035046 	& 56\,246.12					& 207.77 					& 1.4   			& $1.46\pm0.03$\\
\swift\ XRT	& WT	& 00032035049 	& 56\,264.91					& 226.56 					& 3.3  			& $1.34\pm0.02$\\
\swift\ XRT	& WT	& 00032035051 	& 56\,274.47					& 236.12 					& 2.8  			& $1.42\pm0.02$\\
\swift\ XRT	& WT	& 00032035052 	& 56\,292.33					& 253.98 					& 3.2  			& $1.13\pm0.02$\\
\swift\ XRT	& WT	& 00032035055 	& 56\,355.68					& 317.33 					& 3.3  			& $0.66\pm0.01$\\
\swift\ XRT	& WT	& 00032035056 	& 56\,376.48					& 338.13 					& 2.9   			& $1.31\pm0.02$\\
\swift\ XRT	& WT	& 00032035057 	& 56\,397.81					& 359.47 					& 3.4  			& $1.24\pm0.02$\\
\swift\ XRT	& WT	& 00032035058 	& 56\,418.11					& 379.77 					& 3.7  			& $1.30\pm0.02$\\
\swift\ XRT	& WT	& 00032035061 	& 56\,481.50					& 443.16 					& 3.2   			& $1.27\pm0.02$\\
\swift\ XRT	& WT	& 00032035062 	& 56\,502.49					& 464.15 					& 3.5   			& $1.28\pm0.02$\\
\swift\ XRT	& WT	& 00032035065 	& 56\,565.61					& 527.27 					& 1.6  			& $0.94\pm0.02$\\
\swift\ XRT	& WT	& 00032035069 	& 56\,632.89					& 594.55 					& 1.5   			& $1.38\pm0.03$\\
\swift\ XRT	& WT	& 00032035072 	& 56\,692.59					& 654.24 					& 3.2   			& $1.17\pm0.02$\\
\swift\ XRT	& WT	& 00032035073 	& 56\,712.55					& 674.20 					& 3.3  			& $1.25\pm0.02$\\
\swift\ XRT	& WT	& 00032035076 	& 56\,776.29					& 737.95 					& 3.4  			& $1.18\pm0.02$\\
\swift\ XRT	& WT	& 00032035077 	& 56\,797.59					& 759.25 					& 2.7   			& $1.22\pm0.02$\\
\swift\ XRT	& WT	& 00032035079 	& 56\,839.51					& 801.16 					& 2.9  			& $1.11\pm0.02$\\
\swift\ XRT	& WT	& 00032035083 	& 56\,902.72					& 864.38 					& 3.5    			& $1.25\pm0.02$\\
\swift\ XRT	& WT	& 00032035084 	& 56\,923.61					& 885.26 					& 2.0   			& $1.24\pm0.03$\\
\swift\ XRT	& WT	& 00032035087 	& 56\,965.22					& 926.88 					& 3.3   			& $1.16\pm0.02$\\
\swift\ XRT	& WT	& 00032035088 	& 56\,986.31					& 947.97 					& 4.0   			& $1.07\pm0.02$\\
\swift\ XRT	& WT	& 00032035089 	& 57\,007.77					& 969.43 					& 2.6  			& $1.17\pm0.02$\\
\swift\ XRT	& WT	& 00032035091 	& 57\,028.59					& 990.25 					& 5.5   			& $0.97\pm0.01$\\
\swift\ XRT	& WT	& 00032035092 	& 57\,049.20					& 1010.86 				& 1.3  			& $1.02\pm0.03$\\
\swift\ XRT	& WT	& 00032035093 	& 57\,070.48					& 1032.13 				& 3.9   			& $1.21\pm0.02$\\
\swift\ XRT	& WT	& 00032035096 	& 57\,133.39					& 1095.04 				& 3.9   			& $0.89\pm0.02$\\
\swift\ XRT	& WT	& 00032035097 	& 57\,154.58					& 1116.23 				& 3.3  			& $1.15\pm0.02$\\
\swift\ XRT	& WT	& 00032035101 	& 57\,196.43					& 1158.09					& 1.2  			& $1.10\pm0.03$\\
\swift\ XRT	& WT	& 00032035104 	& 57\,204.93					& 1166.59 				& 0.6  			& $1.17\pm0.05$\\
\swift\ XRT	& WT	& 00032035107 	& 57\,260.71					& 1222.37					& 3.8   			& $0.96\pm0.02$\\
\swift\ XRT	& WT	& 00032035108 	& 57\,286.61					& 1248.26 				& 3.5   			& $0.97\pm0.02$\\
\swift\ XRT	& WT	& 00032035114 	& 57\,343.45					& 1305.11 				& 3.2   			& $1.31\pm0.02$\\
\hline
\end{tabular}
\end{table*}

\begin{table*}
\centering
\caption{Log of \cxo\ observations of \galcen\ following the 2013 April outburst. 
The outburst onset  occurred on MJD 56\,407.80237269 (Barthelmy et al. 2013).
All reported count rates are not corrected for pile-up. Part of these observations 
were already analysed by Rea et al. (2013b) and Coti Zelati et al. (2015a, 2017).}
\label{tab:sgr1745}
\vspace{10pt}
\begin{tabular}{ccccccc}
\hline
Instrument     		& Mode		& Obs. ID 			&  Mid point of observation	& Time since outburst onset 	& Exposure  		& Source net count rate \\
         		     	&          		&       			&  (MJD)					& (d)     			     	& (ks)   			& (counts s$^{-1}$)\\ 
\hline
\cxo/HRC	-S		& TE			& 14701			& 56\,411.70				& 3.90					& 9.7				& $ 0.081 \pm 0.003$\\	
\cxo\ ACIS-S 		& TE			& 14702			& 56\,424.55				& 16.75					& 13.7 			& $ 0.545 \pm 0.006$ \\
\cxo/HETG		& TE			& 15040			& 56\,437.63				& 29.83					& 23.8 			& $ 0.150 \pm 0.003$\\
\cxo\ ACIS-S 		& TE			& 14703			& 56\,447.48				& 39.68					& 16.8			& $ 0.455 \pm 0.005$\\
\cxo//HETG 		& TE			& 15651			& 56\,448.99				& 41.19					& 13.8 			& $ 0.141 \pm 0.003$\\
\cxo/HETG		& TE			& 15654			& 56\,452.25				& 44.45					& 9.0 			& $ 0.128 \pm 0.004$\\
\cxo\ ACIS-S 		& TE			& 14946			& 56\,475.41				& 67.61					& 18.2			& $ 0.392 \pm 0.005$\\
\cxo\ ACIS-S 		& TE			& 15041			& 56\,500.36				& 92.56					& 45.4			& $ 0.346 \pm 0.003$\\
\cxo\ ACIS-S 		& TE			& 15042			& 56\,516.25				& 108.45					& 45.7			& $ 0.317 \pm 0.003$\\
\cxo\ ACIS-S 		& TE			& 14945			& 56\,535.55				& 127.75					& 18.2			& $ 0.290 \pm 0.004$\\
\cxo\ ACIS-S 		& TE			& 15043			& 56\,549.30				& 141.50					& 45.4			& $ 0.275 \pm 0.002$\\
\cxo\ ACIS-S 		& TE			& 14944			& 56\,555.42				& 147.62					& 18.2			& $ 0.273 \pm 0.004$\\
\cxo\ ACIS-S 		& TE			& 15044			& 56\,570.01				& 162.21					& 42.7			& $ 0.255 \pm 0.002$\\
\cxo\ ACIS-S 		& TE			& 14943			& 56\,582.78				& 174.98					& 18.2			& $ 0.246 \pm 0.004$\\
\cxo\ ACIS-S 		& TE			& 14704			& 56\,588.62				& 180.82					& 36.3			& $ 0.240 \pm 0.003$\\
\cxo\ ACIS-S 		& TE			& 15045			& 56\,593.91				& 186.11					& 45.4			& $ 0.234 \pm 0.002$\\
\cxo\ ACIS-S 		& TE			& 16508   			& 56\,709.77				& 301.97 					& 43.4			& $ 0.156 \pm 0.002$\\
\cxo\ ACIS-S 		& TE			& 16211 			& 56\,730.71				& 322.91					& 41.8			& $ 0.149 \pm 0.002$\\
\cxo\ ACIS-S 		& TE			& 16212			& 56\,751.40				& 343.60					& 45.4			& $ 0.135 \pm 0.002$\\
\cxo\ ACIS-S 		& TE			& 16213			& 56\,775.41				& 367.61					& 45.0			& $ 0.128 \pm 0.002$\\
\cxo\ ACIS-S 		& TE			& 16214			& 56\,797.31				& 389.51					& 45.4			& $ 0.118 \pm 0.002$\\
\cxo\ ACIS-S 		& TE			& 16210			& 56\,811.24				& 403.44					& 17.0			& $ 0.110 \pm 0.003$\\
\cxo\ ACIS-S 		& TE			& 16597			& 56\,842.98				& 435.18					& 16.5			& $ 0.097 \pm 0.002$\\
\cxo\ ACIS-S 		& TE			& 16215			& 56\,855.22		 		& 447.42					& 41.5			& $ 0.090 \pm 0.001$\\
\cxo\ ACIS-S 		& TE			& 16216			& 56\,871.43				& 463.63					& 42.7 			& $ 0.085 \pm 0.001$\\
\cxo\ ACIS-S 		& TE			& 16217			& 56\,899.43				& 491.63					& 34.5 			& $ 0.079 \pm 0.002$\\
\cxo\ ACIS-S 		& TE			& 16218			& 56\,950.59				& 542.79 					& 36.3			& $ 0.071 \pm 0.001$\\	
\cxo\ ACIS-S 		& TE			& 16963			& 57\,066.18				& 658.38					& 22.7			& $ 0.056 \pm 0.002$\\
\cxo\ ACIS-S 		& TE			& 16966			& 57\,156.53				& 748.72 					& 22.7			& $0.045  \pm 0.001$\\
\cxo\ ACIS-S 		& TE			& 16965			& 57\,251.60				& 843.80 					& 22.7 			& $0.035  \pm 0.001$\\
\cxo\ ACIS-S 		& TE			& 16964			& 57\,316.41				& 908.60					& 22.6			& $0.026  \pm 0.001$\\
\cxo\ ACIS-S 		& TE			& 18055 			& 57\,431.53				& 1023.73					& 22.7			& $0.0133  \pm 0.0008$\\
\cxo\ ACIS-S 		& TE			& 18056  			& 57\,432.76				& 1024.96					& 21.8			& $0.0146  \pm 0.0009$\\
\cxo\ ACIS-S 		& TE			& 18731 			& 57\,582.27				& 1174.47					& 78.4			& $0.0102  \pm 0.0004$\\
\cxo\ ACIS-S 		& TE			& 18732			& 57\,588.00				& 1180.20					& 76.6			& $0.0118  \pm 0.0004$\\
\cxo\ ACIS-S 		& TE			& 18057 			& 57\,669.95				& 1262.15					& 22.7			& $0.0130  \pm 0.0008$\\
\cxo\ ACIS-S 		& TE			& 18058  			& 57\,675.61				& 1267.80					& 22.7			& $0.0135  \pm 0.00081$\\
\hline
\end{tabular}
\end{table*}

\begin{table*}
\centering
\caption{Log of all X-ray observations of \sgrm\ following the 2014 July outburst.
The outburst onset  occurred on MJD 56\,843.39777778 (Stamatikos et al. 2014).
Part of these observations were already analysed by Israel et al. (2016).} 
\label{tab:sgr1935}
\vspace{10pt}
\begin{tabular}{ccccccc} \hline
Instrument     	& Mode	& Obs. ID 			&  Mid point of observation		& Time since outburst onset 	& Exposure  	& Source net count rate \\
         		&          	&       			&  (MJD)						& (d)     			     	& (ks)   		& (counts s$^{-1}$)\\ 
\hline 
\swift\ XRT	& PC		& 00603488000	& 56\,843.44					& 0.04					& 3.4			& $ 0.033 \pm 0.003$	\\
\swift\ XRT	& PC		& 00603488001	& 56\,844.72					& 0.32					& 9.9			& $ 0.027 \pm 0.002$	\\
\swift\ XRT	& PC		& 00603488003	& 56\,845.36					& 1.96					& 3.9			& $ 0.019 \pm 0.002$	\\
\swift\ XRT	& PC		& 00603488006	& 56\,846.77					& 3.37					& 3.7			& $ 0.028 \pm 0.003$	\\
\swift\ XRT	& PC		& 00603488007	& 56\,847.67					& 4.27					& 3.6			& $ 0.018 \pm 0.002$	\\
\swift\ XRT	& PC		& 00603488009	& 56\,851.39					& 7.99					& 3.0			& $ 0.025 \pm 0.003$	\\
\swift\ XRT	& PC		& 00603488008	& 56\,851.62					& 8.23					& 5.3			& $ 0.022 \pm 0.002$	\\
\cxo\ ACIS-S	& TE		& 15874			& 56\,853.66					& 10.26 					& 9.1			& $ 0.110 \pm 0.003$	\\
\swift\ XRT	& PC		& 00603488010	& 56\,854.51					& 11.11					& 7.1			& $ 0.024 \pm 0.002$	\\
\swift\ XRT	& PC		& 00603488011		& 56\,858.54					& 15.14					& 2.9			& $ 0.019 \pm 0.003$	\\
\cxo\ ACIS-S	& CC	& 15875			& 56\,866.48					& 23.08					& 75.1		& $ 0.115 \pm 0.001$	\\
\swift\ XRT	& PC		& 00033349001	& 56\,869.76 					& 26.36					& 2.1			& $ 0.019 \pm 0.003$	\\
\swift\ XRT	& PC		& 00033349002	& 56\,876.70					& 33.30					& 2.2			& $ 0.022 \pm 0.003$	\\
\swift\ XRT	& PC		& 00033349003	& 56\,883.75					& 40.35					& 1.5			& $ 0.023 \pm 0.004$	\\
\swift\ XRT	& PC		& 00033349004	& 56\,890.41					& 47.01					& 1.7			& $ 0.019 \pm 0.003$	\\
\swift\ XRT	& PC		& 00033349005	& 56\,894.65					& 51.25					& 3.7			& $ 0.020 \pm 0.002$	\\
\swift\ XRT	& PC		& 00033349006	& 56\,897.55					& 54.15					& 1.7			& $ 0.022 \pm 0.004$	\\
\cxo\ ACIS-S	& CC	& 17314			& 56\,900.21					& 56.81					& 29.0		& $ 0.107 \pm 0.002$	\\	 
\swift\ XRT	& PC		& 00033349007	& 56\,904.79					& 61.39					& 1.1			& $ 0.016 \pm 0.004$	\\
\swift\ XRT	& PC		& 00033349008	& 56\,906.51					& 63.11					& 1.5			& $ 0.022 \pm 0.004$	\\
\swift\ XRT	& PC		& 00033349009	& 56\,911.27					& 67.88					& 1.5			& $ 0.023 \pm 0.004$	\\
\swift\ XRT	& PC		& 00033349010	& 56\,914.31					& 70.91					& 2.4			& $ 0.024 \pm 0.003$	\\
\swift\ XRT	& PC		& 00033349011		& 56\,918.57					& 75.17					& 1.4			& $ 0.019 \pm 0.004$	\\
\swift\ XRT	& PC		& 00033349012	& 56\,924.50					& 81.10					& 2.3			& $ 0.019 \pm 0.003$	\\
\xmm\ EPN	& FF		& 0722412501		& 56\,927.06					& 83.66					& 16.9		& $ 0.190 \pm 0.003$	\\
\xmm\ EPN	& FF		& 0722412601		& 56\,928.32					& 84.92					& 17.8		& $ 0.189 \pm 0.003$	\\
\xmm\ EPN	& FF		& 0722412701		& 56\,934.36					& 90.96					& 16.1		& $ 0.197 \pm 0.004$	\\
\xmm\ EPN	& FF		& 0722412801		& 56\,946.17					& 102.77					& 8.6			& $ 0.194 \pm 0.005$	\\
\xmm\ EPN	& FF		& 0722412901		& 56\,954.19					& 110.79 					& 6.4			& $ 0.201 \pm 0.006$	\\
\xmm\ EPN	& FF		& 0722413001		& 56\,958.03					& 114.63					& 11.2		& $ 0.189 \pm 0.004$	\\
\xmm\ EPN	& FF		& 0748390801		& 56\,976.27					& 132.87 					& 9.5			& $ 0.194 \pm 0.005$	\\
\swift\ XRT	& PC		& 00632158000	& 57\,075.59 					& 232.19 					& 7.3			& $ 0.039 \pm 0.002$	\\
\swift\ XRT	& PC		& 00632158001	& 57\,075.84					& 232.44					& 1.8			& $ 0.042 \pm 0.005$	\\
\swift\ XRT	& PC		& 00632158002	& 57\,076.59					& 233.19					& 5.9			& $ 0.028 \pm 0.002$	\\
\swift\ XRT	& PC		& 00033349014	& 57\,078.48					& 235.08					& 3.1			& $ 0.030 \pm 0.003$	\\
\swift\ XRT	& PC		& 00033349015	& 57\,080.35					& 236.95					& 5.9			& $ 0.022 \pm 0.002$	\\
\swift\ XRT	& PC		& 00033349016	& 57\,085.51					& 242.11					& 3.9			& $ 0.025 \pm 0.003$	\\
\swift\ XRT	& PC		& 00033349017	& 57\,092.69					& 249.29					& 3.9			& $ 0.023 \pm 0.002$	\\
\xmm\ EPN	& FF		& 0764820101		& 57106.59					& 263.19					& 26.5		& $ 0.250 \pm 0.003$	\\
\swift\ XRT	& PC		& 00033349020	& 57\,127.83					& 284.43 					& 3.0			& $ 0.014  \pm 0.002$	\\
\swift\ XRT	& PC		& 00033349023	& 57\,134.62 					& 291.22 					& 1.4			& $ 0.029  \pm 0.005$	\\	
\swift\ XRT	& PC		& 00033349024	& 57\,221.00					& 377.60					& 2.0			& $ 0.025 \pm 0.004$	\\
\xmm\ EPN	& FF		& 0764820201		& 57\,303.04					& 459.65					& 11.4		& $ 0.207 \pm 0.004$	\\
\swift\ XRT	& PC		& 00033349025	& 57\,377.77					& 534.37					& 3.9 		& $ 0.017 \pm 0.002$	\\	
\swift\ XRT	& PC		& 00686761000	& 57\,526.42					& 683.02					& 1.6 		& $0.058 \pm 0.006$		\\	
\swift\ XRT	& PC		& 00033349026	& 57\,527.81					& 684.41					& 2.9  		& $0.020 \pm 0.003$		\\
\swift\ XRT	& PC		& 00687123000	& 57\,529.84					& 686.45					& 1.2  		& $0.026 \pm 0.005$		\\	
\swift\ XRT	& PC		& 00033349027	& 57\,534.44					& 691.04 					& 2.3  		& $0.017 \pm 0.003$ 	\\
\swift\ XRT	& PC		& 00033349028	& 57\,539.99 					& 696.51 					& 2.8  		& $0.015 \pm 0.002$		\\
\swift\ XRT	& PC		& 00033349030	& 57\,548.89 					& 705.49 					& 1.7  		& $0.014 \pm 0.003$		\\
\swift\ XRT	& PC		& 00033349031	& 57\,554.23					& 710.83 					& 2.6 		& $0.020 \pm 0.003$		\\
\swift\ XRT	& PC		& 00033349032	& 57\,561.42					& 718.02					& 1.6  		& $0.020 \pm 0.004$		\\
\swift\ XRT	& PC		& 00033349033	& 57\,567.59					& 724.19					& 2.0 		& $0.023 \pm 0.003$		\\
\swift\ XRT	& PC		& 00033349034	& 57\,569.82					& 726.42					& 2.4 		& $0.019 \pm 0.003$		\\
\swift\ XRT	& PC		& 00033349035	& 57\,576.88					& 733.48					& 2.8 		& $0.020 \pm 0.003$		\\
\swift\ XRT	& PC		& 00033349036	& 57\,586.36					& 742.96					& 2.5  		& $0.020 \pm 0.003$		\\
\swift\ XRT	& PC		& 00033349037	& 57\,597.51					& 754.11					& 2.8 		& $0.024 \pm 0.003$		\\
\hline
\end{tabular}
\end{table*}

\begin{table*}
\centering
\caption{Log of the X-ray observations of 1E\,1048.1$-$5937\ following the 2016 July 
outburst. No burst signalling the outburst onset was detected in this case. The outburst onset 
is thus considered to be occurred on $\sim$ MJD 57\,592, when an increase in the X-ray flux was 
measured (the previous \swift\ observation was carried out on MJD 57\,588; Archibald et al. 2016b). 
The source is currently being regularly monitored by \swift.} 
\label{tab:1e1048_2016}
\vspace{10pt}
\begin{tabular}{ccccccc} 
\hline
Instrument     	& Mode	& Obs. ID 			&  Mid point of observation		& Time since outburst onset 	& Exposure  	& Source net count rate \\
         		&          	&       			&  (MJD)						& (d)     			     	& (ks)   		& (counts s$^{-1}$)\\ 
\hline 
\swift\ XRT	& WT	& 00032923248	& 57\,592.49 					& 0.49  					& 1.2  	& $0.85  \pm 0.03$	\\
\swift\ XRT	& WT	& 00032923249	& 57\,598.47					& 6.47					& 1.4 	& $0.95  \pm 0.03$	\\
\swift\ XRT	& WT	& 00032923250	& 57\,598.80					& 6.80					& 1.1 	& $0.81  \pm 0.03$	\\
\swift\ XRT	& WT	& 00032923251	& 57\,605.97					& 13.97					& 0.3		& $0.55  \pm 0.04$	\\
\swift\ XRT	& WT	& 00032923252	& 57\,607.53					& 15.53					& 0.9		& $0.78  \pm 0.03$	\\
\swift\ XRT	& WT	& 00032923253	& 57\,608.27					& 16.27					& 0.6		& $0.56  \pm 0.03$	\\
\swift\ XRT	& WT	& 00032923254	& 57\,608.38					& 16.38					& 1.0		& $0.65  \pm 0.03$	\\
\swift\ XRT	& WT	& 00032923255	& 57\,612.02					& 20.02					& 0.6		& $0.69  \pm 0.03$	\\
\swift\ XRT	& WT	& 00030912012	& 57\,612.65					& 20.65					& 1.4		& $0.87  \pm 0.02$	\\
\swift\ XRT	& WT	& 00030912013	& 57\,613.61					& 21.61					& 1.4		& $0.47  \pm 0.02$	\\
\swift\ XRT	& WT	& 00030912016	& 57\,627.91					& 35.91					& 0.1		& $0.59  \pm 0.07$	\\
\swift\ XRT	& WT	& 00030912019	& 57\,628.11					& 36.11					& 0.9		& $0.51  \pm 0.02$	\\
\swift\ XRT	& WT	& 00030912017	& 57\,629.45					& 37.45					& 1.2		& $0.63  \pm 0.02$	\\
\swift\ XRT	& WT	& 00030912020	& 57\,633.13					& 41.13					& 1.5		& $0.47  \pm 0.02$	\\
\swift\ XRT	& WT	& 00030912022	& 57\,634.82					& 42.82					& 0.6		& $0.67  \pm 0.03$	\\
\swift\ XRT	& WT	& 00030912025	& 57\,640.20					& 48.20					& 1.6		& $0.65  \pm 0.21$	\\
\swift\ XRT	& WT	& 00030912023	& 57\,640.50					& 48.50					& 1.4		& $0.51  \pm 0.02$	\\
\swift\ XRT	& WT	& 00030912024	& 57\,641.40					& 49.40					& 1.5		& $0.54  \pm 0.02$	\\
\swift\ XRT	& WT	& 00030912026	& 57\,654.15					& 62.15					& 1.5		& $0.58  \pm 0.02$	\\
\swift\ XRT	& WT	& 00030912027	& 57\,654.39					& 62.39					& 1.3		& $0.53  \pm 0.02$	\\
\swift\ XRT	& WT	& 00030912028	& 57\,655.15					& 63.15					& 1.4		& $0.52  \pm 0.02$	\\
\swift\ XRT	& WT	& 00030912029	& 57\,667.27					& 75.27					& 1.5		& $0.49  \pm 0.02$	\\
\swift\ XRT	& WT	& 00030912030	& 57\,667.41					& 75.41					& 1.1		& $0.44  \pm 0.02$	\\
\swift\ XRT	& WT	& 00030912031	& 57\,668.20					& 76.20					& 1.4		& $0.52  \pm 0.02$	\\
\swift\ XRT	& WT	& 00030912032	& 57\,677.45					& 85.45					& 1.1		& $0.52  \pm 0.02$	\\
\swift\ XRT	& WT	& 00030912033	& 57\,677.85					& 85.85					& 1.6		& $0.53  \pm 0.02$	\\
\swift\ XRT	& WT	& 00030912035	& 57\,687.12					& 95.12					& 1.4		& $0.53  \pm 0.02$	\\
\swift\ XRT	& WT	& 00030912036	& 57\,687.58					& 95.58					& 1.5		& $0.42  \pm 0.02$	\\
\swift\ XRT	& WT	& 00030912037	& 57\,688.17					& 96.17					& 1.5		& $0.52  \pm 0.02$	\\
\swift\ XRT	& WT	& 00030912038	& 57\,697.08					& 105.08					& 1.6		& $0.55  \pm 0.02$	\\
\swift\ XRT	& WT	& 00030912040	& 57\,698.84					& 106.84					& 1.4		& $0.35  \pm 0.02$	\\
\swift\ XRT	& WT	& 00030912041	& 57\,707.35					& 115.35					& 1.0		& $0.38  \pm 0.02$	\\
\swift\ XRT	& WT	& 00030912042	& 57\,707.60					& 115.60					& 1.5		& $0.53  \pm 0.02$	\\
\swift\ XRT	& WT	& 00030912043	& 57\,708.80					& 116.80					& 1.4		& $0.33  \pm 0.02$	\\
\swift\ XRT	& WT	& 00030912044	& 57\,717.34					& 125.34					& 1.4		& $0.39  \pm 0.02$	\\
\swift\ XRT	& WT	& 00030912045	& 57\,717.73					& 125.73					& 1.5		& $0.48  \pm 0.02$	\\
\swift\ XRT	& WT	& 00030912046	& 57\,718.87					& 126.87					& 1.4		& $0.43  \pm 0.02$	\\
\swift\ XRT	& WT	& 00030912047	& 57\,727.02					& 135.02					& 1.4		& $0.39  \pm 0.02$	\\
\swift\ XRT	& WT	& 00030912048	& 57\,727.55					& 135.55					& 1.5		& $0.42  \pm 0.02$	\\
\swift\ XRT	& WT	& 00030912049	& 57\,728.45					& 136.45					& 1.5		& $0.42  \pm 0.02$	\\
\swift\ XRT	& WT	& 00030912050	& 57\,737.22					& 145.22					& 1.1		& $0.49  \pm 0.02$	\\
\swift\ XRT	& WT	& 00030912051	& 57\,737.45					& 145.45					& 1.3		& $0.43  \pm 0.02$	\\
\swift\ XRT	& WT	& 00030912052	& 57\,738.08					& 146.08					& 1.5		& $0.51  \pm 0.02$	\\
\swift\ XRT	& WT	& 00030912053	& 57\,747.15					& 155.15					& 1.4		& $0.46  \pm 0.02$	\\
\swift\ XRT	& WT	& 00030912054	& 57\,747.61					& 155.61					& 1.6		& $0.42  \pm 0.02$	\\
\swift\ XRT	& WT	& 00030912055	& 57\,748.77					& 156.77					& 1.5		& $0.27  \pm 0.01$	\\
\swift\ XRT	& WT	& 00030912057	& 57\,758.77					& 166.77					& 1.4		& $0.41  \pm 0.02$	\\
\swift\ XRT	& WT	& 00030912058	& 57\,759.49					& 167.49					& 1.3		& $0.39  \pm 0.02$	\\
\swift\ XRT	& WT	& 00030912059	& 57\,778.18					& 186.18					& 1.5 	& $0.46  \pm 0.02$	\\
\swift\ XRT	& WT	& 00030912060	& 57\,778.57					& 186.57					& 1.4		& $0.45  \pm 0.02$	\\
\swift\ XRT	& WT	& 00030912061	& 57\,779.68					& 187.68					& 1.4		& $0.111  \pm 0.009$ \\
\swift\ XRT	& WT	& 00030912062	& 57\,785.06					& 193.06					& 1.6		& $0.40  \pm 0.02$	\\
\swift\ XRT	& WT	& 00030912063	& 57\,785.89					& 193.89					& 1.4		& $0.40  \pm 0.02$	\\
\swift\ XRT	& WT	& 00030912064	& 57\,786.09					& 194.09					& 1.5		& $0.41  \pm 0.02$	\\
\swift\ XRT	& WT	& 00030912065	& 57\,793.04					& 201.04					& 0.3		& $0.46  \pm 0.04$	\\
\swift\ XRT	& WT	& 00030912066	& 57794.92 					& 202.92					& 1.5		& $0.39  \pm 0.02$	\\
\swift\ XRT	& WT	& 00030912067	& 57\,796.87					& 204.87 					& 0.6		& $0.36  \pm 0.02$	\\
\swift\ XRT	& WT	& 00030912068	& 57\,797.82					& 205.82 					& 1.1		& $0.38  \pm 0.02$	\\
\swift\ XRT	& WT	& 00030912069	& 57\,803.21					& 211.21					& 1.4		& $0.38  \pm 0.02$	\\
\swift\ XRT	& WT	& 00030912070	& 57\,803.86					& 211.86					& 1.5		& $0.43  \pm 0.02$	\\
\hline
\end{tabular}
\end{table*}

\begin{table*}
\centering
\caption{Log of all X-ray observations of \psr\ following the 2016 July outburst. The outburst onset  occurred on MJD 57\,597.06100694 (Kennea et al. 2016).
Part of these observations were already analysed by Archibald et al. (2016a).} 
\label{tab:psr1119}
\vspace{10pt}
\begin{tabular}{ccccccc} \hline
Instrument     	& Mode	& Obs. ID 			&  Mid point of observation	& Time since outburst onset 	& Exposure  	& Source net count rate \\
         		&          	&       			&  (MJD)					& (d)     			     	& (ks)   		& (counts s$^{-1}$)\\ 
\hline 
\swift\ XRT	& PC		& 00706396000	&  57\,597.08				&   0.01				& 2.2			& $0.544  \pm 0.02$	\\
\swift\ XRT	& WT	& 00034632001	&  57\,597.93				&   0.87 				& 9.6			& $0.510  \pm 0.008$	\\
\swift\ XRT	& WT	& 00034632002	&  57\,601.02				&   3.95 				& 4.8			& $0.53  \pm 0.01$	\\
\swift\ XRT	& PC		& 00034632003	&  57\,603.52				&   6.46 				& 3.0			& $0.202  \pm 0.008$	\\
\swift\ XRT	& PC		& 00034632005	&  57\,606.54				&   9.48 				& 3.0			& $0.212  \pm 0.008$	\\
\swift\ XRT	& WT	& 00034632007	&  57\,609.56				&  12.50 				& 5.5			& $0.49  \pm 0.01$	\\
\swift\ XRT	& WT	& 00034632008	&  57\,610.26				&  13.20 				& 1.3			& $0.39  \pm 0.02$	\\
\swift\ XRT	& PC		& 00034632009	&  57\,612.32				&  15.25 				& 2.3			& $0.179  \pm 0.009$	\\
\swift\ XRT	& WT	& 00034632010	&  57\,627.32				&  30.26 				& 3.0			& $0.40  \pm 0.01$	\\
\swift\ XRT	& WT	& 00034632011		&  57\,630.41				&  33.34 				& 3.2			& $0.40  \pm 0.01$	\\
\swift\ XRT	& WT	& 00034632013	&  57\,637.45				&  40.38 				& 2.1			& $0.30  \pm 0.01$	\\
\swift\ XRT	& WT	& 00034632014	&  57\,641.73				&  44.66 				& 2.4			& $0.28  \pm 0.02$	\\
\swift\ XRT	& WT	& 00034632015	&  57\,647.07				&  50.01 				& 2.1			& $0.33  \pm 0.01$	\\
\swift\ XRT	& WT	& 00034632016	&  57\,647.42				&  50.36 				& 1.7			& $0.23  \pm 0.01$	\\
\swift\ XRT	& WT	& 00034632017	&  57\,648.07				&  51.01 				& 1.9			& $0.40  \pm 0.02$	\\
\swift\ XRT	& WT	& 00034632018	&  57\,657.10				&  60.04 				& 1.9			& $0.36  \pm 0.01$	\\
\swift\ XRT	& WT	& 00034632019	&  57\,657.70				&  60.63 				& 1.9			& $0.31  \pm 0.01$	\\
\swift\ XRT	& WT	& 00034632020	&  57\,658.05				&  60.98 				& 2.1			& $0.34  \pm 0.01$	\\
\swift\ XRT	& WT	& 00034632022	&  57\,667.67				&  70.61 				& 1.6			& $0.25  \pm 0.01$	\\
\swift\ XRT	& WT	& 00034632023	&  57\,668.37				&  71.30 				& 2.0			& $0.24  \pm 0.01$	\\
\swift\ XRT	& WT	& 00034632024	&  57\,679.05				&  81.98 				& 2.0			& $0.17  \pm 0.01$	\\
\swift\ XRT	& WT	& 00034632025	&  57\,679.62				&  82.55 				& 2.0			&$0.31  \pm 0.01$	\\
\swift\ XRT	& WT	& 00034632026	&  57\,680.17				&  83.11 				& 1.8			&$0.19  \pm 0.01$	\\
\swift\ XRT	& WT	& 00034632027	&  57\,687.21				&  90.15 				& 1.9			&$0.22  \pm 0.01$	\\
\swift\ XRT	& WT	& 00034632028	&  57\,687.42				&  90.36 				& 1.9			&$0.34  \pm 0.01$	\\
\swift\ XRT	& WT	& 00034632029	&  57\,688.21				&  91.14 				& 1.7			&$0.22  \pm 0.01$	\\
\swift\ XRT	& WT	& 00034632030	&  57\,693.19				&  96.12 				& 2.1			&$0.23  \pm 0.01$	\\
\swift\ XRT	& WT	& 00034632037	&  57\,707.27				& 110.21				& 2.5			&$0.18  \pm 0.01$	\\
\swift\ XRT	& WT	& 00034632039	&  57\,708.31				& 111.25				& 2.0			&$0.18 \pm 0.01$	\\
\swift\ XRT	& WT	& 00034632040	&  57\,709.87				& 112.80				& 2.1			&$0.16  \pm 0.01$	\\
\swift\ XRT	& WT	& 00034632041	&  57\,714.19				& 117.12				& 1.9			&$0.22  \pm 0.01$	\\
\swift\ XRT	& WT	& 00034632042	&  57\,714.59				& 117.52				& 1.9			&$0.19  \pm 0.01$	\\
\swift\ XRT	& WT	& 00034632043	&  57\,715.18				& 118.12				& 2.1			&$0.146  \pm 0.009$	\\
\swift\ XRT	& WT	& 00034632044	&  57\,729.22				& 132.16				& 2.6			&$0.175  \pm 0.009$	\\
\swift\ XRT	& WT	& 00034632045	&  57\,729.65				& 132.59				& 5.5			&$0.211  \pm 0.007$	\\
\swift\ XRT	& WT	& 00034632046	&  57\,730.14				& 133.08				& 4.9			&$0.142  \pm 0.006$	\\
\hline
\end{tabular}
\end{table*}

\clearpage

\begin{table*}
\section{\swift\ XRT light curves}
\label{xrtcurves}

This section reports a series of figures showing the cooling curves for magnetar outbursts as observed 
by the X-ray Telescope on board \swift\ and in terms of the count rate. The 0.3--10~keV light curves 
were created by exploiting both PC- and WT-mode data, and using the online \swift\ XRT data products 
generator (\url{http://www.swift.ac.uk/user_objects/}). This tool corrects for instrumental artefacts such 
as pile up and bad columns on the CCD (see Evans et al. 2007, 2009 for more details).
\end{table*}

\begin{figure*}
\begin{center}
\includegraphics[width=7.7cm]{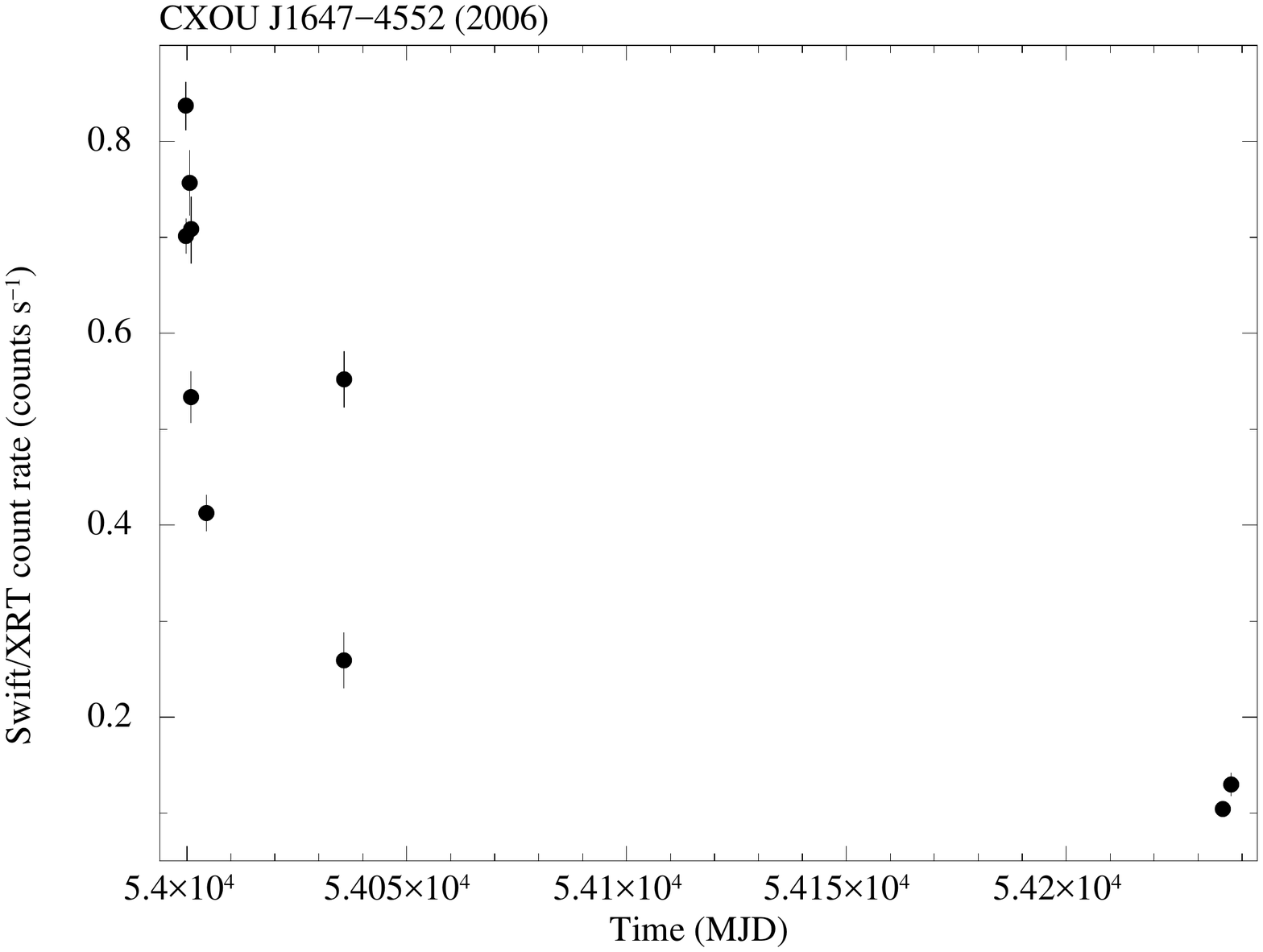}
\includegraphics[width=7.7cm]{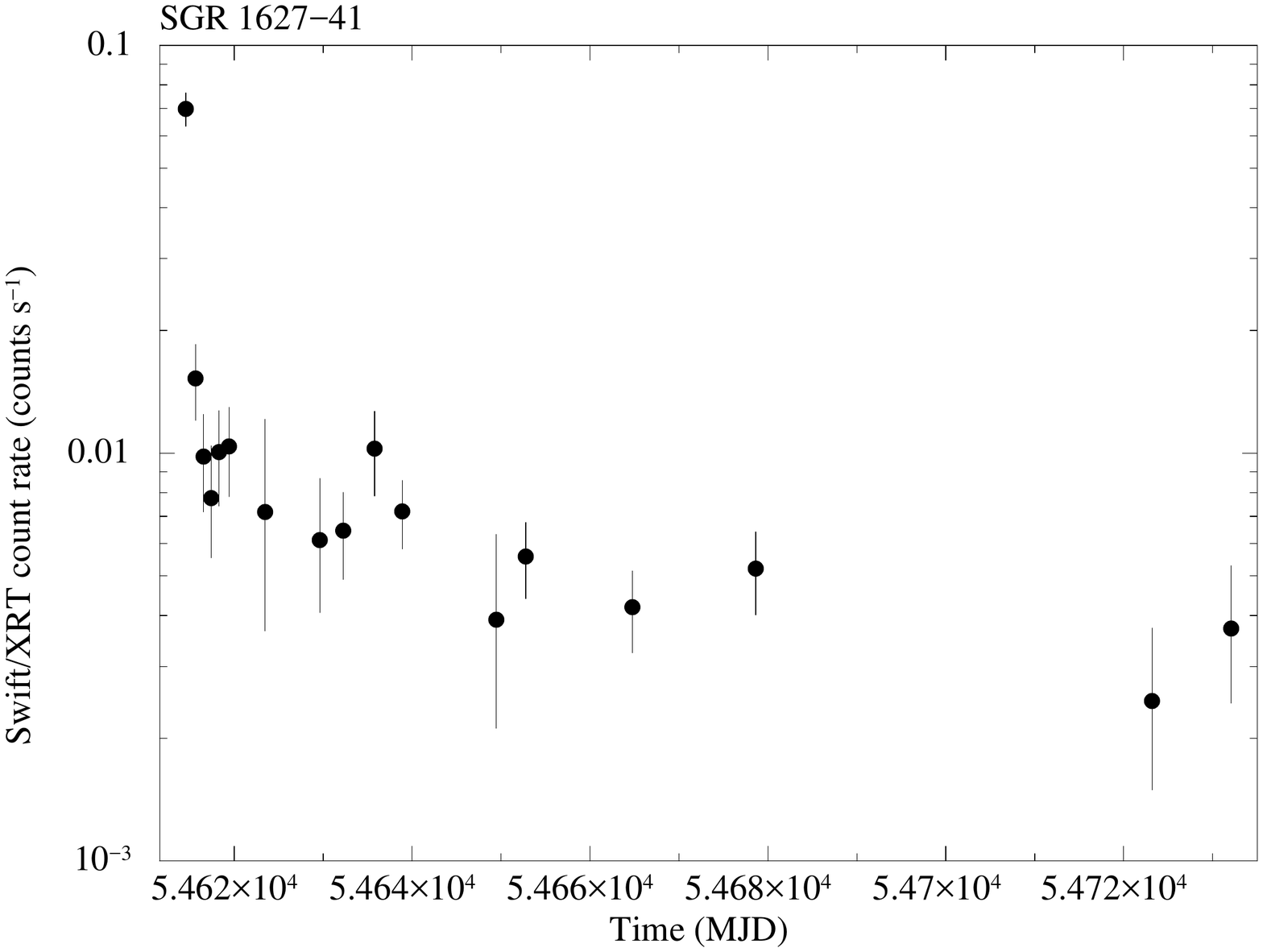}\\ \vspace{-1.2cm}
\includegraphics[width=7.7cm]{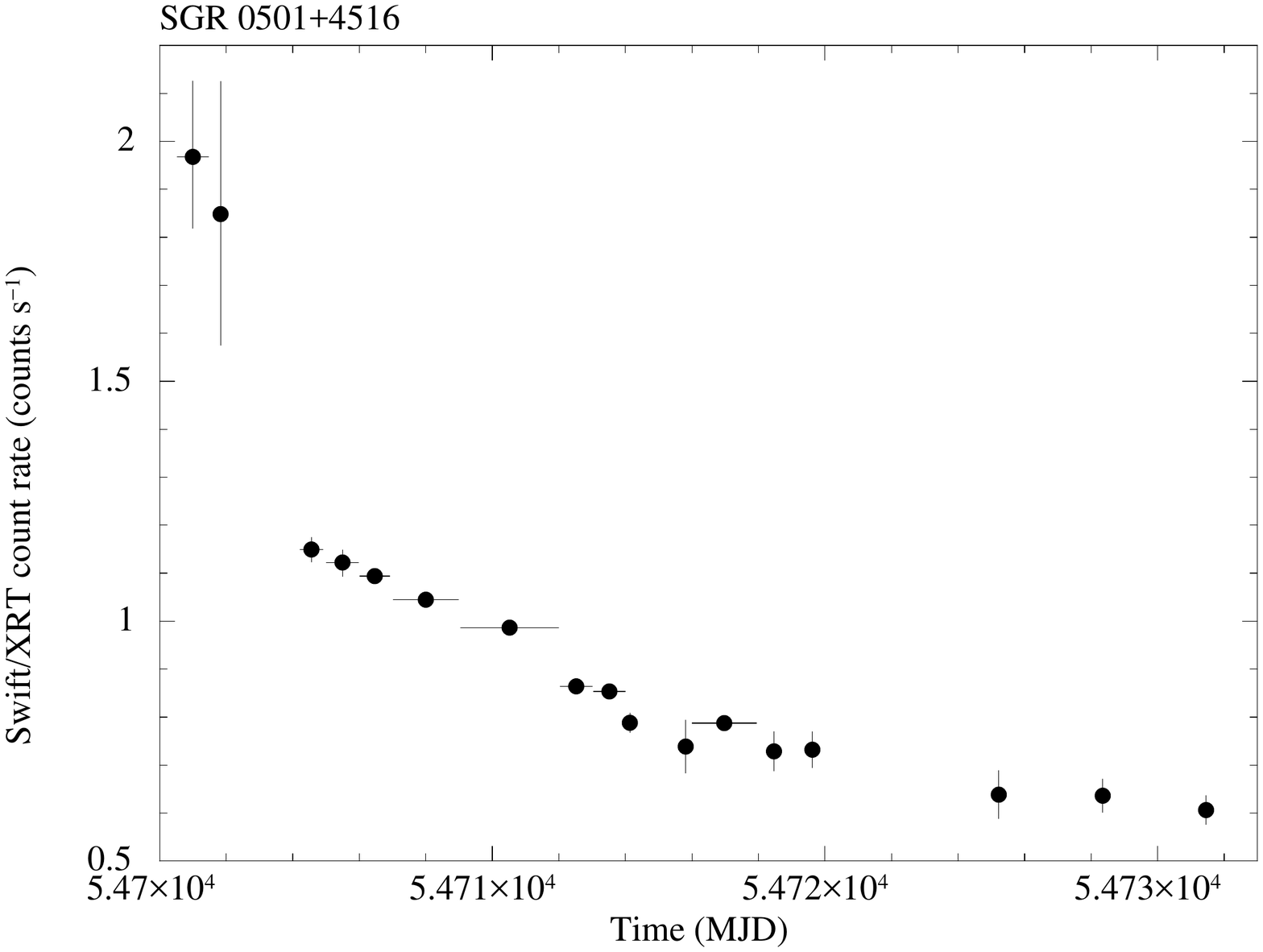}
\includegraphics[width=7.7cm]{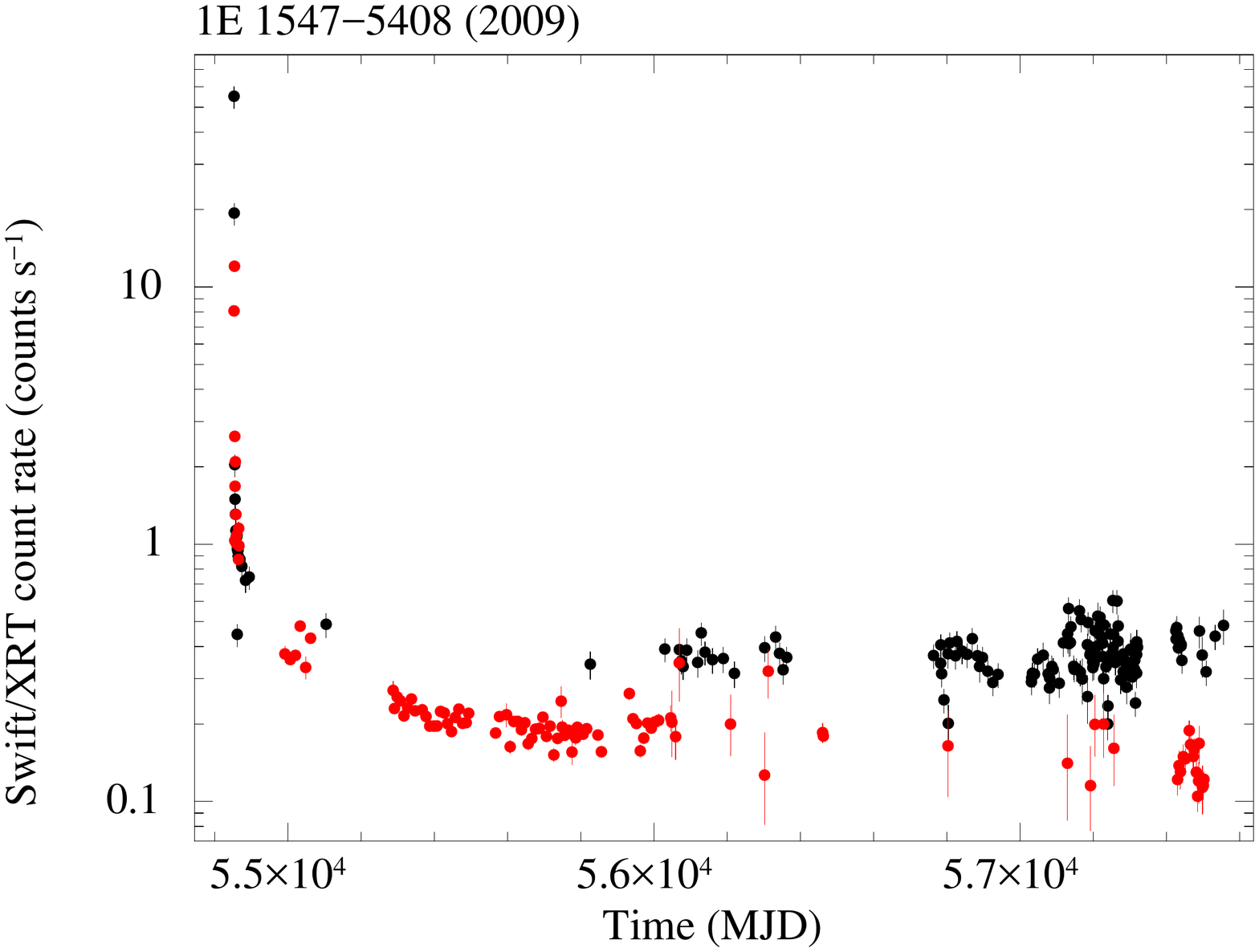}\\ \vspace{-1.2cm}
\includegraphics[width=7.7cm]{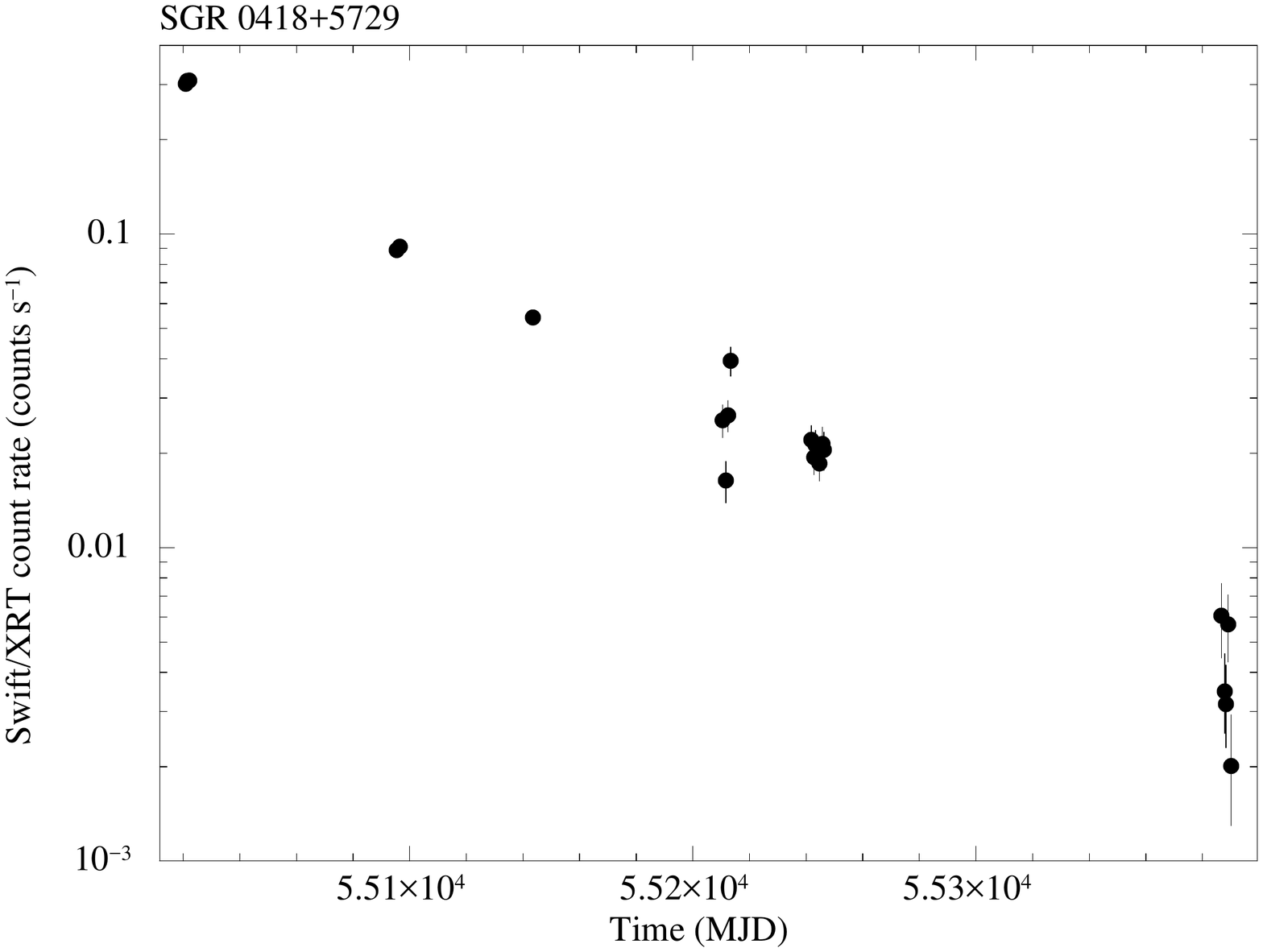}
\includegraphics[width=7.7cm]{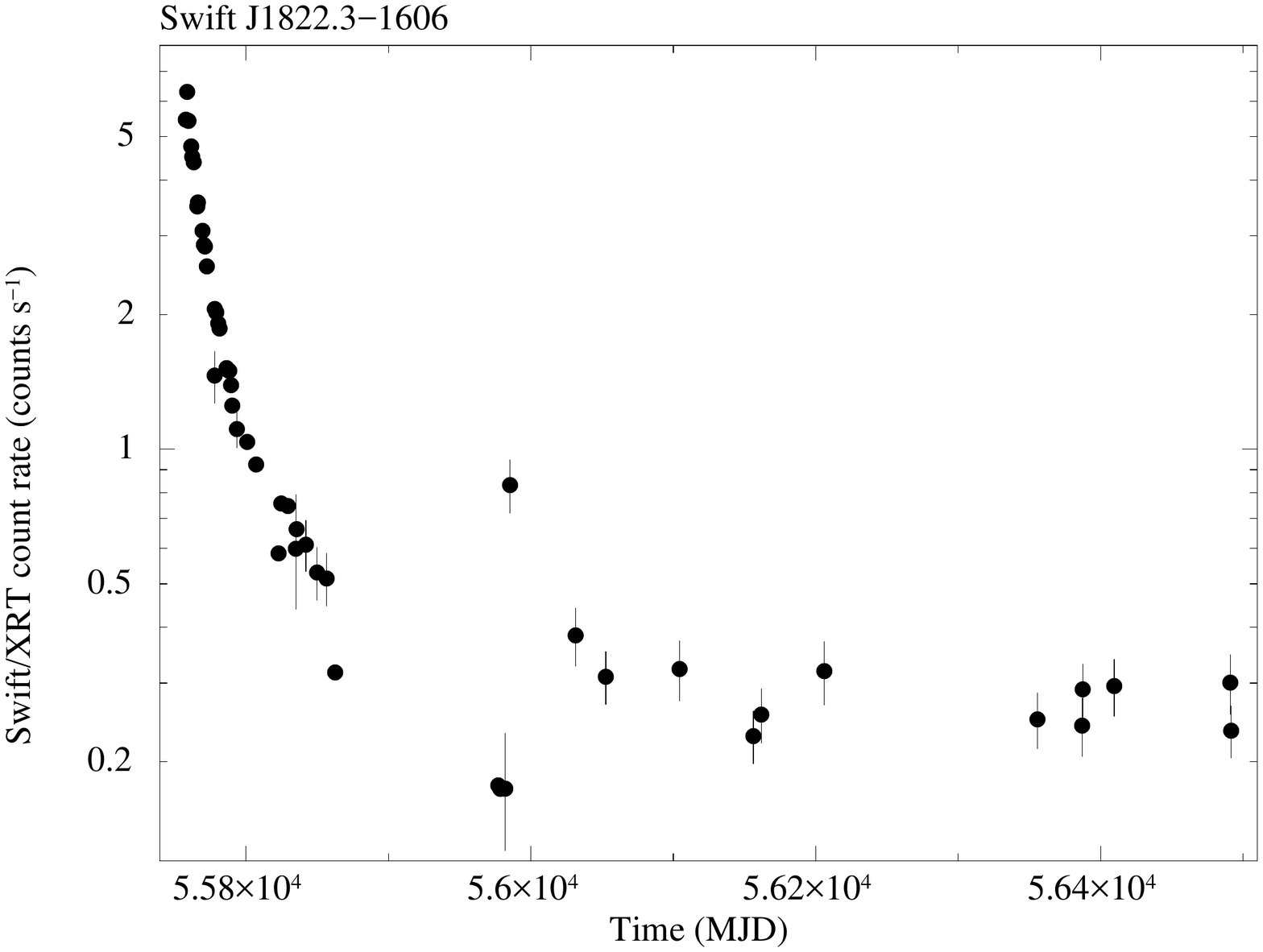}\\ \vspace{-1.2cm}
\includegraphics[width=7.7cm]{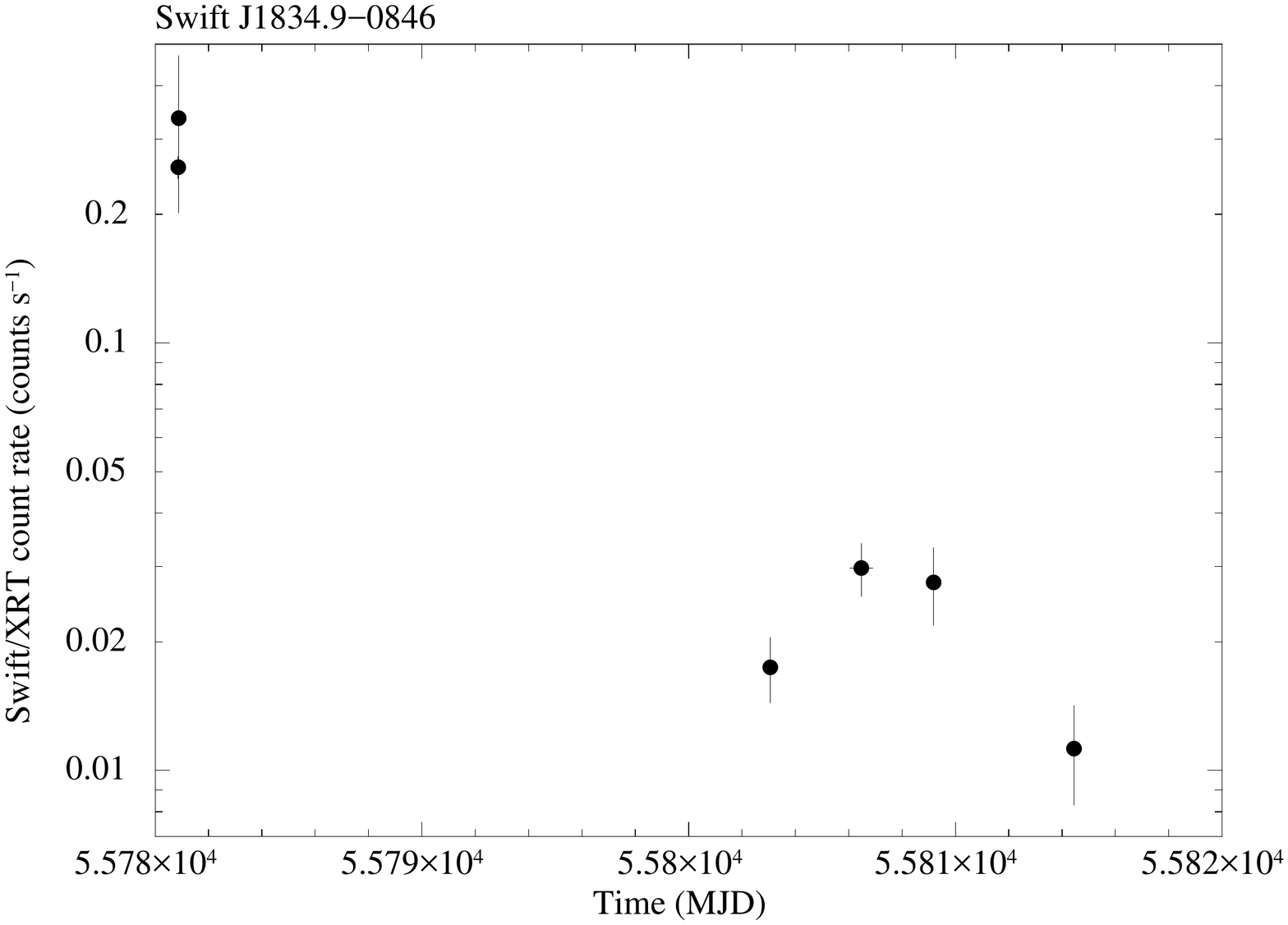}
\includegraphics[width=7.7cm]{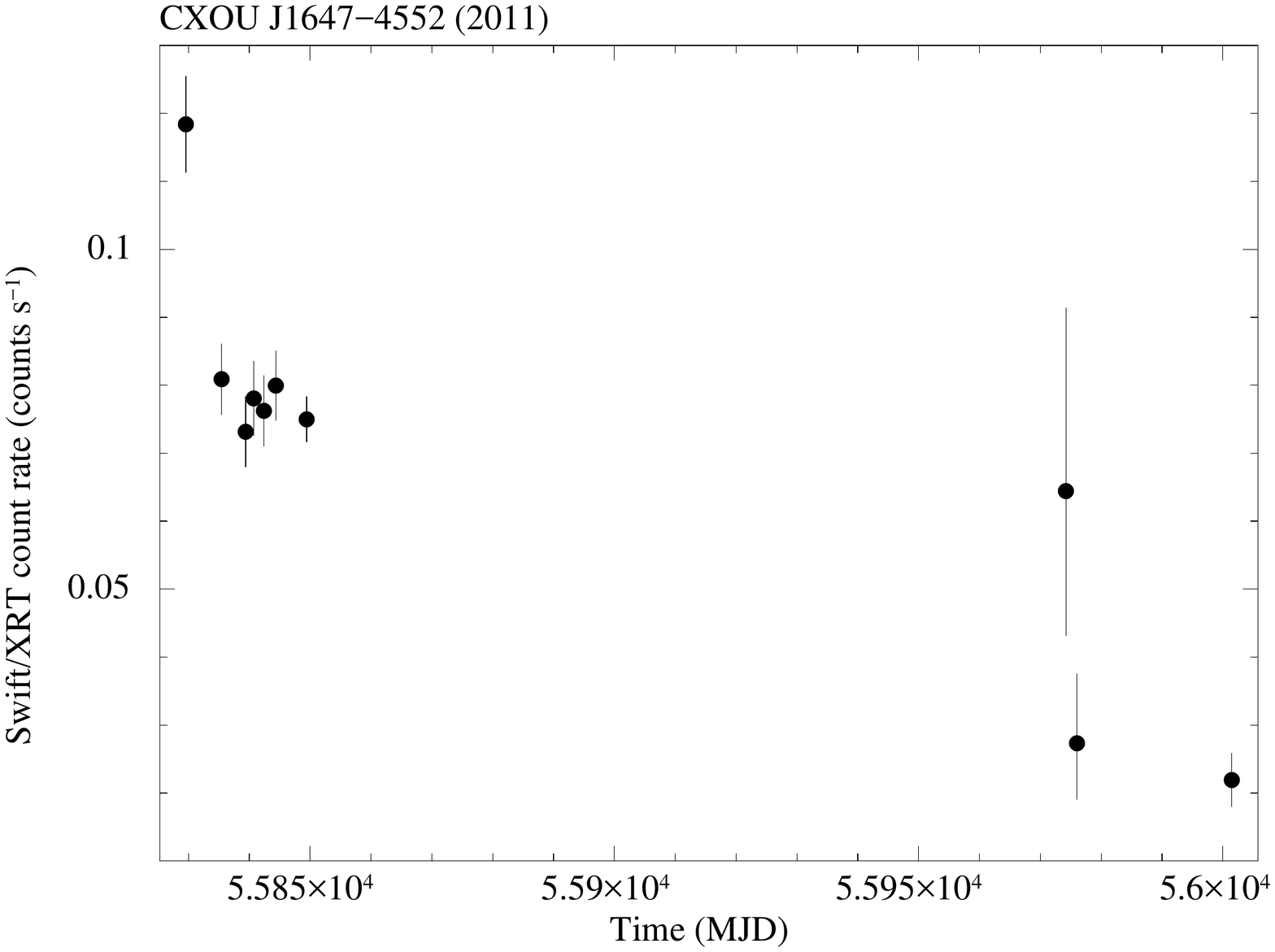}
\vspace{-0.5cm}
\caption{\swift\ XRT long-term 0.3--10~keV light curves of densely monitored magnetar outbursts, created using the online \swift\ XRT data products generator (Evans et al. 2009).
In case both PC- and WT-mode data are available, black (red) dots refer to data acquired with the XRT set in WT (PC) mode.} 
\label{fig:flux_outbursts}
\end{center}
\end{figure*}

\setcounter{figure}{0}
\begin{figure*}
\begin{center}
\includegraphics[width=7.7cm]{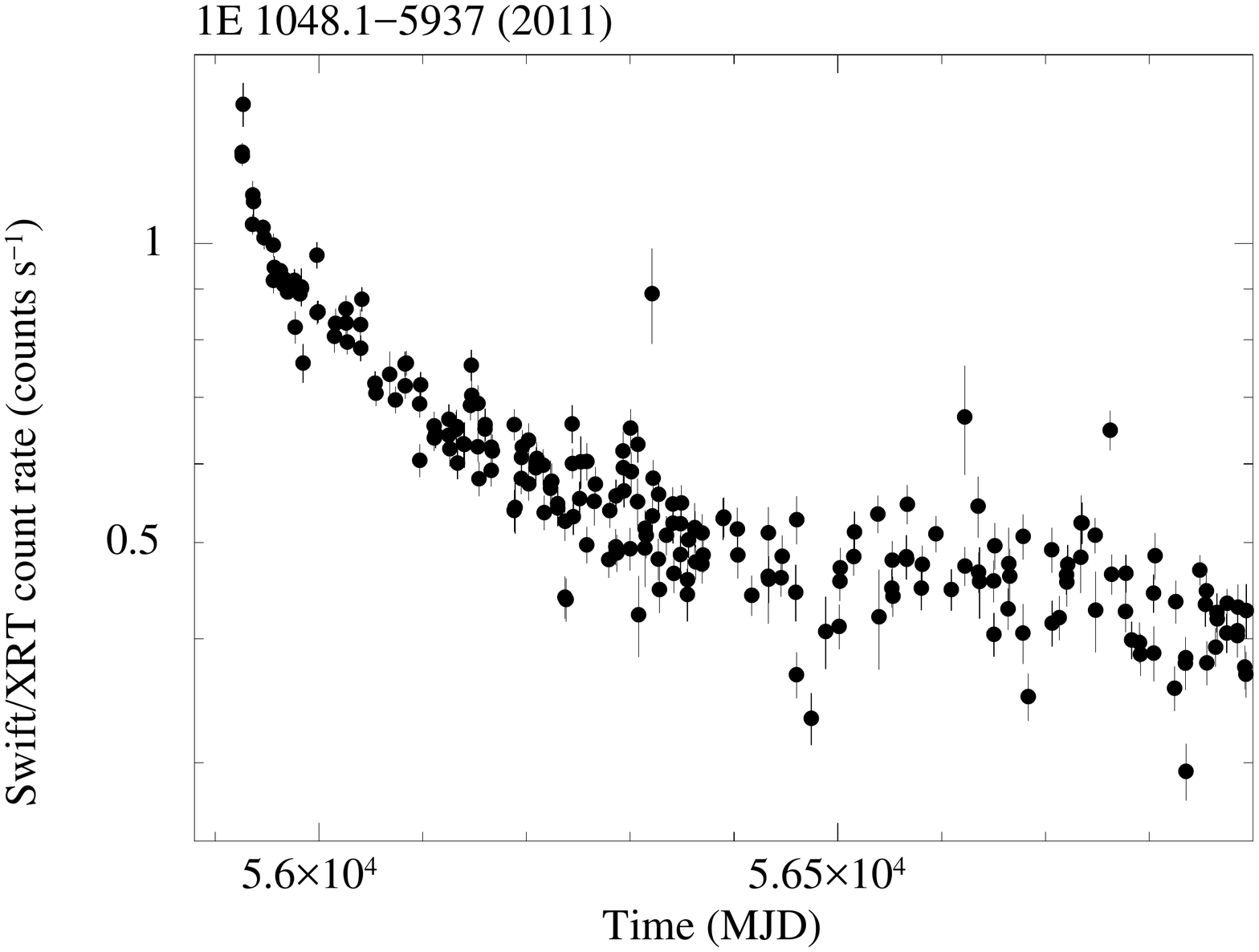}
\includegraphics[width=7.7cm]{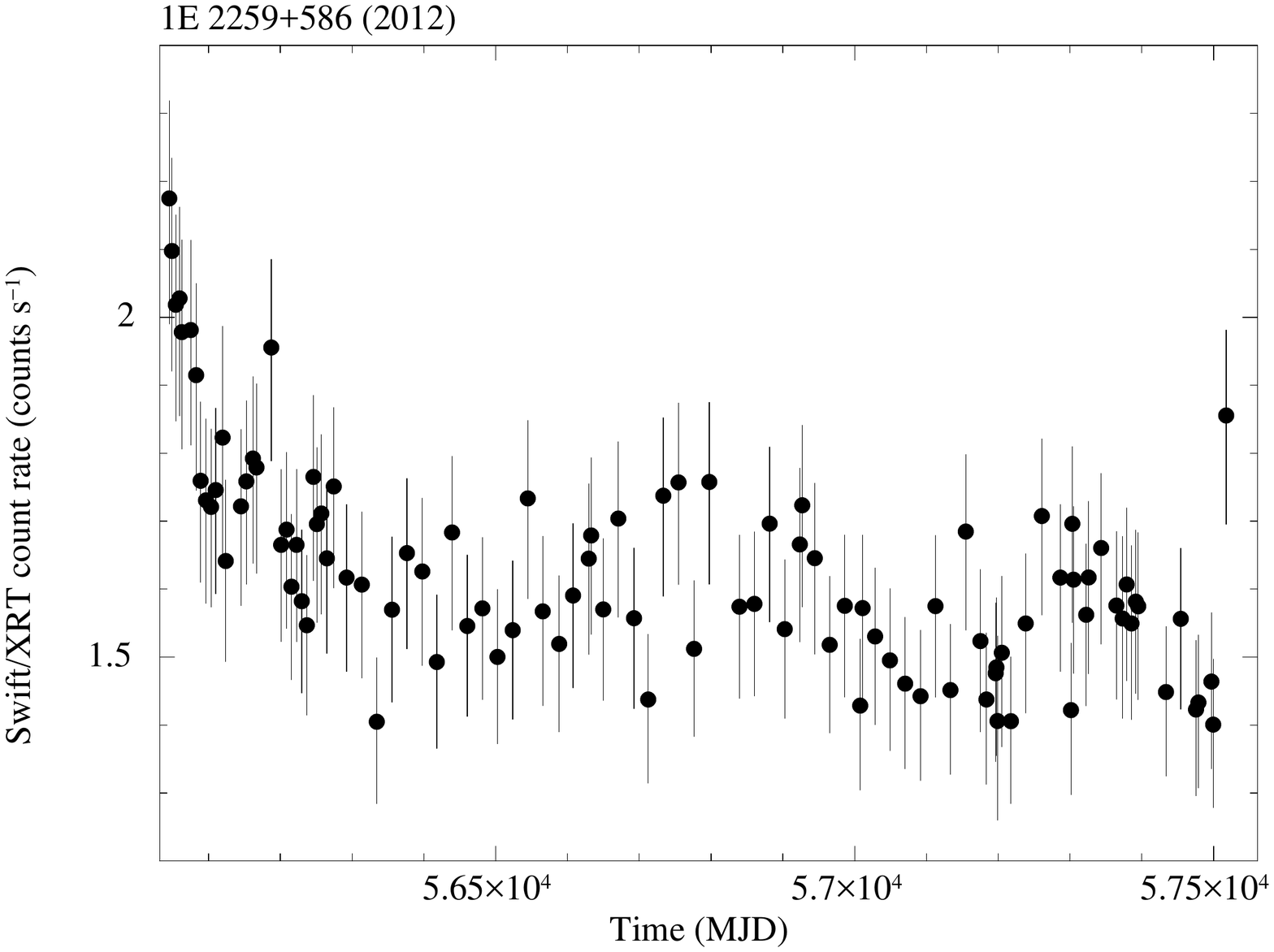}\\ \vspace{-1.2cm}
\includegraphics[width=7.7cm]{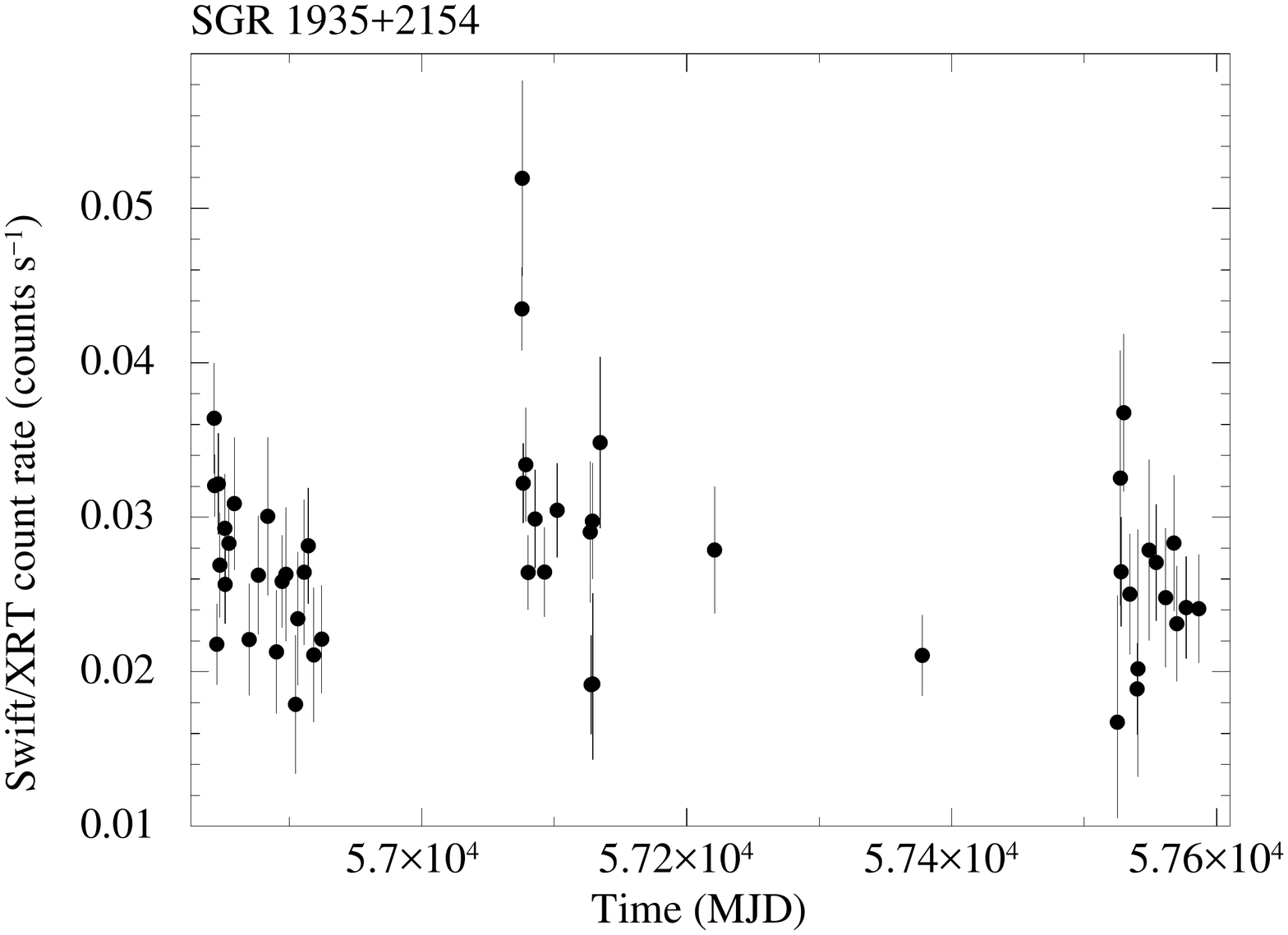}
\includegraphics[width=7.7cm]{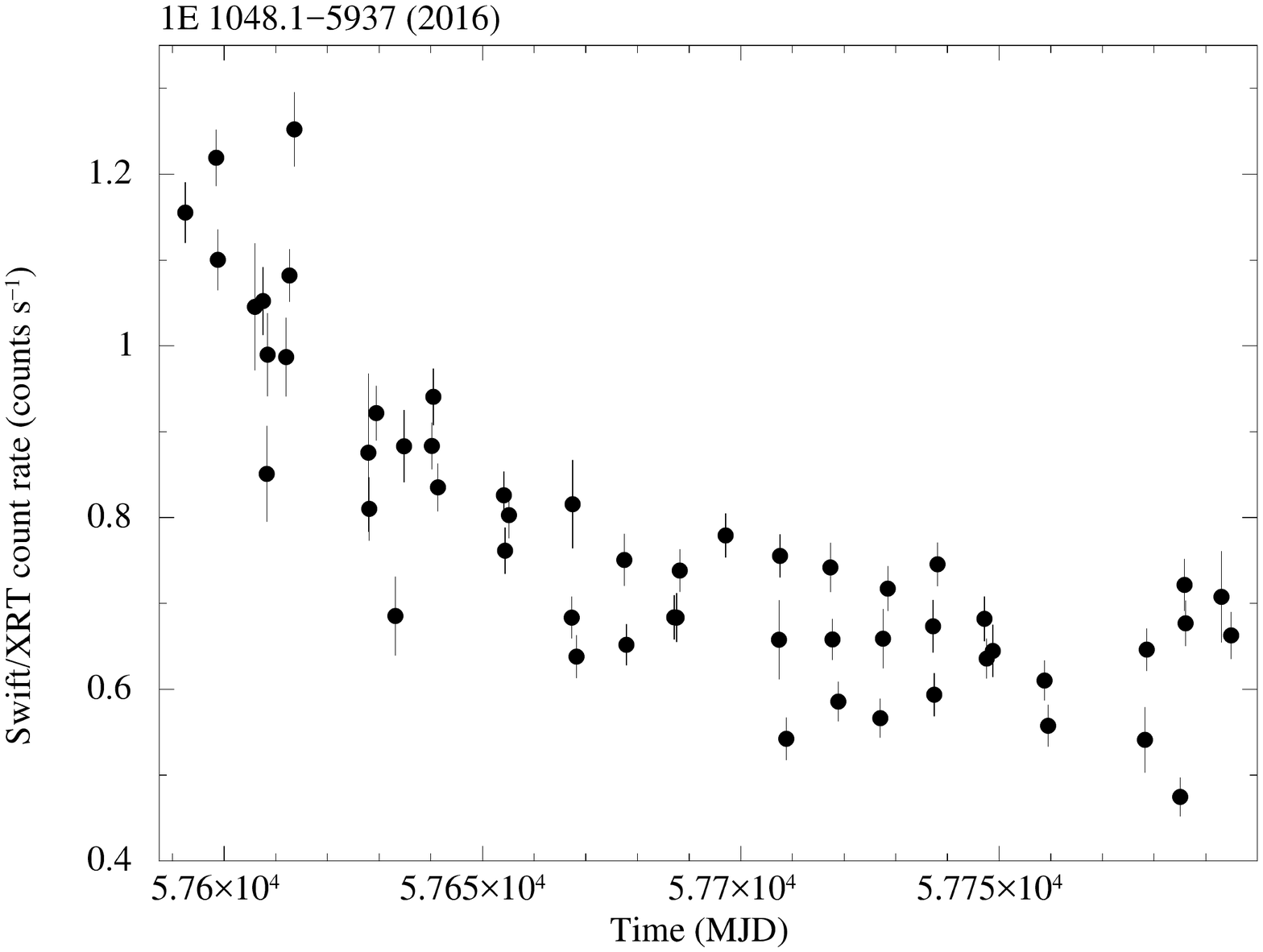} \vspace{-1.2cm}
\includegraphics[width=7.7cm]{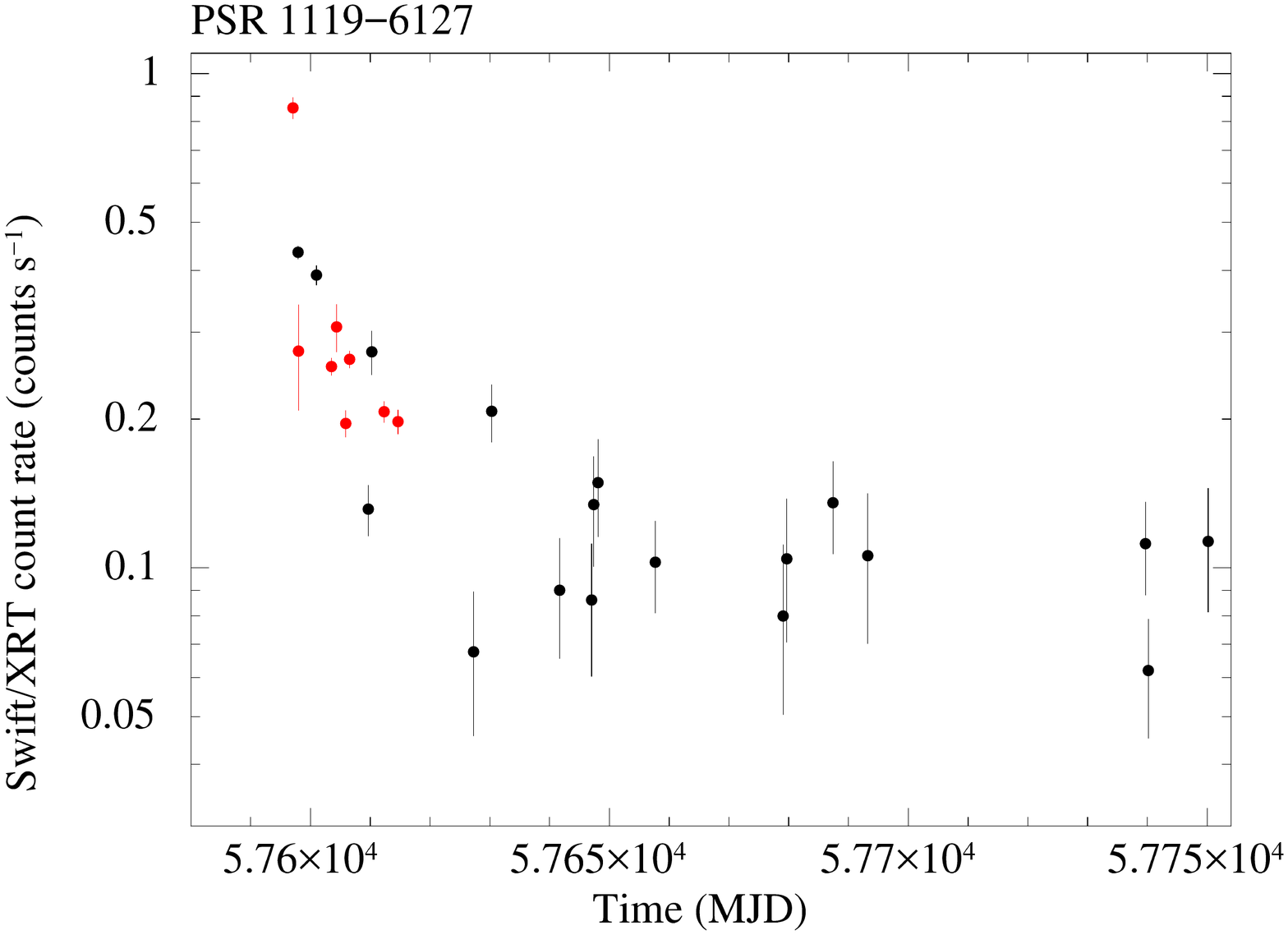}
\includegraphics[width=7.7cm]{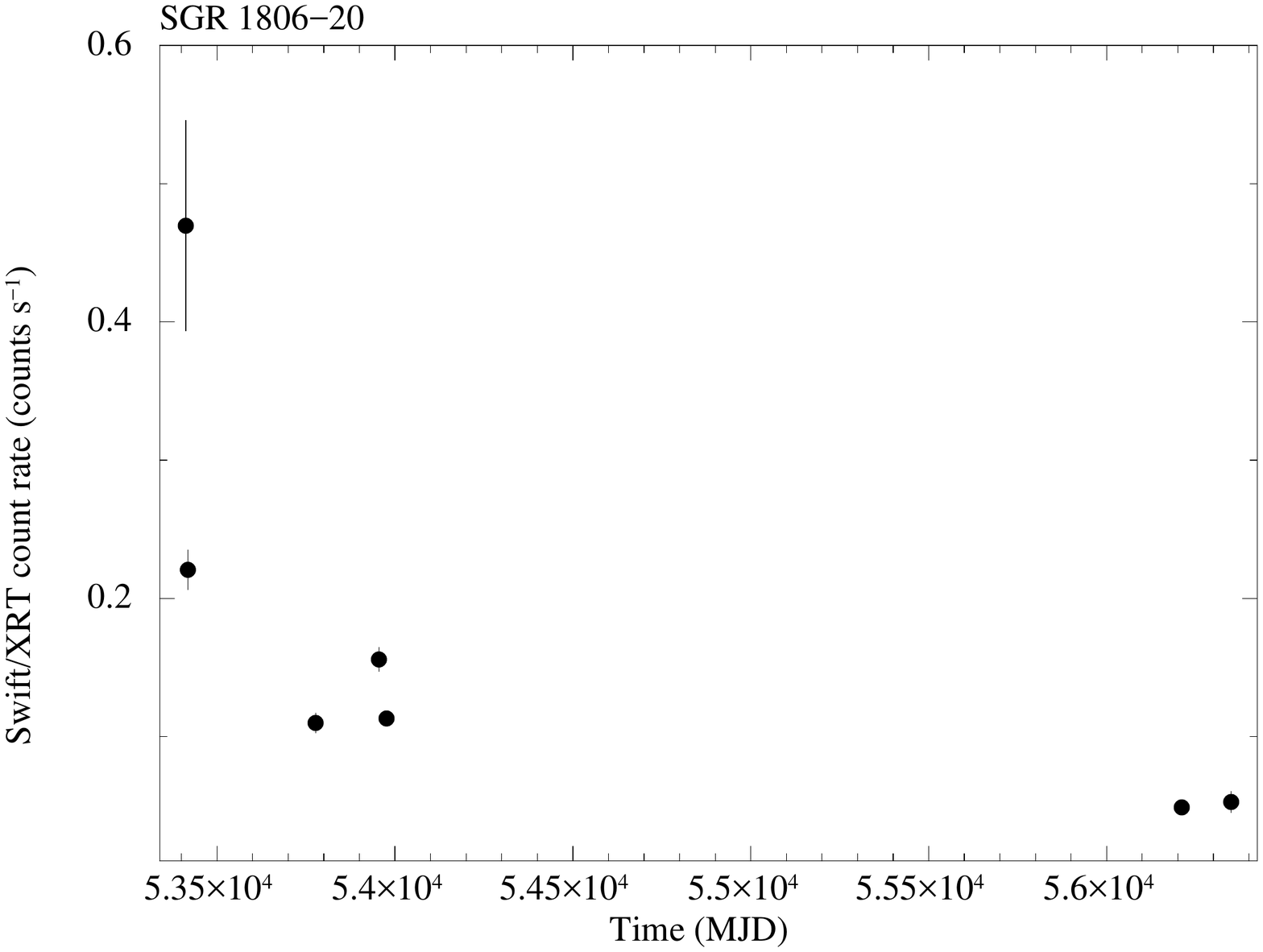} \vspace{-1.2cm}
\includegraphics[width=7.7cm]{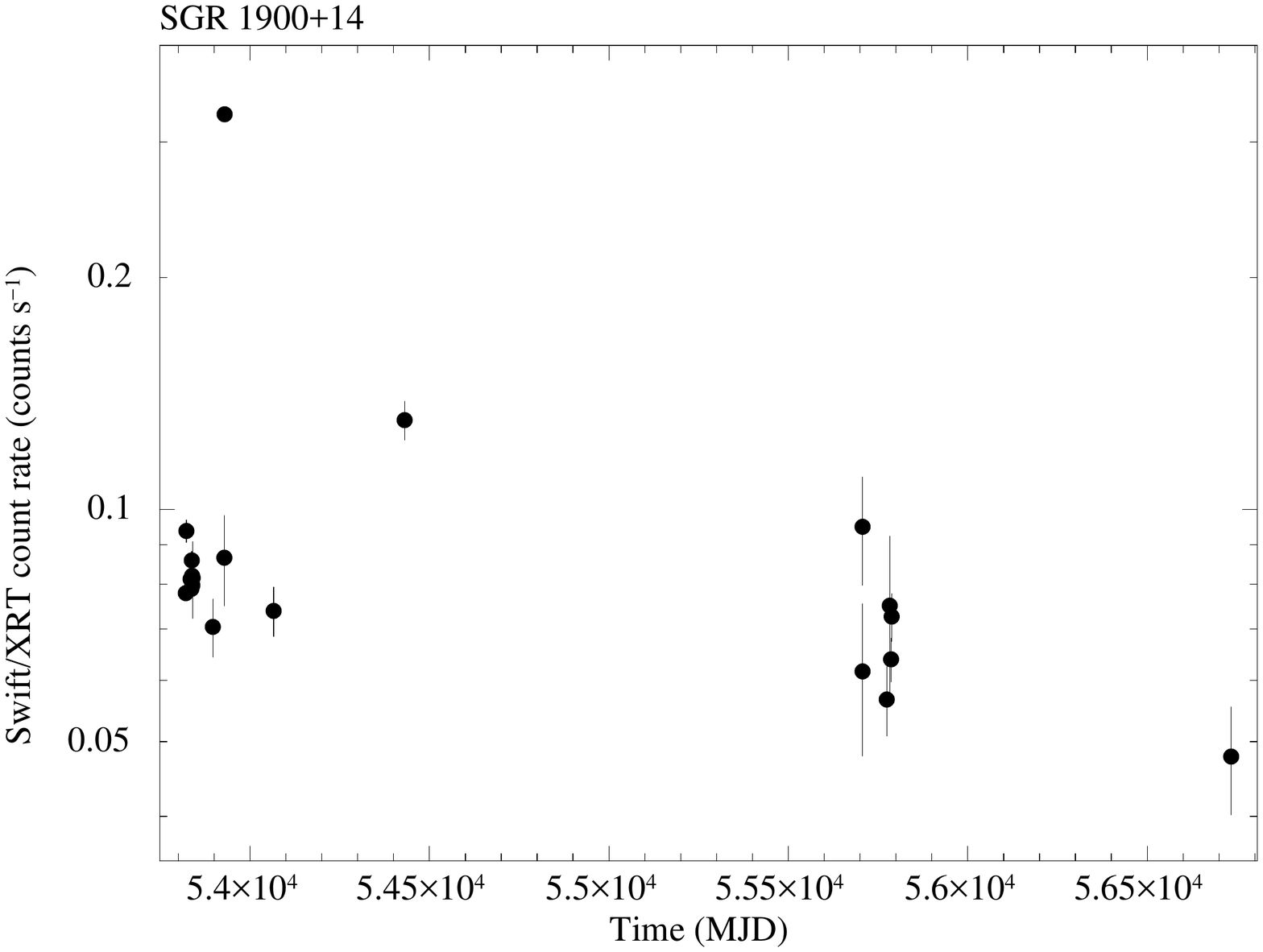}
\vspace{1cm}
  \caption{ -- \emph{continued}}
\label{fig:flux_outbursts}
\end{center}
\end{figure*}

\clearpage
\begin{table*}
\section{High-statistics quality X-ray spectra and fitted models}
\label{goodspectra}

This section reports a series of figures showing several high-quality spectra and the best-fitting empirical 
models (see Table~\ref{tab:results}) for the outbursts that were repeatedly monitored by the \xmm\ 
or \cxo\ observatories. In each case we plot the $E \times F(E)$ unfolded spectra and the models, 
to highlight the contributions of the different spectral components to the total X-ray emission (i.e. 
multiple blackbodies or blackbody plus power law; see the dotted lines in the figures) as a function 
of time. Post-fit residuals in units of standard deviations are also plotted at the bottom of each panel. 
In all cases, the data points were re-binned for plotting purpose, to better visualize the trend in 
the spectral residuals. The colours are associated with the chronological order of the observations 
according to the following code: black, red, green, blue, light blue, magenta, yellow, orange, 
yellow+green, green+cyan, blue+cyan, blue+magenta, red+magenta, dark grey, light grey.
\end{table*}

\begin{figure*}
\begin{center}
\includegraphics[width=8cm]{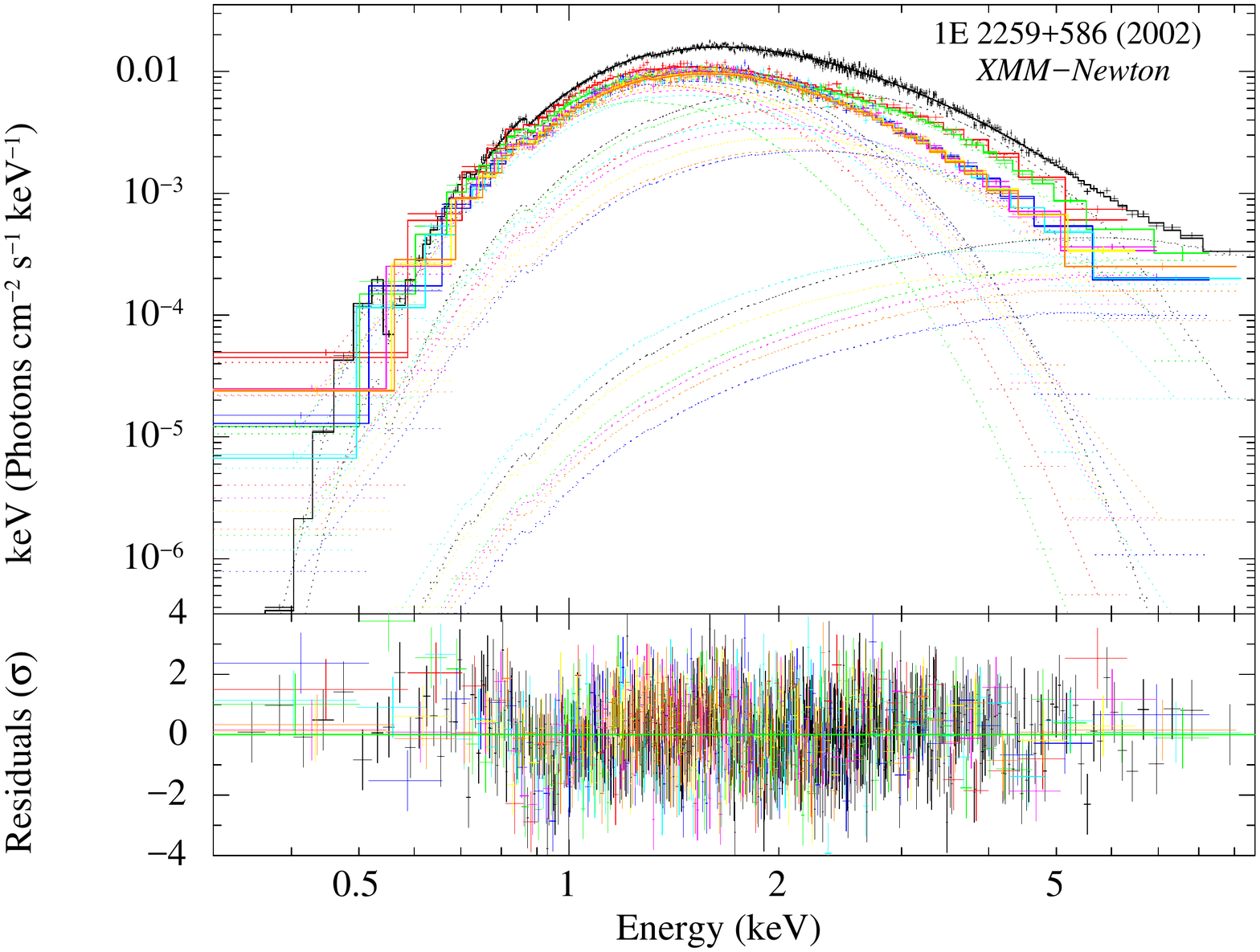}
\includegraphics[width=8cm]{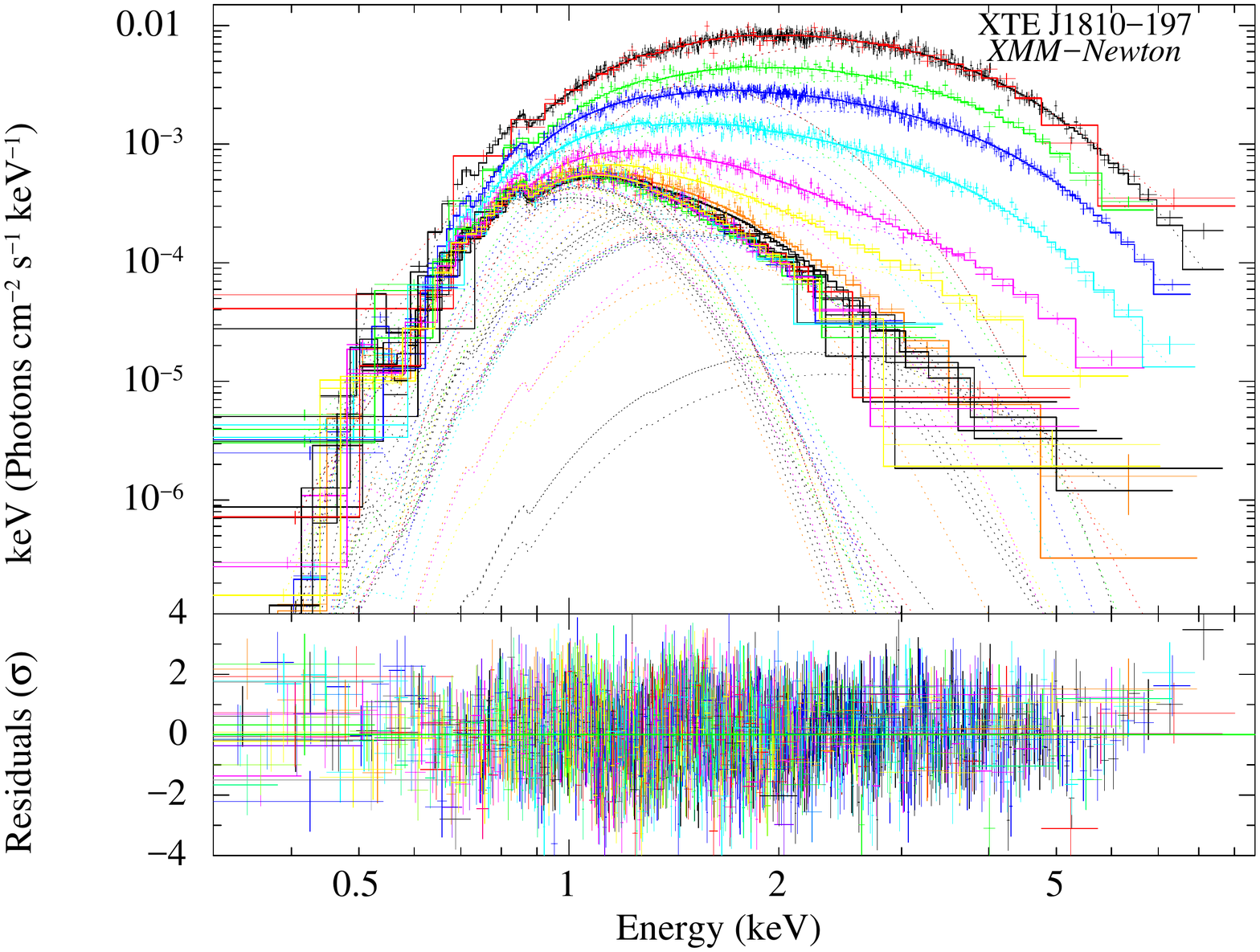}\\
\includegraphics[width=8cm]{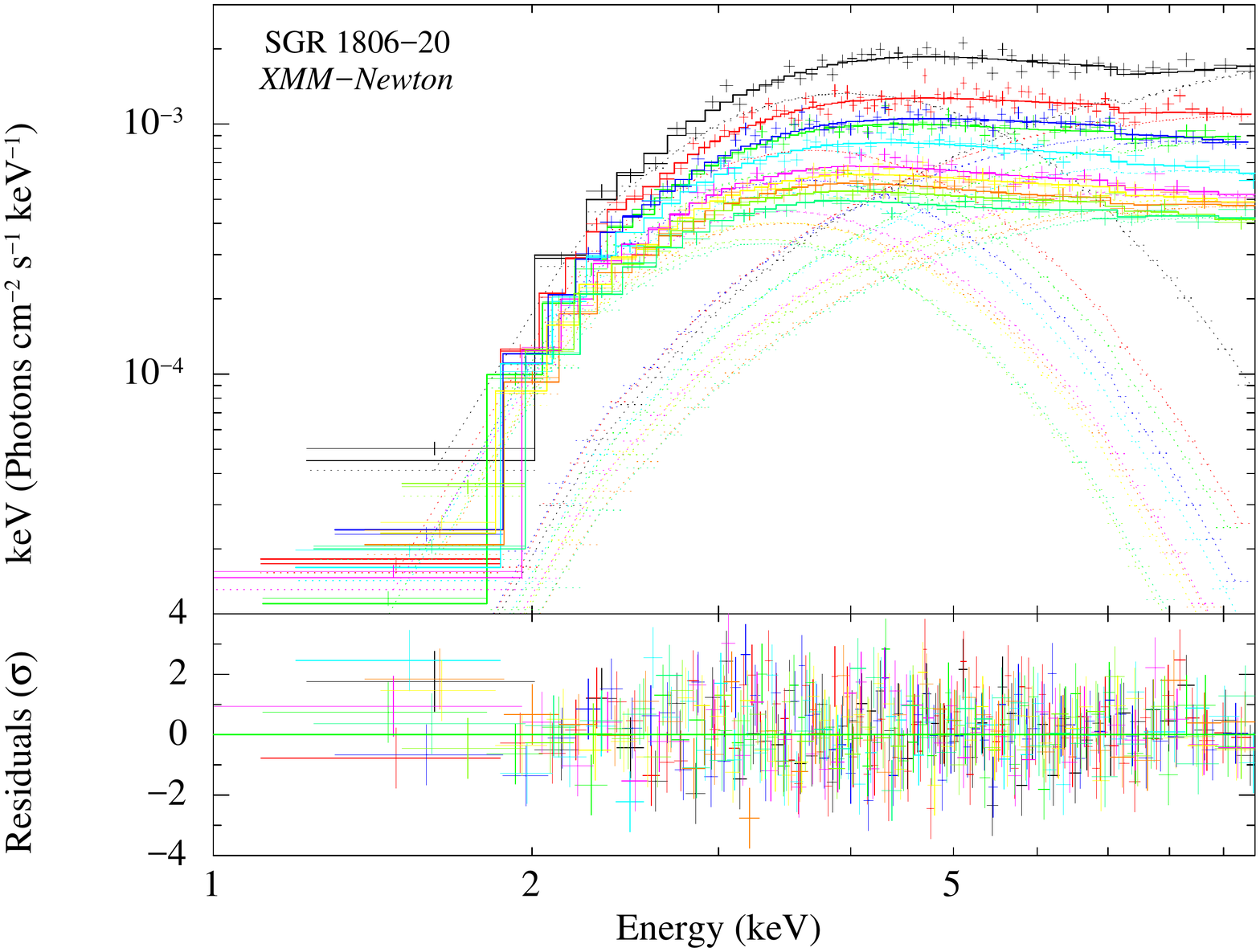}
\includegraphics[width=8cm]{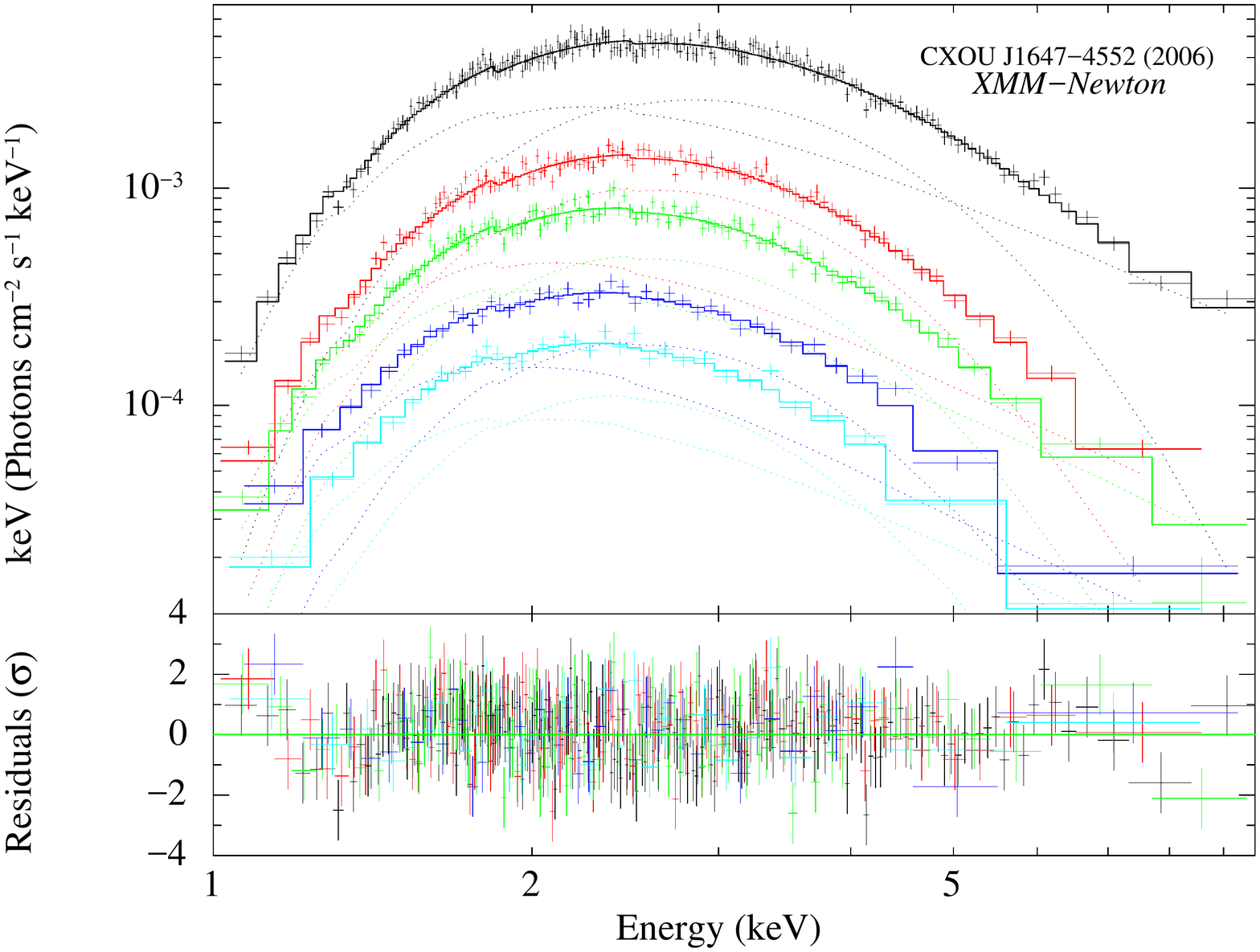}\\
\includegraphics[width=8cm]{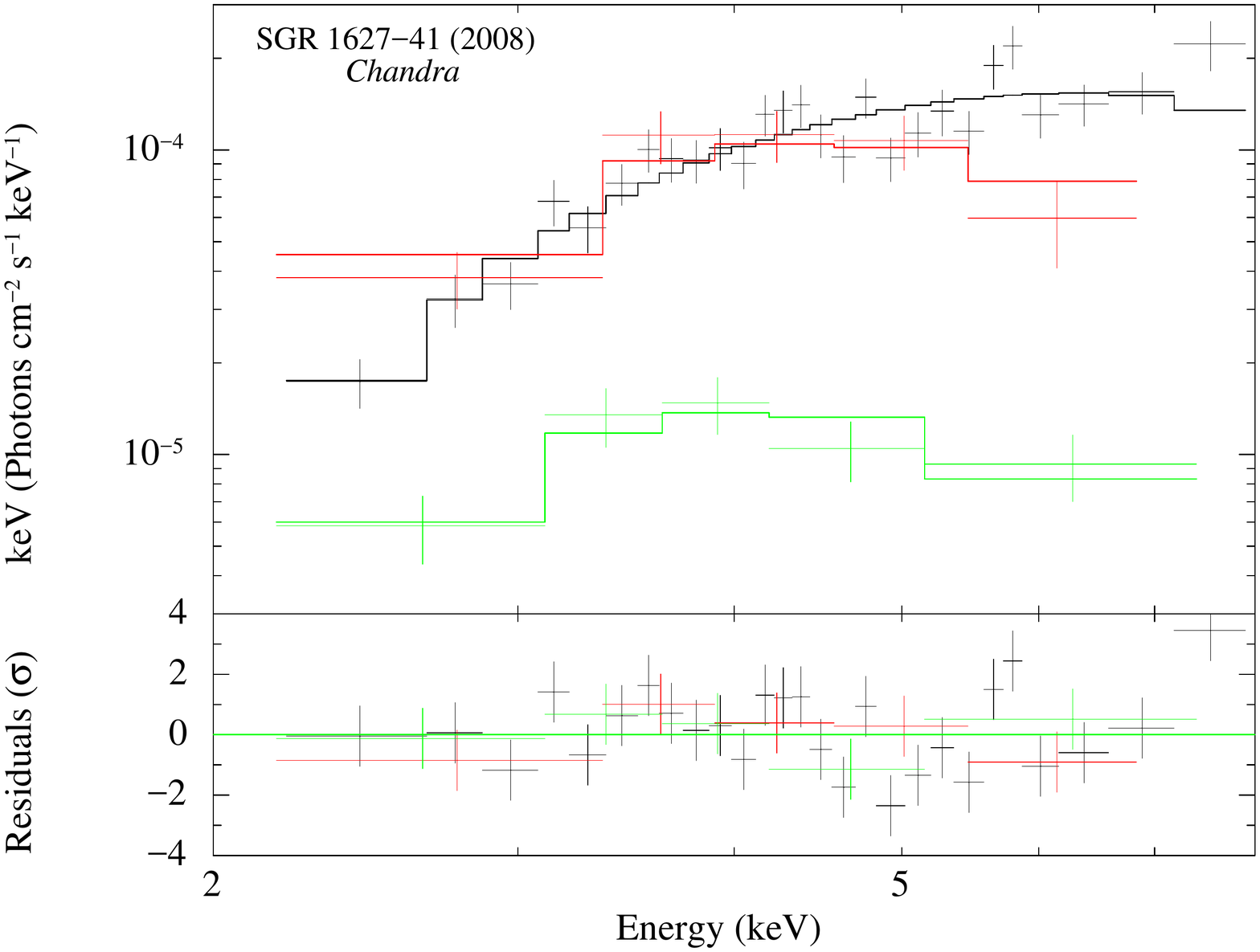}
\includegraphics[width=8cm]{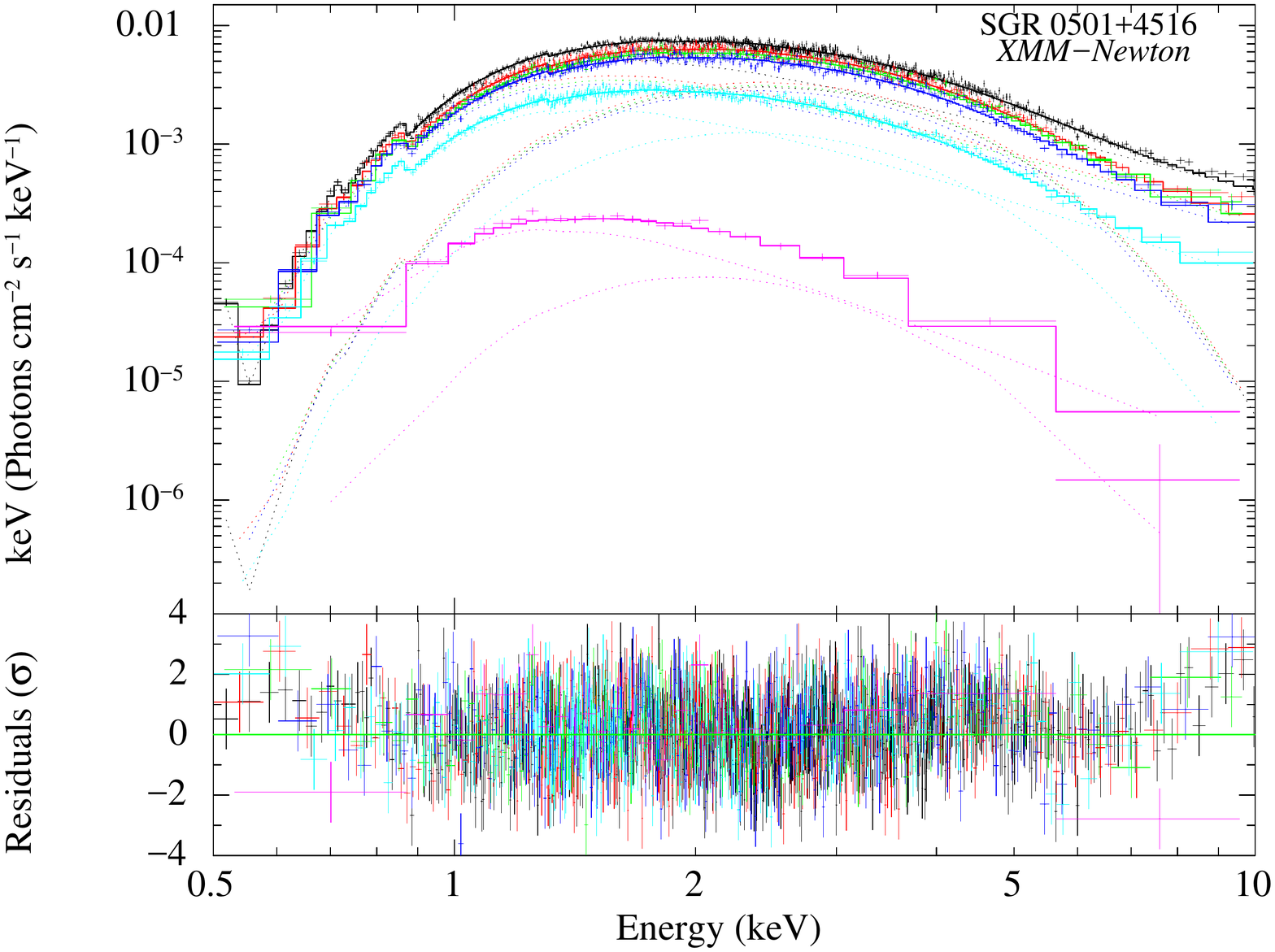}\\
\caption{High quality unfolded spectra for magnetar outbursts that were repeatedly monitored with the \xmm\ or \cxo\ observatories. 
Best-fitting models are marked by the solid lines, whereas the contributions of the different spectral components are marked by the 
dotted lines (see Table~\ref{tab:results} for more details on the models employed).}
\label{fig:spectra_outbursts}
\end{center}
\end{figure*}

\setcounter{figure}{0}
\begin{figure*}
\begin{center}
\includegraphics[width=8cm]{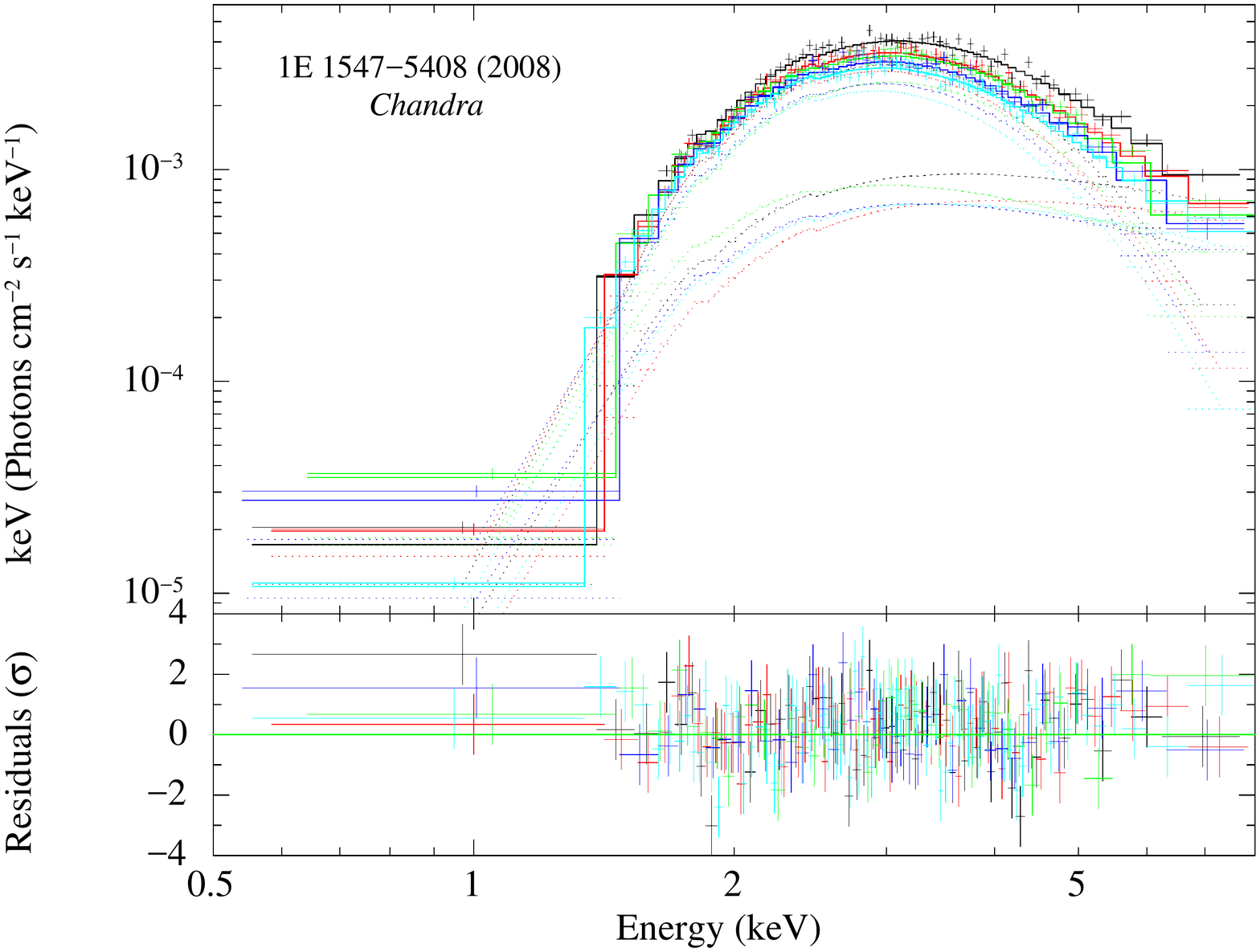}
\includegraphics[width=8cm]{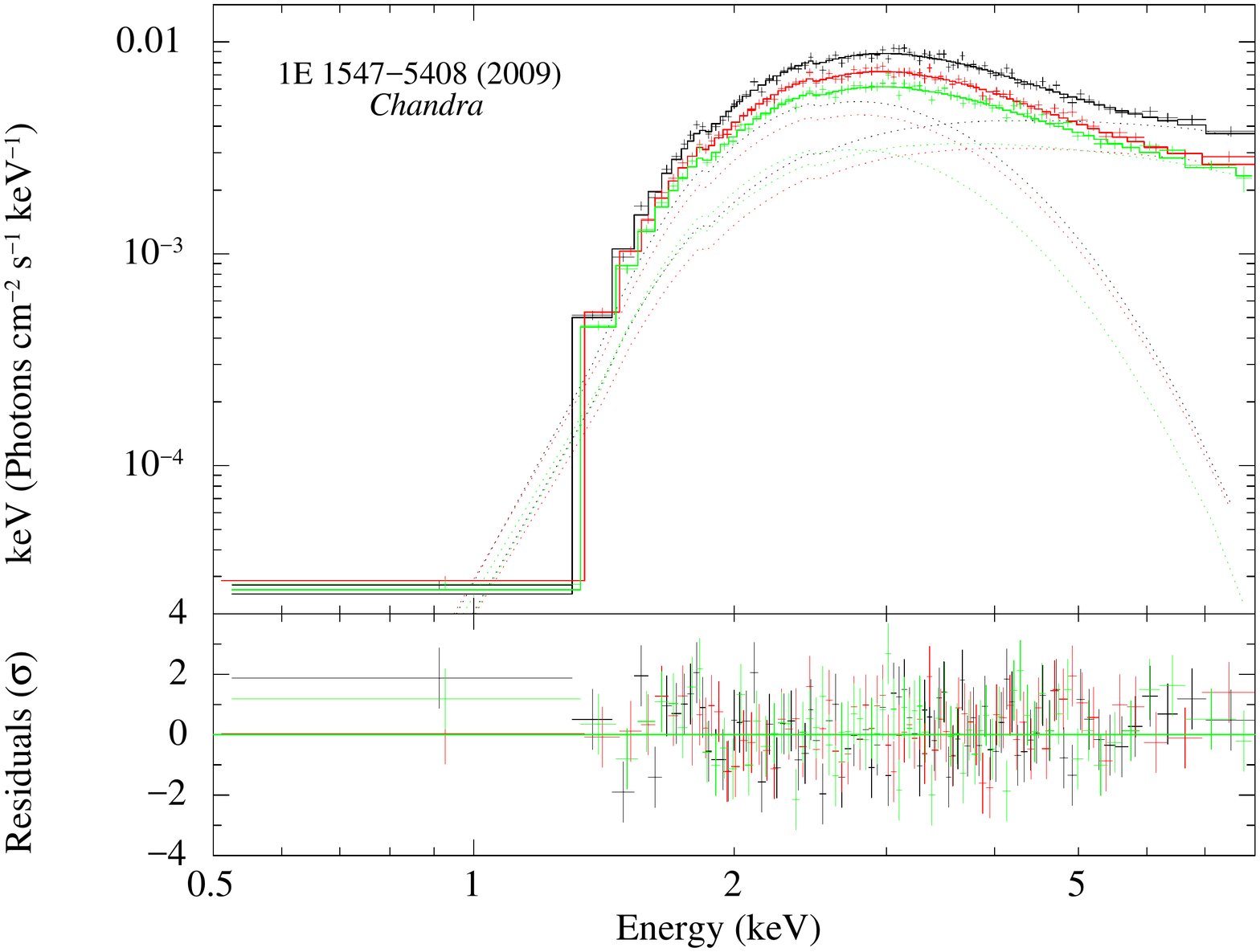}\\
\includegraphics[width=8cm]{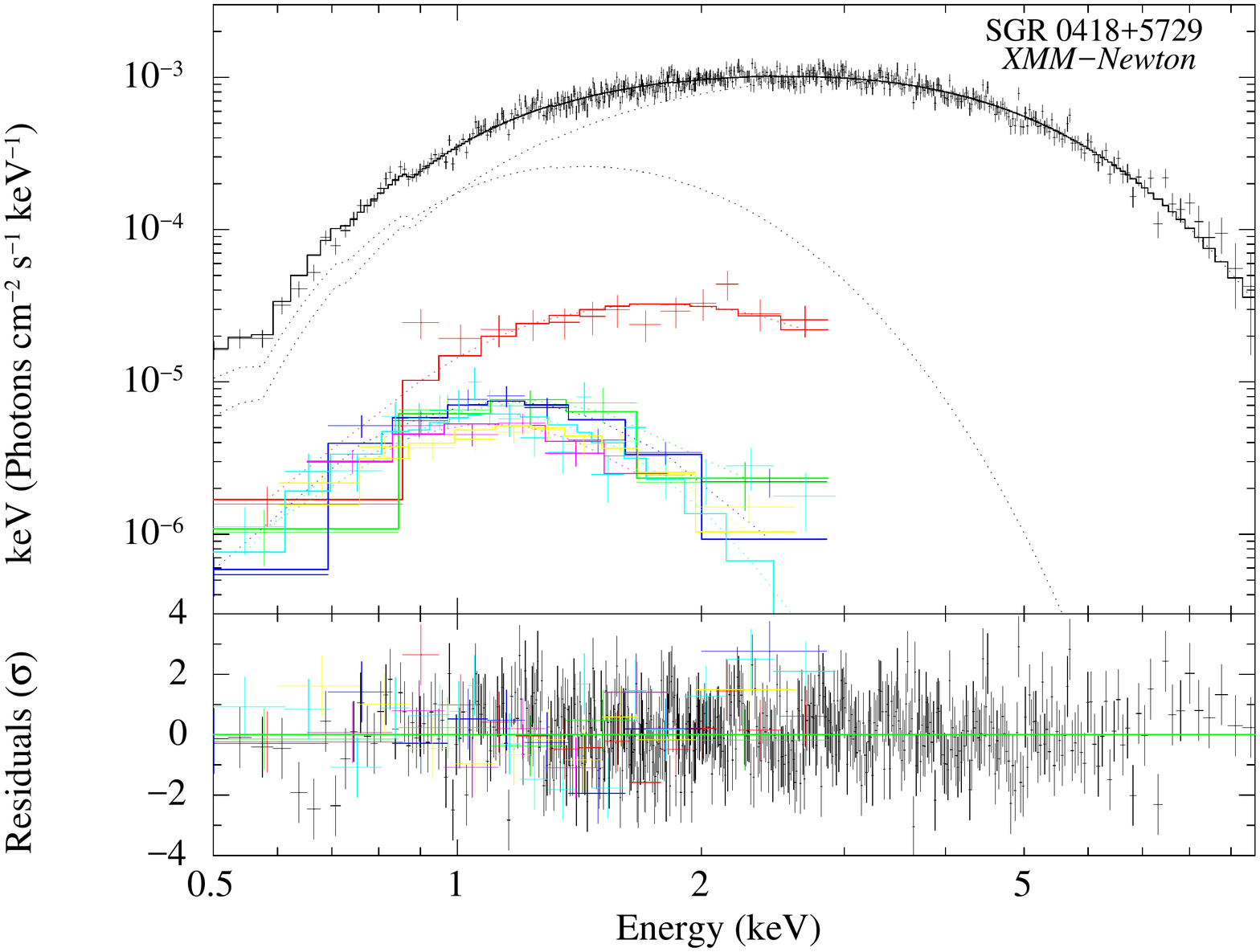}
\includegraphics[width=8cm]{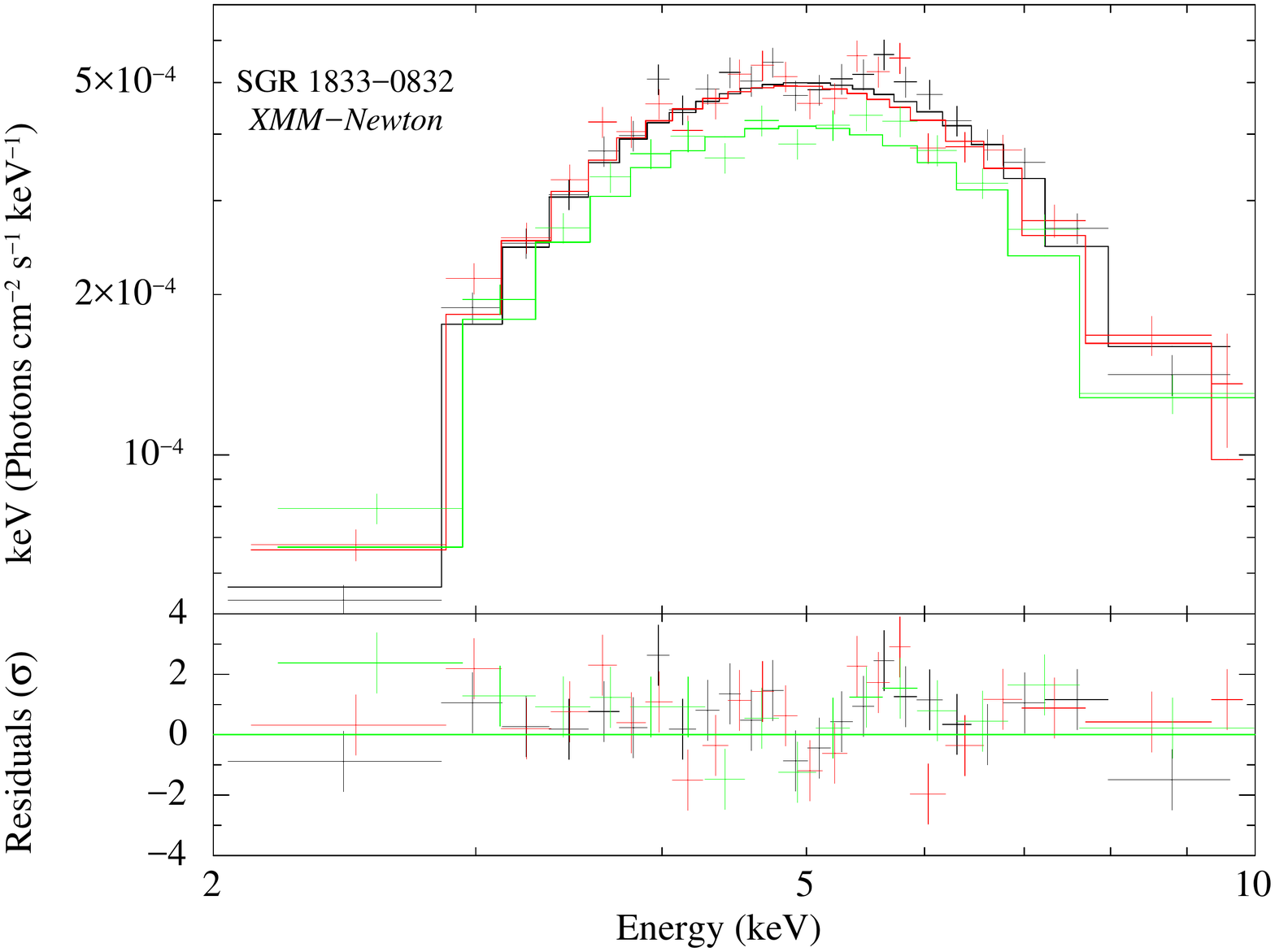}\\
\includegraphics[width=8cm]{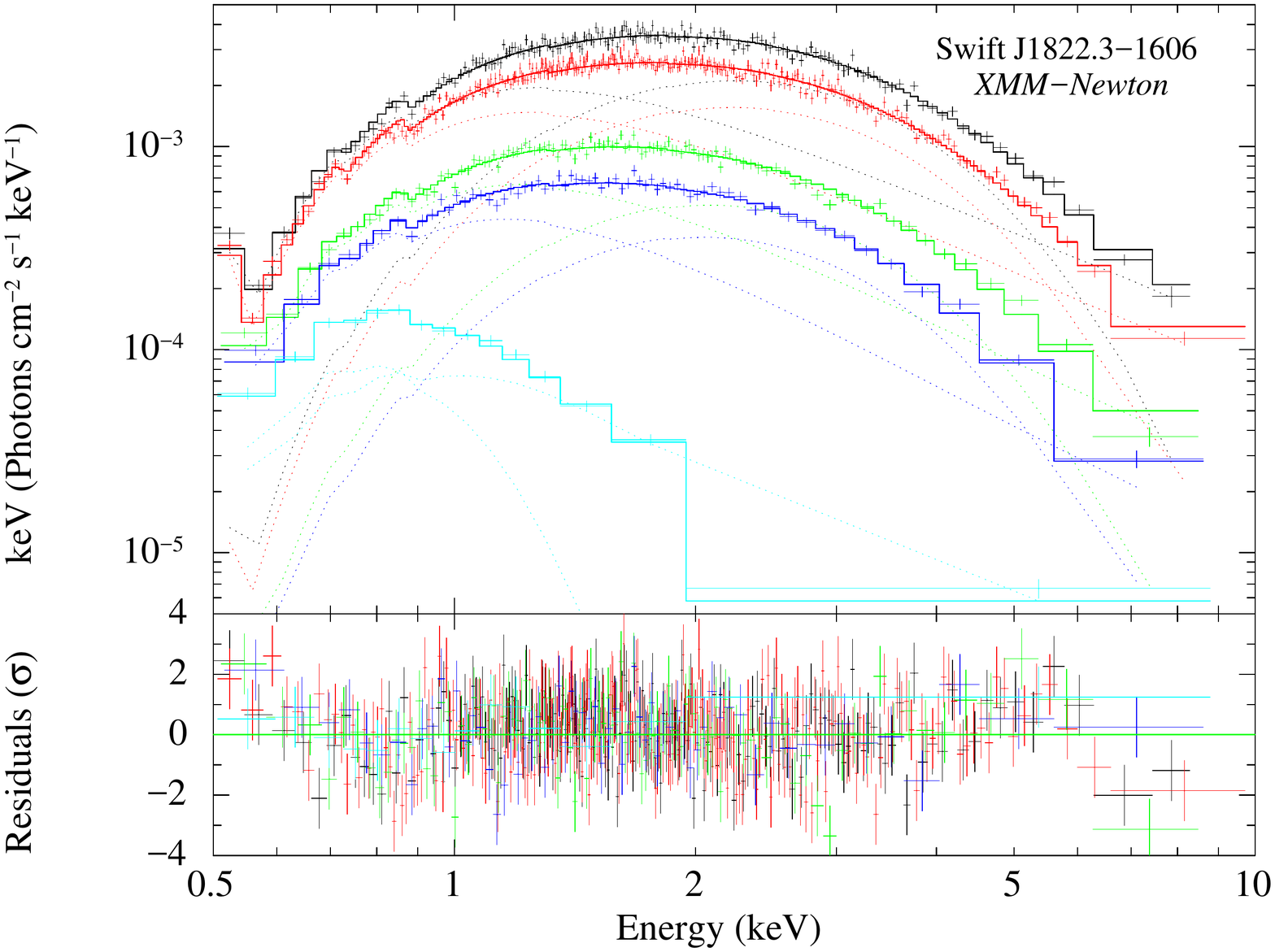}
\includegraphics[width=8cm]{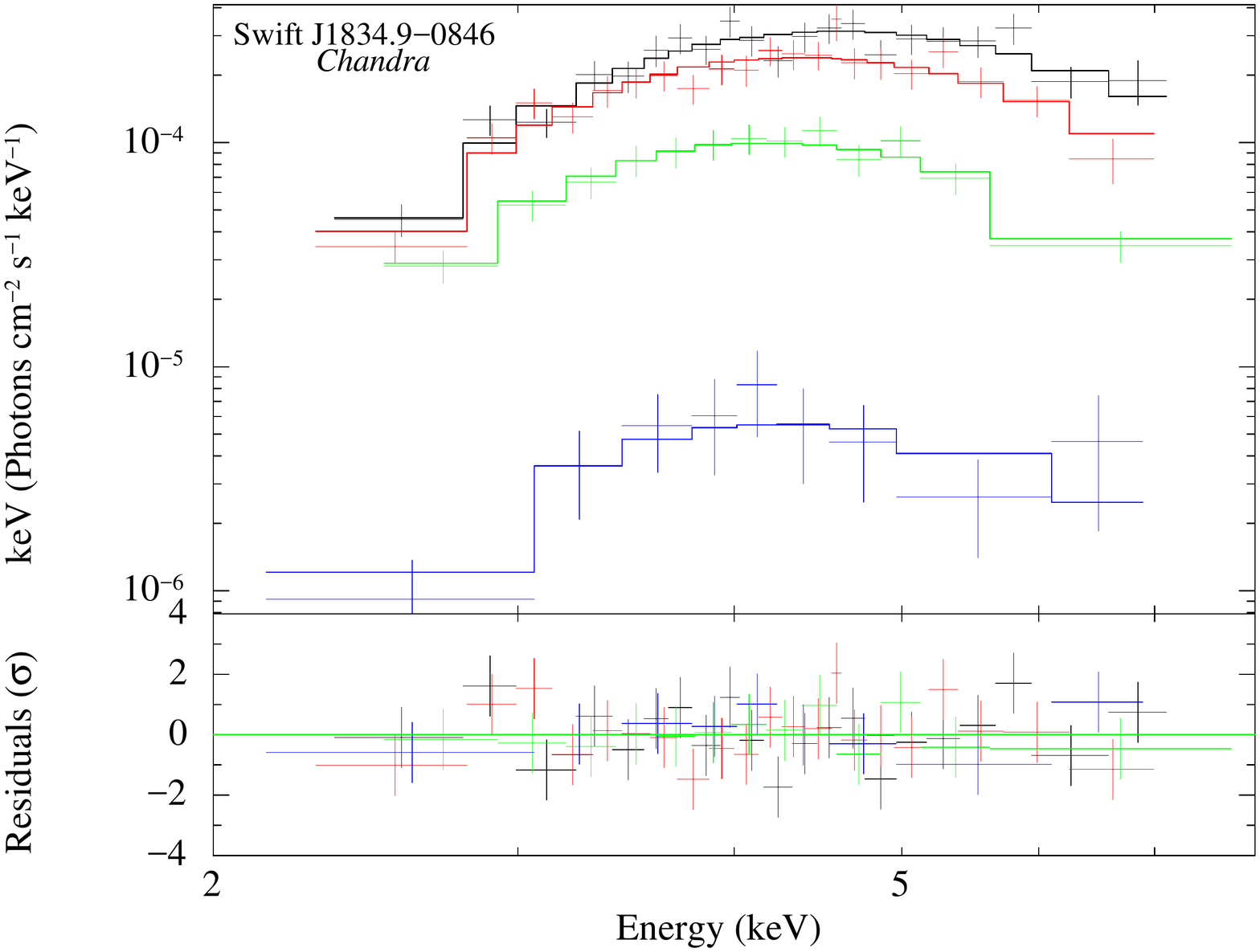}\\
  \caption{ -- \emph{continued}}
\label{fig:spectra_outbursts}
\end{center}
\end{figure*}

\setcounter{figure}{0}
\begin{figure*}
\begin{center}
\includegraphics[width=8cm]{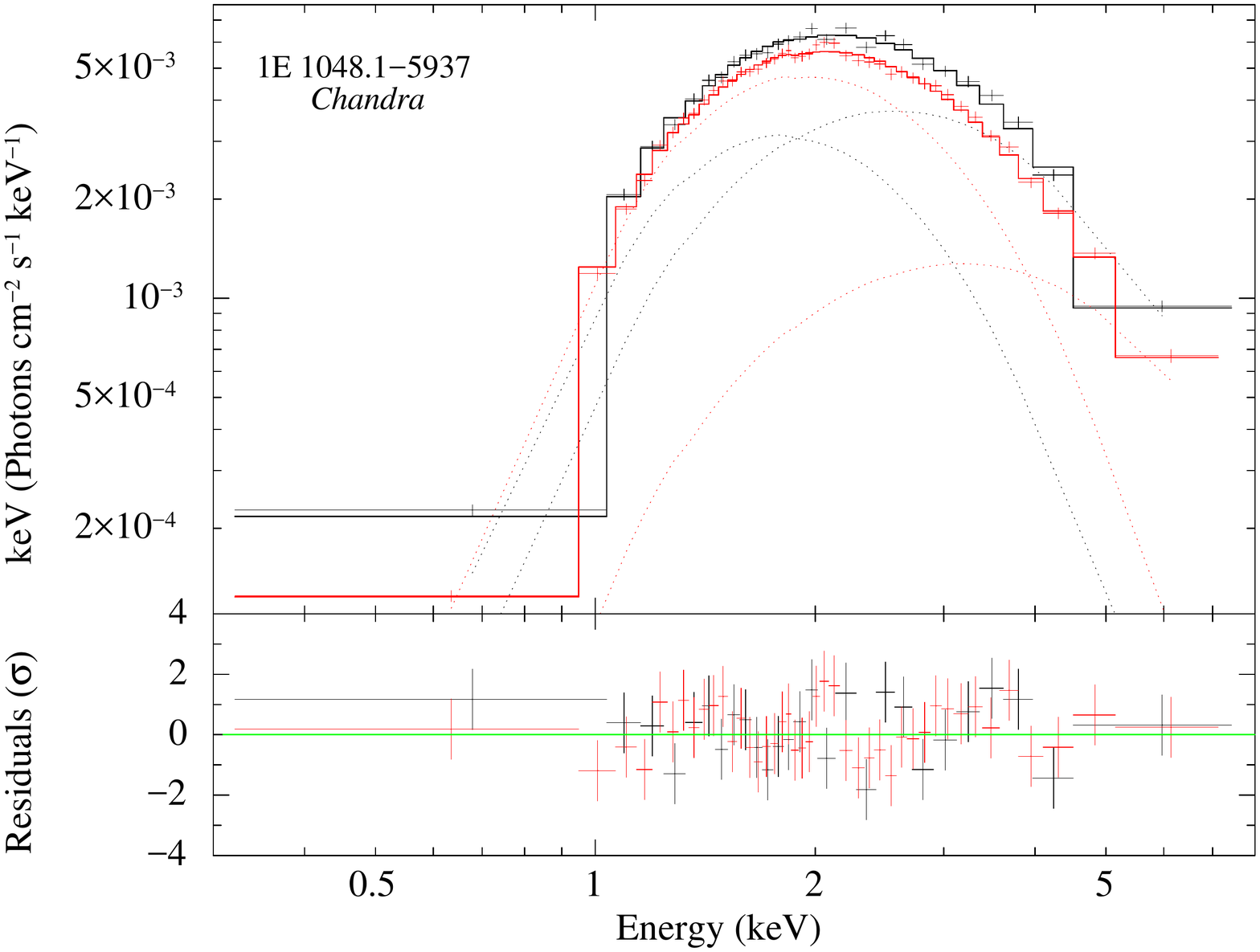}
\includegraphics[width=8cm]{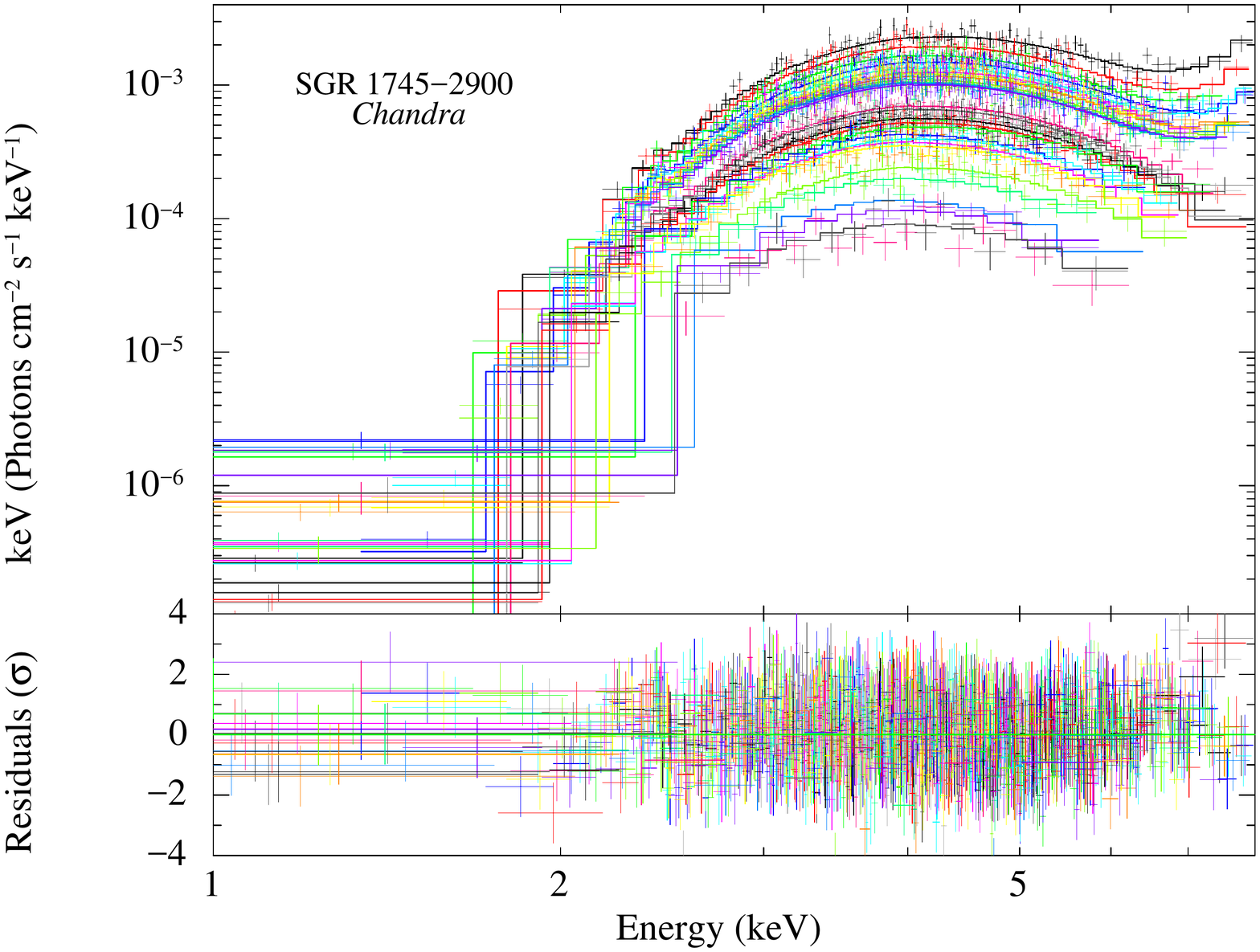}\\
\includegraphics[width=8cm]{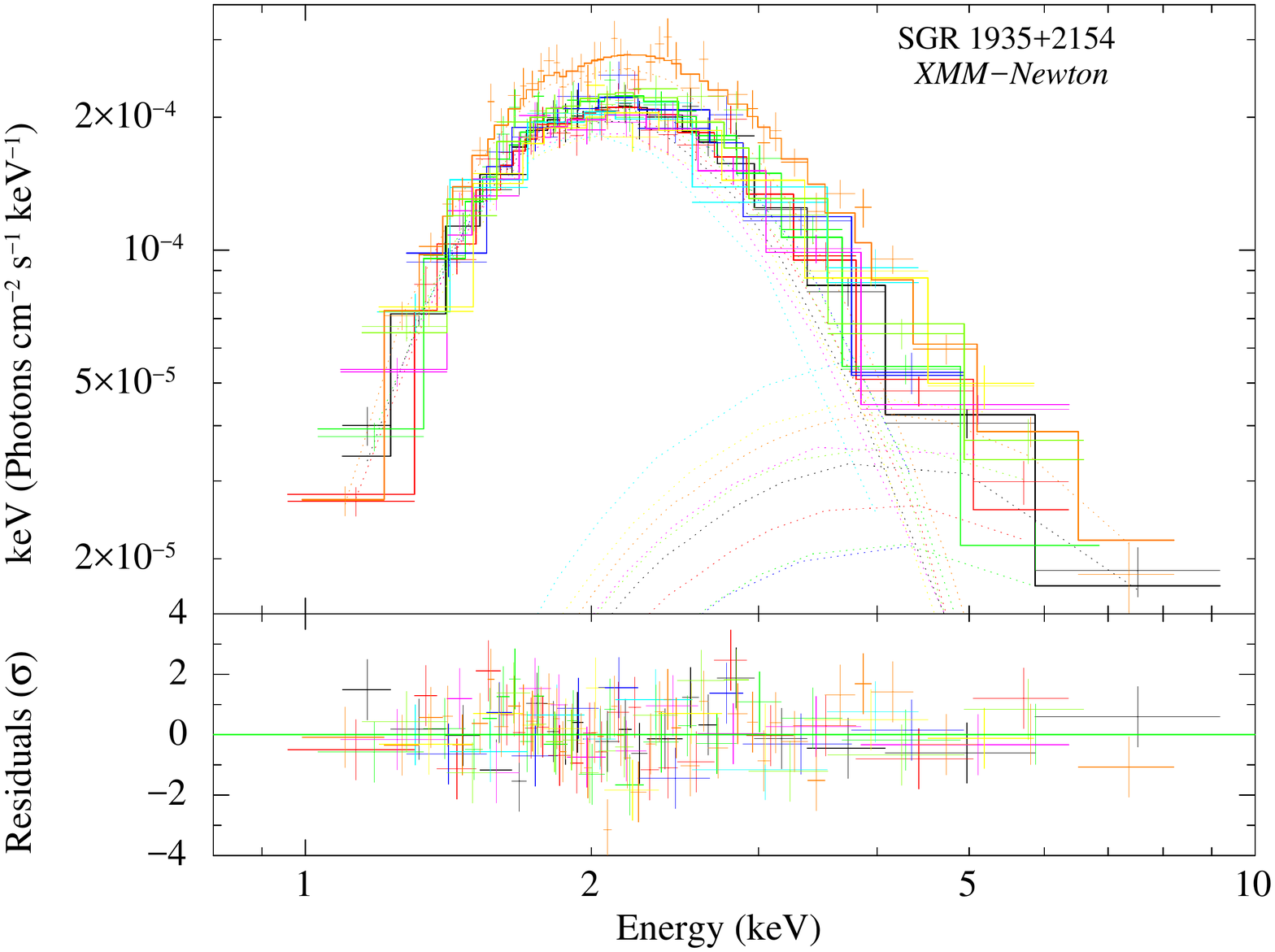}
\caption{Continued.}
\label{fig:spectra_outbursts}
\end{center}
\end{figure*}

\clearpage
\begin{table*}
\section{Outburst light curves}
\label{out_lcurves}

This section shows the cooling curves for all magnetar outbursts re-analysed in this study.
\end{table*}

\begin{figure*}
\begin{center}
\includegraphics[width=2\columnwidth]{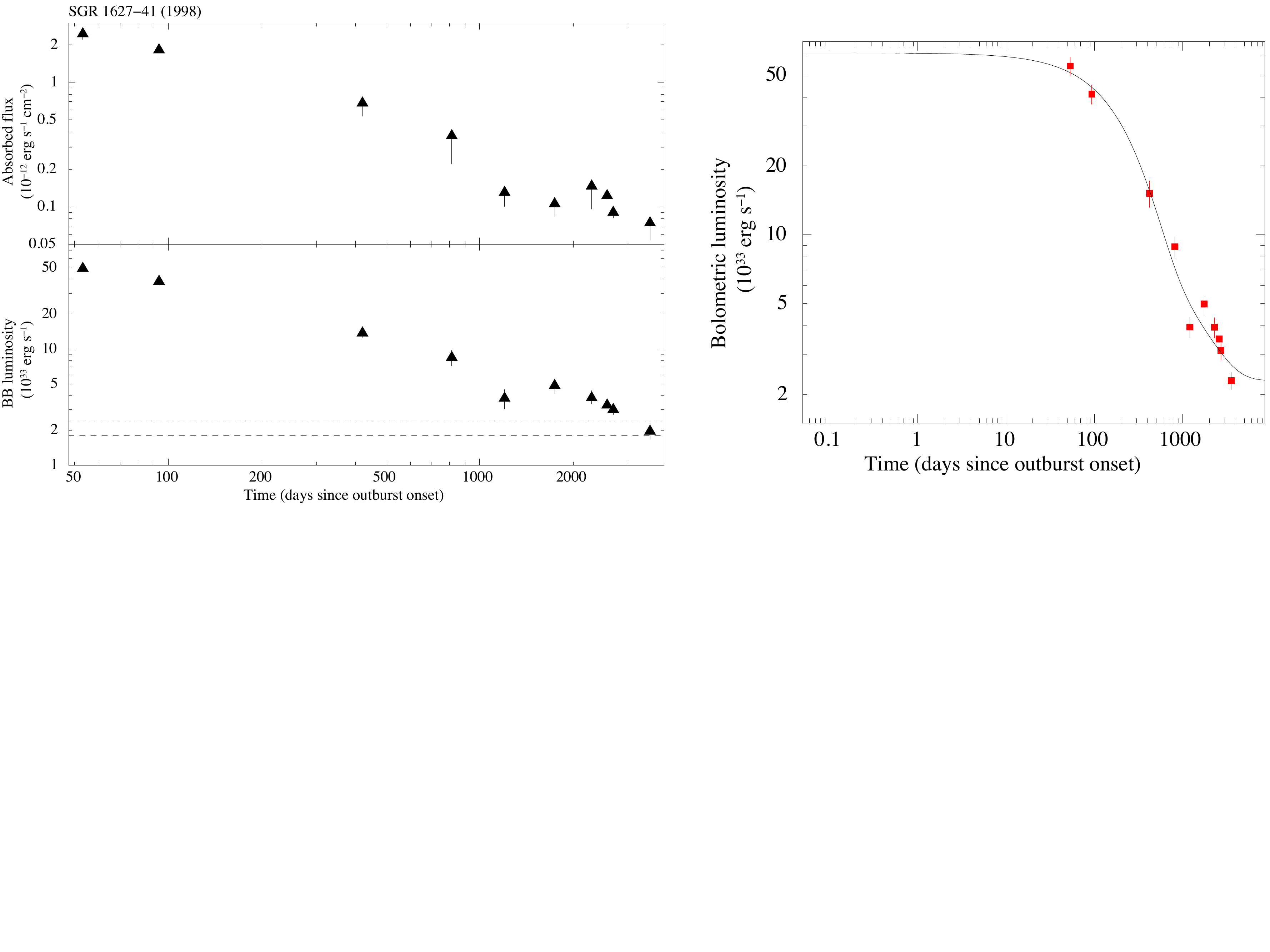}
\vspace{-5.5cm}
\caption{Left-hand panel: temporal evolution of the fluxes and luminosities for the BB model applied to the X-ray data of the 1998 outburst of \sgrd.  
The dashed lines mark the 1$\sigma$ c.l. range for the quiescent luminosity (see Table~\ref{tab:quiescence}). A distance of 11~kpc 
was assumed. Right-hand panel: temporal evolution of the bolometric luminosity with the best-fitting decay model superimposed.}
\label{fig:sgr1627_1998_parameters_outbursts}
\end{center}
\end{figure*}

\begin{figure*}
\begin{center}
\includegraphics[width=2\columnwidth]{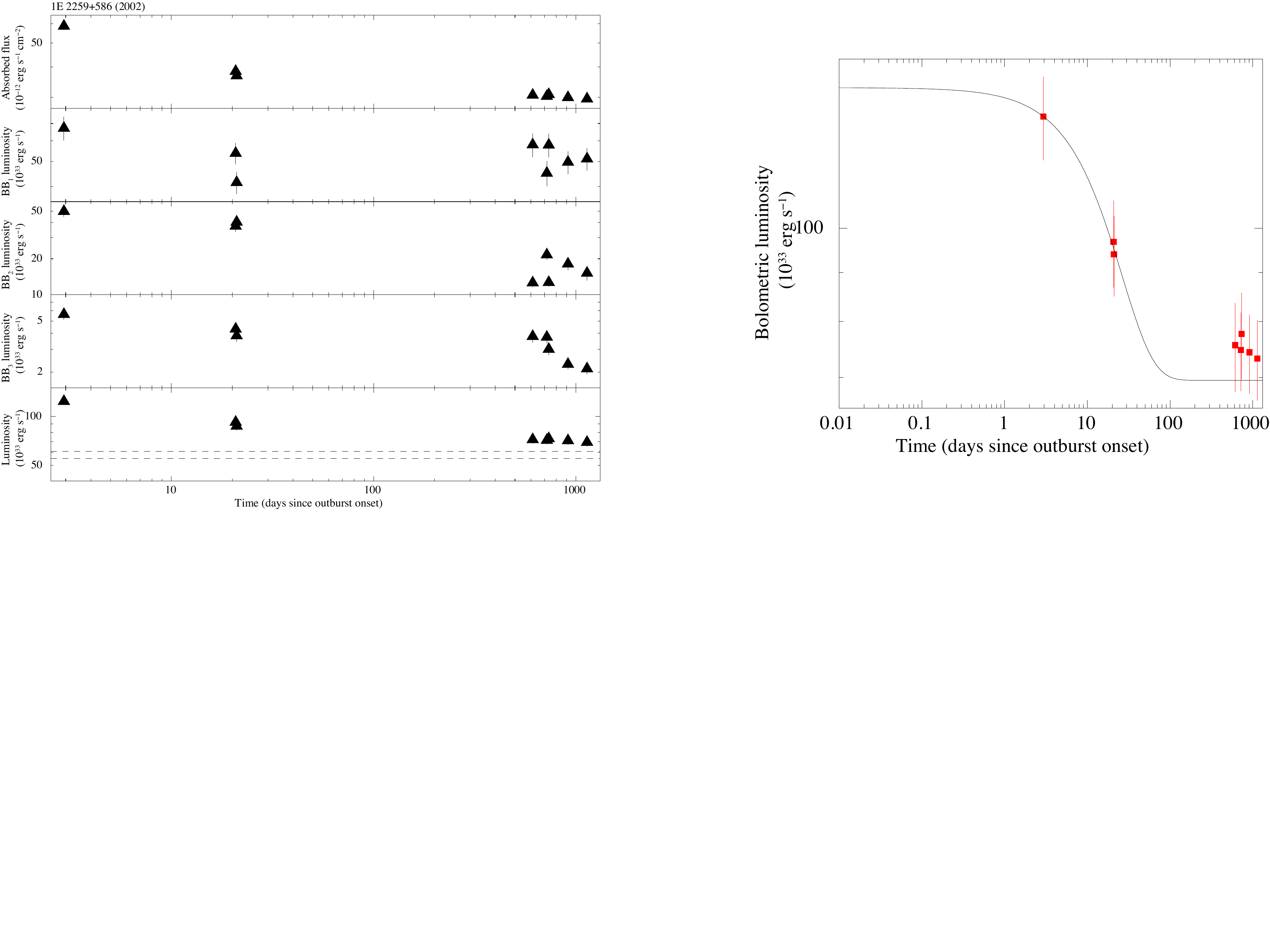}
\vspace{-5cm}
\caption{Left-hand panel: temporal evolution of the fluxes and luminosities for the 3BB model applied to the \xmm\ data of the 2002 outburst of \antgli.  
The dashed lines mark the 1$\sigma$ c.l. range for the quiescent luminosity (see Table~\ref{tab:quiescence}).  
A distance of 3.2~kpc was assumed. Right-hand panel: temporal evolution of the bolometric luminosity with the best-fitting decay model superimposed.}
\label{fig:1e2259_2002_parameters_outbursts}
\end{center}
\end{figure*}

\begin{figure*}
\begin{center}
\includegraphics[width=2\columnwidth]{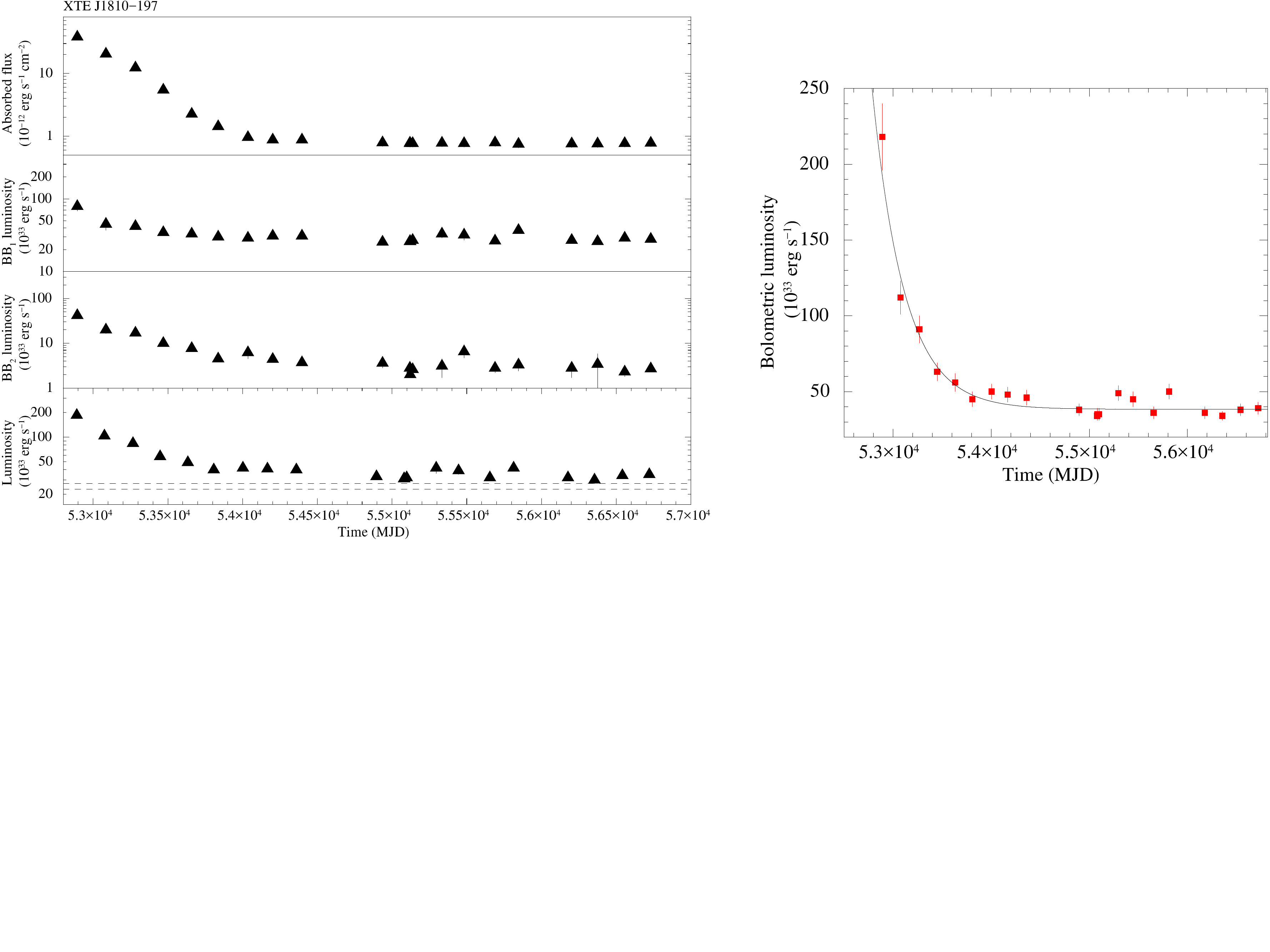}
\vspace{-5cm}
\caption{Left-hand panel: temporal evolution of the fluxes and luminosities of the cold and warm blackbody components for the 3BB+2BB model applied to the \xmm\ data of the 2003 outburst of \xte.  
The dashed lines mark the 1$\sigma$ c.l. range for the quiescent luminosity (see Table~\ref{tab:quiescence}).  
A distance of 3.5~kpc was assumed. Right-hand panel: temporal evolution of the bolometric luminosity with the best-fitting decay model superimposed.}
\label{fig:xte1810_parameters_outbursts}
\end{center}
\end{figure*}

\begin{figure*}
\begin{center}
\includegraphics[width=2\columnwidth]{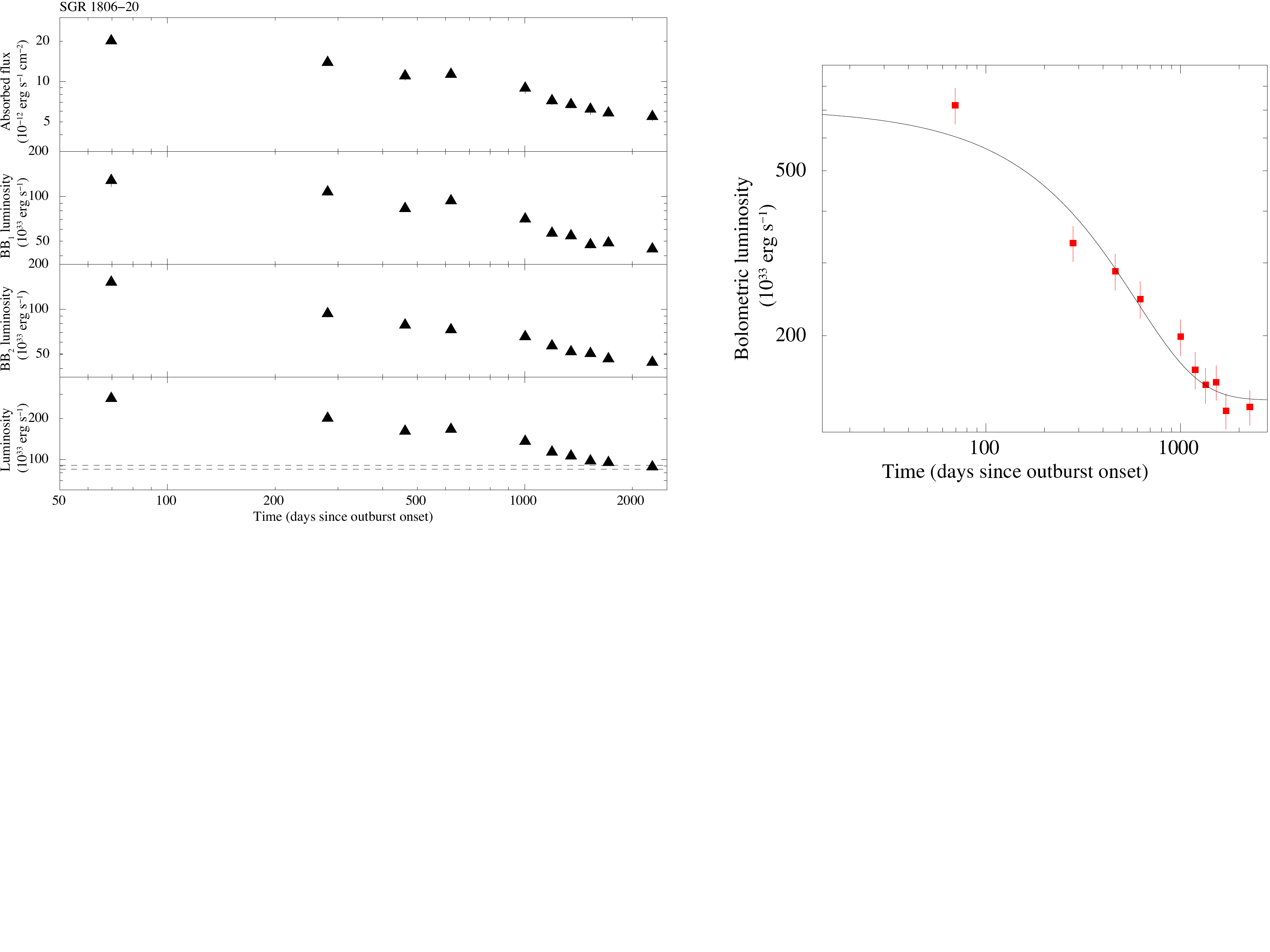}
\vspace{-5.5cm}
\caption{Left-hand panel: temporal evolution of the fluxes and luminosities for the 2BB model applied to the \xmm\ data of SGR\,1806$-$20.  
The dashed lines mark the 1$\sigma$ c.l. range for the quiescent luminosity (see Table~\ref{tab:quiescence}). A distance of 
8.7~kpc was assumed. Right-hand panel: temporal evolution of the bolometric luminosity with the best-fitting decay model superimposed.}
\label{fig:sgr1627_1998_parameters_outbursts}
\end{center}
\end{figure*}

\begin{figure*}
\begin{center}
\includegraphics[width=2\columnwidth]{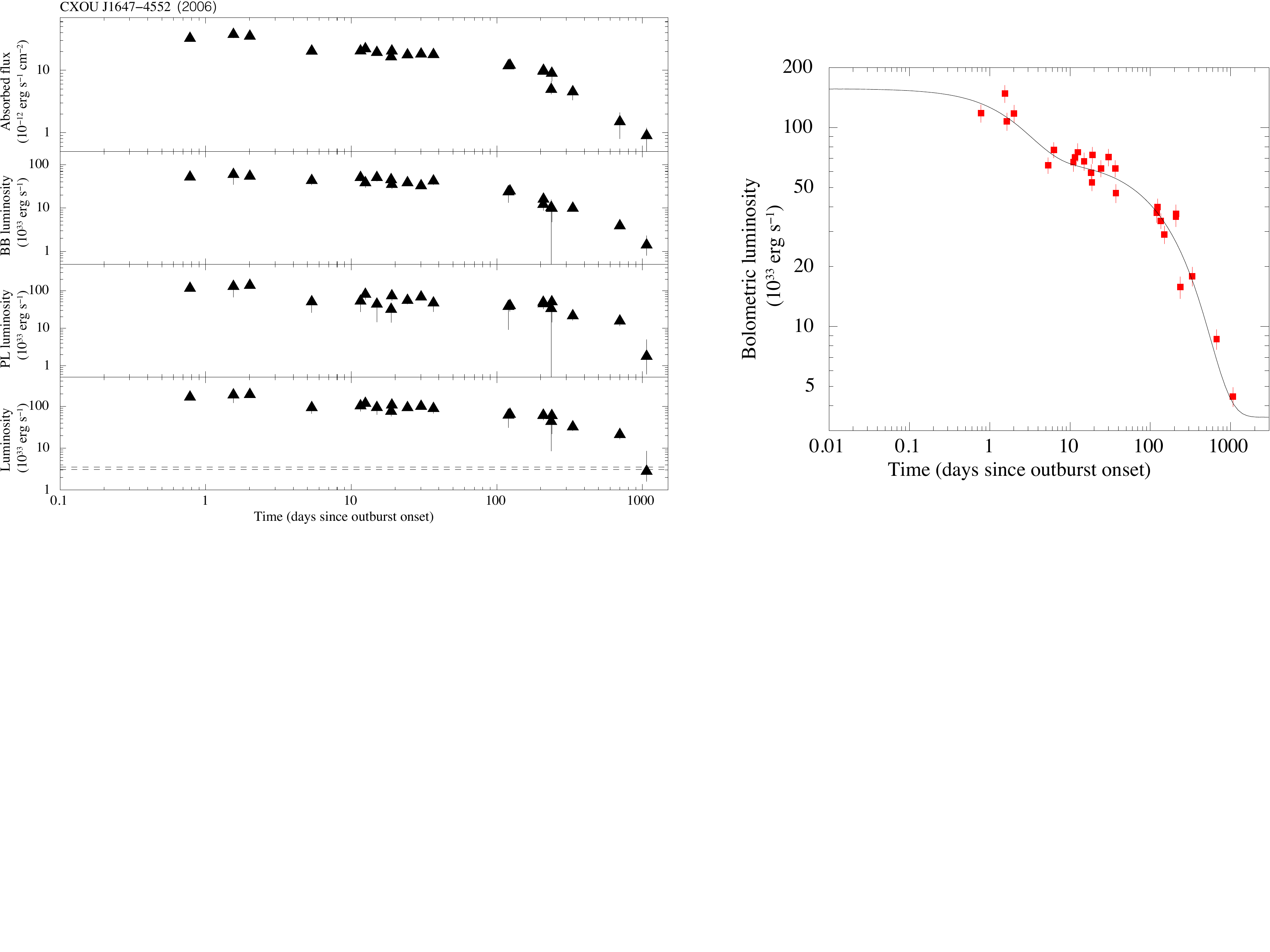}
\vspace{-5.5cm}
\caption{Left-hand panel: temporal evolution of the fluxes and luminosities for the BB+PL model applied to the X-ray data of \wes.  
The dashed lines mark the 1$\sigma$ c.l. range for the quiescent luminosity (see Table~\ref{tab:quiescence}).  
A distance of 4~kpc was assumed. Right-hand panel: temporal evolution of the bolometric luminosity with the best-fitting decay model superimposed.}
\label{fig:cxou2006_parameters_outbursts}
\end{center}
\end{figure*}


\begin{figure*}
\begin{center}
\includegraphics[width=2\columnwidth]{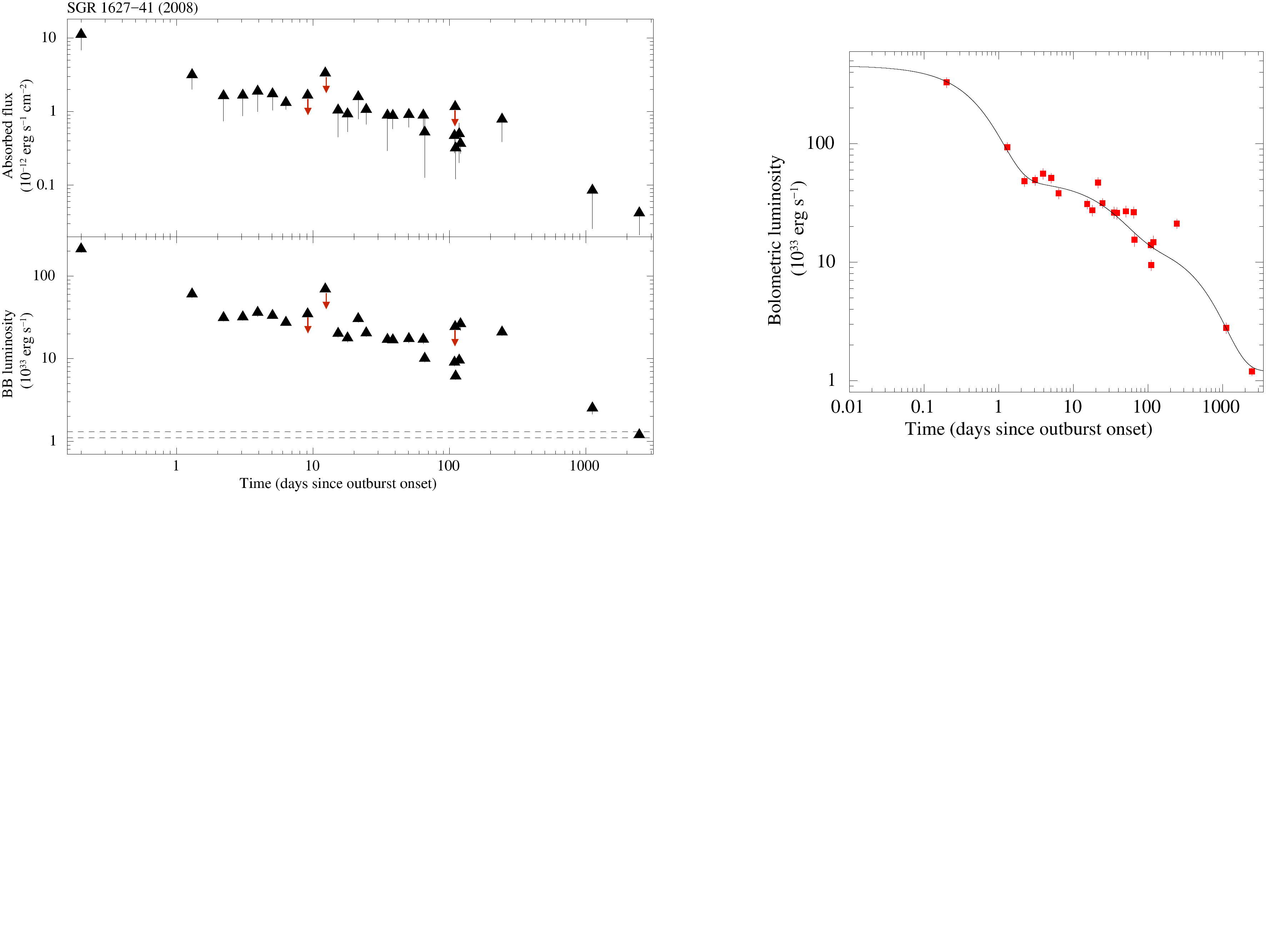}
\vspace{-5.5cm}
\caption{Left-hand panel: temporal evolution of the fluxes and luminosities for the BB model applied to the X-ray data of the 2008 outburst of \sgrd.  
The dashed lines mark the 1$\sigma$ c.l. range for the quiescent luminosity (see Table~\ref{tab:quiescence}). 
The red downward arrowheads indicate the 3$\sigma$ upper limits. A distance of 11~kpc was assumed. Right-hand panel: temporal evolution of the bolometric 
luminosity with the best-fitting decay model superimposed.}
\label{fig:sgr1627_2008_parameters_outbursts}
\end{center}
\end{figure*}

\begin{figure*}
\begin{center}
\includegraphics[width=2\columnwidth]{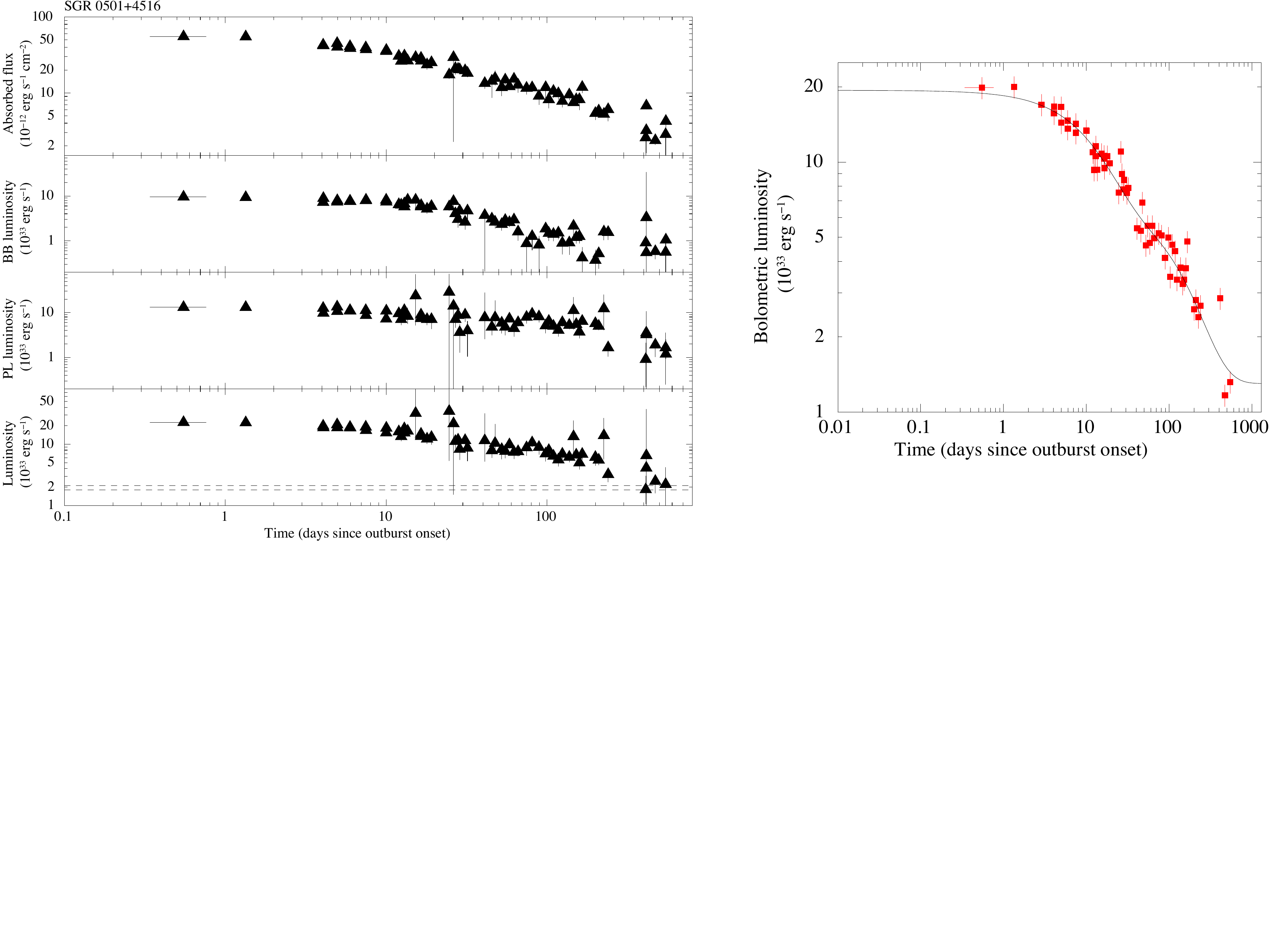}
\vspace{-5.5cm}
\caption{Left-hand panel: temporal evolution of the fluxes and luminosities for the BB+PL model applied to the \swift\ XRT data of \sgre.  
The dashed line marks the approximate value for the quiescent luminosity (see Table~\ref{tab:quiescence}).  
A distance of 1.5~kpc was assumed. Right-hand panel: temporal evolution of the bolometric luminosity with the best-fitting decay model superimposed.}
\label{fig:sgr0501_parameters_outbursts}
\end{center}
\end{figure*}

\begin{figure*}
\begin{center}
\includegraphics[width=2\columnwidth]{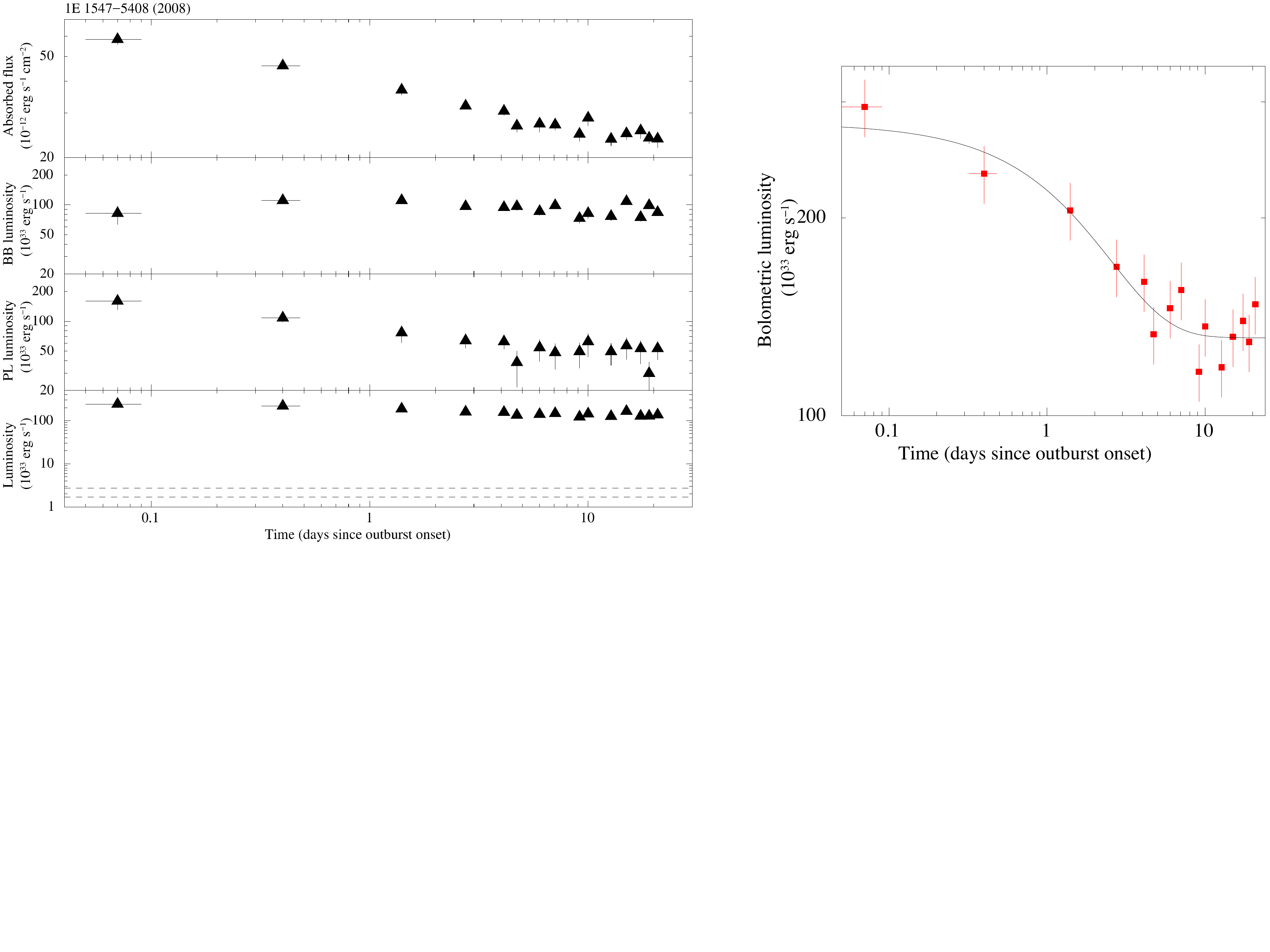}
\vspace{-5.5cm}
\caption{Left-hand panel: temporal evolution of the fluxes and luminosities for the BB+PL model applied to the \swift\ XRT data of the 2008 
outburst of \aa. The dashed lines mark the 1$\sigma$ c.l. range for the quiescent luminosity (see Table~\ref{tab:quiescence}). 
A distance of 4.5~kpc was assumed. Right-hand panel: temporal evolution of the bolometric luminosity with the best-fitting decay model superimposed.}
\label{fig:1e1547_2008_parameters_outbursts}
\end{center}
\end{figure*}

\begin{figure*}
\begin{center}
\includegraphics[width=2\columnwidth]{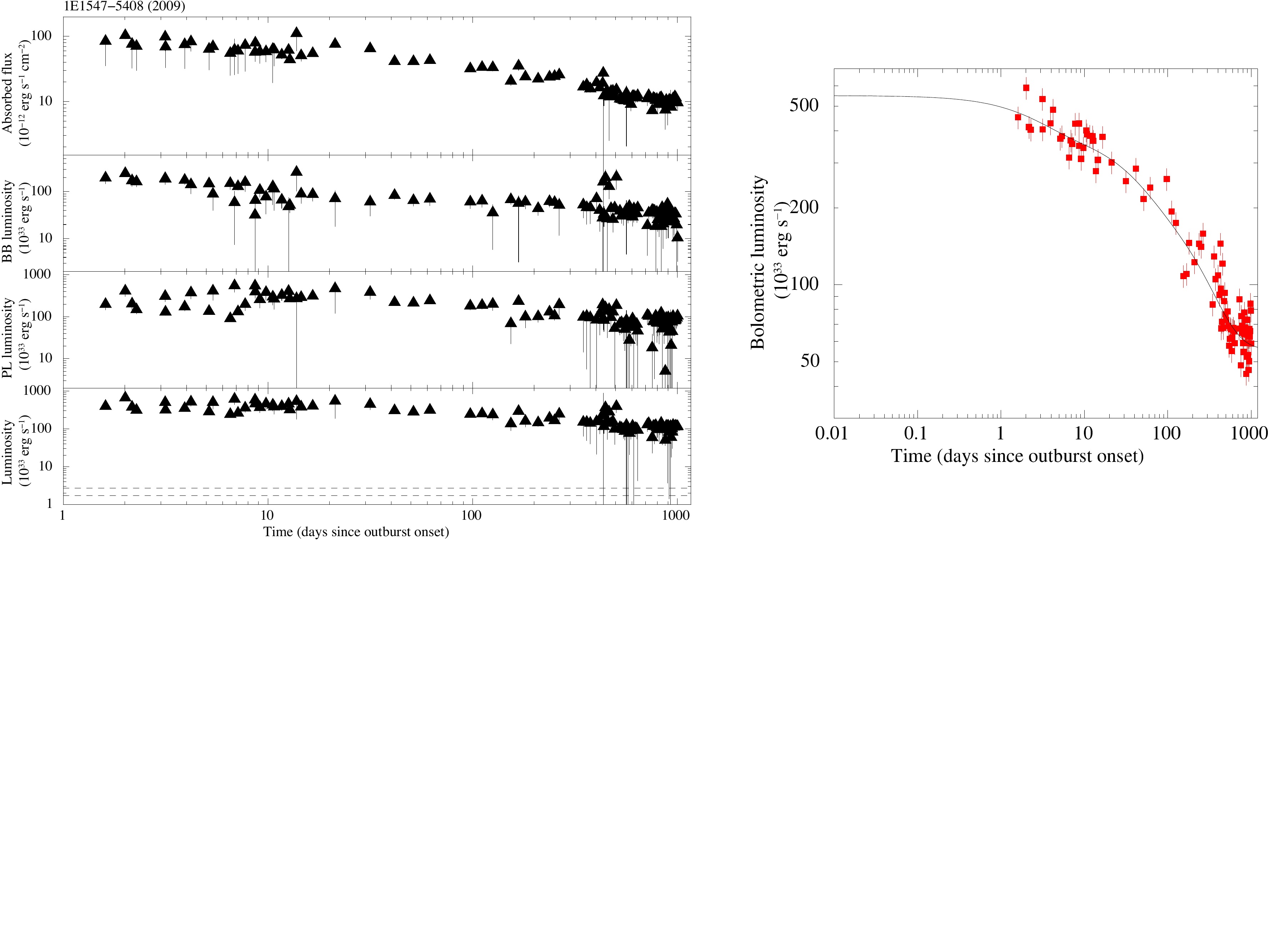}
\vspace{-5.5cm}
\caption{Left-hand panel: temporal evolution of the fluxes and luminosities for the BB+PL model applied to the \swift\ XRT data of the 2009 
outburst of \aa.  The dashed lines mark the 1$\sigma$ c.l. range for the quiescent luminosity (see Table~\ref{tab:quiescence}). 
A distance of 4.5~kpc was assumed. Right-hand panel: temporal evolution of the bolometric luminosity with the best-fitting decay model superimposed.}
\label{fig:1e1547_2009_parameters_outbursts}
\end{center}
\end{figure*}

\begin{figure*}
\begin{center}
\includegraphics[width=2\columnwidth]{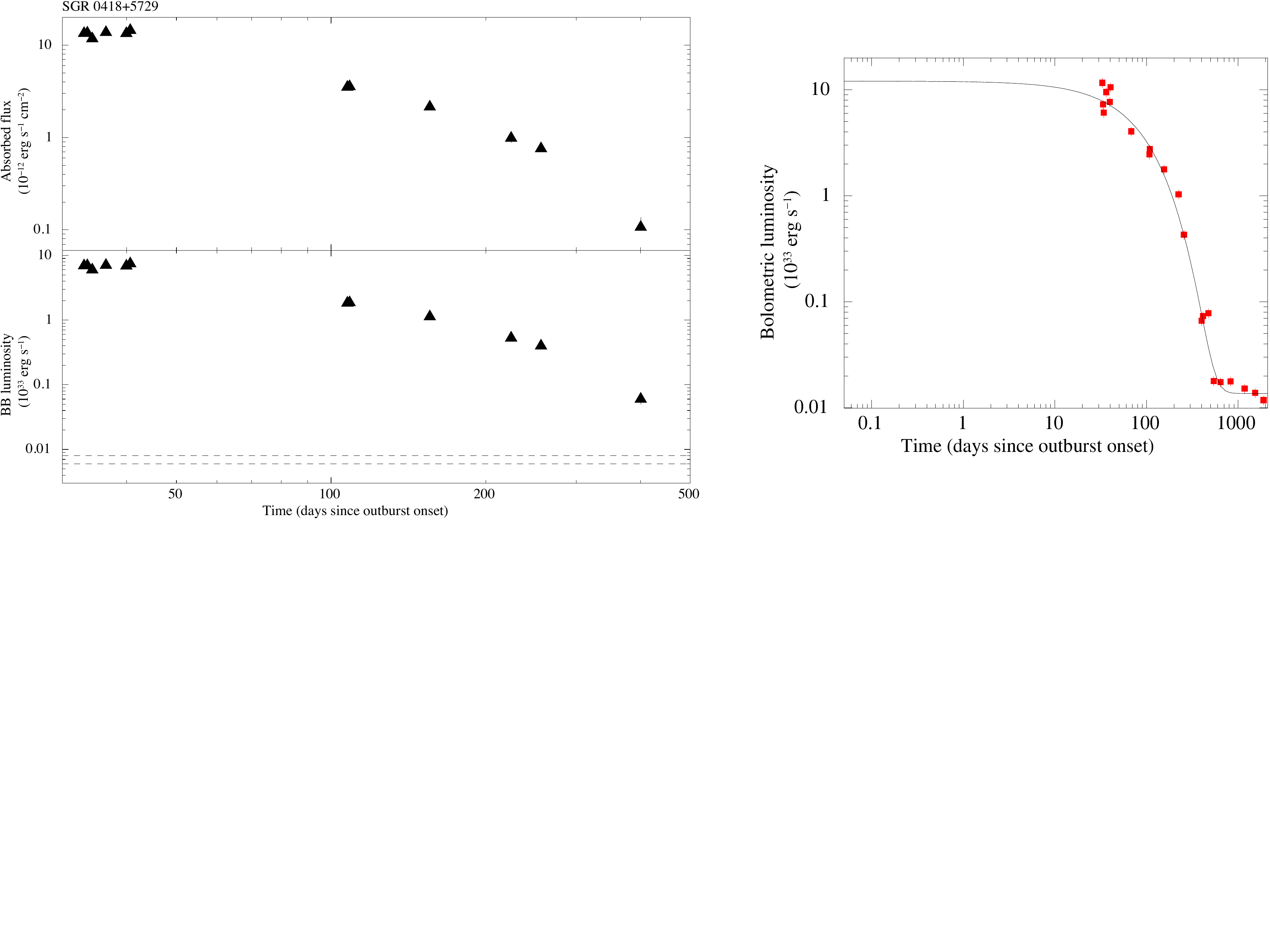}
\vspace{-5.5cm}
\caption{Left-hand panel: temporal evolution of the fluxes and luminosities for the BB model applied to the \swift\ XRT data of the
outburst of \lowba. The dashed lines mark the 1$\sigma$ c.l. range for the quiescent luminosity (see 
Table~\ref{tab:quiescence}).  A distance of 2~kpc was assumed. Right-hand panel: temporal evolution of the bolometric luminosity with the best-fitting decay model superimposed.}
\label{fig:sgr0418_parameters_outbursts}
\end{center}
\end{figure*}

\begin{figure*}
\begin{center}
\includegraphics[width=2\columnwidth]{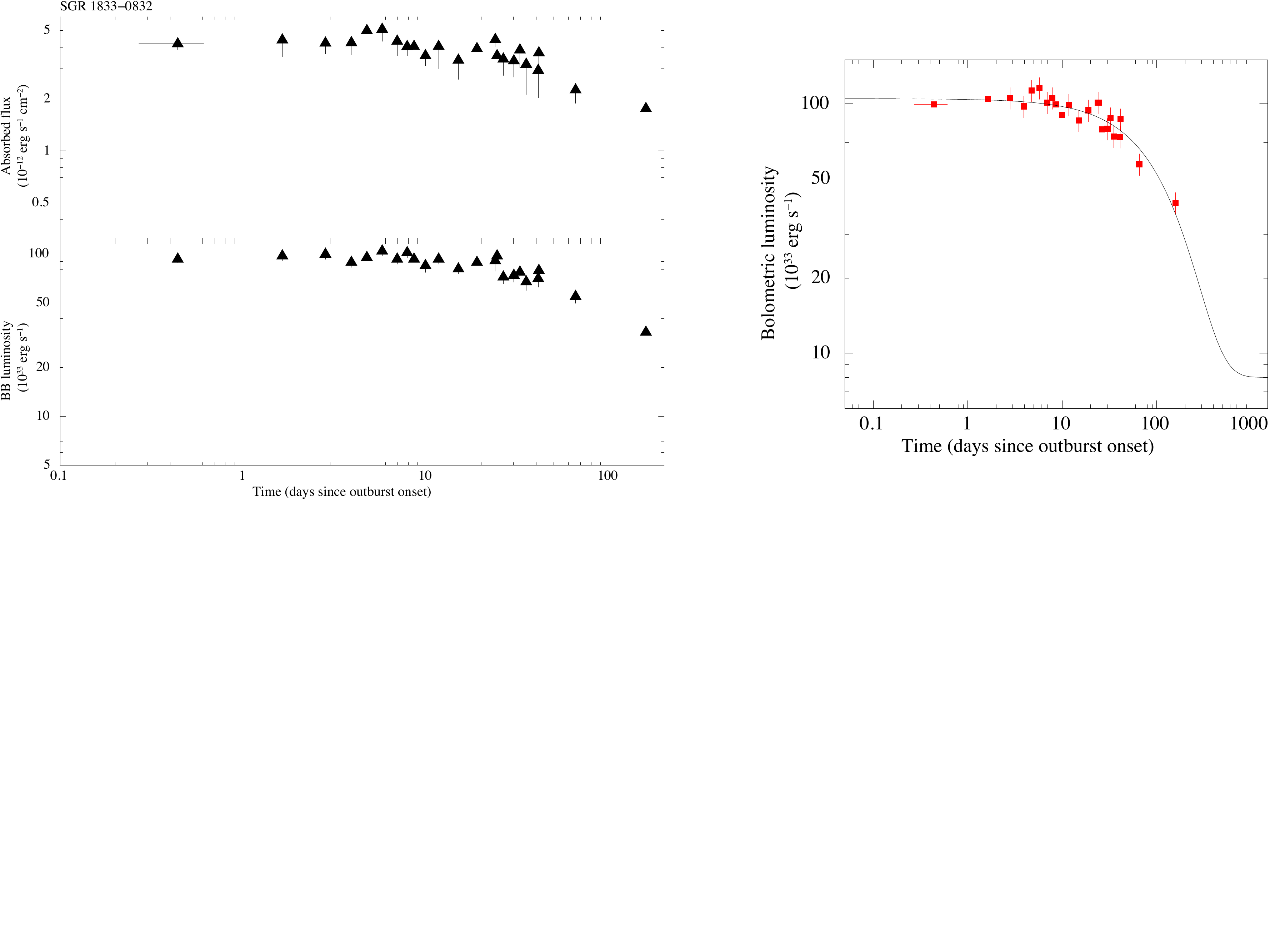}
\vspace{-5.5cm}
\caption{Left-hand panel: temporal evolution of the fluxes and luminosities for the BB+PL model applied to the \swift\ XRT data of the
outburst of SGR\,1833$-$0832. The dashed line marks the upper limit (at the 3$\sigma$ c.l.) for the 
quiescent luminosity (see Table~\ref{tab:quiescence}). A distance of 10~kpc was assumed. Right-hand panel: temporal evolution 
of the bolometric luminosity with the best-fitting decay model superimposed.}
\label{fig:sgr1833_parameters_outbursts}
\end{center}
\end{figure*}

\begin{figure*}
\begin{center}
\includegraphics[width=2\columnwidth]{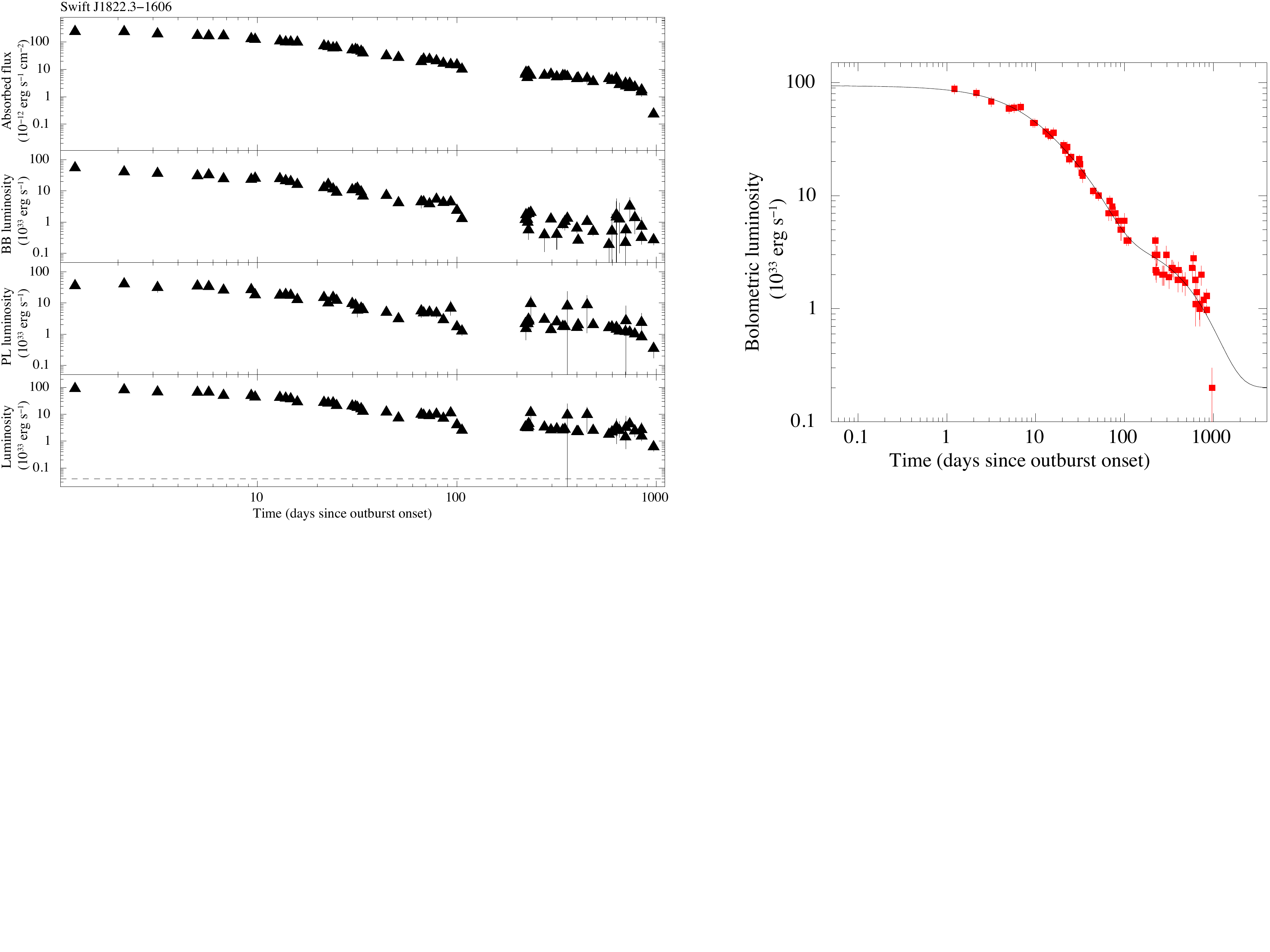}
\vspace{-5.5cm}
\caption{Left-hand panel: temporal evolution of the fluxes and luminosities for the BB+PL model applied to the outburst of \lowbb. 
 The dashed line marks the value for the quiescent luminosity (see Table~\ref{tab:quiescence}). 
A distance of 1.6~kpc was assumed. Right-hand panel: temporal evolution of the bolometric luminosity with the best-fitting decay model superimposed.}
\label{fig:swift1822_parameters_outbursts}
\end{center}
\end{figure*}

\begin{figure*}
\begin{center}
\includegraphics[width=2\columnwidth]{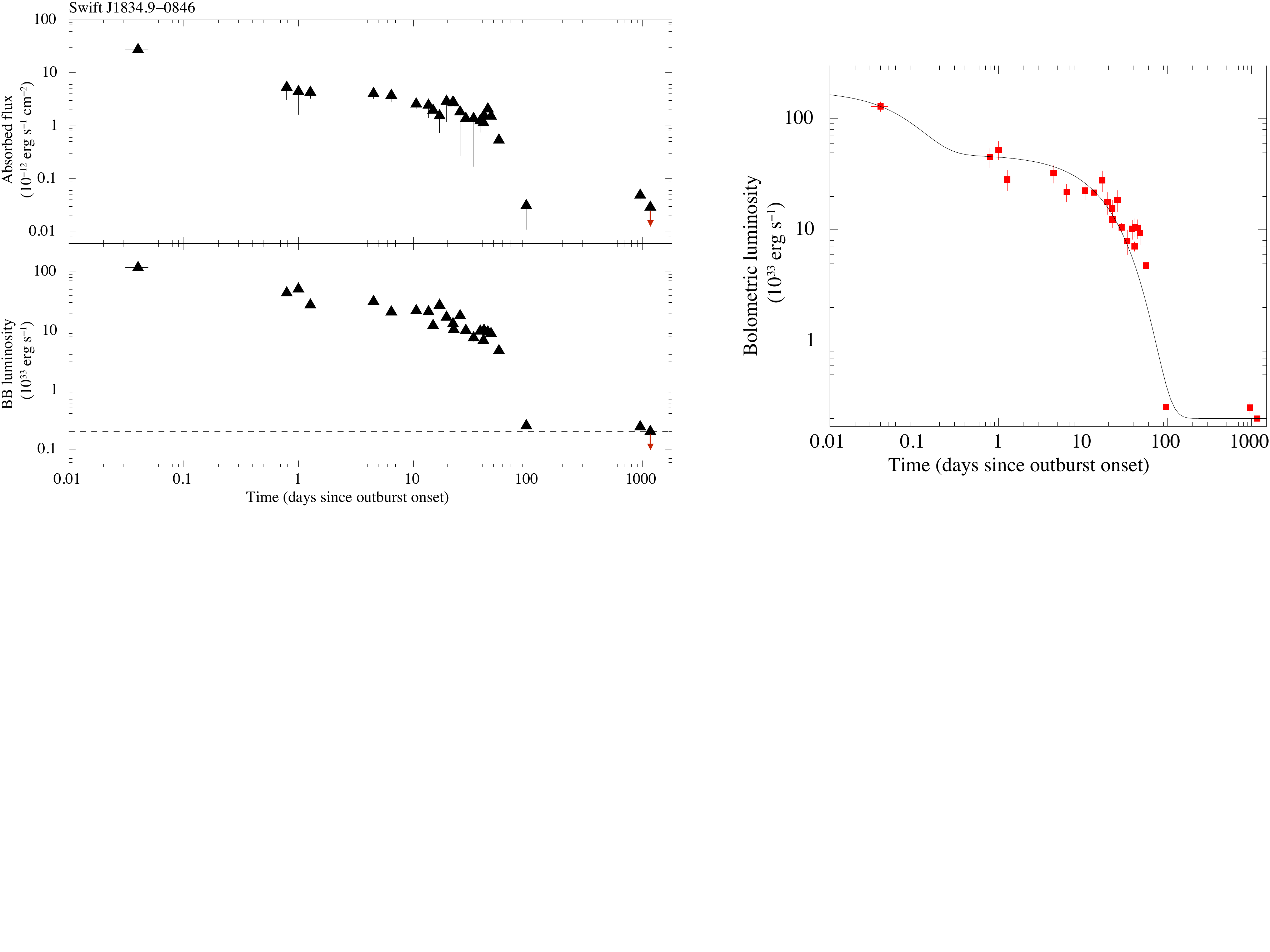}
\vspace{-5.5cm}
\caption{Left-hand panel: temporal evolution of the fluxes and luminosities for the BB model applied to the X-ray data of the outburst 
of Swift\,J1834.9$-$0846.  The dashed line marks the upper limit (at the 3$\sigma$ c.l.) for the quiescent 
luminosity (see Table~\ref{tab:quiescence}). The red downward arrowheads indicate the 3$\sigma$ upper limits. A distance of 4.2~kpc 
was assumed. Right-hand panel: temporal evolution of the bolometric luminosity with the best-fitting decay model superimposed.}
\label{fig:swift1834_parameters_outbursts}
\end{center}
\end{figure*}

\begin{figure*}
\begin{center}
\includegraphics[width=2\columnwidth]{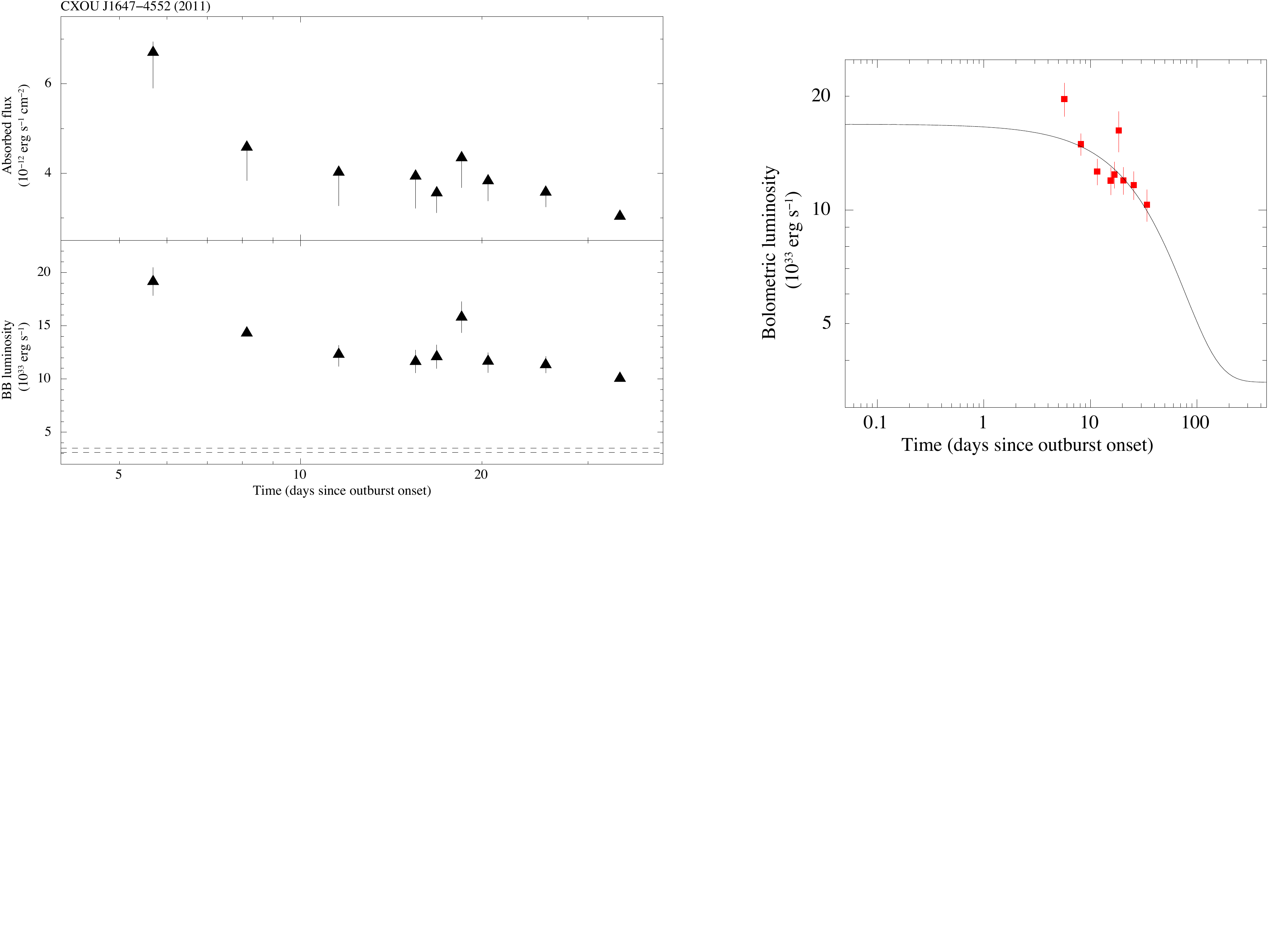}
\vspace{-5.5cm}
\caption{Left-hand panel: temporal evolution of the fluxes and luminosities for the BB model applied to the X-ray data of the 2011 outburst 
of \wes.  The dashed lines mark the 1$\sigma$ c.l. range for the quiescent luminosity (see Table~\ref{tab:quiescence}). 
A distance of 4~kpc was assumed. Right-hand panel: temporal evolution of the bolometric luminosity with the best-fitting decay model superimposed.}
\label{fig:cxou2011_parameters_outbursts}
\end{center}
\end{figure*}

\begin{figure*}
\begin{center}
\includegraphics[width=2\columnwidth]{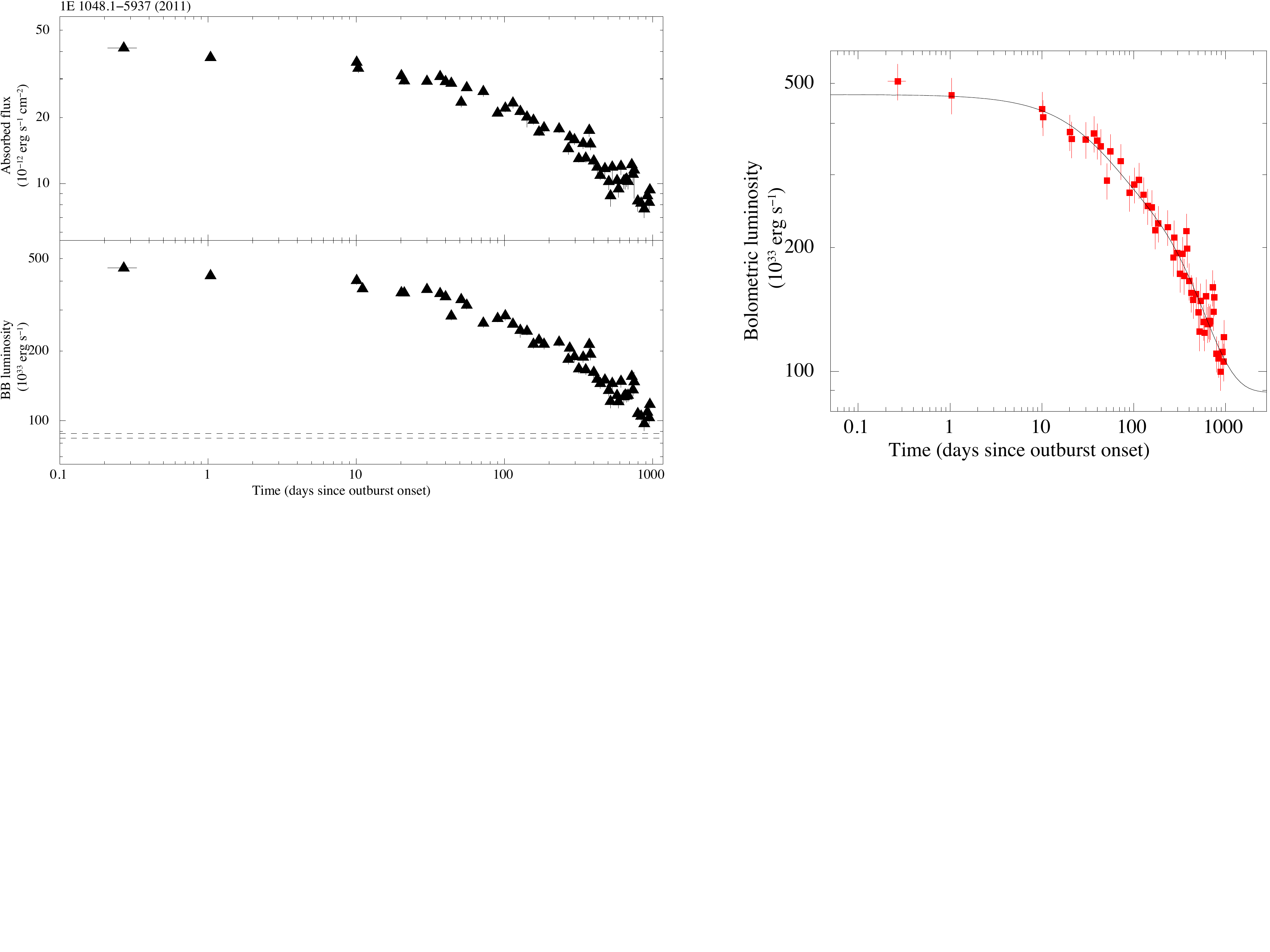}
\vspace{-5.5cm}
\caption{Left-hand panel: temporal evolution of the fluxes and luminosities for the BB model applied to the \swift\ XRT data of the 2011 outburst 
of 1E\,1048.1$-$5937.  The dashed lines mark the 1$\sigma$ c.l. range for the quiescent luminosity (see 
Table~\ref{tab:quiescence}). A distance of 9~kpc was assumed. Right-hand panel: temporal evolution of the bolometric luminosity with the best-fitting decay model superimposed.}
\label{fig:1e1048_parameters_outbursts}
\end{center}
\end{figure*}

\begin{figure*}
\begin{center}
\includegraphics[width=2\columnwidth]{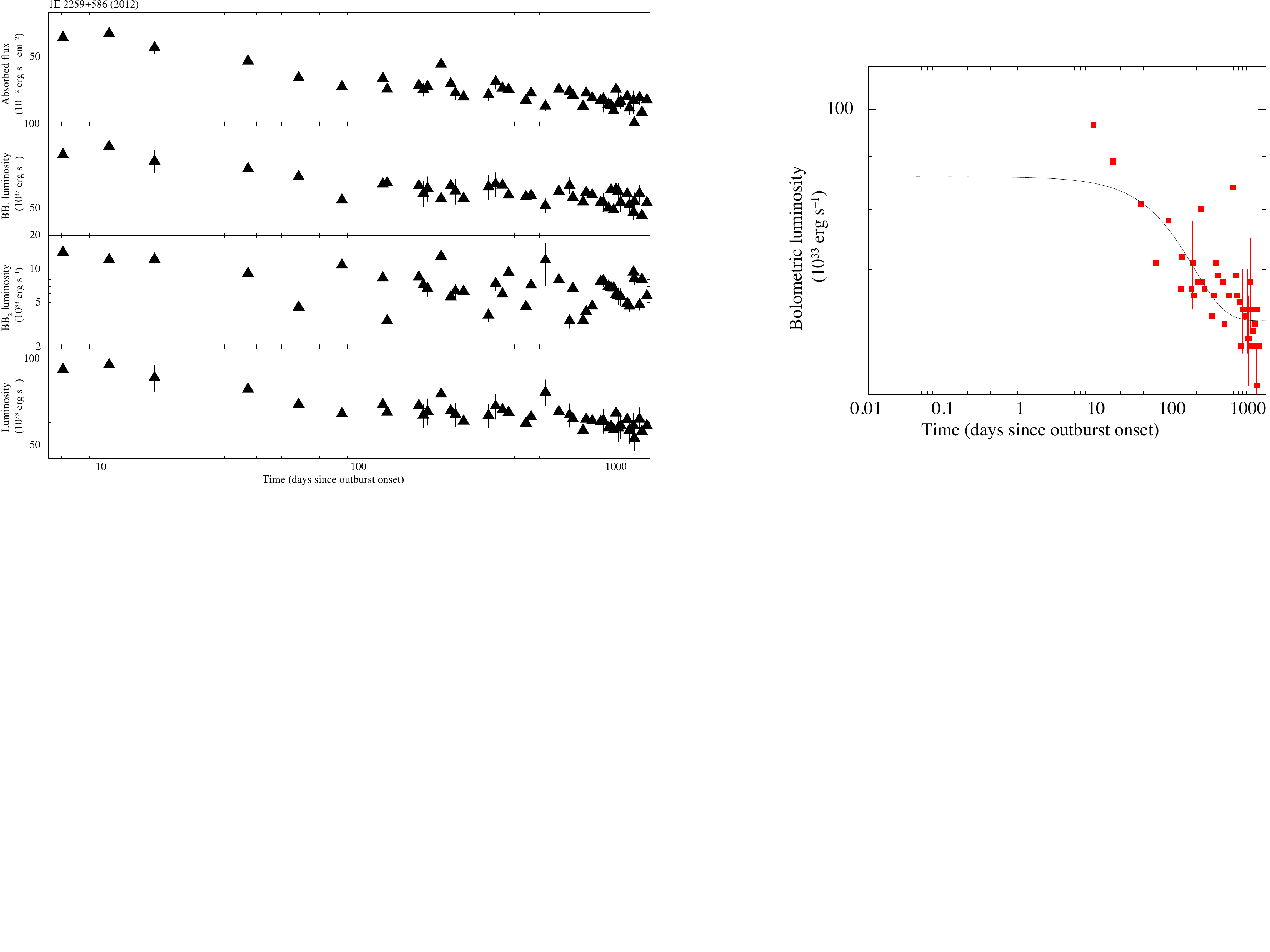}
\vspace{-5.5cm}
\caption{Left-hand panel: temporal evolution of the fluxes and luminosities for the 2BB model applied to the \swift\ data of the 2012 outburst of \antgli.  
The dashed lines mark the 1$\sigma$ c.l. range for the quiescent luminosity (see Table~\ref{tab:quiescence}).  
energy range. A distance of 3.2~kpc was assumed. Right-hand panel: temporal evolution of the bolometric luminosity with the best-fitting decay model superimposed.}
\label{fig:1e2259_2012_parameters_outbursts}
\end{center}
\end{figure*}

\begin{figure*}
\begin{center}
\includegraphics[width=2\columnwidth]{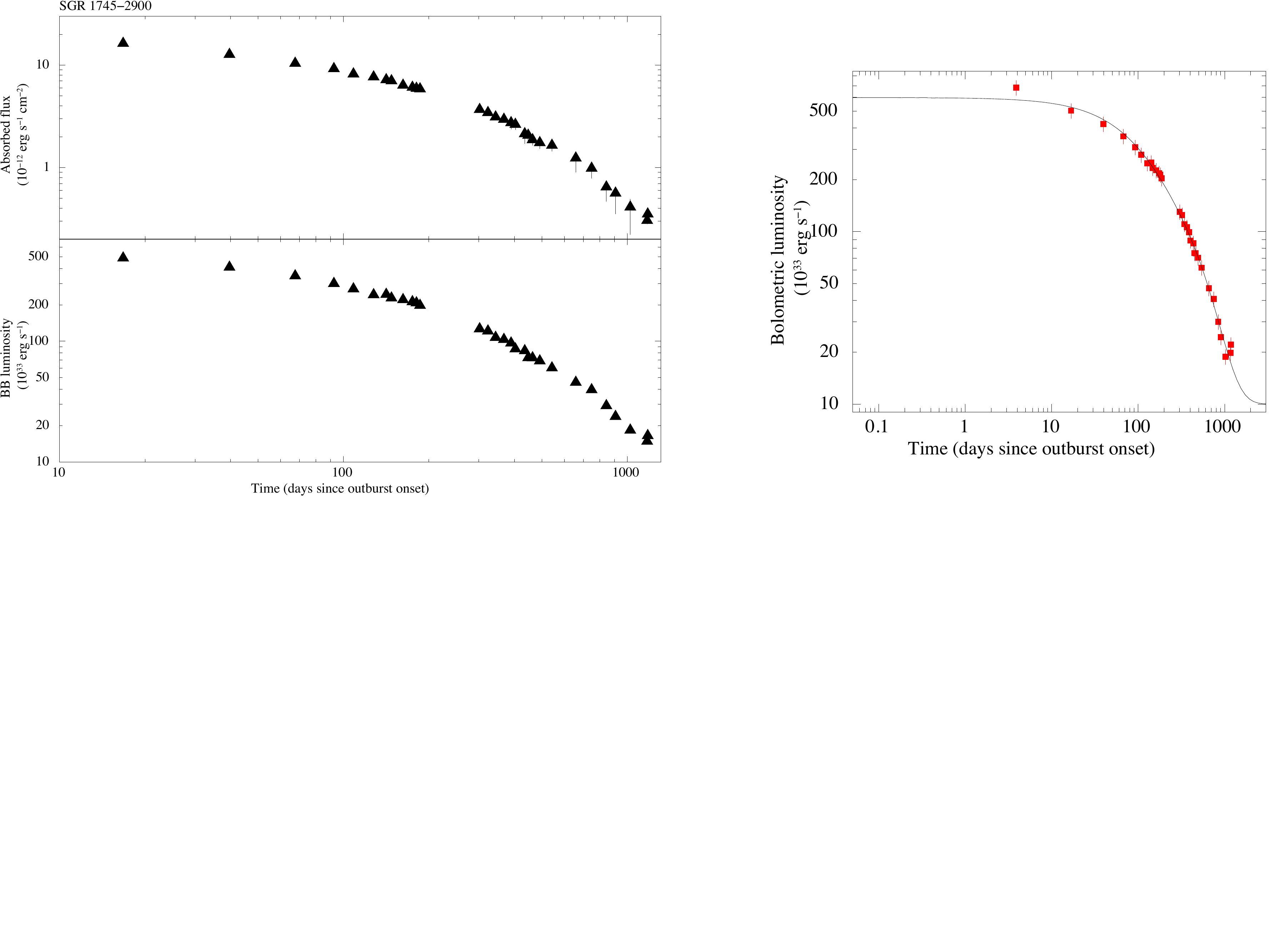}
\vspace{-5.5cm}
\caption{Left-hand panel: temporal evolution of the fluxes and luminosities for the BB+PL model applied to the \cxo\ data of the
outburst of \galcen. 
A distance of 8.3~kpc was assumed. Right-hand panel: temporal evolution of the bolometric luminosity with the best-fitting decay model superimposed.}
\label{fig:sgr1745_parameters_outbursts}
\end{center}
\end{figure*}

\begin{figure*}
\begin{center}
\includegraphics[width=17cm]{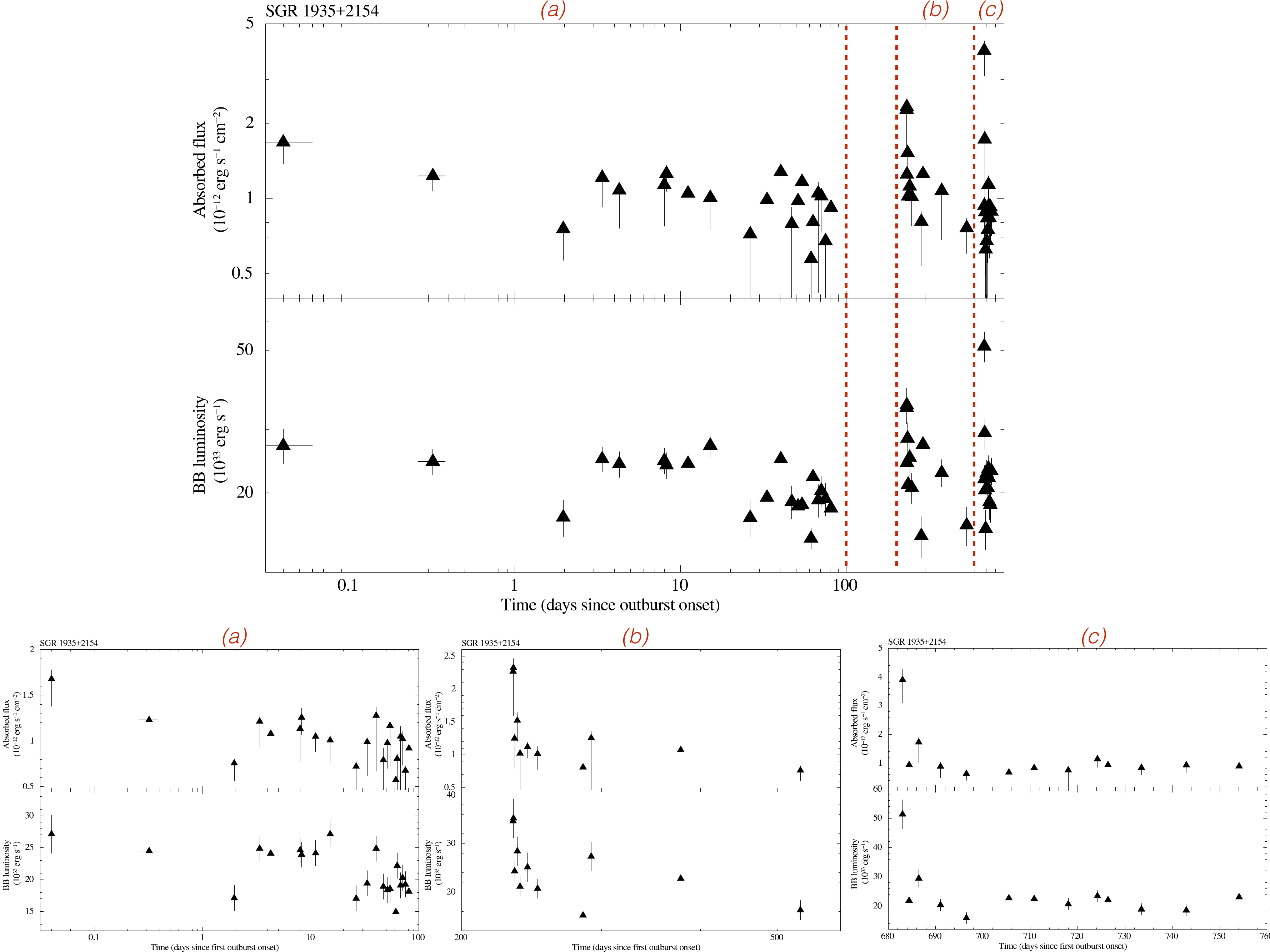}
\caption{Temporal evolution of the fluxes and luminosities for the BB model applied to the \swift\ XRT data of the
outbursts of \sgrm\ (see the bottom panels for a zoom on the individual outbursts). A distance of 9~kpc was assumed (see Israel et al. 2016). The quiescent level is unknown.}
\label{fig:sgr1935_parameters_outbursts}
\end{center}
\end{figure*}

\begin{figure*}
\begin{center}
\includegraphics[width=2\columnwidth]{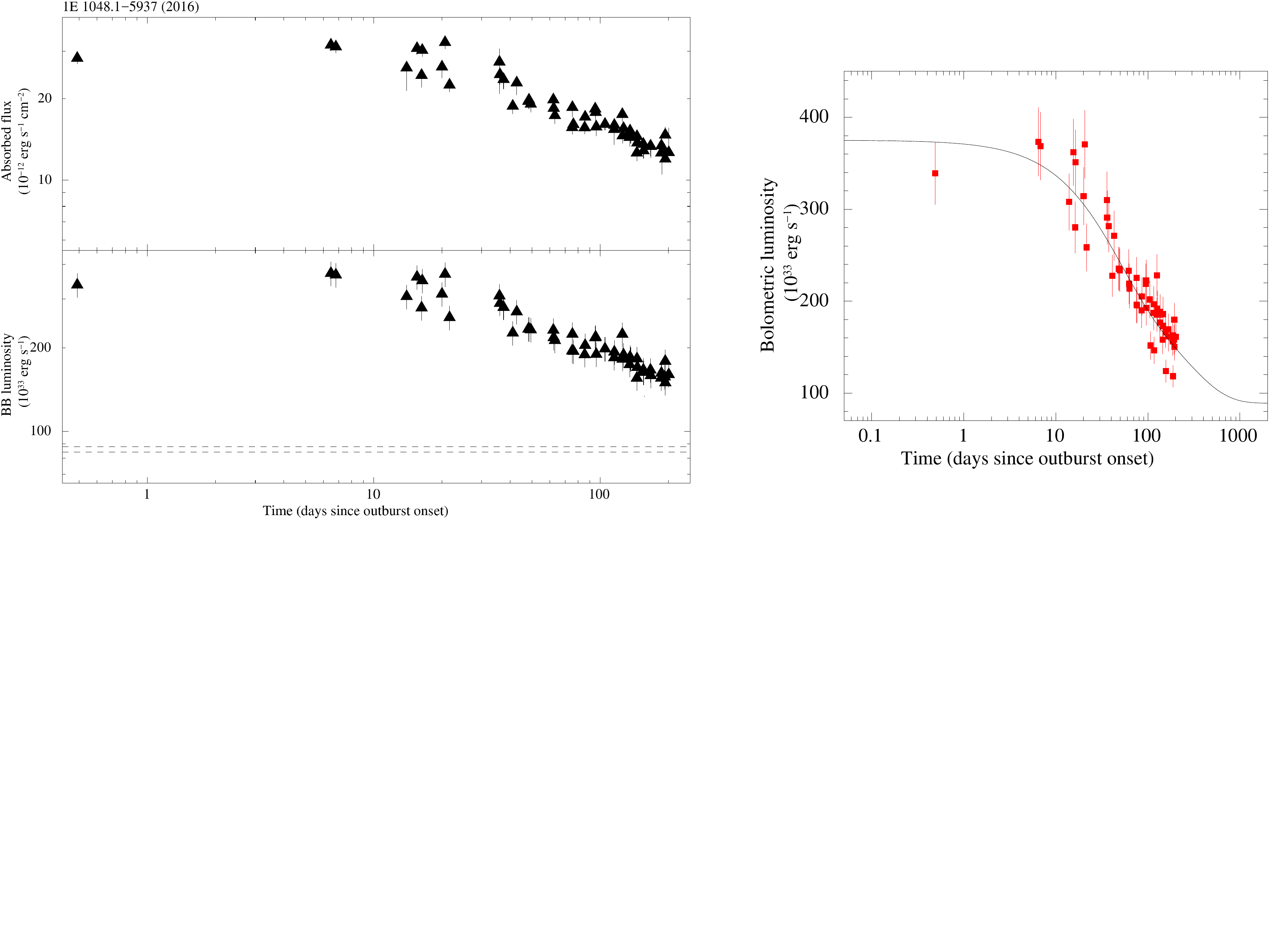}
\vspace{-5.5cm}
\caption{Left-hand panel: temporal evolution of the fluxes and luminosities for the BB model applied to the \swift\ data of the
2016 outburst of 1E\,1048.1$-$5937. The dashed lines mark the 1$\sigma$ c.l. range for the quiescent luminosity 
(see Table~\ref{tab:quiescence}). A distance of 9~kpc was assumed. Right-hand panel: temporal evolution of the bolometric 
luminosity with the best-fitting decay model superimposed.}
\label{fig:1e1048_2016_parameters_outbursts}
\end{center}
\end{figure*}

\begin{figure*}
\begin{center}
\includegraphics[width=2\columnwidth]{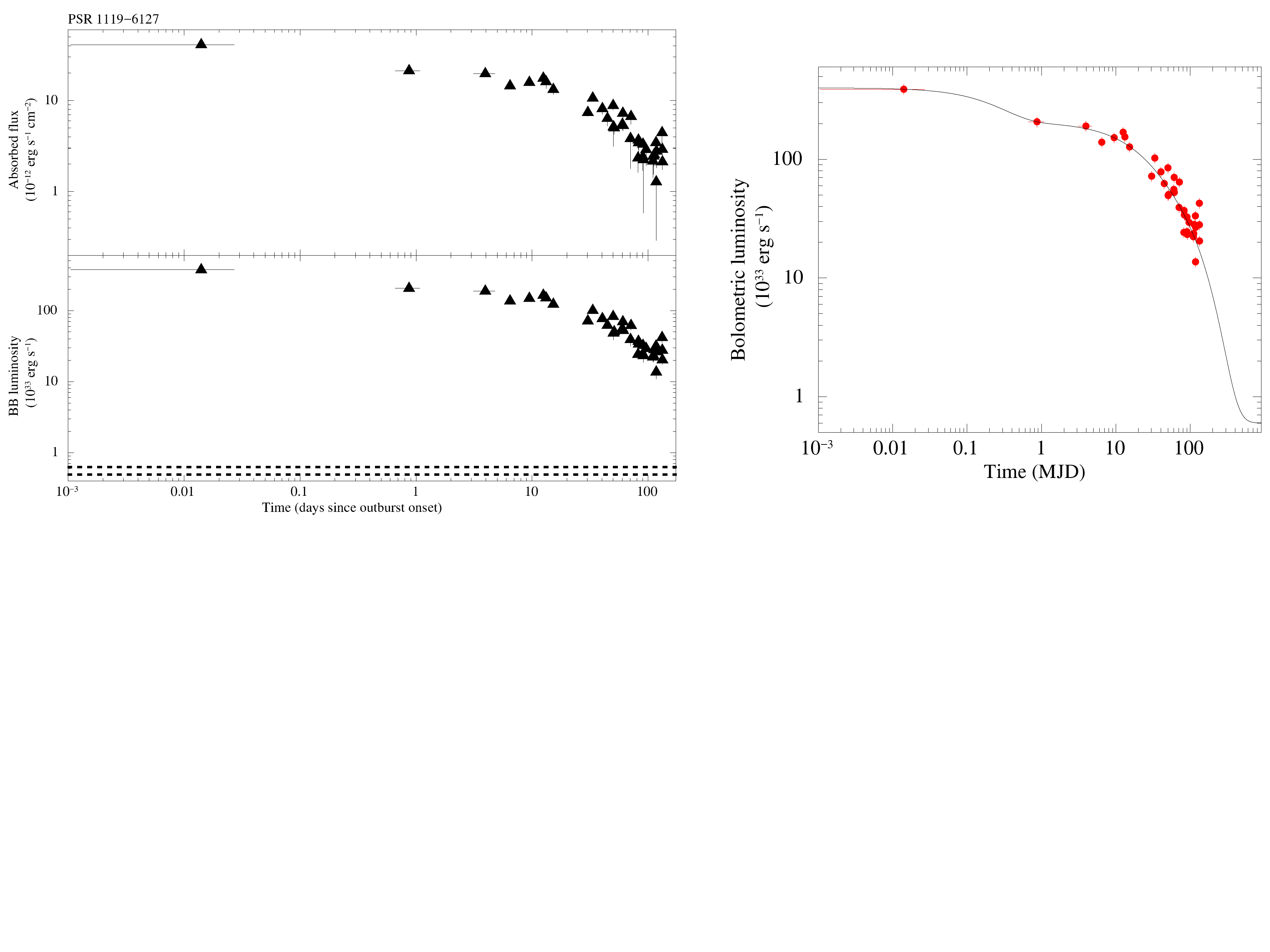}
\vspace{-5.5cm}
\caption{Left-hand panel: temporal evolution of the fluxes and luminosities for the BB model applied to the \swift\ XRT data of the
outburst of \psr. The dashed lines mark the 1$\sigma$ c.l. range for the quiescent luminosity 
(see Table~\ref{tab:quiescence}). A distance of 8.4~kpc was assumed. Right-hand panel: temporal evolution of the bolometric 
luminosity with the best-fitting decay model superimposed.}
\label{fig:psr1119_parameters_outbursts}
\end{center}
\end{figure*}

\clearpage

\bsp
\label{lastpage}
\end{document}